\documentclass[10pt]{article}
\usepackage{amsmath,amssymb,graphicx}
\usepackage{amsthm}
\usepackage{euscript}
\usepackage{eufrak}

\batchmode

\newtheorem{defn}{Definition}
\newtheorem{lem}{Lemma}
\newtheorem{thm}{Theorem}

\newtheorem{cor}{Corollary}

\newcommand{\pr}{\noindent{\bf Proof}. }
\newcommand{\rem}{\noindent{\bf Remark}. }
\newcommand{\rems}{\noindent{\bf Remarks}. }

\newcommand{\pa}{\partial}
\newcommand{\one}{\cO(1)}
\newcommand{\bpsi}{\bar \psi}
\newcommand{\bPsi}{\bar \Psi}
\newcommand{\const}{\textrm{const}}

\newcommand{\supp}{ \mathrm{ supp  }}
\newcommand{\hs}{ \hspace{1cm}}

\newcommand{\Tr}{\textrm{ Tr }}

\newcommand{\un}{\underline}

\newcommand{\Vol}{\textrm{Vol}}
\newcommand{\nat}{\natural}
\newcommand{\tk}{\bbT^{-k}_{N -k}}
\newcommand{\tz}{\bbT^0_{N -k}}

\newcommand{\re}{\textrm{Re } } 

\newcommand{\B}{\Big}
\newcommand{\blan}{\Big  \langle} 
\newcommand{\bran}{\Big  \rangle} 
 
\newcommand{\sgn}{\textrm{sgn}}
\newcommand{\crit}{\textrm{crit}}
\newcommand{\sq}{\square}
\newcommand{\bd}{\boxdot}

\newcommand{\be}{\begin{equation}}
\newcommand{\ee}{\end{equation}}

\newcommand{\bom}{\mbox{\boldmath $\Om$}}

\newcommand{\ba}{\mathbf{a}}

\newcommand{\bh}{\mathbf{h}}

\newcommand{\sZ}{\mathsf{Z}}

\newcommand{\sN}{\mathsf{N}}
\newcommand{\sS}{\mathsf{S}}
\newcommand{\sB}{\mathsf{B}}
\newcommand{\sx}{\mathsf{x}}

\newcommand{\sD}{\mathsf{D}}

\newcommand{\sC}{\mathsf{C}}

\newcommand{\al}{\alpha}
\newcommand{\De}{\Delta}
\newcommand{\de}{\delta}
\newcommand{\ga}{\gamma}
\newcommand{\Ga}{\Gamma}
\newcommand{\ka}{\kappa}
\newcommand{\La}{\Lambda}
\newcommand{\la}{\lambda}
\newcommand{\Om}{\Omega}
\newcommand{\om}{\omega}

\newcommand{\ep}{\epsilon}

\newcommand{\Up}{\Upsilon}
\newcommand{\vep}{\varepsilon}

\newcommand{\cA}{ \EuScript{A} }
\newcommand{\fD}{\EuFrak{D}} 
\newcommand{\fS}{\EuFrak{S}}

\newcommand{\cB}{{\cal B}}
\newcommand{\cC}{{\cal C}}
\newcommand{\cD}{{\cal D}}

\newcommand{\cO}{{\cal O}}

\newcommand{\cH}{{\cal H}}
\newcommand{\cS}{{\cal S}}

\newcommand{\cR}{{\cal R}}

\newcommand{\cK}{{\cal K}}
\newcommand{\cG}{{\cal G}}
\newcommand{\cM}{{\cal M}}
\newcommand{\cN}{{\cal N}}
\newcommand{\cW}{{\cal W}}
\newcommand{\cQ}{{\cal Q}}
\newcommand{\cL}{{\cal L}}

\newcommand{\cJ}{{\cal J}}
\newcommand{\cX}{{\cal X}}

\newcommand{\cZ}{{\cal Z}}

\newcommand{\bcW}{{\bar  \cW}}

\newcommand{\bbR}{{\mathbb{R}}}
\newcommand{\bbZ}{{\mathbb{Z}}}
\newcommand{\bbC}{{\mathbb{C}}}

\newcommand{\bbT}{{\mathbb{T}}}

\begin{document}

\title{Ultraviolet regularity  for     QED in d=3}
\author{ 
J. Dimock
\thanks{dimock@buffalo.edu}\\
Dept. of Mathematics \\
SUNY at Buffalo \\
Buffalo, NY 14260 }
\maketitle

\begin{abstract}
We  study the ultraviolet problem  for  QED in d=3    using Balaban's formulation of the renormalization group.
The model is defined on a fine  toroidal     lattice and  we  seek  control as the lattice spacing goes to zero.
As a first step  we  take a bounded field approximation  and  solve the renormalization problem.    Namely we  show that the bare  energy density   and the bare fermion   mass  can  be chosen to depend on the lattice spacing, so that under the renormalization group flow they  take preassigned values on  a macroscopic   scale.    This is accomplished by a  nonpertubative technique  which   is 
insensitive to whether the renormalizations are finite or infinite.  
\end{abstract}



\section{Introduction}    

\subsection{overview} 
Constructive quantum field theory is concerned with    the construction of mathematically rigorous quantum field theory models.   It has had some  notable successes   in the   case  of super-renormalizable models in dimension $d \leq 3$,  see the surveys  \cite{GlJa87},   \cite{Sum12}.     
However quantum electrodynamics (QED) in  $d=3$  has  so far resisted  analysis.    In this paper  we  start a program to gain control over this model.   The   immediate  task   is   to control  the ultraviolet  (short distance)    singularities in a finite volume.    In  a subsequent paper  \cite{Dim17}
the goal is to prove an ultraviolet stability  bound.        

The  problem is formulated in a renormalization group language for lattice gauge theories.  Originally due to Wilson,  a precise version    was   developed by Balaban. \cite{Bal82a} - \cite{Bal89b}.    The idea is  to   perform a series of block averaging  renormalization group   (RG)
 transformations on the action.    Each transformation      integrates out  some   short distance modes.   
    One  tracks the effective actions  and   hopes   to control the flow by specifying that 
parameters like fermion mass and energy density   end at certain specified values  on a macroscopic  scale.   
This involves  renormalization:  bare  parameters are to  be chosen  to depend  on the lattice spacing.
In    \cite{Dim15}    we  studied weakly coupled   scalar QED in $d=3$  and accomplished this   using a   non-perturbative technique.
In the present we  work we   accomplish the same result for   weakly coupled fermion QED in $d=3$.

 Generally  in quantum field theory     infinite    renormalizations  are  required,   i.e.   some   bare parameters  must   diverge as the lattice spacing goes to zero.   The  
 present  technique is    insensitive to whether the renormalizations are finite or  infinite, and in fact  we  do  not  decide  whether  they are 
 finite or infinite.   However  taking some wisdom from  the perturbative analysis of continuum theories it is quite  likely that the
 energy density  requires an infinite renormalization  and   the fermion mass does not.

In  Balaban's approach  each  RG transformation  features a division of the lattice into regions where the field is small  (i.e bounded)
and a region where the field is large.   The large field region makes  a tiny contribution  to the effective actions and does not require 
renormalization.   The renormalization problem  is  confined to the small field region,  and that is what we study here.    For simplicity  we consider the case that the fields are small on the whole torus,  anticipating that   the analysis  will work in arbitrary 
regions  and     provide the central  
ingredient for the full problem.   A paradigm  for the complete  analysis   is   given  for the scalar  $\phi^4$ interaction in $d=3$ 
in   \cite{Dim11} - \cite{Dim13}.

Some preliminary work  on QED$_3$ was done by Balaban, O'Carroll, and Schor \cite{BOS89}, \cite{BOS91}   and also
by the author   \cite{Dim02}, \cite{Dim04}  (the latter  with massive photons, something we avoid here).  We also mention some
work on the infrared problem  with  massive  fermions integrated out   \cite{DiHu92}.
\bigskip

\subsection{the model}
 
 We  work on the toroidal lattices     
\be
   \bbT^{-N}_M   = ( L^{-N} \bbZ /  L^M \bbZ )^3
\ee
where  $L$  is  a    (large)  positive odd   integer.  
To  begin    we take the         lattice  $\bbT^{-N} _0$  with   unit volume and spacing   $\ep  = L^{-N}$
On  this    lattice the fermi    fields   are elements  of  a Grassman algebra    $\bpsi_{\al} (x),  \psi_{\al} (x) $  indexed  by   points $x  \in  \bbT^{-N}_0  $
  and  components  $1 \leq \al  \leq  4$ and satisfying  
 \be
 \psi_{\al}(x) \psi_{\beta}(y) =   -\psi_{\beta}(y)  \psi_{\al}(x)
 \ee
 and so forth.  Alternatively   we  can  take  $x  \in   L^{-N} \bbZ^3$   and   make the identification  
$ \psi  ( x  +  e_{\mu}  ) = \psi(x) $ where the  $e_{\mu}$ are the unit basis vectors.  We also want to consider anti-periodic boundary conditions and 
then   impose  instead    that    $  \psi  ( x  +  e_{\mu}  ) =  -   \psi(x) $.   In the latter case elements    of the algebra with an   even number  of  
fields  (which is all we consider)   have no change in sign and can be  considered as indexed by points in      $ \bbT^{-N}_0  $
as well.

There  is  also   an   abelian gauge field  (electromagnetic potential,  connection, one-form)   $\cA$ mapping   bonds in   $ \bbT^{-N}_0 $  to  $\bbR$.   A bond from $x$ to a nearest neighbor  $x'$   is the ordered pair    $b= [x,x']$.   We  require  that  
  $\cA(b) =  \cA(x,x')  = -\cA(x',x)$.       Oriented bonds   have the form
$[x,x'] =  [x,  x+ \ep e_{\mu}]$,  and    we  sometimes    write   $A_{\mu}(x)  =  A(x, x+  \ep  e_{\mu})$

    A   covariant derivative   with  charge  $e$  
is    defined     by   
\be
(\pa_{\cA, \mu} f)(x)    =   \B( e^{ie\ep \cA(x, x+ \ep e_{\mu} )} f(x+ \ep  e_{\mu})  - f(x)\B) \ep^{-1} 
\ee
  This  is  a forward derivative.  The transpose  is given by   
  \be
(\pa^T  _{-\cA, \mu} f)(x)     =   \B( e^{ie\ep \cA(x, s- \ep e_{\mu} )} f(x- \ep  e_{\mu})  - f(x)\B) \ep^{-1} 
\ee
and    is  a backward derivative. We  also  consider   the  symmetric derivative   
  \be
    \nabla_{\cA, \mu}   =  \frac12  \B(  \pa_{\cA, \mu} -  \pa^T_{-\cA, \mu}  \B)   
 \ee 
and the    covariant    Laplacian 
\be  \De_{\cA}  =  - ( \pa_{-\cA})^T \pa_{\cA}  \ee

 Let  $\{ \ga_{\mu},  \ga_{\nu} \} = \de_{\mu \nu}$  be  a  representation of the Clifford algebra   for     $d=4$.  
 We  only use   $\ga_0, \ga_1, \ga_2$ in the three dimensional    Dirac operator,  but  $\ga_3$ will also play a roll since it anti-commutes with the
 others.  
 The  Dirac    operator  on spinors  is    
\be   
  \fD_{ \cA} =   \ga \cdot \nabla_{\cA}  -   \frac12   \ep  \De_{\cA}   =    \sum_{\mu=0}^2   
\ga_{\mu}  \nabla_{\cA, \mu}  -   \frac12   \ep  \De_{\cA}  
  \ee
   The extra term  $\frac12   \ep  \De_{\cA}   $ was added by Wilson to prevent  doubling of 
  fermion species.  
  The operator     can also be written 
 \be   
 \begin{split} 
 &(\fD_{ \cA} f)(x) \\
    =   & -     \ep^{-1}    \sum_{\mu} \left[  \left( \frac{ 1 - \ga_{\mu}}{2} \right)   e^{ie\ep \cA(x, x + \ep e_{\mu} )  } f( x + \ep e_{\mu})  \       
 +    \left( \frac{ 1   + \ga_{\mu}}{2} \right)   e^{ie\ep \cA(x, x - \ep e_{\mu} )  } f( x - \ep e_{\mu})  \    -f(x)  \right] \\
 \end{split}
 \ee

 The gauge field   $\cA$ has field strength $d\cA$ defined on plaquettes  (squares)  by  
\be          d\cA(p) = \sum_{b \in \pa p}    \cA (b)\   \ep^{-1}  \ \ \   \textrm{ or }   \ \ \
(d \cA)_{\mu \nu} (x)   =  d \cA  \B(x, x + \ep e_\mu,     
    x + \ep e_\mu + \ep e_{\nu}, x + \ep e_\nu,  x\B)
\ee

The action  with  fixed bare     fermion mass  $0  \leq    \bar m \leq  1$  
\be
\sS(\cA,  \bpsi,   \psi)   =    \frac12   \| d \cA  \|^2 +  <  \bar \psi,   (\fD _{\cA}    +  \bar   m) \psi >    +   m^N  <  \bar \psi,   \psi > 
+   \vep^N   
\ee   
where   
 \be
 \begin{split}
 <  \bar \psi,  \psi >  = &   \int     \bar \psi  (x)  \psi(x)  dx =  \sum_{\al}  \sum_x  \ep^3      \bar \psi_{\al} (x)     \psi_{\al} (x)  \\
   \|   d \cA \|^2  = &  \int | d \cA (p) |^2  dp \equiv     \sum_{\mu < \nu} \sum_x  \ep^3  |  (d \cA)_{\mu \nu} (x) |^2
   \end{split}
      \ee      
The vacuum energy  density   $ \vep^N $   and  the    mass    $ m^N$ are counter terms and    will  be chosen to depend on  $N$. 
The $N \to \infty  $  limit   formally   gives the standard continuum theory.    We   are interested   in bounds uniform in  $N$  on  things   like   the partition function
\be    \int    \exp (  -  \sS(\cA, \bpsi,    \psi)    ) \   D  \bpsi \      D  \psi      \ D  \cA       \hs      D  \cA  =  \prod_{b}  d  (A(b) )   \ee
where  the fermion integral is the standard Grassman integral.   
The     integral will need gauge fixing  to enable convergence.

\subsection{symmetries}
The action is invariant under    lattice symmetries, that is translations, reflections, and rotations by multiples of   $\pi/2$.
Indeed let  $a$ be a lattice point,  let  $r$  be such a   a reflection or rotation,  and let $S$ be  a  corresponding element of $Spin(3)$ so  
that  $S^{-1} \ga_{\mu}  S =  \sum_{\nu} r_{\mu \nu} \ga_{\nu}$.
Then with  
\be    \psi_{a,r}  (x) =   S \psi( r^{-1} (x-a))   \hs    \bpsi_{a,r}  (x) =  ( S^{-1} )^T \bpsi( r^{-1} (x-a))  \hs  A_{a,r} (b)  = A\B(r^{-1}(b-(a,a))\B)
\ee
we have   
\be  
\sS( \cA_{a,r}, \bpsi_{a,r},  \psi_{a,r})    =    \sS( \cA,  \bpsi,    \psi)
\ee
  
 The  action  is     also      gauge invariant.
 For   $\la:  \bbT^{-N}_0  \to  \bbR$   a  gauge transformation is defined by   
 \be   \label{snort}
    \psi^{\la} (x)  =   e^{ie \la(x)}  \psi(x)    \hs       \bpsi^{\la} (x)  =   e^{ -   i e \la(x)}  \bpsi(x)    \hs   \cA^{\la} (x,x')  =  \cA(x,x')  -  \pa  \la(x,x') 
 \ee  
Then $ \fD_{\cA^{\la}}  \psi^{\la}  =  ( \fD_{\cA}  \psi)^{\la} $  and    
\be    \sS( \cA^{\la},  \bpsi^{\la},  \psi^{\la}) =  \sS(\cA,  \bpsi,  \psi)
\ee

Another  symmetry  is charge conjugation invariance.   We  define a charge conjugation matrix  $\sC$ to satisfy
  $ - \ga_{\mu} ^T  =  \sC^{-1} \ga_{\mu}  \sC $  and can choose a representation  such that       $\sC^T  =  \sC^{-1}= -\sC$
  \cite{Wei95}.  
  Then since   $(\nabla  _{\cA})^T =  - \nabla  _{-\cA}$  and  $\De_{\cA}^T = \De_{- \cA }$ 
  \be  \label{lamb1}
      (  \fD_{\cA}   + m  )^T   =   \sC^{-1}       (  \fD_{-\cA}   + m  )  \sC
  \ee 
   Charge conjugation  on the Grassman algebra   is defined by      
  \be   
 \cC  \psi   =  \sC \bpsi     \hs      \cC  \bpsi   =  - \sC \psi   
 \ee 
  and on the gauge field by  $\cA \to - \cA$.  
 Then 
 \be 
     \blan   \cC \bpsi , ( \fD_{-\cA}    +  \bar m)   \cC  \psi   \bran        
 =   -   \blan    \psi , (  \fD_{\cA} +  \bar  m )^T  \bpsi   \bran  
 =      \blan    \bpsi , (  \fD_{\cA} +\bar  m )  \psi   \bran        
  \ee
where we used   $\psi  \bpsi = - \bpsi \psi$.   It follows that the entire action has the symmetry
 \be 
\sS(- \cA, \cC \bpsi ,  \cC \psi  )    =  \sS( \cA, \bpsi ,  \psi  )  
\ee

Note also that since    $\ga_{\mu}^* = \ga_{\mu}$ we also have     $ -  \bar \ga_{\mu}   =  \sC^{-1} \ga_{\mu}  \sC $
which implies      $   \bar \ga_{\mu}   = (\ga_3 \sC)^{-1} \ga_{\mu} (\ga_3 \sC) $.   Assuming   $\cA$ is real we have  
$ \overline{    \nabla_{\cA} }    = \nabla_{-\cA}$ and  $ \overline{    \De_{\cA} }    = \De_{-\cA} $  and therefore with $\bar m$ real
we have the complex conjugation
\be    \label{lamb2}
\overline{     (  \fD_{\cA}   + \bar m  ) } =  ( \ga_3 \sC) ^{-1}       (  \fD_{-\cA}   +  \bar  m  )  (\ga_3 \sC  )  
 \ee
It follows  from   (\ref{lamb1}) and (\ref{lamb2}) that
\be
  \det \     (  \fD_{\cA}   + \bar m  )^*   = \det  \   (  \fD_{\cA}   + \bar m  )
\ee   
from which one can deduce that  (after gauge fixing)   the partition function real and not zero.

\subsection{the scaled model}    The model has been formulated on   a    fine  lattice with unit volume   $\bbT^{-N} _0$. 
But we     immediately   scale up  to  the  large   unit lattice  $\bbT^0_N$.  Then    the ultraviolet  problem  is recast as in  infrared problem,  the natural home of the renormalization group.       Let $\Psi_{\al}(x), \bPsi_{\al}(x)$  be elements of    a Grassman algebra indexed by    $x \in  \bbT^0_N$
and   $1 \leq \al \leq  4$,      and   let    $A:   \{\textrm{ bonds in }  \bbT^0_N   \}  \to  \bbR$ be  a gauge   field    on this  lattice.  These  scale down  to fields   on the 
original lattice   $\bbT^{-N}_0$   by 
\be
       A_{L^{-N}}(b)  =  L^{N/2} A(L^N b)   \hs  \hs      \Psi_{L^{-N}}(x)  =  L^{N} \Psi(L^N x)  \hs     \bPsi_{L^{-N}}(x)  =  L^{N} \bPsi(L^N x)
\ee
The  action on the  new  lattice   is   $\sS_0( A, \bPsi,    \Psi)  =   \sS(A_{ L^{-N}} , \bPsi_{L^{-N}}, \Psi_{L^{-N}} ) $
which is  
\be   \label{snow1}
\sS_0(A, \bPsi,   \Psi)   =   \frac12   \|  d A  \|^2   +  \blan  \bar \Psi,    (\fD _{A}    +  \bar   m^N_0)      \Psi \bran    +   m^N_0  \blan  \bar \Psi,   \Psi   \bran 
+   \vep^N_0   \Vol(\bbT^0_N)  \ee
Now lattice sums are  unweighted   and derivatives are unit lattice derivatives such as
\be
(\pa_{A, \mu }  \Psi)(x)   =  e^{ie^N_0A(x,x + e_{\mu} )}  \Psi(x  + e_{\mu} )  -  \Psi(x)  
\ee 
The   scaled coupling constants are  now tiny and given by   
\be    
   e^N_0  =   L^{-\frac12  N} e  \hs     \bar m_0  =  L^{-N}   \bar m 
  \ee
The  scaled  counter terms are
\be
 \vep^N_0 =L^{-3N}  \vep^N   \hs    m^N_0  =  L^{-N}  m^N  
 \ee 
  In the following we omit the superscript $N$  writing  $e_0,  \bar   m_0 $
and    $\vep_0,      m_0$.

As  we  proceed with the RG analysis the volume will shrink back down.  After $k$ steps the torus  will be $\tz$ or    $\tk$.   The coupling constants  will    scale up to    
\be    
    e_k    =   L^{\frac12 k  } e_0  =   L^{-\frac12(N-k)  } e  \hs    \bar  m_k    =  L^{k} \bar     m_0  = L^{-(N-k)} \bar   m    
  \ee
 The   counterterms    $\vep_k,  m_k$  will     evolve in a more complicated manner.
 \bigskip

  \noindent
\textbf{Conventions}: 
\begin{itemize}
\item
Throughout the paper    $\one$  stands for  a constant   independent of all parameters.  Also  
$c, C, \ga$  are  constants   ($C \geq 1, c, \ga \leq 1$)    which may depend on $L$ and which    may change from line to line.
\item
 Distances are taken in a $\sup$ metric
\be
d(x,y) = |x-y| = \sup_{\mu} |x_{\mu} - y_{\mu}|
\ee
\end{itemize}

\section{RG transformation for fermions}
 
We  explain how  the RG     transformation  is   defined  for  fermions   with a gauge field background.   The analysis is originally due to 
Balaban,  O'Carroll, and Schor  \cite{BOS89},    \cite{BOS91}.   See also  \cite{Dim02},    \cite{Dim04},       \cite{Dim15}.   
 
  \subsection{Grassman variables}
 
 We first   review  some  facts  and  conventions about Grassman variables,  see Appendix  \ref{A}  for more details.   General references
 are  \cite{Sal99},  \cite{FKT02}.

 We consider the  Grassman algebra generated by   $\Psi_{\al} (x), \bPsi_{\al}(x)$
where  $(x,\al)$ are spacetime  and spinor  indices,  $x \in  \bbT_N^0$. 
 Let    $\xi$ stand for   $(x, \al, \om)$  with  $\om = (0, 1)$.   Combine the two by defining  
\be
\Psi(\xi)    = \begin{cases}  
 \Psi_{\al}(x)     &    \xi =    (x, \al, 0)  \\
 \bPsi_{\al}(x)     &    \xi =    (x, \al, 1)  \\
\end{cases}
\ee
  Then the 
$\Psi(\xi)$  satisfy  $\Psi(\xi) \Psi(\xi')    =  - \Psi(\xi') \Psi(\xi) $ and generate the algebra.   
Take some fixed ordering 
for the index set which is all such   $\xi$.   Any element of the algebra    can be uniquely written  
 as
 \be     F(\Psi)  = \sum_{n=0}^{\infty} \    \sum_{\xi_1 < \dots  < \xi_n }  F_n(\xi_1,  \dots,  \xi_n)   \Psi(\xi_1 )  \cdots  \Psi(\xi_n) 
 \ee 
 If the kernel   $ F_n(\xi_1,  \dots,  \xi_n) $  is extended to be an anti-symmetric function of  unordered collections    $(\xi_1,  \dots,  \xi_n)$
 then   
  \be     F(\Psi)  = \sum_{n=0}^{\infty} \frac{1}{n!}     \sum_{\xi_1,  \dots,  \xi_n }  F_n(\xi_1,  \dots,  \xi_n)   \Psi(\xi_1 )  \cdots  \Psi(\xi_n) 
 \ee   
 
 The size of the $F(\Psi)$ is measured by  a norm  of the kernel  depending on a parameter  $h>0$ and defined by
 \be   \label{truenorm}
 \begin{split}  
 \|  F  \|_h   =  &  \sum_{n=0}^{\infty} \ h^n   \sum_{\xi_1 < \dots  < \xi_n } | F_n(\xi_1,  \dots,  \xi_n)|  \\
 =   &   \sum_{n=0}^{\infty} \frac{h^n}{n!}     \sum_{\xi_1,  \dots,  \xi_n } | F_n(\xi_1,  \dots,  \xi_n) | \\
 \end{split}
 \ee

 The integral of an element  of   the Grassman algebra is the projection onto the element  of maximal degree which 
 is identified with the complex numbers       
\be   \int     F(\Psi) \    D   \Psi    =    F_{\max}     
\ee
A  Gaussian   integral  with non-singular covariance $\Ga = \sD^{-1}$  is   defined by 
\be
\begin{split} 
\int F(\Psi )      d\mu_{\Ga}( \Psi  )  =  &    \sZ^{-1}   \int    F (\Psi)        e^{-<\bPsi,\sD\Psi>}  D\Psi    \\
\sZ  =   &      \int        e^{-<\bPsi,\sD\Psi>}  D\Psi   =    \det \sD   \\
\end{split}
\ee
In this  formula  $\Psi$   stands for the pair  $\Psi_{\al} (x),  \bPsi_{\al} (x)$ in   $  F (\Psi)  $,  but just   
$\Psi_{\al}(x) $ in  $<\bPsi,\sD\Psi>$.   This     ambiguity  shows up       throughout the paper, but it should be clear from the context what is meant.   
If   $\bar J_{\al}(x),  J_{\al}(x)$  are  additional  Grassman variables  the Gaussian measure can be characterized by   
\be
\label{generating} 
\int   e^{< \bar J, \Psi>   +   < J, \bPsi>}   d \mu_{\Ga} (\Psi )   =   e^{  < \bar J,  \Ga   J  > }
\ee

 \subsection{block averages}  \label{singer}
 
 Let   $A$  be  a background gauge   field  
 and  let   $\Psi   $   be  a spinor  valued function    on  a unit lattice   $\bbT^0_N$ or a generator of the Grassman algebra.  
 We     define    a  covariant   block averaging operator  $Q(A)$  taking  $\Psi  $ to    
    $Q(A) \Psi  $  defined     on the  $L$-lattice   $\bbT^1_N$.    
 In any  square lattice  let  $B(y)$  be  a cube with    $L$  sites   on a side  centered on  a point $y$.   
 Here  on  the lattice   $\bbT^0_N$      for     $y \in  \bbT^1_N$    we have    
 \be
  B(y) =  \{x  \in  \bbT^0_N: |x-y| < L/2  \}    \ee
  The $B(y)$ partition the lattice.  
 For   $x \in B(y)$  let   $\pi$ be a permutation of  $(1,2,3)$ and  let  $\Ga^{\pi}(y,x)$ be that path  from  $y$ to $x$ obtained 
 by  varying each coordinate to its final value in the order  $\pi$.     There are $3!$ of these.
  For  any     path  $\Ga$    let    $ 
  A(\Ga )  =  \sum  _{b \in  \Ga}   A(b) $
 and define an average over the various  paths  from  $y$ to $x$    by 
 \be   \label{wombat} 
  (\tau A)(y,x)  =      \frac{1}{3!}  \sum_{\pi }    A(\Ga^{\pi}(y,x) )
 \ee
   Then we  define the averaging operator
    \be    \label{dice1}
(Q(A)\Psi  )(y)    =   L^{-3}   \sum_{x \in B(y)} e^{ ie_0  ( \tau A)(y,x) } \Psi  (x)  \hs   y \in \bbT^1_N
 \ee
  and  similarly on     $\bPsi$.   
  
  The   definition is       
   covariant under    symmetries of the lattice   $\bbT_N^1$.  In particular    if  $r$ is     a rotation by a multiple of   $\pi/2$  or  a reflection 
\be
Q(A_r) \Psi  _r  =    (Q(A) \Psi  )_r      
\ee
It is also      is  constructed  to be gauge covariant:
    \be      \label{study} 
 Q(A^{\la})\Psi  ^{\la}  =  (Q(A) \Psi  )^{\la^{(1)}}   \ee
   where  $\la^{(1)}$  is  $\la  $  restricted to  the  lattice    
 $\bbT^1_{N}$.     For    the conjugate field  we  need to take $ Q(-A) \bPsi $   to preserve the coveriance.

    The  transpose   operator  $Q^T(\cA) \equiv (Q(\cA))^T$   maps  functions   $ \Psi $    on   $\bbT^1_N$ to functions      on $\bbT^0_N$.  It is computed with
    sums on  $\bbT^1_N$  weighted by  $L^d$   and   is  given by  
 \be
   (Q^T(A) \Psi )(x)   =   e^{ ie_0(\tau A)(y,x)} \Psi (y)    \hs       x  \in B(y)  
    \ee
Then we  have  
\be      Q(A)  Q^T(- A)     = I 
 \ee
 while   
 \be     P(A)    = Q^T(-A)Q(A)  
 \ee
 is a   projection which satisfies $P^T(\cA) = P(-\cA)$.

 \subsection{the  transformation}
 
 Suppose we start with a density  $\rho (\cA,  \psi  ) $ with fermion    field   $\psi$ and  background gauge field $\cA$   on 
   $\bbT^{-N}_0$.   It scales up
 to a density   
 \be
    \rho_{0}(A,  \Psi_0)  \equiv   \rho_{L^N}  ( A,  \Psi_0)   \equiv       \rho(A_{L^{-N}}  \Psi_{0, L^{-N}} ) 
 \ee
 where   $A, \Psi_0$  are defined      on  $\bbT^0_N $.    
Starting with  $\rho_0( \cA, \Psi_0)$ we  create   a sequence of densities  $\rho_k(  \cA,  \Psi_k)$
 defined for   $\cA$  on   $\tk$ and  $\Psi_k$    on      $ \bbT^0_{N-k}$.      They are defined recursively    
first by 
 \begin{equation}
\begin{split}  \label{kth}
 \tilde  \rho_{k+1} ( \cA,   \Psi_{k+1}) 
=  & \int    \de_G\Big( \Psi_{k+1} -  Q(\cA ) \Psi_k  \Big)  \rho_{k} (\cA,  \Psi_k)   D \Psi_k \\
 \end{split}
\end{equation}
where  $\Psi_{k+1}$   are  new Grassman  variables    defined on the coarser lattice  $\bbT^1_{N-k}$.
 The   $\de_G$ is a Gaussian approximation to  
the delta function.  For a constant  $b = \one$  it is defined by  
\be  \label{spring2}
\begin{split}
\de_G\Big( \Psi_{k+1} -  Q(\cA) \Psi_k   \Big)  
 =   &  N_k
   \exp  \B(  -    \frac{b}{L}   \B \langle   \bPsi_{k+1} -  Q(-\cA) \bPsi_k,    \Psi_{k+1} -  Q(\cA) \Psi_k  \B  \rangle   \B)       \\
\end{split}
\ee
Here   $ N_k  = (bL^2 )^{- 4 s_{N-k-1}} $
where     $s_N =  L^{3N}$ is the number of sites in a 3 dimensional lattice with  $L^N$ sites on a side.  
Also     $< \bPsi_{k+1}, \Psi_{k+1}> =  \sum_x  L^3  \bPsi_{k+1}(x) \Psi_{k+1}(x)$, etc.     The averaging operator  $Q(\cA)$    is  taken to be a modification of  (\ref{dice1}):
  \be    \label{dice2}
(Q(\cA) \Psi_k)(y)    =   L^{-3}   \sum_{x \in B(y)} e^{ ie_k \eta (\tau\cA)(y,x) }  \Psi_k(x)  
 \ee
Here    $(\tau\cA)(y,x)$   is still defined by  (\ref{wombat}),  but  now in $\cA(\Ga)$ the sum is over bonds of length  $\eta  =L^{-k}$, hence the weighting 
factor  $\eta$ in the exponent.   
 The normalization  factor  $N_k$  is      chosen  so that  $ \int  d \Psi_{k+1} \ \de_G\Big( \Psi_{k+1} -  Q(A) \Psi_k  \Big)  =1  $.
 Therefore    
\be     \label{bell}
 \int  \tilde   \rho_{k+1} (\cA,  \Psi_{k+1})\    D \Psi_{k+1}  =     \int     \rho_{k} (\cA,  \Psi_{k} )\   D \Psi_k
\ee

Next     one  scales  back to the unit lattice.   
 If   $\cA$ is a field  on   $\bbT^{-k-1}_{N-k-1}$   and  $ \Psi_{k+1}$ is a field   on $\bbT^{0}_{N-k-1}$   then
   then   
\be
\cA_{L}(b)  = L^{-1/2}   \cA( L^{-1}b) \hs  \Psi_{k+1,L}(x) =  L^{-1}\Psi_{k+1} (L^{-1}x)
\ee
are  fields  on  $\tk$   and  $\bbT^1_{N-k}$ respectively,   and we  define     
  \begin{equation}   \label{scaleddensity}
 \rho_{k+1 } ( \cA,  \Psi_{k+1})  =  \tilde  \rho_{k+1} (\cA_L, \Psi_{k+1,L}) L^{-8(s_N  - s_{N-k-1}) }   
 \end{equation}
If   $\Psi'_{k+1} =  \Psi_{k+1,L}$       we have   by  (\ref{bell})     
 \begin{equation}  \label{preserve}
 \begin{split}
 & \int       \rho_{k+1} (\cA,   \Psi_{k+1})   D \Psi_{k+1}  =  L^{-8(s_N  -  s_{N-k-1}) }  \int     \tilde  \rho_{k+1} (\cA_L, \Psi_{k+1,L})     D \Psi_{k+1}   \\ 
 & =  L^{-8s_N}    \int     \tilde  \rho_{k+1} (\cA_L, \Psi'_{k+1})     D \Psi'_{k+1}
    =   L^{-8s_N} \int       \rho_{k} ( \cA_L,  \Psi_{k})    D \Psi_k
 \\
\end{split} 
\end{equation}
Here the  second step follows by picking out the element of maximal degree, taking account that there are  8 variables per site.    (Formally         $D  \Psi'_{k+1}  =   L^{8s_{N-k-1} }   D \Psi_{k+1} $  goes the opposite way  from  functions ).

Now we claim that   for    $\cA$ on  $\tk$  and  $\Psi_k$ on   $\tz$
\be   \label{lincoln0}
   \int  \rho_{k } ( \cA,  \Psi_k)  D \Psi_k   =      \int   \rho_{0} (\cA_{ L^k}, \psi _{L^k} )   \  D \psi    
\ee
where the integral is over  $\psi $ on   $\tk$. 
  It is true for $k=0$;  suppose it is true for  $k$.   If   $\psi   =  \psi '_{L}$    then  $D \psi  =  L^{ 8s_N} D \psi ' $  and so     by  (\ref{preserve}) 
\be  
\begin{split}
&  \int  \rho_{k+1 } ( \cA,  \Psi_{k+1})  D \Psi_{k+1} 
=        L^{-8s_N} \int       \rho_{k} ( \cA_L,  \Psi_{k})    D \Psi_k  \\
=  &      L^{-8s_N} \int   \rho_{0} (\cA_{ L^{k+1}}, \psi _{L^k} )   \  D \psi    
=       \int   \rho_{0} (\cA_{ L^{k+1}}, \psi '_{L^{k+1}} )   \  D \psi '    \\
\end{split}
\ee
Hence it is true for  $k+1$.

For   $k=N$   (\ref{lincoln0})   says     that for $\cA$ on $\bbT^{-N}_0$ and $\Psi_N$ on $\bbT^0_0$
 \begin{equation}  \label{preserve2}
\int       \rho_{N} (\cA,   \Psi_{N})   D \Psi_{N} =  \int       \rho (\cA,   \psi  )   D \psi 
 \end{equation}
 where the integral is over $\psi \in \bbT^{-N}_0$.
 We   are back to the original integral.      
 The right side  is the integral over a space  with an unbounded number of dimensions,  but   can be computed as the left   side  which    is the integral over a low dimensional space.    This is the point of the renormalization group approach.

\subsection{composition of averaging operators}

 To  investigate   the sequence      $  \rho_{k} ( \cA,   \Psi_{k})$     we  first  study how averaging operators compose.
Define   
\be
Q_k(\cA)  =  Q(\cA) \circ \cdots  \circ  Q(\cA) \hs   ( k   \textrm{ times } )
\ee
 This maps 
fields   on    $\bbT^{-k}_{N-k} $  to  fields  on  $\bbT^0_{N-k}$. 
We assume here that         $\cA $  is defined on   $\tk$   and each $Q(\cA)$ has coupling constant  $e_k$.    The identity     $Q_k(\cA) Q^T_k(-\cA) =I$  
is satisfied and  
\be   
   P_k(\cA)  =    Q^T_k(-\cA) Q_k(\cA)
\ee
is a projection operator.

Later we will   need an  explicit representation for  $Q_k(\cA)$.    In general let       $B_k(y)$  be a cube  with  $L^k$ sites  on a side 
centered on  $y$.     
  Suppose    $x \in \tk$  and  $y  \in  \bbT^0_{N-k}$  satisfy  $x \in B_k(y)$, which is the same as   
  $|x-y| < \frac12$.    There is an associated sequence
$x=y_0,  y_1, y_2,  \dots  y_k = y$ such that  $y_j  \in   \bbT^{-k+j}_{N-k}$  and   $x  \in  B_j(y_j)$. Define
\be    \label{sum}
(\tau_k\cA)(y,x)  =  \sum_{j=0}^{k-1}   (\tau \cA) (y_{j+1}, y_{j}  )  
\ee
Then one can show  \cite{Dim15}   
that for     $\cA,  \psi $ on   $\tk$    and    $\Psi$  on  $ \bbT^0_{N-k}$
\be
\begin{split}
(Q_k(\cA)  \psi  )  (y)  =  &   \int_{ |x-y| <  \frac12}   e^{ ie_k \eta (\tau_k\cA)(y,x) }\psi(x) \ dx \\
 (Q_k^T(\cA)\Psi)(x)  =   &     e^{i e_k \eta(\tau_k\cA)(y,x) }   \Psi (y)    \hs   x \in B_k(y)
  \\  
\end{split}
\ee

Next we show that the individual RG transformations compose to give a transformation with averaging operator  $Q_k(\cA)$:

\begin{lem}  
\label{lucy}   For   $\cA, \psi$  on   $\tk$  and  $\Psi_{k}$  on $\bbT^0_{N-k}$ the density
$\rho_{k, \cA}(\Psi_k)$ can be written  
  \begin{equation}    
 \label{second}
 \rho_{k} ( \cA,  \Psi_k)  
 =  \cN_k  \ \int     \exp  \B(  -  b_k      \B \langle   \bPsi_{k} -  Q_k(-\cA) \bpsi,    \Psi_{k} -  Q_k(\cA) \psi  \B  \rangle   \B)    
    \rho_0   (\cA_{L^k}, \psi _{L^k}) \ \  D  \psi    
\end{equation}
where     
\begin{equation}  \label{oprah} 
b_{k+1}   = \frac{ b b_k   }{b_k + a  }    \hs  \textrm{ or }  \hs 
b_k  =   b \left( \frac{1-L^{-1}}{1- L^{-k}}\right)
\end{equation}
  and  $\cN_k=  b_k^{-4s_{N-k}}$ is a normalizing  constant.  
\end{lem}
\bigskip

 \pr   It holds for  $k=1$ by   (\ref{kth}),(\ref{scaleddensity}) at  $k=0$.   Suppose it is true for  $k$.   Then for  $k+1$ we have
 \be    \label{long} 
 \begin{split}
 \tilde \rho_{k+1} ( \cA,\Psi_{k+1})  =  &  \const   \int   
  \exp  \B(  -    \frac{b}{L}   \B \langle   \bPsi_{k+1} -  Q(- \cA) \bPsi_k,    \Psi_{k+1} -  Q( \cA) \Psi_k  \B  \rangle   \B)   \\
&   \exp  \B(  -  b_k      \B \langle   \bPsi_{k} -  Q_k(- \cA) \bpsi,    \Psi_{k} -  Q_k( \cA) \psi  \B  \rangle   \B)     \rho_{0} (\cA_{L^k}, \psi _{L^k})  
D  \Psi_{k}     D  \psi     \\
\end{split}  
 \ee
 We evaluate the integral by expanding around the critical point of the quantity in the exponential  in $\Psi_k$.   To find the critcal point we treat the fields as real-valued rather
 than elements of a Grassman algebra.  
 Taking derivatives in $\bPsi_k, \Psi_k$ yields  equations for the critical point
 \be 
\begin{split}
- bL^{-1} Q^T(-\cA)\B( \Psi_{k+1}-Q(\cA)\Psi^{\crit}_k \B)  + 
 b_k\B( \Psi^{\crit}_{k}-Q_k(\cA)\psi \B) =&  0 \\
 -bL^{-1} Q^T( \cA) \B(\bPsi_{k+1}-Q(-\cA)\bPsi^{\crit}_k \B)   
+b_k \B( \bPsi^{\crit}_{k}-Q_k(-\cA)\bpsi \B) =&  0 \\
\end{split}
\ee
The first equation can be rewritten as
\be
\B( b_k + bL^{-1}P(\cA) \B)\Psi^{\crit}_{k} =  bL^{-1} Q^T(-\cA)  \Psi_{k+1} +b_k Q_k(\cA)\psi 
 \ee
But  
  \be     \label{lunchmeat}
 \begin{split}  
  \B(    b_k  + bL^{-1}P( \cA)     \B) ^{-1}  
 =   &
    b_k^{-1} (1- P(\cA))    +( b_k + bL^{-1} )^{-1} P( \cA)     \\
 =   &    b_k^{-1}    +     \B( ( b_k + bL^{-1} )^{-1}  -  b_k^{-1}\B)   P( \cA)   \\   
  =   &    b_k^{-1}    -     \frac{bL^{-1}}{b_k + bL^{-1}}   P( \cA)     \\
\end{split}
\ee
We have       $P(\cA)Q^T(-\cA)  =    Q^T(-\cA)$ so
the    $ (1- P(\cA))$  do not  survive  on $bL^{-1} Q^T(-\cA)  \Psi_{k+1}$.  Also   $  P( \cA)  Q_k(\cA) =Q_k^T(-\cA) Q_{k+1}(\cA)$
so applying the inverse gives
\be
\begin{split}
\Psi^{\crit}_{k} = &  \Psi^{\bullet} (\Psi_{k+1}, \psi)\\
\equiv & Q_k(\cA)\psi
+ \frac{bL^{-1}}{b_k + bL^{-1}} Q^T(-\cA)  \Psi_{k+1} -      \frac{bL^{-1}}{b_k + bL^{-1}}  Q_k^T(-\cA) Q_{k+1}(\cA) \psi \\
\end{split}
 \ee
 There is a similar expression for $\bPsi^{\crit}_{k} $. 
We expand the quantity in the exponential around the critical point by 
$\Psi_k =  \Psi^{\crit}_{k}  + W$ and $\bPsi_k =  \bPsi^{\crit}_{k}  +  \bar W$ and then integrate
over $W$ rather than $\Psi_k $.  The linear term must vanish  and the  term quadratic in $W$ just contributes an overall 
constant. Thus we have
 \be    \label{long2} 
 \begin{split}
& \tilde \rho_{k+1} ( \cA,\Psi_{k+1}) =  \const   \int    D  \psi  \  \rho_{0} (\cA_{L^k}, \psi _{L^k})    \\
&  
  \exp  \B(  -    bL^{-1}   \B \langle   \bPsi_{k+1} -  Q(- \cA) \bPsi^{\crit}_k,    \Psi_{k+1} -  Q( \cA) \Psi^{\crit}_k  \B  \rangle   -  b_k      \B \langle   \bPsi^{\crit}_{k} -  Q_k(- \cA) \bpsi,    \Psi^{\crit}_{k} -  Q_k( \cA) \psi  \B  \rangle   \B)   
    \\
\end{split}  
 \ee
However
\be   Q(- \cA) \bPsi^{\crit}_k =  Q_{k+1}(\cA) \psi
+ \frac{bL^{-1}}{b_k + bL^{-1}}   \Psi_{k+1} -      \frac{bL^{-1}}{b_k + bL^{-1}}  Q_{k+1}(\cA)\psi 
\ee
and so
\be
\begin{split}
 \Psi_{k+1}  -   Q(- \cA) \bPsi^{\crit}_k  =  &\B( 1- \frac{bL^{-1}}{b_k + bL^{-1}} \B) ( \Psi_{k+1} -    Q_{k+1}(\cA)\psi ) \\
=  & \frac{b_k}{b_k + bL^{-1}}  ( \Psi_{k+1} -    Q_{k+1}(\cA)\psi ) \\
\end{split}
\ee
also
\be 
 \Psi^{\crit}_{k} -  Q_k( \cA) \psi  =  \frac{bL^{-1}}{b_k + bL^{-1}} Q^T(- \cA)  ( \Psi_{k+1} -    Q_{k+1}(\cA)\psi ) 
\ee
Combining these and using 
\be  bL^{-1} \left ( \frac{b_k}{b_k + bL^{-1}} \right)^2  + b_k \left(   \frac{bL^{-1}}{b_k + bL^{-1}} \right)^2
=   \frac{b_k bL^{-1}}{b_k + bL^{-1}} = b_{k+1}L^{-1}
\ee
we have
\be \label{long3}
 \begin{split}
& bL^{-1}  \B \langle   \bPsi_{k+1} -  Q(- \cA) \bPsi^{\crit}_k,    \Psi_{k+1} -  Q( \cA) \Psi^{\crit}_k  \B  \rangle  
 + b_k      \B \langle   \bPsi^{\crit}_{k} -  Q_k(- \cA) \bpsi,    \Psi^{\crit}_{k} -  Q_k( \cA) \psi  \B  \rangle  \\
& =    b_{k+1}L^{-1}
  \B \langle   \bPsi_{k+1} -  Q_{k+1}(- \cA) \bpsi,    \Psi_{k+1} -  Q_{k+1}( \cA) \psi  \B  \rangle \\
\end{split}
\ee
Make this substitution in  (\ref{long2}). Then  scale with    $\cA  \to  \cA_L,  \Psi_{k+1}  \to   \Psi_{k+1,L},  \psi  \to \psi_L$ 
and use    $Q_{k+1}(\cA_L)  \psi_L  =   (Q_{k+1}(\cA)  \psi )_L$  (now with coupling constant  $e_{k+1}$).  We obtain    
  \be    \label{long4} 
 \begin{split}
 \rho_{k+1} ( \cA,\Psi_{k+1})  =  &  \const   \int   
  \exp  \B(  - b_{k+1}      \B \langle   \bPsi_{k+1} -  Q_{k+1}(-\cA) \bpsi,    \Psi_{k+1} -  Q_{k+1}(\cA) \psi  \B  \rangle   \B)  
 \rho_0(  \cA_{L^k}, \psi _{L^{k+1}})   D \psi \\
\end{split}  
 \ee
 But the constant    must be   $\cN_{k+1}$  in order that  the identity  (\ref{lincoln0}) 
hold,   so we have the result for  $k+1$.

 \subsection{free flow}  \label{minimizers1}

Now consider   an initial density which is a perturbation of the free  fermion   action:    
   \be 
    \rho_{0}(A,   \Psi_0)  = F_0( \Psi_0)    \exp \B(  - \blan  \bPsi,  ( \fD_{\cA} +\bar  m_0)    \Psi   \bran   \B)
\ee
     Insert this in (\ref{second}) 
and use    for   $\cA, \psi   $   on   $\bbT^{-k}_{N - k}$ 
\be 
 \blan  \bpsi_{L^k},  ( \fD_{\cA_{L^k} }+ \bar  m_0)    \psi_{L^k}   \bran =
  \blan  \bpsi,  ( \fD_{\cA} +\bar  m_k)    \psi   \bran    \ee
where     now    $ \fD_{\cA}$    is defined with       coupling constant  $e_k$.
Then   with    $ F_{0, L^{-k}}( \psi  ) =   F_{0}( \psi _{L^k} )   $ we have from (\ref{second})
 \be
 \label{something}
 \rho_{k} (\cA,  \Psi_k)    = 
\cN_k  \int     F_{0, L^{-k}}( \psi  )    
  \exp  \B(   -  b_k      \B \langle   \bPsi_{k} -  Q_k(-\cA) \bpsi,    \Psi_{k} -  Q_k(\cA) \psi  \B  \rangle   -    \blan\bpsi , ( \fD_{\cA}+\bar  m_k  ) \psi \bran     \B)    D  \psi   
\ee
We expand around the critical point in $\psi$  for the expression in the exponential.   The critical point satisfies the equations
\be 
\begin{split}
b_k Q_k^T(-\cA)\B(\Psi_{k}-Q_k(\cA)\psi^{\crit} \B)  
 - (\fD_{\cA} + \bar m_k ) \psi^{\crit} = & 0 \\
 b_kQ_k^T( \cA)\B( \bPsi_{k}-Q_k(-\cA)\bpsi^{\crit} \B) 
 -     (\fD_{\cA} + \bar m_k ) ^T \bpsi^{\crit}= & 0 \\
\end{split}
\ee
These are solved by $\psi^{\crit}=  \psi_k(\cA )  $ and $\bpsi^{\crit}=   \bpsi_k(\cA ) $ where on $\tk$
\be   \label{seventyfour}
\begin{split}
 \psi_k(\cA )   =   \psi_k(\cA, \Psi_k ) \equiv   &  \   b_k  S_k(\cA) Q^T_k(-\cA)   \Psi_k   \\
 \bpsi_k(\cA )      =   \bpsi_k(\cA, \bPsi_k )     \equiv \  &    b_k  S^T_k(\cA) Q^T_k(\cA)   \bPsi_k   \\
\end{split}
 \ee
 and  the  propagator    (Green's function)  is defined as 
\be  \label{Sk}
   S_k(\cA)   =   \Big( \fD_{\cA} +\bar  m_k   +  b_k   P_k(\cA)  \Big)^{-1} 
\ee
This pair of equations (\ref{seventyfour}) is abbreviated as
 \be  
  \psi_k(\cA )   = \cH_k(\cA) \Psi_k
  \ee

We expand around the critical point introducing  by  $\psi  =   \psi_k(\cA )    +  \cW $ and    $  \bpsi  =   \bpsi_k(\cA )    +  \bcW   $  
and integrating over
  new   Grassman  variables
   $\bcW, \cW$   instead of  $\psi, \bpsi$.   The linear term must vanish and 
 we have
\be
 \begin{split}
& b_k      \B \langle   \bPsi_{k} -  Q_k(-\cA) \bpsi,    \Psi_{k} -  Q_k(\cA) \psi  \B  \rangle   +   \blan\bpsi , ( \fD_{\cA}+\bar  m_k  ) \psi \bran \\    
& =   \fS_k(\cA, \Psi_k, \psi_k(\cA))  +  \blan\bcW , \B( \fD_{\cA}+\bar  m_k + b_kP_k(\cA) \B) \cW \bran   \\
\end{split}
\ee
where
\be   \label{freefermi}
 \begin{split}
  \fS_k(\cA, \Psi_k, \psi_k(\cA))  = &  b_k      \B \langle   \bPsi_{k} -  Q_k(-\cA) \bpsi_k(\cA),    \Psi_{k} -  Q_k(\cA) \psi_k(\cA) \B  \rangle    +    \blan\bpsi_k(\cA) , ( \fD_{\cA}+\bar  m_k  ) \psi_k(\cA) \bran  \\ 
 = & \blan \bPsi_k, \B[    b_k -  b_k^2 Q_k(\cA)S_k(\cA)Q^T_k(-\cA) \B] \Psi_k \bran \\
 \equiv  & \blan \bPsi_k,    D_k(\cA)  \Psi_k\bran   \\
\end{split}
\ee
Our expression  becomes
 \be   \label{spinit}
   \rho_{k} (\cA,  \Psi_k)    =   \cN_k    \sZ_k(\cA)  F_k (  \psi_k(\cA) )   
      \exp  \Big( -    \fS_k(\cA, \Psi_k, \psi_k(\cA))    \Big)
  \ee
where
\be  \label{route66}
\begin{split}
    F_k ( \psi ) 
    =  &   \sZ_k(\cA)  ^{-1}  \int      F_{0, L^{-k}}(  \psi  +  \cW ) 
       \exp \B(   -    \blan  \bcW, \B( \fD_{\cA} +\bar     m_k   +  b_k  P_k(\cA)  \B)  \cW   \bran  \B)    D  \cW     \\
 \sZ_k(\cA)   =  & \int  \exp \B(   -    \blan  \bcW, \B( \fD_{\cA} +\bar     m_k   +  b_k  P_k(\cA)  \B)  \cW   \bran  \B)   D  \cZ 
 = \det ( S_k(\cA)  )^{-1}
  \\
 \end{split}
 \ee

\subsection{the next step}   \label{single}

If   we  start  with  the   expression  (\ref{spinit})  for  $\rho_k$   and apply another renormalization transformation we 
again        get  $\rho_{k+1}$.   Working out the details  will give us some useful  identities.   
We have first   
\be   \label{manx}
\begin{split}
&   \tilde  \rho_{k+1 } (\cA,  \Psi_{k+1}) =   \cN_k   N_{k} \sZ_k(\cA) \\  
&   \int      \exp \left(-  bL^{-1}
\blan   \bPsi_{k+1}-Q(-\cA)\bPsi_k, \Psi_{k+1}-Q(\cA)\Psi_k \bran  -   \fS_k(\cA, \Psi_k, \psi_k(\cA))  \right) 
 F_k (  \psi_{k}(\cA) ) \ D   \Psi_k     \\
\end{split}
\ee 
To evaluate the integral
we want    expand around the critical point for this quadratic form in the exponential.   Using the representation (\ref{freefermi}) one can argue as in the previous section the critical point is 
\be   \label{min1}
\begin{split}
  \Psi^{\crit}_k    =     &   bL^{-1} \Ga_k (\cA)  Q^{T}(-\cA)  \Psi_{k+1}  \\
   \bPsi^{\crit}_k    =     &    bL^{-1}\Ga^T_k (\cA)  Q^{T}(\cA)    \bPsi_{k+1}     \\
\end{split}
\ee
 where
 \be
\Ga_k (\cA)  = \B(D_{k}(\cA) +  bL^{-1}P(\cA)\B)^{-1}  
\ee
However it is useful to find another expression for these critical points.

We define on $\tk$ the operator 
 \be 
S^0_{k+1}(\cA) =  \B(\fD_{\cA} + \bar m_k +L^{-1} b_{k+1}P_{k+1}(\cA)\B)^{-1}
\ee
  and the field 
\be 
\psi^0_{k+1} (\cA)  = L^{-1}  b_{k+1}S^0_{k+1}(\cA) Q_{k+1}^T( -\cA ) \Psi_{k+1} 
\ee
These   scale to $ S_{k+1}(\cA),  \psi_{k+1} (\cA) $ on $\bbT^{-k-1}_{N-k-1}$.

\bigskip
\begin{lem} 
\be \label{longo2}
\begin{split}
  \Psi^{\crit}_k(\cA)  = &  \Psi^{\bullet} \B( \Psi_{k+1},  \psi^0_{k+1}(\cA)  \B)  \\
 \psi^0_{k+1}(\cA)    = &  \psi_k\B( \cA,  \Psi^{\crit}_{k}(\cA) \B)  \\
\end{split}
\ee 
\end{lem}
\bigskip

\pr These are identities between operators and it suffices to treat the fields as real-valued functions rather than elements of the 
Grassman algebra. 

The quadratic form in the exponential in (\ref{manx}) is $ J(\Psi_{k+1}, \Psi_k, \psi_k(\cA, \Psi_k))$ is
\be    \label{lady}
\begin{split}
  &  J(\Psi_{k+1}, \Psi_k, \psi) =  bL^{-1} 
\blan   \bPsi_{k+1}-Q(-\cA)\bPsi_k, \Psi_{k+1}-Q(\cA)\Psi_k \bran  \\
& +
b_k\blan \bPsi_{k}-Q(-\cA)\bpsi, \Psi_{k}-Q(\cA)\psi \bran   + \blan \bpsi_k, (\fD_{\cA} + \bar m_k ) \psi \bran  \B) \\
\end{split}
 \ee
The critical point $ \Psi^{\crit}_k(\cA)  $  is the unique solution of $\pa / \pa  \bPsi_k(x) \B[J(\Psi_{k+1}, \Psi_k, \psi_k(\cA, \Psi_k))\B]=0$
or 
\be
\begin{split}
 \frac {\pa  J}{ \pa  \bPsi_k(x)} \B(\Psi_{k+1}, \Psi_k, \psi_k(\cA, \Psi_k)\B) 
+ \int  \cH_k(\cA)(x,y) \frac {\pa  J}{ \pa  \psi(y)} \B(\Psi_{k+1}, \Psi_k, \psi_k(\cA, \Psi_k)\B) dy  =0 \\
\end{split}
\ee

  To get at this  we study the critical point of $ J(\Psi_{k+1}, \Psi_k, \psi)$ in both variables  $\Psi_k, \psi$.
Taking derivatives in  $\bPsi_{k},\Psi_{k} ,\bpsi, \psi$ gives the equations for the critical 
point of $ J(\Psi_{k+1}, \Psi_k, \psi) $
\be  \label{spitfire}
\begin{split}
 \frac{\pa J}{ \pa \bPsi_k}  =  &- bL^{-1} Q^T(-\cA)\B( \Psi_{k+1}-Q(\cA)\Psi'_k \B)  + 
 b_k\B( \Psi'_{k}-Q_k(\cA)\psi' \B) =  0 \\
 \frac{\pa J}{ \pa \Psi_k} = & -bL^{-1} Q^T(\cA) \B(\bPsi_{k+1}-Q(-\cA)\bPsi'_k \B)   
+b_k \B( \bPsi'_{k}-Q_k(-\cA)\bpsi' \B) =  0 \\
 \frac{\pa J}{ \pa \bpsi} =&- b_k Q_k^T(-\cA)\B(\Psi'_{k}-Q_k(\cA)\psi' \B)  
 +  (\fD_{\cA} + \bar m_k ) \psi' =  0 \\
 \frac{\pa J}{ \pa \psi}  =  & -b_kQ_k^T(\cA)\B( \bPsi'_{k}-Q_k(-\cA)\bpsi' \B) 
 -     (\fD_{\cA} + \bar m_k ) ^T \bpsi'=  0 \\
\end{split}
\ee
We have seen these equations before, but now we have to solve then simultaneously.  The first  and third equations  have the solutions
 \be   \label{longo4}
 \begin{split}
   \Psi'_k  = &  \Psi^{\bullet} ( \Psi_{k+1}, \psi')  \\
 \psi'   = &  \psi_k(\cA,  \Psi'_k) \\
 \end{split}
 \ee
Substitute the  expression for $ \Psi'_k $  into the third equation which can be written
 \be \label{equine}
  \B(\fD_{\cA} + \bar m_k + P_k(\cA)\B) \psi' =   b_k Q_k^T(\cA)\Psi'_{k} 
 \ee
 We compute 
\be
 \begin{split}
b_k Q_k^T(-\cA)\Psi'_{k} 
=  & b_k Q_k^T(-\cA) \B( Q_{k} (\cA ) \psi '     + \frac{ bL^{-1}} {b_k + bL^{-1} }  Q^T( -\cA ) \Psi_{k+1}  
     -  \frac{bL^{-1}}{  ( b_k + bL^{-1})   }   Q^T(-\cA)      Q_{k+1} (\cA ) \psi  \B) \\
     =  & b_k  P_{k} (\cA ) \psi '     + \frac{b_k bL^{-1}} {b_k + bL^{-1} }  Q_{k+1}^T( -\cA ) \Psi_{k+1}  
     -  \frac{b_kbL^{-1}}{ ( b_k + bL^{-1})   }       P_{k+1} (\cA ) \psi'  \\
\end{split}
\ee
Substitute this in (\ref{equine}) and use
$
   b_k b( b_k + bL^{-1} )^{-1}   =   b_{k+1}
$
to obtain
\be
  \B(\fD_{\cA} + \bar m_k + L^{-1}b_{k+1}P_{k+1}(\cA) \B) \psi' =L^{-1} b_{k+1} Q_{k+1}^T( -\cA ) \Psi_{k+1}  
   \ee
which has the solution
$\psi'  =  \psi^0_{k+1}(\cA) $.    With this identification the identities (\ref{longo4}) become
\be   \label{longo3}
 \begin{split}
   \Psi'_k  = &  \Psi^{\bullet} ( \Psi_{k+1}, \psi^0_{k+1}(\cA) )  \\
\psi^0_{k+1}(\cA)  = &  \psi_k(\cA,  \Psi'_k) \\
 \end{split}
 \ee
Since $\psi'  = \psi_k(\cA,  \Psi'_k)$   the first  and third equations in  (\ref{spitfire}) now read
\be  
   \frac{\pa J}{ \pa \bPsi_k} \B(\Psi_{k+1} \Psi'_k ,    \psi_k(\cA,  \Psi'_k)\B)  = 0  \hs   \frac{\pa J}{ \pa \psi} \B(\Psi_{k+1} \Psi'_k ,    \psi_k(\cA,  \Psi'_k)\B)  = 0
\ee
Thus  $ \Psi_k = \Psi' _k$ solves the equation  
which which identifies     $ \Psi^{\crit}_k(\cA)  =  \Psi' _k$.
With this  the identities (\ref{longo3}) become the identities (\ref{longo2}) of the theorem. 
This completes the proof.
\bigskip

We expand around the critical point  by
\be  \label{diag}
\Psi_k =  \Psi^{\crit}_{k}(\cA) + W   \hs   \bPsi_k =\bPsi^{\crit}_{k}(\cA) + \bar  W
\ee
  This also  entails  
 \be \label{diag2}
 \psi_k(\cA)=   \psi^0_{k+1}(\cA) + \cW_k(\cA)   \hs     \bpsi_k(\cA)= \bpsi^0_{k+1}(\cA) + \bcW_k(\cA)  
 \ee
where 
\be
  \cW_k(\cA)  =   \psi_k(\cA,W) = \cH_k(\cA) W  
\ee
We introduce
 \be
\begin{split}
& \fS^0_{k+1} \B(\cA, \Psi_{k+1}, \psi \B)\\
= 
 &
b_{k+1}L^{-1}\blan  \bPsi_{k+1}-Q_{k+1}(-\cA)\bpsi, \Psi_{k+1}-Q_{k+1}(\cA)\psi \bran  +  \blan \bpsi , (\fD_{\cA} + \bar m_k ) \psi \bran   \\
\end{split}
\ee
which scales to $ \fS_{k+1} \B(\cA, \Psi_{k+1}, \psi \B)$

\begin{lem} 
\label{samsung}
Under the transformation (\ref{diag}), (\ref{diag2}) the quadratic form 
\be bL^{-1}
\blan   \bPsi_{k+1}-Q(-\cA)\bPsi_k, \Psi_{k+1}-Q(\cA)\Psi_k \bran  +  \fS_k(\cA, \Psi_k, \psi_k(\cA))
\ee
becomes   
\be
 \fS^0_{k+1}\B(\cA, \Psi_{k+1}, \psi^0_{k+1}(\cA)\B)    +     \blan\bar  W,  \B(  D_k(\cA)  +  bL^{-1}   P(\cA)\B) W\bran  
 \ee
\end{lem} 
\bigskip

\pr  Since we are at the critical point the cross terms vanish.   The terms in  $\bPsi^{\crit}_{k}(\cA) $ and $  \bpsi_k(\cA)$
are identified using  (\ref{long3}) as 
\be
\begin{split}
 & bL^{-1} 
\blan   \bPsi_{k+1}-Q(-\cA)\bPsi^{\crit}_{k}(\cA), \Psi_{k+1}-Q(\cA)\Psi^{\crit}_{k}(\cA) \bran 
 +    \fS_k\B(\cA, \bPsi^{\crit}_{k}(\cA), \psi^0_{k+1}(\cA) \B)   \\
  = & 
 \fS^0_{k+1}\B(\cA, \Psi_{k+1}, \psi^0_{k+1}(\cA)\B)    \\
\end{split}
\ee
The terms in $W, \cW_k(\cA)$ are identified  using (\ref{freefermi}) as
\be
  bL^{-1} 
 \blan  Q(-\cA) \bar W,  Q(\cA)W \bran  +    \fS_k\B(\cA, W, \cW_k(\cA)\B)   
  =  \blan\bar  W,  \B(  D_k(\cA)  +  bL^{-1}   P(\cA)\B) W\bran   
\ee
This completes the proof.
\bigskip 

Now in  (\ref{manx})  we make the change of variables  and  integrate over the new Grassman variables 
   $\bar W, W$   instead of  $\Psi_k, \bPsi_k$
   This gives 
   \be   \label{manx2}
\begin{split}
&   \tilde  \rho_{k+1 } (\cA,  \Psi_{k+1}) =   \cN_k   N_{k} \sZ_k(\cA)   \exp \left(-  \fS^0_{k+1}\B(\cA, \Psi_{k+1}, \psi^0_{k+1}(\cA)\B)      \right) 
 \\  
&   \int      \exp \left(   -     \blan\bar  W,  \B(  D_k(\cA)  +  bL^{-1}   P(\cA)\B) W\bran   \right) 
 F_k \B(  \psi^0_{k+1}(\cA)+  \cW_k(\cA) \B) \ D W     \\
\end{split}
\ee 
Next 
identify the  Gaussian integral   $\int [ \cdots ] d \mu_{\Ga_k(\cA)}$  where  
\be
\begin{split}
  d \mu_{\Ga_k(\cA)} (W)= &   \sZ_k(\cA) ^{-1} \exp \B(   -    \blan \bar   W,  \B( D_k(\cA)  + bL^{-1}   P(\cA)\B)  W\bran       \B) DW \\
\de  \sZ _k(\cA)   =  &  \int   \exp \B(   -    \blan \bar  W,  \B(  D_k(\cA)  +  bL^{-1}   P(\cA)\B) W\bran       \B)  D  W   
= \det( \Ga_k (\cA))^{-1}
 \\
\end{split}
\ee
Then  (\ref{manx2}) is rewritten as 
 \be   \label{spinit2}
 \begin{split}
  \tilde    \rho_{k+1} (\cA,  \Psi_k) 
  =  &   \cN_k   N_{k}
   \sZ_k(\cA)  \de\sZ_k(\cA)    \exp \left(-  \fS^0_{k+1}\B(\cA, \Psi_{k+1}, \psi^0_{k+1}(\cA)\B)      \right)  \\   &  
    \int    F_k\B(  \psi^0_{k+1}(\cA)    + \cW_k(\cA) \B)   d \mu_{\Ga_k(\cA)} (W) \\
\end{split}
  \ee
  Next we  scale by (\ref{scaleddensity}).   Using $\psi^0_{k+1}(\cA_L, \Psi_{k+1,L}) = [\psi_{k+1}(\cA)]_L$
  and
\be    \label{sundry}
  \fS^0_{k+1}\B(\cA_L, \Psi_{k+1,L}, [\psi_{k+1}(\cA)]_L \B) 
=\fS_{k+1}\B(\cA, \Psi_{k+1}, \psi_{k+1}(\cA)\B)   
  \ee
  we have
 \be   \label{kingly}
 \begin{split}
 & \rho_{k+1} (\cA,  \Psi_{k+1})  
   =       \cN_k   N_{k}
   \sZ_k(\cA_L) \de   \sZ_k(\cA_L)   \  L^{-8(s_N   -s_{N-k-1})}  \\
  &     \exp  \B( -  \fS_{k+1}\B(\cA, \Psi_{k+1}, \psi_{k+1}(\cA)\B)     \B) \int    F_k\B(    [\psi_{k+1}(\cA)]_L + \cW_k(\cA_L) \B)   d \mu_{\Ga_k(\cA_L)} (W)\\
\end{split}   
 \ee
Comparing  
this  expression     with   (\ref{spinit}) for  $k+1$ we find that 
\be    \label{z}
\cN_{k+1}\sZ_{k+1} (\cA)   =  \cN_k   N_{k}     \sZ_k(\cA_L) \de  \sZ_k(\cA_L)   \  L^{-8(s_N   -s_{N-k-1})}
\ee
and that 
\be  \label{toast0}
F_{k+1}(\psi_{k+1} (\cA) ) =  \int    F_k  \B(     [\psi_{k+1}(\cA)]_L + \cW_k(\cA_L) \B)   d \mu_{\Ga_k(\cA)} (W)
\ee
The latter is   fluctuation integral of a type  we  investigate  in  detail for special $F_k$.  

  We also note  that   $\Ga_k(\cA) $  has the alternate representation:   
\be     \label{lamp}
 \begin{split}
 \Ga_k(\cA)  = & \sB_k(\cA)  + b_k^2 \sB_k(\cA)Q_k(\cA)  S^0_{k+1}(\cA) Q_k^T(-\cA)\sB_k(\cA)   \\
 \end{split}
 \ee
 where  $\sB_k(\cA)$ is the operator  (\ref{lunchmeat})  and   
 \be  
  S^0_{k+1}(\cA)  =     \B(   \fD_{\cA} +    \bar m_k +    b_{k+1}L^{-1}   P_{k+1}(\cA)  \B)^{-1} 
 \ee
  is the operator on  $\tk$ which scales to  $S_{k+1} (\cA)$ on $\bbT^{-k-1}_{N-k-1}$.  
See the Appendix  \ref{ID} with  $y=0$  for  derivation of this representation .

\bigskip

 \subsection{propagators}

\label{green}
 
 We  study  the  propagator   
   \be 
    S_k(\cA)= 
    \B(\fD_{ \cA}   + \bar  m_k   +    b_k   P_k (\cA) \B)^{-1}
    \ee        an operator   on functions on    $\tk$ defined for   a background field    $\cA$ on $\tk$. 
  We first list some general properties. 
  \bigskip         

\noindent
(a.)   With gauge transformation    $\la$    on  $\tk$    defined as in  (\ref{snort})  we have   
\be   \label{buddy}
   \fD_{\cA^{\la}}\psi^{\la}     =   ( \fD_{\cA}\psi  )^{\la}   \hs   
  Q_k(\cA^{\la})\psi^{\la}   =   ( Q_k(\cA)\psi)^{\la^{(0)}}  \ee
where   $\la^{(0)} $ is the restriction to the unit lattice $\tz$.  
It follows that  
\be   \label{sync}
  S_k (   \cA^{\la})\psi^{\la}   =   ( S_k (   \cA)\psi)^{\la} 
\hs      \cH_{k} (   \cA^{\la}) \Psi_k^{\la^{(0)}}   =    ( \cH_{k} (   \cA) \Psi_k  ) ^{\la}  
\ee
Under    charge conjugation    
\be   \label{superduper} 
  \sC^{-1}  S_k( - \cA) \sC    =   S^T_k(\cA)    \hs    \cC   \cH_{k} (   \cA) \Psi_k   =     \cH_{k} (  - \cA) \cC   \Psi_k   
\ee
and if   $\cA$ is real   
\be 
 (\ga_3 \sC)   ^{-1}  S_k( - \cA)(  \ga_3\sC )    = \overline{  S_k(\cA) }    
\ee
\bigskip
   
   \noindent
(b.)  
At some points we will want to change the background field    on $\tk$   from  $\cA + \cZ$    to a  background field $\cA$   
by    
\be   \label{resolvent}  
  S_k(\cA ) -  S_k(\cA+ \cZ)   =  S_k(\cA  + \cZ )   V_k(  \cA,  \cZ)  S_k(  \cA) 
\ee
where
 \be
    V_k(  \cA,  \cZ)  \equiv   \Big( \fD_{\cA+ \cZ }  +  \bar m_k +     b_k  P_k (\cA+ \cZ )   \Big) -  \Big( \fD_{\cA} + \bar  m_k  
    +  b_k  P_k (\cA)   \Big)
\ee
We  have explicitly      with  $\eta  =  L^{-k}  $ 
 \be  \label{V_k}
 \begin{split}
   ( V_k(  \cA,  \cZ)f)(x)  = & -  \sum_{\mu}  \left( \frac{ 1  - \ga_{\mu}}{2} \right)   e^{ie_k\eta \cA(x, x + \eta e_{\mu} )  }   
\left( \frac  {  e^{ie_k\eta \cZ(x, x + \eta  e_{\mu} )  }  -1}{\eta}      \right)            f( x + \eta e_{\mu})  \\  
 & -  \sum_{\mu}  \left( \frac{ 1  + \ga_{\mu}}{2} \right)   e^{ie_k\eta \cA(x, x - \eta e_{\mu} )  }   
\left( \frac  {  e^{ie_k\eta   \cZ(x, x - \eta  e_{\mu} )  }  -1}{\eta}     \right)             f( x - \eta e_{\mu})  \\    
   & +   b_k \B( P_k (\cA+ \cZ ) - P_k (\cA) \B)    \\     
\end{split} 
\ee
\bigskip

\noindent
(c.)  Now  some comments about boundary conditions.     The  propagator $S_k(\cA)$    on the torus   $\tk$ is the same
as the propagator   on  the cube $  [- \frac12 L^{N-k}, \frac12 L^{N-k} ]^3$ with periodic boundary conditions.  In the latter form it can
be  
obtained  from  propagator    $S_{k, \eta \bbZ^3}(\cA)$
on  $\eta \bbZ^3$   (defined with  a periodic extension of     $\cA$)    by the statement that the kernels satisfy  
 \be  \label{louie} 
 S_k(\cA,  x,y)   =    \sum_{n \in \bbZ^3}    S_{k,\eta  \bbZ^3} (\cA,  x,y  + nL^{N-k}  )
\ee
For anti-periodic boundary conditions  this is modified to    
 \be   \label{lux2} 
 S_k(\cA,  x,y)   =    \sum_{n \in \bbZ^3}   (-1)^{n_0+n_1+ n_2}   S_{k,\eta  \bbZ^3} (\cA,  x,y  + nL^{N-k}  )
\ee
This  satisfies  for all $x,y$    
\be 
 S_k(\cA,  x,y + L^{N-k} e_{\mu})  =  - S_k(\cA,  x,y)= S_k(\cA,  x + L^{N-k} e_{\mu},y)
\ee
Hence when  restricted to the cube it satisfies  the  anti-periodic  boundary conditions in either variable.

\subsection{local  propagators}

We  develop    a random walk expansion for    $  S_k(\cA)  $ following Balaban,  O'Carroll, and Schor  \cite{BOS91}. 
The first  step is to find local inverses for the   Dirac operator.

Partition the  lattice    $\tk$  into  large   cubes $\square$  of linear  size  $M=L^m$ centered on  points in   $\bbT^m_{N-k}$ for some integer $m>1$.    Let  $\tilde \square$ be cubes of linear  size $3M$  centered on  the same points,  and more  generally let   
$\tilde \square^{n}$  be the cubes of linear size  $(2n+1)M$ centered on the same points.  

We   seek     local   propagators  $S_k( \square, \cA)$  localized   near   $\tilde  \square$     with   the property that  
for  $x \in   \tilde   \square$
\be   \label{free} 
  \B( \B( \fD_{ \cA}   +   \bar m_k   +    b_k   P_k (\cA) \B)S_k( \square, \cA)f\B)(x)  = f(x)
\ee
We  also   want   bounds on   $S_k( \square, \cA)$    and a certain Holder derivative   
 $ \de_{\al, \cA }  S_k( \square, \cA  ) $.
The    Holder derivative for  $0< \al < 1$  is defined   for  $0< |x-y|< 1$  by    
\be 
  (\de_{\al,\cA} f) (x,y)   =   \frac{e^{ie_k \eta \cA(\Ga_{xy} )} f(y)  -  f(x) }{ |x-y|^{\al} }   
\ee
where      $\Ga_{xy}$ is one of the standard paths from $x$ to $y$.  Or one can replace $\cA(\Ga_{xy} )$ by
the average over paths  $(\tau \cA )(x,y)$.
  There is an associated  
 norm  
\be  
 \|  \de_{\al, \cA} f  \|_{\infty}  =  \sup_{0< |x-y| <   1}    | (\de_{\al, \cA} f) (x,y) |
  \ee

\begin{lem}  \label{sweet2} 
 Let  $e_k$ be sufficiently small  depending on  $L,M$.   Let   $\cA$  on  $\tilde \square ^{5}$    be real-valued and  gauge
 equivalent  to a field satisfying    
     $|   \cA|   <    e_k^{-1 + \ep}$  for some small positive constant  $\ep$. 
       Then   there is an operator   $ S_k( \square, \cA)$  on  functions on    $ \tilde \square^{5} $  
     satisfying   (\ref{free})      and    
     \begin{equation}  \label{sycamore1}
| S_k( \square, \cA) f |, \ 
\|  \de_{\al, \cA }  S_k( \square, \cA  ) f \|_{\infty}  
 \leq       C  \|f\|_{\infty} 
 \end{equation}
Furthermore  let    $  \De_y,  \De_{y'}$  be unit squares centered on unit lattice points     $   y,y'  \in  \tilde \square^5$,  let  $\tilde   \De_y$
be the enlargement of $\De_y$ by a layer of unit cubes,     and  let 
 $\zeta_y$  be   a smooth partition on unity with $\supp\  \zeta_y \subset  \tilde  \De_{y} $.   
 Then    
  \begin{equation}  \label{sycamore2}
\begin{split}
&|1_{\De_y} S_k( \square, \cA)1_{\De_{y'}} f  |, \ 
\| \de_{\al, \cA }   \zeta_y S_k \square, \cA  )1_{\De_{y'}} f \|_{\infty}  
   \leq      C    e^{  -   \ga  d(y,y') } \|f\|_{\infty}    \\
\end{split}
 \end{equation}
 \end{lem}
 \bigskip
 
 The condition on $\cA$  is  satisfied  if, for example,   $|\pa    \cA|    <    e_k^{-1 +2 \ep}$.    
  Indeed   by subtracting a constant   the field $\cA$ is gauge equivalent to  a  field  satisfying     
  \be 
    | \cA|   \leq    \one \textrm{ diameter }(  \tilde  \square^{5})   \| \pa   \cA \|_{\infty}   \leq     \one M    e_k^{-1 +2 \ep}  \leq   e_k^{-1 + \ep}
 \ee
With a little   more  work  this  can be established with only a bound on the field strength      $|  d \cA|    <    e_k^{-1 +2 \ep}$.

  For the proof  of the lemma     see     \cite{BOS91},    \cite{Dim04}.   A candidate   for  $ S_k( \square, \cA)$ would be to restrict the operator
  $\fD_{\cA}   +\bar  m_k   +    b_k    P_{k}(\cA )  $  to   a  neighborhood on  $\tilde \square$, say with Dirichlet boundary conditions,
  and then invert it.     This would satisfy  (\ref{free}) but it is difficult to get good estimates.
  Instead one uses a rather complicated construction involving soft boundary conditions  implemented by a multi-scale  random walk 
  expansion.

The    previous result  is  extended     to  complex fields   $\cA$   on $\tk$  of the form  
 \be  \label{listless} 
 \begin{split} 
   & \cA=  \cA_0  + \cA_1   \\
    & \cA_0  \textrm{ is  real  and  on each  }    \tilde \square ^{5}   \textrm{ is gauge equivalent to a field satisfying  }     |   \cA_0    |  <    e_k^{-1 + \ep},  \\ 
&   \cA_1 \textrm{ is complex and  satisfies  }     |  \cA_1 |    <    e_k^{-1 + \ep}      \\
\end{split}  
\ee  
Then   
 $S_k( \square, \cA)$  has  an analytic  extension to   the region   (\ref{listless}),  and for such fields     $S_k( \square, \cA)$ 
 again   satisfies bounds of the form  (\ref{sycamore1}), (\ref{sycamore2}).
 This follows by expanding   $S_k( \square,\cA)  =   S_k( \square,\cA_0  + \cA_1)$   around $\cA_1 =0$ and using the   bounds   for    $  S_k( \square,\cA_0)$.

 \subsection{random walk expansion}  \label{random}
The random walk expansion is  based on   the operators  
 $S_k(   \square, \cA)$  just    discussed.   We assume that  $\cA$ is in the domain  (\ref{listless}) so  that  these have good estimates by lemma  \ref{sweet2}.    
Let   $h^2_{\sq}$  be  a partition of unity with  $\sum_{\sq} h_{\sq}^2 =1$  and   $\supp\  h_{\sq}$ well inside  $ \tilde \square$.  We  define a parametrix
\be  S_k^*(\cA)  =   \sum_{\sq}  h_{\sq}  S_k(   \square, \cA) h_{\sq}  \ee
On   $\supp \  h_{\sq}$  the identity (\ref{free}) is applicable     and  so 
\be     
  \B(\fD_{ \cA}   + m_k   +    a_k   P_k (\cA)\B) S_k^*(\cA)=    I  -  \sum_{\sq}   K_{\sq}(\cA) S_k(  \square, \cA) h_{\sq}  \equiv  I -K
  \ee
  where
  \be    K_{\sq}(\cA)  =   -\B[  \B( \fD_{ \cA}   +\bar  m_k   +    a_k  P_k (\cA)\B) , h_{\sq}\B]  \ee
 Then   
  \be
  S_k(\cA)  =  S_k^*(\cA) ( I - K)^{-1}   =   S_k^*(\cA) \sum_{n=0}^{\infty}  K^n
  \ee    
  if  the series   converges.  
  This can be written   as    the random walk expansion
   \begin{equation}  \label{g1}
 S_k(\cA) =    \sum_{\om }  S_{k,\om}(\cA)
 \end{equation}
where   a   path  $\om$ is a sequence of cubes $   \om   =  (\sq_0, \sq_1, \dots,   \sq_n )$
 in  $\bbT^m_{N-k}$
such that  $\sq_i, \sq_{i+1}$ are nearest neighbors  (in a sup metric),   and  
\be  S_{k, \om} (\cA)=\B( h_{\sq_0} S_k( \square_{0},\cA )h_{\sq_0}\B)
\B( K_{\sq_1}(\cA) S_k( \square_{1},\cA)h_{\sq_1}\B) \cdots   \B( K_{\sq_n}(\cA) S_k( \square_{n},\cA)h_{\sq_n} \B)
\ee
 Note  that    $ S_{k,\om}(\cA)$  only depends on  $\cA$ in the set  $ \bigcup_{i=0}^n  \tilde \square^{5}_{i} $

 \begin{lem}     \cite{BOS91},    \cite{Dim02}  \label{sweet3}  Let   $M$  be sufficiently large (depending on  $L$),  
  and  $e_k$  sufficiently small (depending on  $L,M$),   and let   $\cA $  be in the domain  (\ref{listless}).  Then
the   random walk  expansion (\ref{g1})  for $S_k(\cA)$  converges to a function analytic in $\cA$   which satisfies  
 \begin{equation}  \label{sycamore3}
| S_k( \cA) f  |, \ 
\|  \de_{\al, \cA }   S_k( \cA  ) f\|_{\infty}  
 \leq       C  \|f\|_{\infty} 
 \end{equation}
Furthermore let   $  \De_y,  \De_{y'}$  be unit squares centered on unit lattice points     $   y,y'  \in  \bbT^0_{N-k}$  and let
 $\zeta_y$   be a smooth partition on unity with $\supp\  \zeta_y \subset  \tilde  \De_{y'} $. 
 Then 
  \begin{equation}  \label{sycamore4}
\begin{split}
&|1_{\De_y} S_k( \cA)1_{\De_{y'}} f |, \ 
\| \de_{\al, \cA }    \zeta_y     S_k( \cA  )1_{\De_{y'}} f \|_{\infty}  
   \leq      C    e^{  -\ga  d(y,y') } \|f\|_{\infty}    \\
\end{split}
 \end{equation}
 \end{lem}
\bigskip

\pr   
We  compute   
 \be  \label{no1}
 \begin{split} 
  \B([    \fD_{\cA},  h_{\sq} ] f\B) (x ) 
  =  & -   \sum_{\mu}  \left( \frac{ 1  - \ga_{\mu}}{2} \right)( \pa  h_{\sq})   (x,x + \eta e_{\mu})    e^{ie_k\eta A(x, x + \eta e_{\mu} )  } f( x + \eta e_{\mu})   \\      
    & - \sum_{\mu}   \left( \frac{ 1  + \ga_{\mu}}{2} \right)( \pa  h_{\sq})(x,x - \eta e_{\mu})   e^{ie_k\eta A(x, x - \eta e_{\mu} )  } f( x - \eta e_{\mu}) 
 \end{split}  
 \ee
and    with  $x \in \De_y$
\be  \label{no2}
  \B( [P_k(\cA),   h_{\sq} ] f\B)(x)     =  \int_{|x'-y| < \frac12}  e^{-ie_k \eta (\tau_k \cA)(y,x)} e^{ie_k \eta (\tau_k \cA)(y,x')}
\B(h_{\sq}(x') - h_{\sq}(x)  \B)  f(x')   dx'
\ee
The functions   $\{ h_{\sq}\}$ can be chosen so that   $| \pa h_{\sq} | \leq   \one M^{-1}$.
Then    the representations   (\ref{no1}),  (\ref{no2})   lead to the bound
\be   | K_{\sq}(\cA) f | \   \leq    \one  M^{-1} \|f\|_{\infty}    \ee
and   therefore  by  lemma  \ref{sweet2}    
\be   \label{spitfire1} | K_{z}(\cA) S_k( \square, \cA)f|   \leq   CM^{-1}   \|f \|_{\infty}
\ee
These  imply  that  
\be       \label{night}
  |S_{k, \om}(\cA) f|  \leq  C (CM^{-1} )^{\om} \|f\|_{\infty} 
 \ee
This is sufficient to establish the convergence  of  the expansion for $M$  sufficiently  large,   since the number of paths with a fixed length $n$ is bounded
by  $\one^n$.   The bound on  the Holder    derivative follows as well.

For the local estimates  use  the locality of $K_{\sq}(\cA)$ and  (\ref{sycamore2})  to obtain
\be      \label{spitfire2} |1_{\De_y}  K_{\sq}(\cA) S_k(\cA, \square)1_{\De_{y'}}f|   \leq   CM^{-1}   e^{  -  \ga  d(y,y') }  \|f \|_{\infty}
\ee
The decay factors combine to give an overall decay factor  (with a smaller  $\ga$) and the result follows
 \bigskip

\rems  (1.) The same bounds    (\ref{sycamore3}), (\ref{sycamore4}) also   hold for  $\cH_k(\cA)$,  for example   
 \be  \label{slavic}
| \cH_k( \cA) f  |, \    \|  \de_{\al, \cA }  \cH_k( \cA  ) f\|_{\infty}     \leq       C  \|f\|_{\infty} 
 \ee
  We  note  that $S^0_{k+1}(\cA) $  also   has a random walk expansion  and satisfies  the same bounds as  $S_k(\cA)$.  Then 
the representation (\ref{lamp})  leads to the bound
\be   \label{clavicle}
| \Ga_k(\cA) f  |  \leq  C  \|f\|_{\infty}
\ee

(2.) The  operator  $\cH_k$ has a kernel   $ \cH_k( \cA,\xi, \sx)  $   where   $\xi=  (x, \beta, \om)$ with $x \in \tk$   and   
 $\sx = (x, \al, \om)$ with $x \in \tz$.   The operator  $\de_{\al, \cA }  \cH_k$  has a kernel      $ (\de_{\al, \cA }  \cH_k)( \cA, \zeta, \sx  )$ with   $ \zeta=  (x,y,  \beta, \om)$   and   $x,y \in \tk$. 
 We  consider the  $\ell^1- \ell^{\infty}$ norms   on the kernels
\be  
\begin{split} 
  \|\cH_k(\cA) \|_{1, \infty}  =   &  \sup_{\xi }    \sum_{\sx }   |\cH_k(\cA, \xi, \sx)| = \sup_{\xi,  \|f \|_{\infty} \leq  1 }    |\B(\cH_k(\cA)f\B)(\xi)|     \\
  \|\de_{\cA, \al} \cH_k(\cA) \|_{1, \infty}  = &   \sup_{\zeta}    \sum_{\sx}   |\de_{\al}\cH_k(\cA, \zeta,  \sx)|
  =  \sup_{\zeta, \|f \|_{\infty} \leq  1}  |\B(\de_{\al}\cH_k(\cA) f \B)( \zeta)|   \\
\end{split}
\ee
The second form for the norms follows  since on   our   finite measure space     $\ell^{\infty}$  is the dual space to  $\ell^1$.      
Then (\ref{slavic}) implies
\be  \label{slavic2}
    \|\cH_k(\cA) \|_{1, \infty}  ,       \|\de_{\cA, \al} \cH_k(\cA) \|_{1, \infty}    \leq  C
\ee

 \subsection{weakened propagators}     \label{weak}
 
 The random walk expansion makes it possible to introduce weakened forms of the propagators.
 Note that if  $|\om| = 0$  then $S_{k, \om} (\cA) = S^*_k(\cA)$ and so 
 \be 
 S_k(s, \cA) =  S^*_k(\cA)+    \sum_{\om: | \om| \geq 1}  S_{k,\om}(\cA)
\ee
 For each  $   \om   =  (\sq_0, \sq_1, \dots,   \sq_n )$ with $n \geq 1$ define
 \be
    X_{\om}   \equiv  \bigcup_{i=1}^n  \tilde \square^5_i  
 \ee
  We     introduce weakening parameters  $\{ s_{\square} \}$ indexed by the  $M$ cubes $\sq$  with  $0 \leq  s_{\square} \leq  1$  and    define 
\be  
s_{\om}  =  \prod_{\square \subset   X_{\om}}  s_{\square} \ee
  Weakened  propagators are defined by  
\be  \label{again}
  S_k(s, \cA) =  S^*_k(\cA)+    \sum_{\om: | \om| \geq 1} s_{\om}  S_{k,\om}(\cA)
\ee
 The  $ S_k(s,\cA)$  interpolate between
$S_k(\cA)= S_k(1,\cA)$   and a strictly local  operator  $ S_{k}(0,\cA)$.
If  $s_{\square}$  is  small  then the coupling through  $\square$ is reduced.   If  $Y$ is a union of $M$ cubes and  $s_{\sq} = 0$ for $\sq \subset Y^c$, then no path with
 $ X_{\om}$ intersecting $Y^c$ contributes to the
sum (\ref{again}).  Then  $S_k(s, \cA)$ only connects points in $Y$ and only depends on $\cA$ in $Y$.

 The   bounds (\ref{sycamore3}), (\ref{sycamore4}) still  hold  for  the weakened propagators   $  S_k(s,\cA)$,  even if we allow $s_{\sq}$ 
 complex and rather large.  
 In fact let $\al_0$ be a small parameter and  take complex $s_{\sq}$ satisfying
    \be   |s_{\square}|  \leq   M^{\al_0} 
    \ee
   Then for $\al_0$ sufficiently small (independent of all parameters)  
    \be 
    |s_{\om}| \leq    \exp \B( \al_0\log M |  X_{\om}|_M\B) \leq  \exp \B( \one \al_0  \log M |\om| \B) 
   = M^{\frac12|\om|}
   \ee
   Still assuming $M$ is sufficiently large this does not affect convergence of the random walk expansion which is driven by the
   factor $M^{-|\om|}$.

  The weakened propagator   $   S_k(s,\cA)$ has  the   analyticity, bounds,  and symmetries  of  $S_k ( \cA)$.
  The weakened propagator  $S_k(s,\cA)$ also gives a weakened operator  $\cH_k(s, \cA)$ which satisfies the same bounds as $\cH_k$.
 Similarly  $S^0_{k+1}(\cA)$  weakens to  $S^0_{k+1}(s, \cA)$  and  $\Ga_k(\cA)$ weakens to  $\Ga_k(s, \cA)$ with
 the same bounds.

 \section{RG transformations for  gauge fields }  \label{rgg} 

For gauge fields  we      follow the  analysis developed  by   Balaban       
 \cite{Bal84a}, \cite{Bal84b},  \cite{Bal85b},      and   Balaban, Imbrie, and Jaffe  
 \cite{BIJ85},  \cite{BIJ88}.   The basic  treatment is identical with    \cite{Dim14},  \cite{Dim15}.

\subsection{axial gauge}

\label{axialgauge} \label{lulu}

We  are concerned  with   formal     integrals   over fields    on    $\bbT^{-N}_0$  
 of  the form   
\be     \label{formal3}   
 \int   f(\cA)   \exp \B( - \frac12 \| d\cA \|^2  \B)  D\cA     
 \ee
Scaling    up by $L^N$  the integral is a constant times
$  \int    \rho_0 (A_0)  DA_0       $  where      for     $A_0$  on   $\bbT_N^0 $
\be   \rho_0 (   A_0 )   =    F_0(A_0)    \exp \B( - \frac12 \| dA_0 \|^2  \B)   
\ee
and  $  F_0(A_0)  =    f_{L^N}(  A_0 )  =   f(  A_{0, L^{-N }}   ) $.
 We   seek       control over   the integral  $ \int    \rho_0 (A_0)  DA_0    $  with  RG transformations,   which   at the same time
 supplies the gauge fixing necessary for convergence.     Specifically   we  want to define  a sequence of well-defined   densities  $\rho_0,  \rho_1,  \dots, \rho_N $  
 where    $\rho_k(A_k) $  is a  function  of     $A_k$  on  $\bbT_{N-k}^0$   and   represents   a   partial integral of $\rho_0$.
 
 Suppose $\rho_k$ is already defined. 
Then 
for $A_k$ on    $\bbT^0_{N-k}$       define  an  averaged field     on   oriented bonds   in      $\bbT^{1}_{N-k} $ by     (for reverse oriented bonds take minus this)  
\be   
\begin{split}
(\cQ A) (y,  y + L e_{\mu} )  
= & \sum_{x \in B(y)  } L^{-4}    A(  \Ga_{x,  x +  L e_{\mu} } ) \\
\end{split}
\ee
where   $  \Ga_{x,  x +  L e_{\mu} }$ is the straight line between the indicated points.   
First consider  for $A_{k+1}$ on $\bbT^1_{N-k}$
\be    \label{funny0} 
\tilde   \rho_{k+1} (A_{k+1})  =  \int   \de ( A_{k+1} - \cQ A_k)   \rho_k(A_k)  \  DA_k   
\ee
This does not converge as it stands.  
For convergence we introduce  an axial gauge fixing function    (justified by a formal   Fadeev-Popov argument)
\be 
\de  (\tau A_k)    =  \prod_{y  \in   \bbT^{1}_{N-k}  }  \prod_{x  \in B(y),  x \neq  y}    \de \B((\tau   A_k)(y, x) \B)   
\ee
where    $(\tau A_k)(y,x)$ is defined in  (\ref{wombat}). 
Instead  of  (\ref{funny0}) we        define  $\tilde  \rho_{k+1} (A_{k+1})$    for     $A_{k+1}$ on     $ \bbT^1_{N-k}$   
by  
\be    \label{funny1} 
\tilde   \rho_{k+1} (A_{k+1})   = \int   \de ( A_{k+1} - \cQ A_k)      \de ( \tau A_k  )  \rho_k(A_k)  \  DA_k   
\ee
Then we return to a unit lattice defining     
 $   \rho_{k+1}(A_{k+1}) $  for     $A_{k+1}$ on    $ \bbT^0_{N-k-1}$        
by 
\be  
\label{scaleddensity2}
    \rho_{k+1}(A_{k+1}  )  =     \tilde  \rho_{k+1} (A_{k+1,L})  L^{\frac12 (b_N- b_{N-k-1}   -\frac12 (s_N- s_{N-k-1}) }
\ee
Here   $b_n =  3L^{3N}$ is the number of bonds in a three dimensional   toroidal lattice with $L^N$ sites on a side, and again
$s_N = L^{3N}$ is the number of sites.

The result of the iteration can  be computed explicitly as
\be  \label{fourfour} 
\begin{split}
  \rho_k (A_k)  =     &    \int    \de (A_k -   \cQ_k  \cA )  \de (\tau_k \cA)    \rho_{0,L^{-k}} ( \cA)    D\cA  \\
  =  &        \int    \de (A_k -   \cQ_k  \cA )  \de (\tau_k \cA)    F_{0,L^{-k}}  (\cA)    \exp   \B( - \frac 12 \|  d \cA \|^2    \B)    D\cA  \\
  \end{split}  
\ee
where now   $\cA$ is defined on bonds in     $\tk$    and   the $k$-fold averaging operator is defined by     $\cQ_k  =   \cQ  \circ  \cdots  \circ  \cQ$.  Then       $ \cQ_k \cA$ is   given   on    oriented bonds  in  $\bbT^0_{N-k}$  by     
\be   
(  \cQ_k \cA) (y,  y +  e_{\mu} )  =  \int_{|x -y|  < \frac12  }  L^{-k}   \cA  ( \Ga_{x,  x +   e_{\mu}})\ dx
\ee
and the   gauge fixing function is   now   
 \be 
  \de (\tau_k \cA)   =      \prod_{j=0}^{k-1}    \de  (\tau  \cQ_j \cA )   
 \ee
One  can show  that  if    $F_0$   is exponentially bounded then   $ \rho_k (A_k)$    is well-defined.   
Furthermore    the final density   $\rho_N$ is a constant  given by   
\be  
\rho_N   =         \int    \de (  \cQ_N  \cA )  \de (\tau_N \cA)     f  (\cA)    \exp   \B( - \frac 12 \|  d \cA \|^2    \B)    D\cA
\ee
This is  a    gauge fixed version of the original integral  (\ref{formal3}).

To  evaluate   $ \rho_k (A_k)$  as given by  (\ref{fourfour})  we expand around the minimum of  $\| d \cA \|^2$ subject
to the constraints of the delta functions.  We define $\cA^{\sx}_k = \cA^{\sx}_k (A_k)$ on $\tk$ by
\be
\cA^{\sx}_k  \equiv      \cH^{\sx}_kA_k  = \textrm{ minimizer of } \| d \cA \|^2  \textrm{ subject to } \cQ_k \cA = A_k, \tau_k \cA =0
\ee
where the linear operator  $\cH^{\sx}_k$ has a specific representation in terms of Green's functions. 
 Expanding around the minimizer by  $\cA  =  \cA^{\sx}_k    + \cZ$  the linear term vanishes and  one finds 
\be  \label{sex} 
      \rho_k(A_{k})   =    \sZ_k      F_k (   \cA^{\sx}_k )     \exp \B(   - \frac12   \|  \cA^{\sx}_k   \|^2    \B)   
 \ee
where 
\be  \label{route76}
\begin{split}
 F_k (\cA^{\sx}_k  )   
   = & \sZ_k^{-1}    \int    \de (  \cQ_k \cZ  )  \de (\tau_k \cA)    F_{0,L^{-k}}  (\cA^{\sx}_k + \cZ)    \exp   \B( - \frac 12 \|  d \cZ \|^2    \B)  D \cZ
    \\   
\sZ_{k}   =   &   \int    \de (  \cQ_k \cZ  )   \de (\tau_k \cA)     \exp  \B ( - \frac 12 \|  d \cZ \|^2   \B )  D \cZ \\
\end{split}
\ee

Next  consider how one passes from  the representation for  $\rho_k$ to the representation for  $\rho_{k+1}$.     
     Suppose we are starting with the expression  (\ref{sex}) for  $ \rho_k(A_k) $.   In the next step   generated by  (\ref{funny1})  we  have
\be  \label{six}
\begin{split}
\tilde \rho_{k+1} (A_{k+1}) 
 =   & \sZ_{k}      \int    \de (A_{k+1} -    \cQ A_k  )\   \de(  \tau  A_k)  \    F_k (\cA^{\sx}_k  )        \exp \B(   - \frac12 \| d \cA^{\sx} \|^2  \B)   
 DA_k  \\
 \end{split}
\ee
Define the minimizer  $\cA_k^{\min}=\cA_k^{\min}(A_{k+1})$ by 
\be
\cA_k^{\min}  \equiv      H_k^{\sx}A_{k+1}  = \textrm{ minimizer of } \| d \cA^{\sx}_k \|^2  \textrm{ in } A_k \textrm{ subject to } \cQ A_k = A_{k+1}, \tau \cA_k =0
\ee
Expand around the minimizer  by    $A_k =  \cA_k^{\min} + Z$  and integrate over $Z$ instead of $A_k$
Then
\be
\cA^{\sx}_k =  \cA^{0, \sx}_{k+1}+ \cZ_k
\ee
where
\be 
\cA^{0, \sx}_{k+1}= \cH^{\sx}_k \cA_k^{\min} =  \cH^{\sx}_k  H_k^{\sx}A_{k+1}  \hs   \cZ_k =  \cH^{\sx}_kZ
\ee  
On the constrained subspace 
\be  \frac12  \| \cA^{\sx}_k \|^2  =   \frac12  \| d \cA^{0, \sx}_{k+1} \|^2 +  \frac12  \|d \cZ_k  \|^2 
\ee
and we also write with $\de = d^T$ on two-forms (functions on plaquettes)
\be
 \|  d\cZ_k \|^2   = \blan Z, \De_k  Z\bran    \hs  \textrm{ where } \hs     \De_k =   \cH^{\sx, T}_{k} \de d  \cH^{\sx}_{k} 
 \ee
We   find 
\be  \label{spotless} 
  \tilde     \rho_{k+1}(A_{k+1})   =    \sZ_k      \exp \B(   - \frac12 \| d \cA^{0, \sx}_{k+1}\|^2    \B)    \    \int     \de (  \cQ   Z  )   \de(  \tau Z)  \    F_k \B( \cA^{0, \sx}_{k+1}  +  \cZ_k \B)     \exp \B(   - \frac12 \blan Z,  \De_k Z \bran     \B)  D Z 
 \ee
 This scales to 
 \be     \label{spotless2} 
 \begin{split}
 \rho_{k+1}(A_{k+1})   = &  L^{\frac12 (b_N- b_{N-k-1})-\frac12 (s_N- s_{N-k-1}) }     \sZ_k   
    \exp \B(   - \frac12 \| d \cA^{0, \sx}_{k+1}(A_{k+1,L})\|^2    \B)    \\
     &    \int     \de (  \cQ   Z  )   \de(  \tau Z)  \    F_k \B( \cA^{0, \sx}_{k+1}(A_{k+1,L}) +  \cZ_k \B)     \exp \B(   - \frac12 \blan Z,  \De_k Z \bran     \B)  D Z  \\
     \end{split} 
 \ee
 Compare this with $
     \rho_{k+1}(A_{k+1})   =    \sZ_{k+1}      \exp (   - \frac12 \| d \cA^{ \sx}_{k+1}\|^2    )      F_{k+1} ( \cA^{ \sx}_{k+1} ) 
 $
and we have the    identifications    (making special choices of $F$) 
\be    \label{sammy}
\begin{split}
 \sZ_{k+1}  = &  \sZ_k \de  \sZ_k     L^{\frac12 (b_N- b_{N-k-1})-\frac12 (s_N- s_{N-k-1}) }  
      \\
 \cA^{\sx}_{k+1,L}  =   & \cA^{0, \sx}_{k+1}(A_{k+1.L})     \\
F _{k+1}  (\cA^{\sx}_{k+1}  ) 
   = &  \de   \sZ_{k} ^{-1}    \int     \de (  \cQ   Z  )   \de(  \tau Z)  \    F_k \B( \cA^{\sx}_{k+1, L}    +\cZ_k \B)     \exp \B(   - \frac12 \blan Z,  \De_k Z \bran     \B)  D Z 
    \\
  \de   \sZ_{k}   =   &    \int    \de (  \cQ   Z  )   \de(  \tau Z)  \       \exp \B(   - \frac12  \blan Z, \De_k  Z\bran  \B)  D Z
   \\ 
 \end{split}
\ee
Note  that if  $F_0$ is gauge invariant then   $F_k$ is gauge invariant for any $k$.

 The   fluctuation  integral     (\ref{sammy})    
can be parametrized as
\be   \label{springgarden4}
 F_{k+1} ( \cA )  =      \int     F_k\B(\cA_L  + \cH_k  C   \tilde Z   \B)      \exp \B(  - \frac12   \blan C  \tilde  Z , \De_k C \tilde  Z   \bran     \B) 
    \    D    \tilde   Z  
   \Big/  \{  F_k =1 \}        
\ee
where    $ \tilde Z = ( \tilde Z_1, \tilde  Z_2)$.  The field  $\tilde Z_1$ is defined on bonds  within  each block $B(y)$ and satisfies 
   $ \tilde Z_1  \in \ker \tau $.  The field    $\tilde Z_2$   is    defined on bonds joining   $B(y), B(y')$ denoted $B(y,y')$,  but not
the central bond on each face denoted  $b(y,y')$.    The mapping   $Z  =  C \tilde Z  $  is the identity  on all bonds except the central bonds and assigns a
value to the central bonds so that   $ \cQ   Z =0,   \tau Z=0$. 
If  we  define  $  
  C_{k}  =   (  C^T     \De_k  C )^{-1}     
$
then   the  integral can be expressed with  the Gaussian measure  $\mu_{C_k}$  with covariance  $C_k$   as      
\be    \label{sometimes2} 
  F_{k+1}  (\cA)     =        \int           F_k \B( \cA_L +  \cH^{\sx}_k  C  \tilde   Z        \B)    d \mu_{  C_{k}}   ( \tilde  Z )   
\ee      

 Integrals of this type can be explicitly evaluated by choosing a basis for the functions $\tilde Z$.   We mention a particular class of
 bases which will be used in the following.   For  the $\tilde Z_1$ we take   functions  $\{e^y_i\}$ on bonds in  $B(y)$ which are in
 $\ker \tau$ and orthonormal with respect to  the usual inner product.  For the $\tilde Z_2$  we take  delta functions $\de_b$  for   $b \in B(y,y') -b(y,y')$. 
 Together they give an orthonormal  basis 
 \be \label{Ubasis}
 \{ \Up_{\al} \}   = \B( \bigcup_y   \{ e^y_i \}  \B)  \   \bigcup  \   \B( \bigcup_{y,y'}   \{ \de_b \}_{b \in B(y,y') -b(y,y')}. \B)
 \ee
 For such a basis   integrals over $\tilde Z$ are evaluated by
 \be
 \begin{split}
\int f(\tilde Z) d \mu_{  C_{k}}   ( \tilde  Z )   
=  & \int  f\B( \sum_{\al} z_{\al} \Up_{\al}\B) d \mu_{ \hat C_{k}}   ( z )   \\
 = & (2 \pi)^{-n/2}    \det ( \hat C_k^{-1} )^{\frac12} \int    f\B( \sum_{\al} z_{\al} \Up_{\al}\B)      \exp \B( -\frac12 \blan z, \hat C_k^{-1} z \bran \B) \prod_{\al} dz_{\al}\\
 \end{split}
 \ee 
 where $\hat C_{\al \beta} = C(\Up_{\al}, \Up_{\beta} ) $ and $n$ is the number of elements in the basis. The expression is independent of the basis.

 \subsection{Landau gauge}

One also   formulate  the problem in the Landau gauge.    Instead of axial gauge fixing one imposes that the divergence
 vanishes.    Instead of  (\ref{fourfour}) we define for $A_k$ on $\tz$ and $\cA$ on $\tk$
\be 
 \rho_{k} (A_{k}) = \const  \int \de( A_{k} - \cQ \cA) \de (R_k \de \cA )  F_{0,L^{-k}}  (\cA)   \exp( -\frac12 \| d \cA \|^2 \B) D\cA
\ee
Now $\de = d^T$ on one-forms (functions on bonds) is the divergence operator and $\de \cA$ is a scalar.  The operator
 $R_k$ is the projection onto the subspace $\De (\ker Q_k )$ where $Q_k$ is the averaging operator on scalars, and $\de (R_k \de \cA ) $
 denotes the delta function in this subspace.   A Fadeev-Popov argument shows  equivalence with the axial gauge expression.
 
Evaluation of such integrals depends on 
\be   \cA_k = \cH_k A_k = \textrm{ minimizer of } \|d \cA \|^2 \textrm{ subject to  } \cQ \cA = A_k \textrm{ and } R_k  \de \cA =0
\ee
There is an explicit expression for $\cH_k$ in terms of Landau gauge Green's functions  defined  for $a>0$ by 
\be
\cG_k  = \B( \de d + dR_k \de + a \cQ^T_k \cQ_k \B)^{-1}
\ee
It is 
\be    
\cH_k =  \cG_k \cQ^T_k (\cQ_k  \cG_k \cQ^T_k)^{-1}
\ee
  
The minimizer        $\cH_k$   
is gauge equivalent  to  the axial gauge minimizer    in the sense that    
\be  
\label{relation}  \cH^{\sx}_{k}  =    \cH_{k}  +    \pa  \cO_k  \ee
   for some operator  $ \cO_k$.
 This means that  in  gauge invariant  expression we  can  can   replace  $ \cH^{\sx}_{k} $ by     $ \cH_{k}$.  Hence
  we  can make this replacement in  the fluctuation integral  (\ref{sometimes2}),  and in particular $\| d\cA^{\sx}_k \|^2$ can be
 replaced by $\| d\cA_k \|^2$  where
 $\cA_k = \cH_k A_k$.  This is useful because  $\cH_k$ is more regular than  the axial  $\cH^{\sx}_k$. 
Also         $\De_k$  can be expressed in the Landau gauge as
 \be    \label{zee}
 \blan     Z,     \De_k   Z\bran     =   \|  d \cH_k  Z \|^2  
 \ee

There is  a bound below on $\De_k$.   We have   $Z = \cQ_k \cH_k Z$, hence
$dZ = \cQ^{(2)}_k  d\cH_k Z$ for a certain averaging operator  $\cQ^{(2)}_k $ on two forms, and hence 
$\| dZ \|^2 \leq  \| d\cH_k Z \|^2$.  Furthermore  Balaban \cite{Bal84b} shows that there is a constant $c_0$ depending only on $L$ such that 
  $\| dZ \|^2  \geq c_0 \|Z\|^2$
on the constrained surface $\cQ Z =0, \tau Z = 0$.   Thus on the same surface we have
\be
  \blan     Z,     \De_k   Z\bran   \geq   c_0 \|Z\|^2 
  \ee
  This shows that the fluctation integrals  of the previous section are well-defined.

\subsection{random walk expansions}   \label{alvin}

We quote some results on random walk expansions,   essentially all   due to Balaban.

\subsubsection{expansion for $\tilde \cG_k$}

First consider the Green's function $\cG_k$.  With the projection operator $P_k = I  - R_k$ it can also be written
 \be
\cG_{k}= (  \De-  dP_k \de + a \cQ^T_k \cQ_k )^{-1}
\ee

\begin{lem}  \label{random3}  \cite{Bal84a}, \cite{Bal84b}, \cite{Bal85b}  
The   Landau gauge Green's function   $\cG_k$ has      a  random walk expansion $\cG_k = \sum_{\om} \cG_{k, \om}$
 based on blocks  of size $M$, convergent for $M$ sufficiently large.   This yields the bounds   
 \be
\label{oscar1}
 | \cG_k f |, \   | \pa \cG_k f|,   
\| \de_{\al}  \pa  \cG_k f \|_{\infty}       \leq      C     \|f\|_{\infty}  
\ee
  Furthermore let   $  \De_y,  \De_{y'}$  be unit squares centered on unit lattice points     $   y,y'  \in  \bbT^0_{N-k}$  and let
 $\zeta_y$   be a smooth partition on unity with $\supp\  \zeta_y \subset  \tilde  \De_{y'} $. 
 \be    \label{oscar2}
 \begin{split}  
  |1_{\De_y} \cG_k1_{\De_{y'}} f |, \   |1_{\De_y} \pa \cG_k1_{\De_{y'}} f|,   
\| \de_{\al } \zeta_y \pa   \cG_k1_{\De_{y'}} f \|_{\infty}       \leq    &  C    e^{  -\ga  d(y,y') } \|f\|_{\infty}     \\
\end{split}
\ee
\end{lem}
\bigskip

\rem These bounds can be obtained more simply by other methods. By deriving them from a random walk expansion it
gives us freedom to modify the operator without losing the bounds.
\bigskip

\pr  We  sketch the proof.  
We use a  covering of $\tk$ by cubes  $\bd$ of width $2M$ centered on the points of the $M$ lattice $\bbT^m_{N-k}$.  
\footnote{We use a covering by $2M$ cubes $\bd$ for consistency with  \cite{Bal84a}, \cite{Bal84b}.  We could use the partition $\sq$
into $M$ cubes introduced before, but the following discussion would have to be modified}
 Let $\tilde \bd$ be the union of all  such cubes whose distance to $\bd$ is zero.
Similarly define enlargements $ \tilde \bd ^2, \tilde \bd^3, \dots $.

 For each cube $\bd$ one introduces the inverse on
the 3-fold enlargement  $\tilde \bd ^3$ :  
\be
\cG_{k, \bd }= \B[  \De-  dP_{k, \bd} \de + a \cQ^T_k \cQ_k \B]_{\tilde \bd^3}^{-1} 
\ee
Here  we  take the inverse with periodic boundary conditions,    
so we are regarding the  cube $\tilde \bd^3$ as a little torus.   These satisfy  the  bounds 
(\ref{oscar2}). 

 Take  a   partition of unity  $\sum_{\bd} h^2_{\bd} =1$  with  $\supp \ h_{\bd} \subset \bd$,  and define 
 a parametrix
\be
 \cG^*_k  = \sum_{\bd} h_{\bd} \cG_{k, \bd } h_{\bd}
 \ee
 Here we identify $\supp\ h_{\bd} $ with a subset of the torus $\tilde \bd^3$
 so that $ h_{\bd}\cG_{k, \bd} h_{\bd}$ can be regarded as an operator on the full torus $\tk$. 
 Let $\cD_k = \De-  dP_k \de + a \cQ^T_k \cQ_k$ and let  $\cD_{k, \bd} = [ \De-  dP_{k, \bd} \de + a \cQ^T_k \cQ_k ]_{\tilde \bd^3}$
 so that $\cG_k = \cD_k^{-1}$ and $\cG_{k, \bd} =  \cD_{k, \bd} ^{-1}$.
 Then we compute
 \be
 \begin{split}
   \cD_k \cG^*_k  = &  \sum_{\bd}   h_{\bd} \cD_{k, \bd}  \cG_{k, \bd } h_{\bd}
 - \sum_{\bd}  K_{\bd} \cG_{k, \bd } h_{\bd} \\
 & K_{\bd} =    h_{\bd} \cD_{k, \bd}  -   \cD_k h_{\bd} \\
 \end{split} 
\ee
 The first term is the identity operator.   For the second term we write
 \be \label{uno}
   K_{\bd} 
= - \B[ \De+ a \cQ^T_k \cQ_k , \ h_{\bd} \B]  +   \B(    h_{\bd} (dP_{k, \bd}\de)- (dP_k \de) h_{\bd}  \B)  \ee
 Let $\zeta_{\bd}$ be a smooth approximation to the characteristic function of $\bd$  with $\zeta_{\bd}=1$ on $\bd$ and all points
 a distance $\frac13 M$ from $\bd$, and with $\zeta_{\bd} =0$ on points greater than $\frac23 M$ from $\bd$.
 In the second  term  in $K_{\bd}$  multiply by $\zeta_{\bd} + (1- \zeta_{\bd})$.  Since $(1-\zeta_{\bd}) h_{\bd} =0$
 this term can be written
 \be\label{dos}
     \zeta_{\bd}  \B(  (d P_k \de- d P_{k,\bd} \de) h_{\bd} - [ (dP_{k, \bd}\de), h_{\bd} ]  \B) 
 -   (1-\zeta_{\bd} )  (dP_k \de) h_{\bd}  
\ee
 All the terms in  (\ref{uno}), (\ref{dos}) are well localized except the last and we insert  here  $1= \sum_{\bd'} h^2_{\bd'} $. Only $\bd \neq \bd'$  contributes.
 Now we can write 
\be
   \cD_k \cG^*_k  = I - K   = I -   \sum_{\bd' , \bd} K_{\bd', \bd} \cG_{k, \bd } h_{\bd}
\ee 
 where
\be  \label{sitz}
 K_{\bd', \bd}  = \begin{cases}  -\B[ \De+ a \cQ^T_k \cQ_k , \ h_{\bd} \B]   
  + \zeta_{\bd}  \B( [ dP_{k, \bd}\de, h_{\bd} ] - d (P_k - P_{k,\bd}) \de h_{\bd} \B)  &  \hs \bd' = \bd    \\
-  h^2_{\bd'}(1-\zeta_{\bd} )  (dP_k \de) h_{\bd} 
& \hs  \bd' \neq \bd \\
\end{cases}
 \ee
 Then   
  \be \label{sisty}
  \cG_k =  \cG^*_k ( I - K)^{-1}   =   \cG^*_k \sum_{n=0}^{\infty}  K^n  =  \sum_{\om }   \cG_{k, \om} 
  \ee    
where for  a sequence $\om = (\bd_0, \bd_1, \bd_2 \dots, \bd_{2n-1} ,\bd_{2n})$ of $2M$ cubes
\be  \label{sisty2}
  \cG_{k, \om} = \B( h_{{\bd}_0} \cG_{k, \bd_0 } h_{{\bd}_0}\B)
\B( K_{\bd_1, \bd_2} \cG_{k, \bd_2} h_{\bd_2} \B)\cdots  \B(  K_{\bd_{2n-1},\bd_{2n}} \cG_{k, \bd_{2n}}   h_{\bd_{2n}}\B)
\ee

Now we claim that
\be
 \label{borneo1}
  | 1_{\De_y} \cK_{\bd' ,\bd}   \cG_{k,  \bd }1_{\De_{y'}} f | \leq      C M^{-1}e^{  -\ga  d(y,y') } \|f\|_{\infty} 
\ee
For $\bd \neq  \bd'$ the first term $[ \De+ a \cQ^T_k \cQ_k , \ h_{\bd} ]   $ in (\ref{sitz}) is local and can be expressed in terms of derivatives
of $h$ which are $\cO(M^{-1})$.  Combined with the exponential decay for $\cG_{k, \bd}$ and its derivatives this gives a bound of the form
(\ref{borneo1}). 
The other terms require some rather detailed knowledge about $P_k, P_{k, \bd}$ which are given by
\be
\begin{split}
P_k = G_kQ^T_k N_k Q_k G_k \hs \hs  &  N_k = (Q^T_k G_k^2 Q_k )^{-1} \\
P_{k, \bd} = G_{k, \bd}Q^T_{k} N_{k, \bd} Q_{k} G_{k, \bd} \hs \hs  &  N_{k, \bd} = (Q^T_{k} G_{k, \bd}^2 Q_{k} )^{-1} \\
\end{split}
 \ee
 Here $Q_k$ is the averaging operator on scalars and $G_k = ( \De + a Q_k^TQ_k)^{-1}$ on scalars. 
The operators $G_{k,\bd}, N_{k, \bd}$ are defined on the torus $\tilde \bd ^3$.   Both  $G_k, G_{k, \bd}$ 
satisfy bounds of the form  $ | 1_{\De_y}  G_{k }1_{\De_{y'}} f | \leq      C e^{  -\ga  d(y,y') } \|f\|_{\infty}  $
The same is true for $N_{k},N_{k, \bd}$ (pointwise bounds for these unit lattice operators)  and hence for $P_k, P_{k, \bd}$.   
 The term  $ [ dP_{k, \bd}\de, h_{\bd} ] $ in (\ref{sitz})  can
be expressed in derivatives of $h_{\bd}$  and together with the bounds on $P_{k, \bd}$ yields a bound of the form (\ref{borneo1}).
Furthermore both  $G_k, G_{k, \bd}$ have random walk expansions based on the fundamental cubes $\bd$.  The expansions are locally the
same and differ only in the global topology.   In particular the leading terms are the same and so when localized near $\bd$ the difference $G_k - G_{k, \bd}$  is $\cO(M^{-1})$.
The same is true for the pair $N_k, N_{k, \bd}$ and hence for $P_k, P_{k, \bd}$. 
This leads to a bound of the form (\ref{borneo1}) for the term  $ \zeta_{\bd} d (P_k - P_{k,\bd}) \de h_{\bd}$ in (\ref{sitz}). 
\footnote{Note that $\de h_{\bd} = h_{\bd} \de + [\de, h_{\bd}]$ does not necessarily involve derivatives of $h_{\bd}$}
Finally for the term $ h^2_{\bd'}(1-\zeta_{\bd} )  (dP_k \de) h_{\bd} $ in (\ref{sitz})  we use the fact that the supports of $ (1-\zeta_{\bd} )$
and $h_{\bd}$ are separated by $\frac13 M$.  The exponential decay of $P_k$ then gives a factor $\cO(e^{-\frac13\ga M})  $.  Hence this term is
is  $\cO(M^{-1})$ and satisfies (\ref{borneo1}).   All this is a rather long story for which we refer to Balaban \cite{Bal84a}, \cite{Bal84b}    . 

Now we write
\be
\begin{split}
&  1_{\De_y} \cG_k 1_{\De_{y'}} = \sum_{\om} \sum_{y_1, y_3, \dots, y_{2n-1} }  \\
&  1_{\De_y}\B( h_{{\bd}_0} \cG_{k, \bd_0 } h_{{\bd}_0}\B) 1_{\De_{y_1}}
\B( K_{\bd_1, \bd_2} \cG_{k, \bd_2} h_{\bd_2} \B)1_{\De_{y_3}}\cdots    1_{\De_{y_{2n-1}}}\B(  K_{\bd_{2n-1},\bd_{2n}} \cG_{k, \bd_{2n}}   h_{\bd_{2n}}\B) 1_{\De_{y'}} \\
\end{split}
\ee
The sums are restricted by the conditions  $y \in \bd_0$, $y' \in \bd_n$  and  for $i$ odd   $y_i \subset \bd_{i-1} \cap \bd_i \neq \emptyset $.
Then (\ref{borneo1}) gives the bound
\be  \label{borneo3}
\begin{split}
 &|1_{\De_y }  \cG_k 1_{\De_{y'}}f| \\
 & \leq      \sum_{\om}\sum_{y_1, y_3, \dots, y_{2n-1} }   C ( C M^{-1} )^{| \om|}
 \exp\B(  -\ga   \B( d(y,y_1) +  d(y_1,y_3)+ \dots +  d(y_{2n-1},y') \B) \B)    \|f\|_{\infty} \\
\end{split}
\ee
Split the exponent into thirds. 
In  the first third  we use $d(y,y_1) +  d(y_1,y_3)+ \dots +  d(y_{2n-1},y') \geq d(y,y')$.  The second third is used for convergence of the sum over the $y_i$.
For the last third 
let $z_i \in \bbT^m_{N-k}$ be the center of the cube $\bd_i$ which we could then label as $\bd_{z_i}$.  We claim that  for $i$ odd
\be
  d(y_i,y_{i+2})  \geq   \frac 13  d' (z_i,z_{i+2})  
\ee
where $d'(z,z')$ is the usual   distance but set to zero if $\bd_z,\bd_{z'}$ are neighbors, i.e. if $d(z,z') \leq 2M$.
Indeed if  $d(z_i,z_{i+2} ) \geq 3M$ then $d(y_i,y_{i+2}) \geq d(z_i,z_{i+2} ) -2M \geq \frac 13 d(z_i,z_{i+2} )$, while if
$d(z_i,z_{i+2} )  \leq 2M$ the inequality is trivial.   Similarly $d(y_{2n-1},y') \geq \frac13 d'(z_{2n-1},z_{2n})$.
 Now  write the  sum over $\om$ as  a sum over $n$ and  $z_0,z_1, \dots, z_{2n}$   with $z_0,z_{2n}$ constrained by the conditions
 $ \bd_{z_0} \ni y$ and $ \bd_{z_{2n}} \ni y'$ and   for $i$ odd $\bd_{z_{i-1} } \cap \bd_{z_i} \neq \emptyset$.  We have then
\be
\begin{split}
 &|1_{\De_y }  \cG_k 1_{\De_{y'}}f| \\
 & \leq     e^{  - \frac13 \ga  d(y,y') }  \sum_{n=0}^{\infty}  \sum_{z_0, \dots, z_{2n} }   C ( C M^{-1} )^{n}
 \exp\B(  - \frac 19 \ga   \B(  d'(z_1,z_3)+ d'(z_3,z_5)   +\dots +  d'(z_{2n-1},z_{2n}) \B) \B)    \|f\|_{\infty} \\ \end{split}
\ee
The sums over $z_i$ for $i$ even gives a factor $\one^n$, the sum over $z_i$ for $i$ odd converges by the exponential decay, and the
sum over $n$ converges for $M$ sufficiently large by the factor $M^{-n}$.  We  obtain
\be
 |1_{\De_y }  \cG_k 1_{\De_{y'}}f|
\leq       C    e^{  -\frac13 \ga  d(y,y') } \|f\|_{\infty}    
\ee
  which is   the first  estimate in (\ref{oscar2}) with a new $\ga$.  The derivatives are treated similarly.  The bounds (\ref{oscar1}) follow by
summing over $y'$.   The results stated in the lemma now follow.
\bigskip

However because  of the presence of long jumps the expansion is not as local as we would like.   The long jumps arise from the term
$K_{\bd', \bd} = h^2_{\bd'}(1-\zeta_{\bd} )  (dP_k \de) h_{\bd}$ in the case $\bd' \neq \bd$. The remedy is to insert the random walk expansion for  $P_k$.    After some rearrangements one ends with the following result \cite{Bal85b}.  The expansion is not based 
just on cubes but on  small connected unions of cubes $X$, called \textit{localization domains}.  There is a constant $r_0 = \one$ such that the number of $M$ cubes  
in $X$ satisfies $|X|_M \leq r_0$.  A walk is a sequence  of localization domains
$ X_0, X_1, \dots, X_n $ with the property that $X_i \cap X_{i+1} \neq \emptyset$.  For each $X$ there
are  operators  $R_{\al}(X)$  localized in $X$ and indexed by  $\al$ which ranges over a finite set including 0.    If $\al =0$ then the
only localization domain possible is $X= \bd $ and 
\be
R_0(\bd ) = h_{\bd} \cG_{k,\bd} h_{\bd}
\ee
We have the bounds 
 \be    \label{oscar3}
 \begin{split}  
&   |1_{\De_y} R_{\al}(X) 1_{\De_{y'}} f |, \   |1_{\De_y} \pa R_{\al}(X) 1_{\De_{y'}} f|,   
\| \de_{\al } \zeta_y \pa   R_{\al}(X) 1_{\De_{y'}} f \|_{\infty}    \\
&  \hs \hs \hs   \leq  
\begin{cases}     C    e^{  -\ga  d(y,y') } \|f\|_{\infty}   &  \al =0    \\
   C M^{- 1}   e^{  -\ga  d(y,y') } \|f\|_{\infty}    &    \al  \neq 0    \\
\end{cases}
\end{split}
\ee
The expansion has the form $ \cG_k  = \sum \cG_{k, \om}$
where the sum is over indexed walks 
\be \label{crackerjack}
\om = \B( (0, X_0), (\al_1, X_1), \dots, (\al_n, X_n) \B)  
\ee
with $\al_i \neq 0$  if $i \neq 0$ and
\be
\cG_{k, \om}   =  R_0(X_0) R_{\al_1}(X_1) \cdots  R_{\al_n}(X_n)
\ee
The walk can also be written with $|\om| =n$
\be 
\cG_k = \sum_{\bd}  h_{\bd} \cG_{k,\bd} h_{\bd} + \sum_{|\om| \geq 1} \cG_{k, \om}
\ee
The  results stated in the lemma also  follow from this form of the random walk expansion.
\bigskip

Expansions of this form will be called \textit{generalized random walk expansions}.  We do not require that all possible combinations of 
the $(\al, X)$ actually occur in the sum over $\om$.  
\bigskip

In our expansion   we can introduce weakening parameters.
Given  a walk $\om$ of the form (\ref{crackerjack})  let $X_{\om}$ be the connected set $X_{\om} = \cup_{i=0}^n X_i$.  Then $\cG_{k, \om}$ is
localized in $X_{\om}$.  For the  weakened propagator  we   associate  with each $M$ cube $\sq$  a variable $s_{\sq}$ and define
\be  \label{newweak}
 \cG_k(s)  = \sum_{\bd}  h_{\bd} \cG_{k,\bd} h_{\bd} +  \sum_{\om: | \om| \geq 1 } s_{\om}  \cG_{k, \om}  \hs   s_\om = \prod_{\sq \subset   X_{\om} } s_{\sq}
\ee
If $Y$  is a union of $M$ cubes and
$s_{\sq} =0$ for $\sq \subset Y^c$, then no path with $ X_{\om}$ intersecting $Y^c$ contributes to $\cG_k(s)$ and so $\cG_k(s)$
is localized in $Y$.

\begin{lem}  \label{oscar4} There is a  small constant $\al_0 = \one$ such that for
For  $|s_{\square} | \leq M^{\al_0  }$ the weakened propagator $\cG_k(s)$ satisfies  (\ref{oscar1}), (\ref{oscar2}).
\end{lem} 
\bigskip

\pr  We have $|X_i|_M \leq r_0 = \one $ and so for $| \om| = n$
\be 
|X_{\om}| \leq  \sum_{i=0}^n |X_i|_M  \leq r_0 ( | \om | +1 )
\ee
Therefore  if $|s_{\square} | \leq M^{\al_0  }$ and $\al_0$  issufficiently small 
\be 
| s_{\om}|
\leq   \exp \B(    \al_0 \log M  |X_{\om}|_M \B)  
\leq   \exp \B(     \al_0 r_0\log M( | \om | +1 )\B) 
\leq   \one  M^{ \frac12 | \om| } 
\ee
This changes the convergence factor $M^{ -| \om| }$ to $M^{- \frac12 | \om| }$, but for $M$ sufficiently large this is still
small enough to guarantee convergence of the expansion.

\bigskip 

\rems
\begin{enumerate}
\item The operator $\cG_{k, \bd}$ itself has a random walk expansion of the form of the lemma.  The sums over $\bd$  would then
be over  the much smaller set $\tilde \bd^3$ rather than $\tk$.   The expansions are locally the same and only differ with the global topology.
In particular the leading term $\cG_k^*$ is the same in each case.  It follows that  say for  $y,y' \in \bd$ 
\be \label{lumbar}
  |1_{\De_y} (\cG_k - \cG_{k, \bd} ) 1_{\De_{y'}} f |       \leq      C  M^{-1}  e^{  -\ga  d(y,y') } \|f\|_{\infty}     
\ee
\item The operator $\cG_{k+1}$ on $\bbT^{-k-1}_{N-k-1}$ scales up to an operator $\cG^0_{k+1}$ on $\tk$.  It is given for any $a>0$
by 
\be   
\cG^0_{k+1}  = \B(   \de  d   +    d R^0_{k+1} \de   + a \cQ^T_{k+1} \cQ_{k+1}   \B)^{-1}
\ee
Here $R^0_{k+1}$ is the projection onto    $\De (\ker \cQ_{k+1}) $.  All the results stated for $\cG_k$ hold for $\cG^0_{k+1}$ as well. 
\end{enumerate}

\subsubsection{expansion for $\cN_k  =   (\cQ_{k}  \cG_{k} \cQ^T_{k})^{-1}$}

We consider the operators
\be
\begin{split}  \cN_k \equiv  (\cQ_{k}  \cG_{k} \cQ^T_{k})^{-1} & \hs  \textrm{ on  } \bbT_{N-k}^0 \\ 
\cN^0_{k+1}  \equiv  (\cQ_{k+1}  \cG^0_{k+1} \cQ^T_{k+1})^{-1} &\hs  \textrm{ on  } \bbT^1_{N-k} \\
\end{split}
\ee

\begin{lem} \cite{Bal84a}, \cite{Bal84b}, \cite{Bal85b}  The  operator  $\cN_k $
has a random walk expansion  $\cN_k = \sum_{\om} \cN_{k, \om}$ based on blocks of size $M$, convergent for $M$ sufficiently large which yields a bound
\be   \label{Ndecay}
| \cN_k(b,b')| \leq C e^{- \ga d(b,b')}
\ee
The same is true for $\cN^0_{k+1}$.
\end{lem} 
\bigskip

\pr  The proofs are essentially the same and we give the proof for $\cN^0_{k+1}$. First
consider the operator  $\cQ_{k+1}  \cG^0_{k+1,\bd}\cQ^T_{k+1}$ 
defined on $\tilde \bd^3$ again regarded as a small torus.  This is bounded above independent of the volume  by our estimates on $ \cG^0_{k+1,\bd}$.
We claim that it is also bounded below.  This relies on the identity   for $A$ on  the $L$-lattice in $\tilde \bd^3 $
\be  \label{english}
\blan A,\   \cQ_{k+1}\cG^0_{k+1,\bd } \cQ^T_{k+1}A \bran =  \blan [\cQ^TA]^{\sx},\tilde   C_{k,\bd}  [\cQ^TA]^{\sx} \bran
\ee
where  $\tilde   C_{k,\bd} $ is the inverse of  $\De_k + a\cQ^T \cQ $ on the subspace $\ker \tau$ of the unit lattice in $\tilde \bd^3$  and $[\cQ^TF]^{\sx} $ is the
projection of $\cQ^TF$ onto this subspace.  
 For the proof on a general torus see  \cite{Bal84b}   or Appendix \ref{C}.  Now $\De_k + a\cQ^T \cQ $ is bounded above as a
 quadratic  form  and so $ C_{k,\bd} $ is bounded below.  Hence the right side of (\ref{english}) is bounded below by constant times
$ \| [\cQ^TA]^{\sx}\|^2 $.   However we show in appendix \ref{D} that
$ \| [\cQ^TA]^{\sx}\|^2  \geq L^{-1} \|A\|^2$.   Thus we have
\be
c \|A\|^2   \leq   \blan A,\   \cQ_{k+1}\cG^0_{k+1,\bd } \cQ^T_{k+1}A \bran  \leq C \|A \|^2
\ee
Hence the inverse has the same bounds.   But before we invert it we restict from $\tilde \bd^3$  to smaller cube $\bd$, that is to functions
on bonds with at least one end in $\bd$.   The restriction
satisfies the same bounds  and  if we define
\be
 \cN^0_{k+1, \bd } = \B[ \cQ_{k+1}\cG^0_{k+1,\bd } \cQ^T_{k+1} \B]^{-1}_{\bd}
\ee
then with  new constants  for $A$ on $\bd$
\be
c \|A\|^2   \leq   \blan A,\   \cN^0_{k+1, \bd }A \bran  \leq C \|A \|^2
\ee
Furthermore  by (\ref{oscar2}) we have the bound  $|(\cQ_{k+1} \cG^0_{k+1, \bd} \cQ^T_{k+1})(b,b')| \leq  C    e^{  -\ga  d(b,b') }  $.  It  follows by   Balaban's theorem   on unit lattice operators (section 5 in \cite{Bal83b})  that the same is true for the inverse  and so  
\be \label{oliver2} 
|  \cN^0_{k+1,\bd}( b,b') |      \leq      C    e^{  -\ga  d(b,b') }  
\ee

To generate  the  random walk  for    $\cN_k $ we again take the partition of unity $\sum_{\bd} h_{\bd}^2 =1$, identify 
$h_{\bd}  \cN^0_{k+1,\bd} h_{\bd} $ with an operator on $\bbT^1_{N-k}$ and  
 define the parametrix
\be
 \cN^*_{k+1}  = \sum_{\bd} h_{\bd}  \cN^0_{k+1,\bd} h_{\bd} \ee
Then we compute
\be   \label{kingly2}
\begin{split}
\cQ_{k+1}\cG^0_{k+1 } \cQ^T_{k+1} \cN^*_{k+1} 
= &  \sum_{\bd}h_{\bd}  \B[ 1_{ \bd}\cQ_{k+1}  \cG^0_{k+1, \bd} \cQ^T_{k+1} 1_{ \bd}   \cN^0_{k+1,\bd} \B]h_{\bd} \\
& -  \sum_{\bd} \B[ h_{\bd}, 1_{ \bd} \cQ_{k+1} \cG^0_{k+1, \bd} \cQ^T_{k+1} 1_{ \bd}\B]  \cN^0_{k+1,\bd} h_{\bd} \\
 & -  \sum_{\bd}  1_{\tilde \bd} \cQ_{k+1} \B( \cG^0_{k+1, \bd} - \cG^0_{k+1} \B)\cQ^T_{k+1}   h_{\bd}   \cN^0_{k+1,\bd} h_{\bd} \\
&  + \sum_{\bd, \bd' } h^2_{\bd'} 1_{\bd^c}  \cQ_{k+1}  \cG^0_{k+1} \cQ^T_{k+1}   h_{\bd}  \cN^0_{k+1,\bd} h_{\bd} \\
  \equiv &   I - \sum_{\bd} K_{\bd' ,\bd}\cN^0_{k+1,\bd} h_{\bd} \equiv  I - K  \\
\end{split}
 \ee
 Here we have identified the first term as the identity.
 Now we have 
\be
  \cN^0_{k+1}  = \cN^*_{k+1}(I-K)^{-1} =   \cN^*_{k+1}   \sum_{n=0}^{\infty}K^n = \sum_{\om} \cN_{k, \om} 
\ee
where for  a sequence $\om = (\bd_0, \bd_1, \bd_2 \dots, \bd_{2n-1} ,\bd_{2n})$ 
\be 
  \cN^0_{k+1, \om} = \B( h_{{\bd}_0} \cN^0_{k+1, \bd_0 } h_{{\bd}_0}\B)
\B( K_{\bd_1, \bd_2} \cN^0_{k+1, \bd_2} h_{\bd_2} \B)\cdots  \B(  K_{\bd_{2n-1},\bd_{2n}}  \cN^0_{k+1, \bd_{2n}}   h_{\bd_{2n}}\B)
\ee

Now we claim that  
 \be \label{pinko}
|K_{\bd', \bd} \cN^0_{k+1, \bd} h_{\bd}(b,b') | \leq   C  M^{-1}  e^{  -\ga  d(b,b') }
\ee
This is argued as follows.  The second term in (\ref{kingly2}) can be expressed in terms of derivatives of $h_{\bd}$ and so is $\cO(M^{-1})$.
For the  third term in (\ref{kingly2}) we use that $\cG^0_{k+1, \bd} - \cG^0_{k+1}$ satisfies a bound like  (\ref{lumbar}) to get an estimate
$\cO(M^{-1})$ .  For  the fourth  term in (\ref{kingly2})  we can assume  that $ \bd^c$ and $\supp\  h_{\bd}$ are separated by
at least $\frac13 M$.  By  the exponential decay of $\cG_k$ this term has a decay factor  $\cO(e^{- \frac13 \ga M})$ 
and hence   $\cO (M^{-1})$ as well.

The estimate (\ref{pinko}) gives convergence of the expansion and the stated estimates just as in the previous lemma. 
   However again the long jumps are unwelcome.   They come from  the term $K_{\bd' \bd} = h^2_{\bd'} 1_{\bd^c}  \cQ_{k+1}  \cG^0_{k+1} \cQ^T_{k+1}   h_{\bd}  \cN^0_{k+1,\bd} h_{\bd} $ for $\bd \neq \bd'$.  The remedy in this case is to replace 
the  $ \cG^0_{k+1}$ by its random walk expansion.   Again the result is a generalized random walk expansion of the form  
\be 
\cN^0_{k+1} = \sum_{\bd}  h_{\bd} \cN^0_{k+1,  \bd}  h_{\bd} + \sum_{|\om| \geq 1} \cN^0_{k+1, \om}
\ee
where 
\be
\cN^0_{k+1, \om}   = \sum  R_0(X_0) R_{\al_1}(X_1) \cdots  R_{\al_1}(X_1)
\ee
and  for $\al \neq 0$,  $R_{\al} (X)$ satisfies
\be
|R_{\al} (X; b,b') | \leq CM^{-1} e^{- \ga d(b,b') }
\ee

\subsubsection{expansion for $\cH_k$}

From the representation $\cH_k =  \cG_k \cQ^T_k (\cQ_k  \cG_k \cQ^T_k)^{-1}$  and the last two results we have

\begin{lem}  \label{random4}  \cite{Bal84a}, \cite{Bal84b}  , \cite{Bal85b}
The   Landau gauge minimizer   $\cH_k$ has      a   generalized  random walk expansion 
 based on blocks  of size $M$, convergent for $M$ sufficiently large.   This yields the bounds   
 \be
\label{slavic3}
 | \cH_k f |, \   | \pa \cH_k f|,   
\| \de_{\al}  \pa  \cH_k f \|_{\infty}       \leq      C     \|f\|_{\infty}  
\ee
  The local version is   for     $y,y'$ on  $\tk$
 \be    \label{alfie}
 \begin{split}  
  |1_{\De_y} \cH_k1_{\De_{y'}} f |, \   |1_{\De_y} \pa \cH_k1_{\De_{y'}} f|,   
\| \de_{\al } \zeta_y \pa   \cH_k1_{\De_{y'}} f \|_{\infty}       \leq    &  C    e^{  -\ga  d(y,y') } \|f\|_{\infty}     \\
\end{split}
\ee
\end{lem}

One can introduce weakening parameters as before and define $\cH_k(s)$.   For $|s_{\sq}| \leq M^{\al_0}$ and $\al_0$ sufficiently small  these 
satisfy the same bounds as $\cH_k$.

\subsubsection{expansion for $\tilde \cG_k$}

The Green's function $\cG_k$ can be defined by stating  that $\cA = \cG_k  J$ is the  minimizer of the quadratic form 
 $\frac12 < \cA, ( \de  d   +    d R_{k} \de   + a \cQ^T_{k} \cQ_{k}  ) \cA >-<\cA, J>$. 
   We are also interested in a modified Green's function  $\tilde \cG_k$
defined by  stating  that $\cA =\tilde  \cG_k  J$ is the  minimizer of the same  quadratic form subject to the constraint  $\cQ_k \cA = 0$ 
(in which case the term $ a \cQ^T_{k} \cQ_{k}$ is optional).   These turn out to be related by 
\be    \label{lorenzo2}
\tilde \cG_{k} =   \cG_{k} -  \cG_{k}  \cQ_{k}^T  \cN_{k}    \cQ_{k} \cG_{k}  
\ee
On the  small torus $\tilde \bd^3$ this takes the form
\be    \label{lorenzo3}
\tilde \cG_{k, \bd} =   \cG_{k, \bd} -  \cG_{k, \bd}  \cQ_{k}^T  \cN_{k, \bd}    \cQ_{k} \cG_{k, \bd}  
\ee

\begin{lem} \label{ossie0}
The operator $\tilde \cG_{k}$ can  has a generalized random walk expansion of the form 
\be  \label{eggs}
\tilde \cG_{k} =  \sum_{\bd} h_{\bd} \tilde \cG_{k, \bd} h_{\bd} +    \sum_{\om: | \om | \geq 1}  \tilde \cG _{k, \om}
\ee
This yields the bounds (\ref{oscar2}) for $\tilde \cG_{k} $.  The second  term  in (\ref{eggs}) is $\cO(M^{-1})$.   
\end{lem}
\bigskip

  We have seen that the operators $\cG_{k}$ and  $ \cN_{k} $  have  a random walk expansions. 
We can insert these in the definition of $\tilde \cG_k$ and get a random walk expansion.   However this would not have the leading term
we want.    We need to modify the procedure  and we sketch the idea below.   If we had supplied full details in the previous theorems
we would already be familiar with this strategy  taken from \cite{Bal85b}

In (\ref{lorenzo2}) insert the expansion for $\cG_{k} $ in the first term and  the expansion for $ \cN_{k}  $ in the second term.  This yields
\be \label{lorenzo4}
\tilde \cG_{k} =  \sum_{\bd} h_{\bd}\cG_{k, \bd}h_{\bd}  -  \sum_{\bd} \cG_{k}  \cQ_{k}^T  h_{\bd}\cN_{k, \bd}h_{\bd}     \cQ_{k} \cG_{k}  
+ \cO(M^{-1})
\ee
where as before the sum is over a covering of the $M$-lattice by $2M$ cubes. 
For each  fixed $\bd$ we introduce a new cover  defined from the old cover   by replacing   $\bd$ by $ \tilde \bd $, then deleting all
cubes contained in $\tilde \bd $,  and leaving the other cubes in the old cover 
alone.  Denote the cubes of the new cover by $\bd_0$.  Make a random walk expansion for $\cG_k$ based on the new cover .   This will have the form   $\cG_{k}  =     \sum_{\bd_0 } h_{\bd_0}  \cG_{k, \bd_0} h_{\bd_0}  + \cO(M^{-1})$.   When this is inserted in two places in the second term in  (\ref{lorenzo4})  only the  term with $\bd_0 \supset \bd$  gives an $\one$  contribution which was the goal. 
Using also $ h_{\bd_0}  \cQ_{k}^T  h_{\bd}= \cQ_{k}^T  h_{\bd}$  the second term in  (\ref{lorenzo4}) can be written 
with $\bd_0 \supset \bd$
\be 
  \sum_{\bd}  h_{\bd_0}  \cG_{k, \bd_0}  \cQ_{k}^T  h_{\bd}\cN_{k, \bd}h_{\bd}     \cQ_{k}   \cG_{k, \bd_0} h_{\bd_0} 
+ \cO(M^{-1})
\ee
Next move the $h_{\bd}$ to the outside using  $[ h_{\bd} ,    \cQ_{k} ] =   \cO(M^{-1})$ and
$[ h_{\bd} ,   \cG_{k, \bd_0} ] =   \cO(M^{-1})$ and use $ h_{\bd_0}   h_{\bd}=   h_{\bd}$. 
Then the last expression becomes
\be  \label{ouch1} 
  \sum_{\bd}  h_{\bd}  \cG_{k, \bd_0}  \cQ_{k}^T \cN_{k, \bd}     \cQ_{k}   \cG_{k, \bd_0} h_{\bd} 
+ \cO(M^{-1})
\ee
 Now write
\be   \label{ouch2}
 h_{\bd}  \cG_{k, \bd_0} =   h_{\bd}  \cG_{k, \bd} + h_{\bd}\B(  \cG_{k, \bd_0}- \cG_{k, \bd} \B) \zeta_{\bd}
+ h_{\bd}\B(  \cG_{k, \bd_0}- \cG_{k, \bd} \B) (1-\zeta_{\bd} )
\ee
 The last term here is $\cO(e^{-\frac 13 \ga M } )= \cO(M^{-1})$ since $h_{\bd} $ is separated from $1- \zeta_{\bd}$ by at least $\frac13 M$ and both 
 $\cG_{k, \bd_0}$ and   $\cG_{k, \bd}$ have exponential decay. 
 The second term is  localized near $\bd$  and  $\cO(M^{-1})$   as in (\ref{lumbar}).   Insert (\ref{ouch2}) in (\ref{ouch1})  and then back in (\ref{lorenzo4}). 
This gives the desired result
\be \label{lorenzo6}
\begin{split}
\tilde \cG_{k} = &   \sum_{\bd} h_{\bd}\B( \cG_{k, \bd}- \cG_{k, \bd}  \cQ_{k}^T  \cN_{k, \bd}   \cQ_{k}   \cG_{k, \bd} \B) h_{\bd} 
+ \cO(M^{-1}) \\
= & \sum_{\bd} h_{\bd} \tilde \cG_{k, \bd}h_{\bd} + \cO(M^{-1}) \\
\end{split}
\ee

\subsubsection{expansions for $C_k,C_k^{\frac12}$} 
We also need a random walk expansion for $C_k = (C^T\De_k C)^{-1}$ and $  C_k^{\frac12}$.  
We have already noted  that $C^T \De_k C$ is bounded below, and it is also bounded above by the
bound on $\cH_k$.  
 Thus the same is true for the inverse:
\be \label{sudbury0}
c \| \tilde Z \|^2 \leq    \blan  \tilde Z,    C_{k}  \tilde Z \bran   \leq   C \| \tilde Z \|^2
\ee
Here $\tilde Z$ is a restricted unit lattice variable defined as in section \ref{axialgauge}.
 We also consider  $C_{k, \bd} = (C^T\De_{k,\bd} C)^{-1}$  defined on the small torus $\tilde \bd^3$ which satisfies
 \be \label{sudbury1}
c \| \tilde Z \|^2_{\tilde \bd^3}  \leq    \blan  \tilde Z,    C_{k, \bd}  \tilde Z \bran   \leq   C \| \tilde Z \|^2_{\tilde \bd^3}
\ee

\begin{lem} 
\label{ossie1} 
  \cite{Bal84b},  \cite{Bal85b},  \cite{Bal88a}  
The   operators  $     C_k,         C_k^{\frac12} $ have   generalized    random walk expansions 
 based on blocks  of size $M$, convergent for $M$ sufficiently large.  For
 $C_k$ it has the form
 \be \label{licorice2}  
  C_k =     \sum_{\bd}    h_{\bd}  C_{k, \bd}h_{\bd} 
 + \sum_{\om: |\om| \geq 1} C_{k, \om}
\ee
The expansions yield the bounds: 
 \be  \label{succinct0}
|  C_k(\Up, \Up') |, \  |C^{\frac12}_k(\Up, \Up') | \leq    C  e^{ - \ga d(\Up, \Up')  }
\ee
where $d(\Up, \Up') \equiv d (\supp \Up, \supp \Up' )$ and $\Up, \Up'$ are taken from the basis $\{ \Up_{\al} \}$  defined in (\ref{Ubasis}).
The first term  in (\ref{licorice2})  is bounded above and below  (by (\ref{sudbury1})) and the second term in (\ref{licorice2}) is $\cO(M^{-1} )$. 
\end{lem}
\bigskip

\pr  We give  details for $C_k$. 
 The expansion  is based on  the  representation \cite{Bal85b}, \cite{Dim14}.
\be \label{goose}
 C C_k   C^T = (1 + d \cM ) \cQ_k \tilde \cG_{k+1} \cQ^T_k ( 1+ d \cM )^T
\ee
where
\be    \label{lorenzo5}
\tilde \cG_{k+1}  =   \cG^0_{k+1} -  \cG^0_{k+1} \cQ_{k+1}^T  \cN^0_{k+1}   \cQ_{k+1} \cG^0_{k+1}
\ee
The operator  $\tilde \cG_{k+1}  $ has a random walk expansion like $\tilde \cG_k$ and satisfies the same bounds (\ref{oscar2}).
Also  $\cM$ maps one-forms   to scalars by
\be
(\cM Z )(x) = -  (\tau Z )(y,x) + L^{-3} \sum_{x'\in B(y),x' \neq y} (\tau Z )(y,x')
\ee

Insert the expansion (\ref{eggs})  for $\tilde \cG_{k+1}$ in (\ref{goose})  and obtain
\be \label{heat}
\begin{split}
 C C_k C^T =  &  \sum_{\bd} (1 + d\cM ) \cQ_k  h_{\bd} \tilde \cG_{k+1, \bd} h_{\bd}  \cQ^T_k ( 1+ d \cM )^T \\
 + &   \sum_{\om: |\om| \geq 1}  (1 + d \cM ) \cQ_k   \tilde \cG_{k+1, \om} \cQ^T_k ( 1+ d \cM )^T \\
 \end{split}
\ee
In the first term move the $h_{\bd}$ to the outside and identify 
\be  \label{llama} 
 C C_{k, \bd} C^T =  (1 + d \cM ) \cQ_k  \tilde \cG_{k+1, \bd}  \cQ^T_k ( 1+ d \cM )^T  
\ee
 to write it as 
as
\be \label{soused}
 \sum_{\bd}h_{\bd} \B( C C_{k, \bd} C^T  \B) h_{\bd} +  \sum_{\bd}L_{4, \bd}
\ee
Here  $L_4(\bd)$ involves the commutators 
 $[\cQ_k, h_{\bd}] = \cO(M^{-1})$  and $[ d \cM , h_{\bd}]= \cO(M^{-1})$.
Using these and the exponential decay for $\tilde \cG_{k+1, \bd}$ one can show 
\be   \label{ells}
| L_{4, \bd}(\Up, \Up') | \leq C M^{-1} e^{-\ga d(\Up,\Up')} 
\ee
Now in the first term in (\ref{soused})  move the $h_{\bd}$ back inside and write it as
\be
   C  \B(\sum_{\bd}    h_{\bd}  C_{k, \bd}h_{\bd} \B)
 C^T +  \sum_{\bd}L_{5, \bd} \hs 
 \ee
The operator $L_{5, \bd}$ involves the commutator $[C, h_{\bd}] = \cO(M^{-1})$ (see  Appendix A in \cite{Dim15}) and  also satisfies the bound (\ref{ells}). 
Now (\ref{heat}) becomes
\be \label{heat2}
  C C_k C^T=  C \B( \sum_{\bd}    h_{\bd}  C_{k, \bd}h_{\bd} \B)
 C^T + \sum_{\bd}(L_{4, \bd}+ L_{5, \bd}) 
 +  \sum_{\om: |\om| \geq 1}  (1 + d \cM ) \cQ_k   \tilde \cG_{k+1, \om} \cQ^T_k ( 1+ d \cM )^T
\ee

The operator $C^T \De_{\bd} C$  has an exponentially decaying kernel also inherited from the bounds on $\cH_k$.     By (a  slight modification of)   Balaban's theorem   on unit lattice operators (section 5 in \cite{Bal83b})  the same is true for the inverse  and so  
\be \label{oliver3} 
|  C_{k, \bd} (\Up, \Up') |\leq    C  e^{ - \ga d(\Up, \Up'')  }
\ee
The second and third terms satisfy the same bound.    Hence  $C C_k C^T$ has a random walk expansion and satisfies an exponential decay bound.

Now we argue that the results for   $ C C_k C^T $ imply the same for $C_k$.   Recall that $C$ is a map  on $V = V_1  \oplus V_2$ where 
$V_1$ is functions on bonds in $\cup_y B(y)$ in $ \ker \tau$  and $V_2$ is functions on bonds in  $\cup_{y,y'} B(y,y') - b(y,y') $.
and it maps to $V' = V \oplus V_3$ where $V_3$ is  functions on  the central  bonds  $b(y,y')$.  As a map from $V$ to $V$ it is the identity.  The transpose $C^T$  goes from $V'$ to $V$,  but if we restrict it to functions on $V$ it is again  the identity.  (For $Z, \tilde Z \in V$ we have $<C^T Z, \tilde Z> = <Z, C \tilde Z> = <Z, \tilde Z>$.)
 Thus  $C C_k C^T $ on $V \times V$ is the same
 as $C_k$ on $V \times V$.   This gives the result for $C_k(\Up, \Up') = < \Up, C_k \Up'>$.
 (These remarks, which mirror similar observations in \cite{Bal85b} render the discussion of $C^{-1}$ in \cite{Dim15} unnecessary.)

  \bigskip
  
 The result for $C^{\frac12}$ is based on the representation 
 \be 
 C_k^{\frac12} = \frac{1}{\pi} \int_0^{\infty} \frac {dx}{\sqrt{x}}  ( C^T \De_k C + x ) ^{-1} dx
 \ee
 and representation like (\ref{goose}) for  $C ( C^T \De_k C + x ) ^{-1} C^T$. For details see \cite{Bal85b}, \cite{Dim14}.
 
\subsubsection{expansion for $C_k(Y)$}  

\label{loco}
We will also need an expansion for $C_k(Y) =[ C^T\De_k C]^{-1}_{Y} $ 
 where $Y$ is a union of $M$-cubes in $\tz$.  The notation means we  restrict to functions on bonds with at least one end in $Y$ before taking the inverse.

 This is a special case of a multi-scale situation which arises discussed in detail in
\cite{Bal84b}, \cite{Bal85b}.  We  will make extensive use of it in   \cite{Dim17}, here we just sketch the situation. 
There one has a decreasing sequence of regions $\bom =( \Om_1, \Om_2, \dots, \Om_k )$ each $\Om_j$  a union of $L^{-(k-j)}M$-cubes in $\tk$, and associated with it an
operator $\De_{k, \bom}$.   One is interested in 
\be 
 C_{k, \bom}(\Om_{k+1}) \equiv  [ C^T\De_{k, \bom} C ]_{\Om_{k+1}}^{-1}
\ee
where $\Om_{k+1} \subset \Om_k$ is a union of $LM$ or $M$ cubes. 
For this operator one has a representation like (\ref{goose}) which has the form with $\bom^+ = (\bom, \Om_{k+1}) $
\be  \label{omniverous}
C C_{k, \bom}(\Om_{k+1}) C^T= \B[(1 + d \cM ) \cQ_k \tilde \cG_{k+1, \bom^+} \cQ^T_k ( 1+ d \cM )^T \B]_{\Om_{k+1}}
\ee
where 
\be    \label{lorenzo9}
\tilde \cG_{k+1, \bom^+}  =   \cG^0_{k+1, \bom^+} -  \cG^0_{k+1, \bom^+}  \cQ_{k+1}^T   \B(\cQ_{k+1} \cG^0_{k+1, \bom^+} \cQ_{k+1}^T\B)^{-1}   \cQ_{k+1}  \cG^0_{k+1, \bom^+} 
\ee
and where  on $\tk$
\be   
\cG^0_{k+1, \bom^+}  = \B(   \de  d   +    d R^0_{k+1, \bom^+} \de   +  \cQ^T_{k+1, \bom^+} \ba \cQ_{k+1, \bom^+}   \B)^{-1}
\ee
Both   $\cG^0_{k+1, \bom^+} $ and  $ (\cQ_{k+1} \cG^0_{k+1, \bom^+} \cQ_{k+1}^T)^{-1} $ have (multiscale) random walk 
expansions and this generates expansions for $\tilde \cG_{k+1, \bom^+}$ and    $C_{k, \bom}(\Om_{k+1})$ as in lemmas \ref{ossie0}, \ref{ossie1}.

In the case at hand we have $\Om_j = \tk$ for $1 \leq j \leq k$  and $ \Om_{k+1} = Y $.   Then $C_k(Y) =[C^T \De_k C]_Y^{-1}$ and the identity
(\ref{omniverous})  reads
\be \label{kitten}
CC_{k}(Y)C^T= \B[ (1 + d \cM ) \cQ_k  \tilde \cG_{k+1, Y^c,Y} \cQ^T_k ( 1+ d \cM )^T\B]_Y
\ee
where $\tilde \cG_{k+1, Y^c,Y}$ is defined from  $  \cG^0_{k+1, Y^c,Y} $ as in (\ref{lorenzo9}) and the latter is an operator of the form
\be
  \cG^0_{k+1, Y^c,Y}  \equiv     \B(   \de  d   +    d R^0_{k+1, Y^c,Y} \de   +  a \B[\cQ^T_{k} \cQ_{k}\B]_{Y^c}   +  a\B[\cQ^T_{k+1} \cQ_{k+1}\B]_{Y}   \B)^{-1}
\ee
Also for  $ \bd \subset Y$ we define on the small torus $\tilde \bd^3$ the operator $C_{k, \bd}(Y) =[C^T \De_{k, \bd} C]_{Y\cap \tilde \bd^3}^{-1}$.
This  is bounded above and below and satisfies an identity like (\ref{kitten}). 
Then we have   the following variation of lemma \ref{ossie1}.

\begin{lem}    \cite{Bal84b},  \cite{Bal85b},  \cite{Bal88a}  
     \label{ossie2}  
The   operators  $ C_k(Y) $  have    generalized   random walk expansions 
 of    the form
 \be \label{licorice}  
  C_k(Y)  =  \sum_{\bd} h_{\bd}  C_{k, \bd}(Y) h_{\bd}  +   \sum_{\om: |\om| \geq 1} C_{k, \om}(Y)
\ee
The expansions yield the bounds: 
 \be  \label{succinct}
|  C_k(Y; \Up, \Up') | \leq    C  e^{ - \ga d(\Up, \Up')  }
\ee
 The the first term in (\ref{licorice})  is bounded above and below, and the second term in (\ref{licorice}) is $\cO(M^{-1})$. 
\end{lem}
\bigskip

 \subsubsection{resummed random walk}
\label{newly}

 The random walk expansions can be resummed so that any particular union of $M$-blocks can be treated as a unit
 \cite{Bal85b}, \cite{Bal88a}.
 This can be done for any of the random walks discussed so far,  but we discuss the details for $\cG_k$. 
 Recall that the  generalized random walk expansion can be written in the form  
\be \label{dingdong}
  \cG_k  = \sum_{\om}  \cG_{k, \om}
=  \sum_{X_0, \al_1, X_1, \dots, \al_n, X_n} R_0(X_0)  R_{\al_1}(X_1)\cdots R_{\al_n}(X_n)
\ee
 The  sum 
is restricted to sequences where $X_i \cap X_{i+1} \neq \emptyset $,  but we can regard it as an unrestricted
sum since the summand vanishes if the constraint is violated. 

 Let $Y$ be a  connected union of $M$ cubes. Define
\be
\begin{split}
\tilde R_0 (Y) = &   \sum_{X_0, \al_1, X_1, \dots, \al_n, X_n: X_i  \subset  Y  } R_0(X_0)  R_{\al_1}(X_1)\cdots R_{\al_n}(X_n)\\
\tilde R(Y) = &   \sum_{ \al_1, X_1, \dots, \al_n, X_n: X_i  \subset Y  }  R_{\al_1}(X_1)\cdots R_{\al_n}(X_n)  \\
 \end{split}
 \ee
These converge and satisfies the same bounds (\ref{oscar3})  as $R_{\al} (X)$. 
The proof is the same as the proof for the convergence of the overall series.

 For any union of $M$ cubes  $Y$ 
let $\{ Y_{\beta} \}$ be the connected components. 
  We resum  the expansion (\ref{dingdong}) grouping together parts of the walk which stay in the same  $Y_{\beta}$.  Given  a  general sequence
    $ (X_0, \al_1, X_1, \dots, \al_n, X_n)$ replace each 
chain  of  localization domains  staying in  some $Y_{\beta}$ by $Y_{\beta}$. We generate   a new sequence
$ (\cX_0, \cX_1, \dots, \cX_m)$ where  $\cX_0$ is either some  $Y_{\beta} $ or $(0, \bd)$ with $\bd \subset Y^c$,  and  $\cX_i$ for $ i\neq 0$ is either some $Y_{\beta}$ or  a pair $(\al,X)$ with  $\al \neq 0$ and $X \cap Y^c \neq \emptyset$.
   We associate with $\cX_0$ an operator
   \be
   R'_0(\cX_0) =
    \begin{cases}
     \tilde R_0(Y_{\beta}) & \hs \cX_0 = Y_{\beta}\\   R_0(\bd) & \hs \cX_0 = (0, \bd)\\
   \end{cases}
   \ee
   and with $\chi_i$
     \be
   R'(\cX_i) =
    \begin{cases}
     \tilde R(Y_{\beta}) & \hs \cX_i = Y_{\beta}\\   R_{\al}(X) & \hs \cX_i = (\al , X)\\
   \end{cases}
   \ee
These are localized in the associated region and satisfy
 \be    \label{oscar5}
 \begin{split}  
   |1_{\De_y}  R'_0(\cX_0) 1_{\De_{y'}} f|     \leq   &   \    C    e^{  -\ga  d(y,y') } \|f\|_{\infty}      \\
    |1_{\De_y}  R'(\cX_i) 1_{\De_{y'}} f|     \leq   &   \    C M^{-1}   e^{  -\ga  d(y,y') } \|f\|_{\infty}      \\
 \end{split}
\ee
together with the bounds on derivatives as in (\ref{oscar3}).

Classify the terms in the original sum by the sequence  $ (\cX_1, \dots, \cX_m)$ that they generate.  This gives an new expansion
\be \label{dingdong2}
\begin{split}
  \cG_k  = &   \sum_{(\cX_0, \cX_1 \dots \cX_m)} \
 \sum_{(X_0, \al_1, X_1, \dots, \al_n, X_n)  \to (\cX_0, \cX_1 \dots,   \cX_m)} R_0(X_0) R(X_2)\cdots R(X_n) \\
 = &   \sum_{(\cX_0,\cX_1 \dots \cX_m)}  \B( \sum_{(X_0, \al_1, X_1, \dots, \al_n,  X_n)  \to \cX_0} R_0(X_0) R(X_1) \cdots R(X_n)  \B)
 \\
 & \hs \hs \hs
 \prod_{i=1}^m \B(  \sum_{(\al_1, X_1, \dots, \al_m,  X_n)  \to \cX_i}  R_{\al_1}(X_1) \cdots R_{\al_m}(X_n)\B) \\
 = &  \sum_{(\cX_0,\cX_1 \dots \cX_m)}  R'_0(\cX_0) R'(\cX_1) \cdots R'(\cX_m) \\
\end{split}
\ee
We   must have $\cX_i \cap \cX_{i+1} \neq \emptyset$ for a non-zero contribution.  This is our new random walk expansion.
\bigskip

\section{Polymers}

 \subsection{dressed Grassman variables }

As we  track the flow of the RG transformations  the densities will be  expressed in terms  
 of localized  elements   of the Grassman algebra  depending on the fundamental variables 
$\Psi_k$  on $\tz$     through the    fields    $\psi_k(\cA)=  \cH_k(\cA) \Psi_k$  on $ \tk$.         We  assume  $\cA$ is in the domain (\ref{listless})   so we have good estimates  on $\cH_k(\cA)$.

Let     $\xi  =  (x, \beta, \om )   $  with    $x \in \tk$,  $1 \leq  \beta \leq 4$, and  $\om =0,1$.   
We  treat    $\psi, \bpsi$  as a  single field by
\be
\psi_k(\cA,  \xi)    = \begin{cases}  
 \psi_{k,\beta}(\cA, x)     &    \xi =    (x, \beta, 0)  \\
 \bpsi_{k,\beta}(\cA,  x)     &    \xi =    (x, \beta, 1) \\ 
 \end{cases} 
\ee
We also define fields depending on two variables $x,y \in  \tk$ .  Let  $\zeta   =   (x,y, \beta,  \om )$ and define
\be  \chi_k(\cA,   \zeta) = \B( \de_{\al, \cA} \psi_k(\cA)\B)( \zeta)     = \begin{cases}  
|x-y|^{- \al} \B(e^{ie_k \eta ( \tau\cA)(x,y) } \psi_{k,\beta}(y)   -  \psi_{k, \beta}(x) \B)  &  \zeta =   (x,y, \beta,   0)  \\
|x-y|^{- \al} \B(e^{-ie_k \eta ( \tau\cA)(x,y) } \bpsi_{k,\beta}(y)   -  \bpsi_{k, \beta}(x) \B)  &  \zeta =  (x,y,\beta,   1)  \\
\end{cases}
\ee
We  consider  elements  of the Grassman algebra generated by $\Psi_k$    of the form  
  \be  
 \begin{split}  
 E\B(\cA,  \psi_k(\cA), \chi_k(\cA)  \B)
   =  &  \sum_{n,m=0}^{\infty}  \frac{1}{n!m!}  \int  \   E_{nm}(\cA,  \xi_1,  \dots,  \xi_n; \zeta_1, \dots,  \zeta_m)    \\
    &  \psi_k(\cA, \xi_1 )  \cdots  \psi_k(\cA, \xi_n)  \chi_k(\cA, \zeta_1 )  \cdots  \chi_k(\cA, \zeta_n) 
     d\xi_1 \cdots  d\xi_n    d\zeta_1 \cdots  d\zeta_m 
       \\
   \end{split}
\ee
Here  with $\eta = L^{-k}$
\be    \int d \xi   =  \sum_{x,\beta, \om}  \eta^3  \hs   \int  d \zeta       =    \sum_{x,y,  \beta, \om}  \eta^6    
\ee
The  kernel is the collection of functions   $  \{    E_{nm}(\cA)  \} $. The
 $ E_{nm}(\cA,  \xi_1,  \dots,  \xi_n; \zeta_1, \dots,  \zeta_m)  $  are taken to be  anti-symmetric the $\xi_i$ and the $\zeta_j$ separately.  
 Note that the $\psi_k(\cA), \chi_k(\cA) $ are not independent and   different kernels may give the same  algebra element.

We  define a norm on the $E(\cA)$    by  
    \be  \label{thing1}
  \| E_{nm}(\cA)  \|    =      \int    | E_{nm}(\cA,   \xi_1,  \dots,  \xi_n, \zeta_1, \dots,  \zeta_m)  |     d\xi_1 \cdots  d\xi_n    d\zeta_1 \cdots  d\zeta_m 
\ee
and       
for      a pair of   positive   real numbers     $\bh=    (h_1, h_2) $          
\be     \label{thing2}    \| E(\cA) \|_{  \bh }  
      =         \sum_{n,m=0}^{\infty}  \frac{ h_1^n  h_2^m }{n!m!}     \|  E_{nm}(\cA)  \|        
\ee
This is not a norm on the algebra element but rather on the representation,  i.e. on the kernels.

  The norm has the property that if   $G(\cA)   = E(\cA)   F(\cA)   $ then  the kernels  satisfy 
 \be  
      \| G(\cA )  \|_{  \bh }    \leq \|   E(\cA)   \|_{  \bh }   \|  F(\cA)      \|_{   \bh }  
 \ee
This is a special case  of   (\ref{enron})  from   Appendix \ref{A3}.    In the terminology there the two measure spaces  are
$(T_1, \nu_1)$  with  $T_1$  equal to all  $\xi$  and  $  \nu_1(\xi)  =  \eta^3 $ and    $(T_2, \nu_2)$ with $T_2$   equal to all $\zeta$ 
and  $ \nu_2( \zeta)  =  \eta^6$.

We  also      want to bound   the true norm     by the dressed  norm.   This again   is  a special case of a result  
from   in     Appendix  \ref{A3}.    In that terminology       
our   fields are  
\be 
\begin{split}    \psi_1(\xi)  =&   (\cH_1 \Psi)(\xi)  =   (\cH_k(\cA) \Psi)(x, \beta, \om )    \\ 
  \psi_2 (\zeta)  =   &    (\cH_2\Psi) (\zeta)     =  |x-y|^{-\al} \B(    e^{(-1)^{\om} ie_k \eta  (\tau \cA) (x,y) }( \cH_{k}(\cA) \Psi )(y,\beta, \om)     - (\cH_k(\cA)    \Psi)(x, \beta, \om)  \B)  \\
  \end{split}
\ee
By (\ref{slavic2})
\be 
\begin{split}
   \|   \cH_1 \|_{1, \infty}  =   &    \|   \cH_k(\cA)  \|_{1, \infty}  \leq C   \\
    \|  \cH_2 \|_{1, \infty}  \leq      &   \one  \|   \cH_k(\cA)  \|_{1, \infty}  +     \|\de_{\al}     \cH_k(\cA)  \|_{1, \infty}  \leq   C   \\
\end{split}
\ee  
Here for the second estimate we consider separately the two cases  $|x-y |  \leq  1 $  and  $|x-y |  \geq  1 $.
In the case   $|x-y |  \leq  1 $ we  identify the norm   $ \|\de_{\al}     \cH_k(\cA)  \|_{1, \infty}$.  In the case  
  $|x-y |  \geq  1 $  we    bound the parallel translation by  $\one$ 
   and then  estimate each term by    $ \|    \cH_k(\cA)  \|_{1, \infty}$.
It  follows  by   lemma \ref{lunge}     in  Appendix \ref{A3}   that   if $E'(\cA, \Psi)  =   E\B(\cA,  \psi_k(\cA), \chi_k(\cA)  \B)$,
 then the  norm   $\| E'(\cA) \|_h$  as defined in   (\ref{truenorm}) satisfies
\be
  \| E'(\cA) \|_{h }     \leq    \|   E(\cA) \|_{ Ch,  C h }  
   \ee

\subsection{a domain for  the gauge field} 

Before proceeding we  make some further  restrictions on the gauge field.
 Let  $\ep>0$ be a fixed small number.   A domain  our fields is defined by  the following (notation slightly different from  \cite{Dim15} )
 
 \begin{defn}  
  $\cR_k$   is all    complex-valued fields  $\cA$ on $\tk$ satisfying  
 \be  \label{ding}
  | \cA|     <   e_k^{-\frac34 + \ep}  \hs
   |\pa \cA|  <   e_k^{-\frac34 + 2\ep}    \hs 
   |\de_{\al} \pa \cA|      <   e_k^{-\frac34 +3 \ep}   
   \ee 
 \end{defn}

 We  also  define an extended domain  with the property that  the fields are locally gauge equivalent  a field in $\cR_k$.  
 Choose a constant  $c_0  = \one$   and let   $\square^{\dagger}  =\tilde    \square^{(c_0L)}$  be the enlarged union of  $M$-cubes   with 
 $c_0L$  cubes on a side.   ($\square^{\dagger}$ was called  $\square^{\nat} $ in \cite{Dim15})

  \begin{defn}   
$ \tilde   \cR_k$  is all fields     $\cA  = \cA_0  + \cA_1$ on $\tk$
  where  $\cA_0$ is  real and   on      each $\square^{\dagger} $  is    gauge equivalent  to a field   in $\cR_k$  and    $\cA_1$ is complex and in $\cR_k$.   
\end{defn}

Assuming  $c_0L \geq  5$    these  conditions  are stronger   than the conditions (\ref{listless})   which we  needed  for   the treatment of the   $S_k(\cA), \cH_k(\cA)$.   Indeed if $\cA \in \tilde \cR_k$ then $e_k^{-\frac14}\cA$ satisfies these conditions.     Also  in  \cite{Dim15}  it is established 
that in each $\square^{\dagger}$ the real   field  $\cA_k = \cH_k A_k$   is gauge equivalent to a field $\cA$   satisfying   
 \be      \label{localest}
     | \cA|, \    |\pa \cA|, \        |\de_{\al} \pa \cA|   \leq  CM \| d \cA_k \|_{\infty}     
\ee
Thus we only need control  over the field strength  $  d \cA_k $  to conclude that  $\cA_k \in \tilde \cR_k$

\subsection{polymer functions}  \label{polymersection}

Next we  localize.  
A   \textit{polymer}   $X$  in  $\tk$  is defined  to be a   connected union of $M$ cubes,  with the convention that two  cubes are connected   if    they have an entire face in common.  The set of all polymers   is denoted  $\cD_k$. 
A \textit{polymer function} depending on a gauge field $\cA$  in   $\tilde \cR_k$ and a polymer $X$     is an element of the Grassman algebra of the form   
  \be  
 \begin{split}  
 E(X,  \cA,  \psi_k(\cA), \chi_k(\cA)  )  =  &  \sum_{n,m=0}^{\infty}  \frac{1}{n!m!}  \int     E_{nm}(X, \cA,  \xi_1,  \dots,  \xi_n, \zeta_1, \dots,  \zeta_m)    \\
    &   \psi_k(\cA, \xi_1 )  \cdots  \psi_k(\cA, \xi_n) \ \chi_k(\cA, \zeta_1 )  \cdots  \chi_k (\cA, \zeta_m) 
\     d\xi_1 \cdots  d\xi_n    d\zeta_1 \cdots  d\zeta_m  
       \\
   \end{split}
\ee
We require that only terms with equal numbers of  $\Psi, \bPsi$  contribute.

The  kernels    $  E_{nm}(X, \cA,  \xi_1,  \dots,  \xi_n, \zeta_1, \dots,  \zeta_m) $  are required to vanish   unless all $\xi_i   \in X$
 and  all $\zeta_i \cap X \neq \emptyset$, and  to   depend  on  $\cA(b)$ only if $b \cap X \neq \emptyset$.   
We  also     require that   $  E(X, \cA    ) $    is bounded and analytic  on the domain   $\cA  \in \tilde  \cR_k$.
Norms   $\| E_{nm}(X,\cA  ) \| $   and  $\|  E (X, \cA)  \|_{\bh  }$     are defined    as  in (\ref{thing1}), (\ref{thing2})   and now we
define
\be    \| E_{nm}(X  ) \|_{\tilde \cR_k}    = \sup_{\cA \in \tilde  \cR_k}  \| E_{nm}(X,\cA  ) \| \hs 
   \|  E (X)  \|_{\bh,\tilde  \cR_k  }    =  \sup_{\cA \in \tilde  \cR_k} \|  E (X, \cA)  \|_{\bh  }
  \ee
We  have $ \|  E (X)  \|_{\bh,\tilde  \cR_k  }  \leq    \|  E (X)  \|_{\tilde  \cR_k, \bh  } $ where the latter is defined by  
\be   
   \|  E(X) \|_{\tilde \cR_k,   \bh }  
      =         \sum_{n,m=0}^{\infty}  \frac{ h_1^n  h_2^m }{n!m!}     \|  E_{nm}(X)  \|_{\tilde \cR_k}        
\ee

The polymer functions are required to have tree decay in the polymer $X$.    Size    is measured on the $M$-scale and    we  define    $d_M(X)$
by  
\begin{equation}   
  M d_M(X)  =  \textrm{length of the shortest  continuum    tree  joining   the $M$-cubes    in     $X$}  .
\end{equation} 
If  $|X|_M$ is  the
number of $M$ cubes in  $X$,    then  
\be     d_M(X)  \leq   |X|_M  \leq   \one  ( d_M(X)  + 1 )
\ee  
Also  there are constants  $\ka_0, K_0 = \one$ such that  for any  $M$-cube  $\square$ 
\be   \label{lemon} 
  \sum_{X \in \cD_k, X  \supset \square}    e^{- \ka_0 d_M(X)}  \leq  K_0  
  \ee
We assume  $\ka = \one$ and   $\ka \geq  \ka_0$.      We  define an  associated norm
\be  \| E\|_{\tilde \cR_k,  \bh,\ka} =   \sup_{X}   \| E(X  ) \|_{\tilde \cR_k,  \bh} e^{\ka  d_M(X)  }    
\ee

It is useful to let  $\bh$ depend on the running coupling constant  $e_k$.  Pick a fixed 
$\ep>0$ small.   Then   define $h_k =e_k^{-\frac14}$ and
 \be 
    \bh_k  =(h_{k,1}, h_{k,2}) =    \B(  h_k,  e_k^{\ep}h_k \B) = \B( e_k^{-\frac14}, e_k^{-\frac14 + \ep} \B)
     \ee
 The basic norm after  $k$ steps is then  
 \be 
     \| E\|_k   \equiv      \| E\|_{\tilde   \cR_k,  \bh_k,\ka}
    \ee
 The space of all  polymer functions with this norm    is a Banach space called  $\cK_k$.

\subsection{scaling}  
\label{scaling0}

At this point  drop  the reference to the specific fields  $\psi_k(\cA, \xi), \chi_k(\cA, \zeta)$  and  consider  general  Grassman variables 
$\psi(\xi) , \chi(\zeta) $  on $\tk $ of the same type,  but   do not assume any relation between them.    
Let  $ E(X,\cA,  \psi, \chi  )$  be  a polymer function  of these variables as above.

We  will want   to scale the polymer function.   Since $M$-cubes do not scale to $M$-cubes we first need
a blocking operation.   If  $X    \in  \cD_k$ let   $\bar X^L$  be the smallest union of $LM$ blocks  containing  $X$.  Then 
if   $Z$ is  a connected union of $LM$ blocks  we define \be
   ( \cB  E) (Z, \psi ,\chi  )   =   \sum_{X:  \bar X^L  = Z }   E(X, \psi , \chi    )
\ee
We   define  a scaled polymer function 
$(\cB E)_{L^{-1}}$  on    $\bbT^{-k-1}_{N-k-1}$  as follows.   For  $Y, \cA, \psi, \chi$ on       $\bbT^{-k-1}_{N-k-1}$ 
\be
(\cB E)_{L^{-1}}(Y,  \cA,  \psi,  \chi  )   \equiv    
(\cB E)(LY,   \cA_L,   \psi_L,  \chi_L   )    
=   \sum_{X:  \bar X^L  = LY  }  E(X, \cA_L, \psi_L , \chi_L  )
\ee
Then we  have  
\be   
 \sum_{X  \in \cD_k}   E_k(X, \cA_L, \psi_L , \chi_L)   =  \sum_{Y \in \cD_{k+1} }   (\cB E)_{L^{-1}}(Y,  \cA,  \psi,  \chi  ) 
\ee
The scaled fields  on $\tk$  are   
\be       
\begin{split}
   \cA_L(b)  = &  L^{-\frac12} \cA(L^{-1} b)  \\
   \psi_L(\xi) = &   \psi_L(x, \beta, \om)  =  L^{-1}  \psi(L^{-1} x, \beta, \om)  \\
   \chi_L(\xi) = &   \chi_L(x,y,  \beta, \om)  =  L^{-1-\al}  \chi(L^{-1} x,L^{-1}y, \beta, \om)  \\
 \end{split}   
\ee
If we define    $    E(X,   \cA_L,   \psi_L,  \chi_L   ) =   (S_L E)(X, \cA, \psi, \chi) $, then     
the kernel of  $S_L E$  is   
\be   
\begin{split}  &  (\cS_L  E) _{nm} (X, \cA,\xi_1,  \dots,  \xi_n, \zeta_1, \dots,  \zeta_m)   \\
=   &  L^{3n + 6m}  L^{-n}L^{-(1 + \al)m}    E_{nm} (X, \cA_L,L\xi_1,  \dots, L \xi_n, L\zeta_1, \dots, L \zeta_m)   \\
\end{split}
\ee
and it has the norm       
\be  \label{spoon} 
\begin{split}
 &  \| (\cS_L  E) _{nm} (X, \cA)  \|   \\ 
=   &    \int     L^{3n + 6m}  L^{-n}L^{-(1 + \al)m}  | E_{nm} (X, \cA_L,L\xi_1,  \dots, L \xi_n, L\zeta_1, \dots, L \zeta_m)|    d\xi_1 \cdots  d \xi_n \  d\zeta_1 \cdots  d \zeta_m   \\
   =   &    \int   L^{-n}L^{-(1 + \al)m}    | E_{nm} (X, \cA_L,\xi'_1,  \dots,  \xi'_n, \zeta'_1, \dots,  \zeta'_m)|   d\xi'_1 \cdots  d \xi'_n \  d\zeta'_1 \cdots  d \zeta'_m   \\
=     &  L^{-n} L^{-(1+ \al) m}   \|  E _{nm} (X, \cA_L)  \|  \\
\end{split} 
\ee

If       $\cA  \in \cR_{k+1} $  then since $e_{k+1} = L^{\frac12} e_k$
 \be   \label{ding3} 
 \begin{split}  
    | \cA_L|  <   &        L^{-\frac12 }    e_{k+1}^{-\frac34 +   \ep}     <    L^{-\frac78+ \frac12 \ep}  e_k^{-\frac34 + \ep}         \\
      |  \pa    \cA_L |        <   &  L^{- \frac32  }    e_{k+1}^{-\frac34 +2 \ep}      <   L^{-\frac{15}{8}+ \ep}    e_k^{-\frac34 +2 \ep}  \\
        |\de_{\al}  \pa    \cA_L |        <   &  L^{- \frac32 - \al }    e_{k+1}^{-\frac34 +3 \ep}      <   L^{-\frac{15}{8} - \al  + \frac32 \ep}    e_k^{-\frac34 +3 \ep}  \\
\end{split} 
\ee
The $L$ factors are all less than $L^{-\frac34} $ so    $\cA_L  \in  L^{-\frac34}\cR_k $.   It  follows also that    $\cA  \in  \tilde   \cR_{k+1} $  implies 
  $\cA_L  \in L^{-\frac34} \tilde  \cR_k$.
Thus we have
\be   \label{ding4}
      \| (\cS_L  E) _{nm} (X)  \|_{\tilde  \cR_{k+1}}   \leq       L^{-n} L^{-(1+ \al) m}   \|  E _{nm} (X)  \|_{ L^{-\frac34}  \tilde   \cR_k  }
\ee
For  the moment we throw away  the  contracting factors $  L^{-n} L^{-(1+ \al) m}$, 
 enlarge $L^{-\frac34 }\tilde   \cR_k$ to  $\tilde  \cR_k$,
  and take $\bh_{k+1}  < \bh_k$
to get
\be   \label{L3}  
  \|  (\cS_L    E )(X)    \|_{\tilde  \cR_{k+1}, \bh_{k+1}}         \leq      \|     E(X) \|_{ \tilde   \cR_k,    \bh_k } 
  \leq          \| E  \|_k      e^{- \ka   d_M(X)  }  
\ee
and so   
\be 
  \| (\cB    E)_{L^{-1}}(Y)  \|_{\tilde  \cR_{k+1},   \bh_{k+1} }    \leq    
       \| E  \|_k    \sum_{X:   \bar X  = LY}    e^{- \ka   d_M(X)  }    
 \ee
If   $\bar X  =  LY$ then  $  L  d_M(Y)  \leq     d_M(X)$  so 
we  can extract a factor   $e^{-  L(\ka - \ka_0)d_M(Y) }$ leaving    $e^{- \ka_0d_M(X) }$.  
Then using    (\ref{lemon}) the sum over $X$ is  bounded by   $\one   |LY|_M \leq   \one L^3 e^{d_M(Y)}$.
Therefore    
\be   \label{xmas}
 \sum_{X:   \bar X  = LY}    e^{- \ka   d_M(X)  }    \leq  \one L^3 e^{-L(\ka - \ka_0 -1)d_M(Y)}  \leq  \one L^3 e^{- \ka d_M(Y) }
\ee
where the second inequality holds for $L$ sufficiently large.      
Thus we get
the crude bound  
\be   \label{L4}  
  \| (\cB   E)_{L^{-1}}  \|_{k+1}  \leq  \one L^3     \|     E \|_{k}
\ee

\subsection{symmetries}  \label{symmetries}
We  assume the    polymer functions    $E(X,\cA,  \psi, \chi  )$  have the following symmetries:
\begin{itemize} 
\item    Invariance  under   $\tz$     lattice  symmetries
  \be   E(rX +a,  \cA_{a,r},  \psi_{a,r}, \chi_{a,r})  =  E(X, \cA,  \psi, \chi )  \ee
\item
 Gauge invariance
\be
  E(X, \cA^{\la},   \psi^{\la},  \chi^{\la} )  =  E(X,\cA,  \psi,   \chi )
   \ee
\item 
  Charge conjugation invariance  
  \be
    E(X,-\cA,  \cC  \psi  ,  \cC  \chi    )  =  E(X,\cA,   \psi, \chi  ) 
     \ee
\item   Complex conjugation.   If  $\cA$ is real  the kernels satisfy      (now distinguishing  $\psi, \bpsi$ and $\chi, \bar \chi$) 
\be 
\label{bungee} 
 \ker    E\B(X,-\cA,  \ga_3 C  \psi ,  [(\ga_3 C)^{-1}]^T   \bpsi,      \ga_3  C  \chi,[(\ga_3 C)^{-1}]^T   \bar  \chi      \B)
  = \overline{     \ker  E\B(X,\cA, \psi,     \bpsi , \chi,  \bar \chi    \B)  }  
  \ee
 \end{itemize}

For the first three we also require  the   corresponding  transformation  properties for the  the kernels.     In particular  the gauge invariance says that
$  E_{nm}(X, \cA- \pa \la )$  and $  E_{nm}(X, \cA) $ differ by a phase factor.
It follows that  the the norm is gauge invariant:
\be   \label{needit} 
   \|  E_{nm}(X, \cA- \pa \la )\|    =     \|  E_{nm}(X, \cA )\|  
\ee

Here are some consequences for   the piece  $ E_{00}(X, \cA) $ with no fermion fields.     The  $p^{th}$   derivative    in  $\cA$     is  
 the multilinear functional 
\be  
\frac{ \de^p  E_{00}}{ \de \cA^p } \Big (X, \cA ;  f_1,  \dots   ,f_p \Big)
 =   \frac{  \pa^p} { \pa t_1 \dots \pa t_p}  \Big[     E_{00}(X, \cA +   t_1  f_1    +  \dots  + t_p f_p,  0  ) \Big]_{t=0}
 \ee
If    one of the functions has the form   $f_i =  \pa  \la$,  then  by gauge invariance there is no dependence  on   $t_i$
and  the derivative vanishes.   Thus we have the Ward identity
\be   \label{Ward} 
 \frac{ \de^p  E_{00}}{ \de \cA^p } \Big (X, \cA ;  f_1,  \dots,  \pa \la,  \dots    ,f_p \Big)
 = 0
 \ee
Charge    conjugation invariance   gives $ E_{00}(X,-\cA)  =   E_{00}(X,\cA)$  and this  implies   
\be \label{suzy2}
  \frac{  \de^p    E_{00}}{ \de \cA^p }(X,0) =   0  \hs    \textrm{  if  $p$ is odd  }
  \ee

\subsection{normalization}

As we iterate the RG transformations    the scaling operation    can increase the size of   the  polymer functions   by as much as $\cO(L^3)$  as is
evident from (\ref{L4}).    We   have  to watch this carefully  and start by introducing a criterion  to avoid the growth.   The following is similar to the analysis in  \cite{BDH98}, \cite{Dim11},  \cite{Dim15}.

\begin{defn}
A polymer function     $E(X,\cA, \psi_k(\cA),  \chi_k(\cA ) )$ with  kernels  $ E_{nm}(X,\cA )$  satisfying   the stated   symmetries   is said to be {\em normalized}   if in  addition to the   
vanishing  derivatives   (\ref{Ward}), (\ref{suzy2})       we   have 
\begin{equation}  \label{normalization}
 E_{00}(X,0)  = 0    \hs 
\int       E_{20}  \B(X, 0;  (x, \al,1), (y, \beta,0) \B) \   dx dy = 0
 \end{equation}
\end{defn}
\bigskip

We generally only require normalization small polymers. 

\begin{defn}
A  polymer  $X$  is  {\em small} 
if  $d_M(X)  \leq   L$ and   {\em large}  if   $d_M(X) >  L$.  The set of all small polymers   in denoted $\cS$.
\end{defn}
\bigskip

\subsubsection{extraction}

Normalization is  achieved by extracting certain  relevant  terms from     the polymer function.
Given  $E (X, \cA, \psi, \chi)$  with kernels    $ E_{nm}(X, \cA)$ on $\tk$  satisfying lattice, gauge, and charge conjugation   symmetries    we  define    $(\cR E)(X,\cA, \psi, \chi) $   as  follows. 
If  $X$ is large then  $(\cR E)(X,\cA)= E(X,\cA)$.    If   $X$ is small  ($X  \in \cS$) then  $(\cR E)(X, \cA)$  is 
defined by 
\begin{equation}  \label{renorm}
 E( X, \cA,  \psi, \chi)  
=   \al_0( E,X)  \Vol( X )   +    \int_X \bpsi\  [ \al_2( E,X)]\ \psi          +   (\cR E)( X,\cA,  \psi, \chi)    
\end{equation} 
where    
\begin{equation} 
\begin{split}
\al_0( E, X)   =& \frac{1}{ \Vol (X)}   E_{00}(X,0 )  \hs  
[ \al_{2}( E,X)]_{\al  \beta} =     \frac{1}{\Vol(X)}   
\int   E_{20}  \B( X,0;   (x, \al, 1), (y, \beta, 0)       \B)   dx  dy  \\
  \end{split}
\end{equation}

\bigskip

\begin{lem}  
\label{amble}  $\cR  E$  is  invariant under lattice, gauge, and charge symmetries.   $\cR  E$  is normalized for small polymers  and 
satisfies
\begin{equation}     \label{sunup}    
    \|    \cR   E \|_{k}    \leq  \cO(1)    \|    E   \|_{k}   
\end{equation}
\end{lem}
\bigskip

\pr 
The  invariance follows since everything else in  (\ref{renorm})  is invariant.
The derivatives  (\ref{normalization}) match on the left and right  except  for  the  term   $\cR  E$,   hence its derivatives vanish.
The  bound  holds  since everything else in  (\ref{renorm}) satisfies the bound.   We omit the details.

\bigskip

 For   global quantities we   only have to remove  energy and mass terms.

\begin{cor}   
 \begin{equation}  \label{renorm2}
\sum_X E(X) =     -   \vep( E)  \Vol(  \bbT_{ \sN-k} ) -   m( E)  \int \bpsi \psi  
+    \sum_X  \cR E(X)
\end{equation} 
where
\begin{equation}  \label{sundown0}
\begin{split}
\vep( E)  =  &    -   \sum_{X \supset \square,  X \in \cS}   \al_0( E,X) \\
  m ( E) \de_{\al \beta} =    [  m( E) ]_{\al \beta}=        & - \sum_{X \supset \square,  X \in \cS}   [ \al_2( E,X)  ]_{\al \beta}  \\
\end{split}
\end{equation}
are  real and satisfy  
\be  \label{sundown}
\begin{split}
|\vep( E)|  \leq   &   \one \|  E \|_{k}     \\   
| m( E)  |    \leq  &   \one  h_{k}^{-2}   \|  E \|_{k}   =   \one     e_k^{ \frac12 }   \|  E \|_{k}  \\
 \end{split} 
\ee
\end{cor}
\bigskip

\pr      The constant term in $\sum_XE(X)$ is
$ \sum_{X \in \cS}     \al_0( E,X)  \Vol( X )$.  Insert  $\Vol(X)  =  \sum_{ \square \subset  X }  \Vol(\square)$
and change the order of the sums to write this as 
 \be  -  \sum_{\square}  \vep( E, \square) \Vol(\square)    \hs   \textrm{ where }  \hs    \vep ( E, \square)  =  -  \sum_{X \supset \square,  X \in \cS }  \al_0( E,X)
 \ee  
But   $\vep ( E, \square)$ is independent  of  $\square$ and is denoted  $\vep( E)$ to give the first term in (\ref{renorm2}). The bounds on  $\vep( E)$ 
  follows    directly.
 
The  mass term  in $\sum_XE(X)$  is   $  \sum_X \int_X \bpsi\  [ \al_2( E,X)]\ \psi  $.      Write   $\int_X   = \sum_{\square \subset X}  \int_{\square} $  
and change the order of the sums to write this as 
\be 
   \label{oink1}  
  - \sum_{\square}     \int_{\square}   \bpsi\    m(   E, \square  )  \psi  
\hs  \textrm{ where } \hs   m( E, \square )  = - \sum_{X  \supset  \square, X \in \cS}   \al_2( E,X)
\ee
But   $ m( E, \square ) $ is independent of  $\square$    and is denoted    $m( E)$ and  
 the expression becomes    $ \int    \bpsi\    m( E    )  \psi  $.
 
Next we  explain   why the matrix  $m( E)$  is a multiple of the identity.   Our assumption that  $ E$ is  invariant under lattice symmetries implies 
if   $S$ is a spinor representation of a rotation or reflection  $r$ then    $S^{-1} \al( E, rX) S  =  \al( E,  X)$.   Specialize to  
$r$ leaving the center of  $\square$ invariant and sum over  $X  \in \cS,  X \supset \square$  to get  
$S^{-1} m( E) S  =  m( E)$. Take   $S = \ga_{\mu}$ for  $0 \leq \mu \leq  3$ which induce  reflections.  We   conclude 
that    $[\ga_{\mu}, m( E)]=0  $  and hence that  $m( E)$ is a multiple of the identity.               

We  also  need to show that  $m( E)$ is real.  This follows   since   $ \overline{  E_{2,0}(X,0)  }
=  ( \ga_3 C  )^{-1}  E_{2,0}(X,0) \ga_3C$  by (\ref{bungee})  and hence   $\overline{  m( E)}  =  ( \ga_3 C  )^{-1} m( E)\ga_3C 
=  m( E)   $.  Similarly   $\vep( E)$ is real.  

The bounds on $\vep(  E), m( E)$ follow from 
\be 
\begin{split}
 \Vol(X) |\al_0( E, X)| \leq &  e^{-\ka d_M(X)} \| E\|_k \\
\Vol(X)  |[\al_2( E, X)]_{\al, \beta} | \leq &   \|  E_k(X,0)\|
\leq   h_k^{-2}\|  E_k(X,0)\|_{\bh_k}
\leq  h_k^{-2} e^{-\ka d_M(X)} \| E\|_k \\
\end{split}
\ee
and $\sum_{X \supset \square} e^{-\ka d_M(X)} \leq \one$
\bigskip

\rem   For the subsequent paper  we  generalize this
construction to the case where instead of  $\sum_X E(X)$ we have a restricted sum
$\sum_{X \subset  \La}   E(X)$   where $\La$ is   a union of $M$ blocks.
In this  case  $\vep( E, \square)$ and $ m( E, \square )  $  are replaced by    
\be  \label{oink2}  
 \begin{split} 
 \vep_{\La}( E, \square)  =  & -    \sum_{X \in \cS  :   \square \subset X \subset \La  }  \al_0( E,X)  \\
 m_{\La} ( E, \square )  =  &  - \sum_{X \in \cS  :   \square \subset X \subset \La  }   \al_2( E,X)  \\
 \end{split}  
\ee 
which vanishes unless  $\square  \subset  \La $.         
Now  we have  
 \be  \label{renorm3}
\sum_{X  \subset \La}   E(X) =     -   \vep( E)  \Vol( \La ) -   m( E)  \int_{\La }  \bpsi \psi  
+    \sum_{X  \subset \La}    \cR E(X)     +    B_{\La }  
\ee
where the extra term is   
\be 
 B_{\La}     =   - \sum_{\square \subset  \La}  (  \vep_{\La}( E, \square) - \vep ( E)  )  \Vol(\square )
-     \sum_{\square \subset  \La}  \int_{\square}   \bpsi   \B( m_{\La}( E, \square) -   m ( E)  \B) \psi  
\ee
Inserting  the definitions    (\ref{oink2})  only polymers $X$  which intersect both  $\La$ and $\La^c$ contribute,  denoted  
$X \# \La$,   and   this can be rearranged to   
\be  
B_{\La}    =   \sum_{X \in \cS,    X \# \La}    B_{\La }  (X)  
\ee
where  
\be
 B_{\La }  (X)    =  -  \al_0( E,  X) \    \Vol ( \La \cap X )  -  \int_{X \cap \La}  \bpsi  \   \al_2( E,  X) \  \psi    
\ee      
This   correction term is   localized around the boundary of  $\La$ and we have
\be   
  \| B_{\La }   \|_k   \leq   \one   \|    E  \|_k 
\ee
\bigskip

\subsubsection{adjustment}

The next  result shows that if   $E(X,\cA, \psi_k(\cA),  \chi_k(\cA ) )$  is normalized for small $X$ then  we can
adjust the kernels so  that both  $ E_{00}(X,0)=0$    (there is no energy term) 
and      $ E_{20} (X, 0)  =0$  (there is no mass term). 

\begin{lem}  
\label{adj}
If  $E(X,\cA_k,  \psi_k(\cA)   )$  with kernels  $ E_{nm}(X, \cA)$  is normalized for  $X \in \cS$  then  it can be written with  new  kernels  $ E^{\nat}_{nm}(X, \cA)$  modified for  $X \in \cS$ and  $n+m =2$  such that $ E^{\nat}_{20} (X, 0)  =0$.  Furthermore
\begin{enumerate}
\item  For  $\cA  \in \tilde \cR_k$
\be   \label{3bounds}
\begin{split}
  \| E^{\nat} _{20}(X,\cA) \|     \leq    & \one  \|  E_{20}(X,\cA)  \|   \\
 \| E^{\nat} _{11}(X,\cA) \|     \leq  & \one  ML   \|  E_{20}(X,\cA)  \|  +    \|  E_{11}(X,\cA)  \|    \\
          \| E^{\nat} _{02}(X,\cA) \|     \leq   &  \one( ML )^2  \|  E_{20}(X,\cA) \|  +       \|  E_{02}(X,\cA)  \|\\
\end{split}
\ee 
\item
 Let    $e_k$ be  sufficiently small   (depending on  $L,M$) then 
\be
  \|  E^{\nat} (X, \cA)  \|_{\bh}   \leq  \one \|  E(X, \cA) \|_{\bh}
\ee
\end{enumerate}
\end{lem}
\bigskip

\pr  The modification  comes  in the piece     with two   $\psi$   fields which we write in an abreviated notation as
  \be  
    \int_{X   \times  X}   
    \bpsi_k  (\cA, x )    E_{20}  (X, \cA , x, y )      \psi_k   (\cA,y  )  dx dy  
\ee
We introduce    dummy variables  $x_0,y_0$   and write this  as 
  \be  
 \frac{1}{ \Vol (X)^2}   \int_{X   \times  X  \times   X   \times  X}   
    \bpsi_k  (\cA, x )    E_{20}  (X, \cA , x, y )      \psi_k   (\cA,y  )  dx dy  dx_0 dy_0   
\ee
Now make the substitution
\be    
\begin{split} 
 \psi_k   (\cA,y  )       =  &     e^{ie_k\eta (\tau \cA)(y,y_0) }  \psi_k(\cA, y_0)    -   \chi_k( \cA,  y,y_0)|y-y_0|^{\al}   \\
  \bpsi_k   (\cA,x  )       =  &      e^{-ie_k\eta (\tau\cA)(x,x_0) }  \bpsi_k(\cA, x_0)   -   \bar \chi_k( \cA,  x,x_0) |x-x_0|^{\al}  \\
 \end{split}  
\ee
This yields four terms    (all gauge invariant,  all integrals over  $X$ )
 \be   \label{four} 
 \begin{split}
 &  \frac{1}{ \Vol (X)^2}   
\int      \bpsi_k(\cA, x_0) \B[ \int   e^{-ie_k\eta (\tau\cA)(x,x_0) }    E_{20}  (X, \cA,x  ,y )  e^{ie_k\eta (\tau \cA)(y,y_0) } dx dy \B] \psi_k(\cA, y_0)  dx_0 dy_0  \\
 -  &  \frac{1}{ \Vol (X)^2}  
 \int  \bpsi_k(\cA, x_0)  \B[ \int   e^{-ie_k\eta (\tau\cA)(x,x_0) }    E_{20}  (X, \cA,x  ,y )|y-y_0|^{\al}dx  \B]  \chi_k( \cA,  y,y_0)  dx_0 dy_0  dy \\
 - & \frac{1}{ \Vol (X)^2} 
\int     \bar \chi_k( \cA,  x,x_0) \B[ \int  |x-x_0|^{\al}   E_{20}  (X, \cA,x  ,y )   e^{ie_k\eta (\tau \cA)(y,y_0) } dy \B] \psi_k(\cA, y_0) dx   dx_0 dy_0  \\
 + & \frac{1}{ \Vol (X)^2} 
\int     \bar \chi_k( \cA,  x,x_0) \B[   |x-x_0|^{\al}   E_{20}  (X, \cA,x  ,y )|y-y_0|^{\al} \B]  \chi_k( \cA,  y,y_0)  dx_0 dy_0  dx  dy   \\
  \end{split} 
 \ee

For the first term  has  a kernel
\be   E^{\nat} _{20}  (X, \cA,x_0  ,y_0 )  =    \frac{1}{ \Vol (X)^2}  \int   e^{-ie_k\eta (\tau\cA)(x,x_0) }    E_{20}  (X, \cA,x  ,y ) 
  e^{ie_k\eta (\tau \cA)(y,y_0) }    dx dy 
\ee
This  does vanish at  $\cA =0$ by  the normalization   assumption.   We want to bound   the norm
$ \| E^{\nat} _{20}  (X, \cA )\|  $.   Since  $X \in \cS$   it  is contained in  some    $\square^{\dagger}$   and so      the field   $\cA  \in  \tilde  \cR _{k}$   is gauge equivalent to a field     $\cA  \in \cR_{k} $.    Since  $ \| E^{\nat} _{20}  (X, \cA )\|  $ is gauge invariant
it suffices to assume $\cA \in \cR_k$.   Since  $x,x_0  \in X$ we have   $|x-x_0| \leq \one ML$  and so
\be  
|e_k\eta (\tau\cA)(x,x_0)|   \leq \one e_k ML   \|  \cA \|_{\infty}  \leq   \one MLe_k^{\frac14+ \ep}  \leq 1  
\ee 
This yields the desired bound
\be  
\begin{split}
\|  E^{\nat} _{20}  (X, \cA) \|    \leq  & \one   \frac{1}{ \Vol (X)^2}  \int      | E_{20}  (X, \cA,x  ,y )| dx dy dx_0 dy_0 \\
 \leq   &   
 \one \int    | E_{20}  (X, \cA,x  ,y )| dx dy   =  \one  \| E_{20}(X, \cA) \|   \\
\end{split}
  \ee

The  second term has the kernel 
\be 
   E^{(1)}_{11}(X, \cA,   x_0,y_0,y)   =  \frac{1}{ \Vol (X)^2}    \int  e^{-ie_k\eta (\tau\cA)(x,x_0) }   E_{20}  (X, \cA,x  ,y )|y-y_0|^{\al}dx 
    \ee
 and we estimate
\be
\begin{split}  \|    E^{(1)}_{11}(X, \cA) \| 
\leq     &  \one  \frac{1}{ \Vol (X)^2}    ML     \int    |  E_{20}  (X, \cA,x  ,y )  | dx_0 dy_0  dx    dy   
=    \one  ML   \| E_{20}(X, \cA)  \|      \\
\end{split}
\ee
This gives a contribution to $ E^{\nat}_{11}(X, \cA)$  as  does    the third term   in  (\ref{four}), as well as the original  term
   $  E_{11}(X, \cA)  $.   The stated bound on     $ E^{\nat}_{11}(X, \cA)$ follows.

The last    term   has the kernel   
\be  
   E^{(1)}_{02} (X,  \cA, x, x_0, y, y_0)  =  \frac{1}{ \Vol (X)^2} |x-x_0|^{\al}  E_{20} (X, \cA, x,y) |y-y_0|^{\al}
   \ee
with norm bounded by   $ \one  (ML)^2 \| E_{20}(X, \cA)\|$.  This contributes to $ E^{\nat}_{02}(X, \cA)$ as does the original term  
$ E_{02}(X, \cA)$  and the stated bound follows.
This completes the proof of part 1.
 
Part 2 is where we use the fact that $\psi$ and $\chi$  are weighted differently.
 It suffices to look at the terms
$ E^{\nat}_{11}(X, \cA)$  and   $ E^{\nat}_{02}(X, \cA)$  since  $ E^{\nat}_{20}(X, \cA) =0$ and all the others are unchanged.   In the first case we 
   \be
\begin{split}
            h_{k,1}     h_{k,2}       \|    E^{\nat}_{11}  (X, \cA)  \|   
  &      \leq         h_{k,1}     h_{k,2}   \B(  \one  ML   \|  E_{20}(X, \cA)  \|     +    \|  E_{11}(X, \cA)  \|       \B)  \\
      &      \leq        \frac12   h_{k,1} ^2    \|  E_{20}(X,\cA)  \|     +   h_{k,1}     h_{k,2}      \|  E_{11}(X)  \|  \\
       &      \leq        \|   E(X,\cA)   \|_{\bh_k}  \\
     \end{split}
 \ee
Here we used   $h_{k,2}  = h_{k,1} e_k^{\ep}$ and then  $\one ML e_k^{\ep} < \frac12$ for  $e_k$ sufficiently small.
The term    $ E^{\nat}_{02}(X, \cA)$  is treated similarly and the result follows

   \subsubsection{improved scaling}

After the adjustments of the last two sections are made we have improved scaling.
   
    \begin{lem}   
\label{scalinglem}  Let  $L$ be   sufficiently large and  
  $e_k$  sufficiently small   (depending on  $L,M$).   Suppose  $E (X, \cA,  \psi, \chi)   $   has all the symmetries and satisfies
 for small sets  $X$:
   \be
 E_{00}(X,0)=0   \hs    E_{20}(X, 0) = 0 
\ee
Then    
  \be   \label{crude}  
  \| (\cB    E)_{L^{-1}}    \|_{k+1}  \leq         \cO( 1) L^{-\frac14 +2 \ep}      \|    E\|_{k} 
\ee
\end{lem}
\bigskip

\pr       This  follows a similar proof in \cite{Dim15}.   Refer to the previous bound in section  \ref{scaling0}  and
consider separately small  polymers $X \in \cS$ and large polymers $X \notin \cS$.   For  $X \notin \cS$ we have
$d_M(X) >L$ and so in the sum  (\ref{xmas})  restricted to  $X \notin \cS$  we can take 
$e^{- \ka d_M(X)  }  \leq  e^{-L} e^{- (\ka-1) d_M(X) }$.   The rest of the estimate proceeds essentially as before
and we find that these terms  are bounded by   $\one L^3 e^{-L}  \|  E_k \|$ which is more than enough.

Thus    we consider small sets.  It suffices to  show that  
  (\ref{L3}) can be improved for  $X \in \cS$  to  
 \be   \label{snort2}  
     \|  (   S_L  E)  (X )  \|_{\tilde   \cR_{k+1},  \bh_{k+1}}  
     \leq   \one   L^{- \frac{13}{4} +2 \ep}    \|     E(X) \|_{  \tilde  \cR_k,     \bh_k}   \ee
 The extra factor   $L^{-\frac{13}{4}}$  beats   the factor  $L^3$ in the blocking operation and we have the result.   
As noted in the proof of lemma  \ref{adj}, for small polymers we can replace the extended domain $\tilde \cR_k$ by
the more managable  $\cR_k$.  So it suffices to prove for  $X \in \cS$
  \be   \label{snort3}  
     \|  (   S_L  E)  (X )  \|_{   \cR_{k+1},  \bh_{k+1}}  
     \leq   \one   L^{- \frac{13}{4}  +2 \ep}   \|     E(X) \|_{   \cR_k,     \bh_k} 
       \ee

 Taking account the total number of  fermi  fields is even we  write
 \be     \label{lichee} 
 \begin{split}
  &    \| \cS_L   E  (X  ) \|_{\cR_{k+1},   \bh_{k+1}}  =     \|  (\cS_L  E )_{00}(X )   \|_{\cR_{k+1} }   \\
 + & \frac12   h_{k+1,1}^2     \|  ( \cS_L  E) _{20}  (X)   \|_{\cR_{k+1} }         
 +   h_{k+1,1}     h_{k+1,2}   \|   ( \cS_L   E  )_{11}  (X)   \|_{\cR_{k+1} }         
 +  \frac12        h_{k+1,2}^2   \|  ( \cS_L  E )_{02}  (X)   \|_{\cR_{k+1} }         \\
  +  &   \sum_{n+m  \geq 4}   \frac{ h_{k+1,1}^n     h_{k+1,2}^m}{n!m! }     \|   (  \cS_L   E ) _{nm}  (X)     \|_{\cR_{k+1} }     \\
  \end{split}
 \ee
 and look at each term separately.

The last  term  in (\ref{lichee})  is bounded by  (\ref{ding4}) by    
\be  
\begin{split}    \sum_{n+m  \geq 4}  \frac{ h_{k+1,1}^n     h_{k+1,2}^m}{n!m! }   
     L^{-n}L^{-(1+\al)m }     \|      E_{nm}  (X)     \|_{L^{- \frac34 }  \cR_{k}}
\leq  &  L^{-4}  \sum_{n+m  \geq 4}  \frac{ h_{k,1}^n     h_{k,2}^m}{n!m! }   \|      E_{nm}  (X)     \|_{  \cR_{k}} \\
\leq   &  L^{-4} \| E (X)  \|_{\cR_k, \bh_k}  
\\
\end{split}  
\ee
which    suffices.

For the third  term  in (\ref{lichee})    we  have  by  (\ref{ding4})
 and  $ h_{k+1,i}    \leq    L^{-\frac18 +  \frac12 \ep  } h_{k,i} $  
 \be
\begin{split}
   h_{k+1,1}     h_{k+1,2}   \|   (\cS_L    E)_{11}  (X)   \|_{\cR_{k+1} }  
 &      \leq         h_{k+1,1}     h_{k+1,2}    L^{-1}L^{-(1+\al)  }   \|    E_{11}  (X  )  \|_{L^{-\frac34} \cR_k}   \\
  &      \leq        L^{-\frac94  -   \al  + \ep }   h_{k,1}     h_{k,2}     \|    E_{11}  (X)  \|_{\cR_k}    \\
       &      \leq        L^{-\frac{13}{4} +2 \ep   }  \|   E(X)   \|_{\cR_k, \bh_k}  \\
     \end{split}
 \ee
 Here we have assumed         $ 1-  \ep < \al <1  $ so $\al - \ep > 1- 2\ep$.  
 The   fourth term  in  (\ref{lichee}) is even easier since we have  $L^{-2(1 + \al)}$ instead of  $L^{-1}L^{-(1+ \al)}$.
But for the second term   we only have  $L^{-2}$ which is not enough.

To   continue we have to  take advantage of the scaling in $\cA$.  
Let  $x_0$ be a point in   $X$  and for  $\cA  \in \cR_{k+1}$  define   
$  \cA'(x )  =  \cA  (x)  -   \cA(x_0) $. 
Then  this is a gauge transformation with  $\la(x)  =    \cA(x_0) \cdot  (x- x_0)  $.    For  any  fixed $n$  and $x \in X$
we have   for $e_k$ sufficiently small
\be     | \cA'(x)|  <     \one    ML  \| \pa  \cA \|_{\infty}  \leq  \one   ML  e_{k+1}^{- \frac34 +  2 \ep}     <    L^{-n}  e_k^{-\frac34 + \ep}         
\ee
and   the same holds for   $\cA'_L$.  
Also   $  |  \pa    \cA'_L |   <   L^{-\frac74 }   e_k^{-\frac34 +2 \ep} $
and    $   |\de_{\al}  \pa    \cA'_L |,            <   L^{- \frac74- \al }    e_k^{-\frac34 +3 \ep} $  from (\ref{ding3})
and   we   conclude that    $\cA'_L   \in   L^{- \frac74 }   \cR_{k}$  which improves the   original   $\cA_L   \in   L^{-\frac34}   \cR_{k}$. 
Since the gauge transformation is complex we no longer  have that     $ \|  E_{nm}(X, \cA )\|  $ is gauge invariant.   
The  kernels  are    transformed  by  phase factors    $e^{ie_k \la}$.   However $|ie_k\la| \leq \one ML e_k^{\ep}  \leq  1$  and so   there is a constant  $c = \one$ (depending on $n,m$) so
\be      
 c^{-1}    \|  E_{nm}(X, \cA_L    )\|  \leq         \|  E_{nm}(X, \cA' _L  )\|   \leq    c     \|  E_{nm}(X, \cA_L   )\| 
\ee

Now for the second term  in   (\ref{lichee})  we  have     
\be  \label{sump}
  \frac12   h_{k+1}^2   \|  (\cS_L   E)_{20}  (X,  \cA)\|   = \frac12   h_{k+1}^2  L^{-2} \|   E_{20}  (X,  \cA_L)\| 
 \leq  \one    h_{k}^2    L^{-2}  \|  E_{20}  (X,  \cA'_L ) \|
\ee
    Since
 $\cA'_L   \in   L^{-\frac74}   \cR_{k}$ we have  that       $t \to  t \cA'_L$ is  an analytic  function from  
complex  $t$  satisfying      $|t| <   L^{ \frac74}$   to     $\cR_k $.    Hence    $t \to      E_{20}  (X, t \cA'_L)$
is analytic  with norm  bounded  by  $\|   E_{20} (X) \|_{\cR_k} $.     Since $   E_{20}  (X, 0)=0$  we have   
\be      E_{20}  (X,    \cA'_L)
 =  \frac{1}{2 \pi  i}    \int_{|t| = L^{\frac74} }\frac{dt}{t(t-1)}       E_{20} (X, t   \cA'_L)dt     
 \ee
and this gives the estimate for  $\cA \in  \cR_{k+1}$
\be     \label{lotus} 
 \|  E_{20}  (X,    \cA'_L) \|
\leq     L^{- \frac74}   \|   E_{20} (X) \|_{\cR_k} 
\ee
Put this in (\ref{sump}) and  the second term   in   (\ref{lichee}) is bounded by 
\be  
    \one     h_{k}^2     L^{-\frac{15}{4}}   \|  E_{20}(X) \|_{\cR_k} 
  \leq     \one   L^{-\frac{15}{4}}   \|   E(X)   \|_{\cR_k, \bh_k}
\ee
 
 Finally consider the    first  term in  (\ref{lichee}).    Since   $ E_{00}  (X, \cA)$  vanishes at zero and is even in $\cA$  
 the expansion around $\cA =0$  starts with the second order term.   We have for $\cA  \in \cR_{k+1}$, again making 
a gauge transformation to  $\cA'$
 \be   (\cS_L  E )_{00}(X,   \cA  ) =     E_{00}(X,   \cA_L  )  =    E_{00}(X,   \cA'_L  )
 =     \frac{1}{2\pi i}  \int_{|t| =     L^{\frac74} }  \frac{ dt}{t^2(t-1)} E_{00}  (X, t  \cA'_L  )
 \ee
which  gives   for  
    \be     \|   (\cS_L  E )_{00}(X,   \cA  )\| \leq     
  \cO(1)  L^{-\frac72}  \| E_{00}(X) \|_{ \cR_k}   \leq      \cO(1)  L^{-\frac72}   \| E(X) \|_{ \cR_{k, \bh_k} } 
  \ee

 \subsection{polymer  propagators}   \label{localgreen}

We  can also localize   the  fermion    propagators  with polymers  using the random walk expansion (\ref{g1}).   Assume 
$\cA$ is in the domain  (\ref{listless}) or the smaller $\tilde \cR_k$, and  for a walk   $\om = (\om_0, \om_1, \dots,  \om_n)$  define  $ X'_{\om}  = \cup_{i=0}^n  \tilde \square_{\om_i}$.
Then   write  
\be    \label{algae}
  S_k (\cA )  =  \sum_{X  \in \cD_k}   S_k (X, \cA  )   \ee
where   
\be
  S_k (X, \cA  )  = \sum_{\om:  X'_{\om}  =X  }   S_{k, \om} (\cA) 
   =\sum_{n=0}^{\infty} \ \     \sum_{\om: |\om| = n,   X'_{\om}  =X  }   S_{k, \om} (\cA)  
\ee
Then   $S_k (X, \cA  )$ only depends on $\cA$ in $X$, and   the kernel  $S_k (X, \cA, x,y)$   vanishes unless $x,y \in X$. 

  Recall   that   if  $|\om| = n$ then  
$   | S_{k, \om} (\cA)f|   \leq    C(CM^{-1})^n    \| f\|_{\infty}$. 
But     $  d_M(X)  \leq    |X |_M =  |X'_{\om} |_M   \leq  27(n+1)$   so we  can  make the estimate    
\be
       (CM^{-\frac12 })^{n}   \leq  \one   (CM^{-\frac12 })^{d_M(X) / 27}  \leq    \one  e^{-\ka  d_M(X)}
\ee
    for  $M$  sufficiently large.    The remaining factor        
$  (CM^{-\frac12})^{n}$   still gives the overall convergence of the series
and we have  the bound     
\be     \label{short}
| S_k (X, \cA  ) f|   \leq    C  e^{-\ka  d_M(X) } \| f \|_{\infty  } 
\ee

\section{Fermion determinant}

\subsection{a determinant identity}  

In  \cite{Bal96}  Balaban  establishes    the   following identity for  the     determinant     of  a  positive    self-adjoint  matrix   T.   It  is   
\be    \det  T  =  \exp    (\Tr  \log  T )
\ee
where
  for  any  $R_0 >0$  
\be 
  \log T  = T \int_{R_0}^{\infty} \frac{dx}{x}  (T +x )^{-1}   - \int_{0}^{R_0} dx  (T+x)^{-1}   + \log  R_0      
\ee

Here we  prove a generalization  for the case  where  $T$  has  negative spectrum as  well.    We   take the  branch  of the  logarithm
with the cut on the negative   imaginary axis,    so   $\log$ is defined on the entire  real axis except the origin.   Then the identity 
  $ \det  T  =  \exp    (\Tr  \log  T ) $ still  holds and   we have

\begin{lem}    \label{ouch}
Let   $T$  be an invertible  self-adjoint   matrix.    Then    $ \det  T  =  \exp    (\Tr  \log  T ) $  where    for  any  $R_0 >0$  
\be 
  \log T   =       T \int_{R_0}^{\infty} \frac{dy}{y}  ( T+  iy )^{-1}  -   i \int_0^{R_0} dy  ( T+ iy)^{-1}   +  \log  R_0    + \frac{   i\pi  }{2}      
\ee
\end{lem}  
\bigskip

\pr    Consider  simple  closed   curve    $\Ga$  traversed counterclockwise and  made up of the pieces
\begin{itemize}
 \item   $\Ga_R  =    \{ z \in  \bbC  :   |z| =  R, \  -\frac{ \pi}{2} + \ep   \leq  \arg  z    \leq      \frac{3 \pi}{2} - \ep   $   \} 
 \item   $\Ga_{-}  =    \{ z \in  \bbC  :    r  \leq  |z|   \leq    R  ,\   \arg z    =   \frac{3 \pi}{2}  -   \ep    $  \}  
\item  $ \Ga_r  =   \{ z \in  \bbC  :  |z| =  r,\    -\frac{ \pi}{2} + \ep   \leq  \arg  z    \leq      \frac{3 \pi}{2} - \ep    $\}
\item   $\Ga_{+}  =    \{ z \in  \bbC  :    r  \leq  |z|   \leq    R  ,\    \arg  z    =    -\frac{ \pi}{2} +  \ep       $  \}  
  \end{itemize} 
For  $R$ sufficiently large  and  $r $  sufficiently small     this encloses the spectrum   of   $T$.   The function $\log z$ is  analytic inside  $\Ga $  and so  
\be
\log  T   =  \frac {1} {2\pi i}  \int_{\Ga}      dz  \       \log z \    ( z -  T  )^{-1}    
\ee
Now  for any  $r< R_0< R$  split the contour by  $\Ga = \Ga{<}  + \Ga_{>}$   where
$\Ga_{<}  =   \Ga \cap   \{  z: |z| \leq  R_0 \}$  and $\Ga_{>}  =   \Ga \cap   \{  z: |z| \geq  R_0  \}$.  
In  the integral  over   $\Ga_{>} $  we  insert the identity  \be   \label{bolux} 
  ( z -  T  )^{-1}     =  z^{-1}  +  z^{-1}  T   ( z -  T  )^{-1} 
   \ee
 We   take the limit  $\ep \to 0$.  For the first term   the discontinuity  in $\log z$ across the negative imaginary axis    contributes    $- 2 \pi i$ and
we get  
\be   
\begin{split}   
\lim_{\ep  \to 0 }     \frac {1} {2\pi i}  \int_{\Ga_{>}}   \frac{   dz  }{z}  \         \log z     
=    &  - \int_{R_0}^R \frac{ dy }{y}        +     \frac {1} {2\pi i}   \int_{|z| =   R }     \frac{   dz  }{z}  \         \log z         \\
=  &      - \log R  +   \log R_0    +  \frac{1}{2\pi}   \int_{- \frac12 \pi}^{\frac32 \pi}    d \theta  (  \log R    +  i \theta )  \\
=  &          \log R_0    + \frac{   i\pi  }{2}      \\
\end{split}   
\ee
For the integral  of the second term   we  have     
\be  
\lim_{\ep  \to 0 }     \frac {1} {2\pi i}  \int_{\Ga_{>}}     \frac{dz}{z}   \log z \  T   ( z -  T  )^{-1}        
 =      T  \int_{R_0}^R  \frac{dy}{y}   ( iy +  T  )^{-1}  +      \frac {1} {2\pi i}   T \int_{|z| =R }    \frac{  dz  }{z} \log  z\    ( z -  T  )^{-1}   
 \ee
 Take the limit  $R \to \infty$.    
 The second term is  $\cO( R^{-1} \log R )$  and converges to zero.  For the first term  the integrand  is  $\cO (y^{-2})$  and   it   converges to the integral over  $[R_0, \infty)$.    Finally there is the integral  over    $\Ga_{<} $ which is   
\be
 \lim_ {\ep \to 0 }     \frac {1} {2\pi i}  \int_{\Ga_{<} }      dz  \       \log z \    ( z -  T  )^{-1}  
 = -i    \int_r^{R_0}  dy    ( iy  +   T  )^{-1}       +     \frac {1} {2\pi i}   \int_{|z| =   r }      dz       \log z \    ( z -  T  )^{-1}      
 \ee
 Take the limit  $r \to 0$.   Since  zero is not an eigenvalue    the second term  is   $\cO( r \log r)$   and converges to zero.   For   the first term the integrand is bounded    and  it converges to the integral over   $[0, R_0]$.   

\subsection{determinant of the fluctuation operator}

We   want to apply this to  $ D_k(\cA) +  bL^{-1}  P(\cA)     $   where   $  D_k(\cA)    =       b_k - b_k^2  Q_k(\cA) S_k(\cA )Q^T_k(- \cA)   $
and    
\be   S_k (\cA )   =   \B(  \fD _{\cA}    +  \bar   m  +b_k  P_k(\cA)    \B)^{-1}  \hs
     \fD_{ \cA} =   \ga \cdot \nabla_{\cA}  -   \frac12   \eta  \De_{\cA}
\ee
At  first  we assume  $\cA$ is real.        
Then     $ (  \ga \cdot \nabla_{\cA} ) ^*  =   -  \ga \cdot \nabla_{\cA}$  while      $ \De_{\cA}^*  =    \De_{\cA}$,   so 
$  \fD_{ \cA} $ is not  self-adjoint.      However   $ (\ga_3 \   \ga \cdot \nabla_{\cA} ) ^*  =   \ga_3 \   \ga \cdot \nabla_{\cA}$  and        $\ga_3 \De_{\cA}^*  =    \ga_3\   \De_{\cA}$,   so 
$\ga_3  \fD_{ \cA} $ is    self-adjoint.    The same is true  for  $\ga_3P_k(\cA)$   and hence  for   $ S_k(\cA)  \ga_3 $  and  $  D_k(\cA)  \ga_3 $.  
Therefore     $(D_k(\cA) + bL^{-1}  P(\cA) )\ga_3$ is self-adjoint.    Since  $\det \ga_3 =1$   this has the same determinant as  
   $D_k(\cA) +  bL^{-1} P(\cA) $  namely  $  \de \sZ_k(\cA) $.

\begin{lem}    For  $\cA$ real and in the domain  (\ref{listless})  (or  $\tilde \cR_k$ )   
   \be    \label{lankylanky} 
\begin{split}
  \de \sZ_k(\cA)  =   & \det      \B( (D_k(\cA) + bL^{-1} P(\cA)) \ga_3 \B)   \\
= &   \exp  \B (     4    | \tz  |  \B(   ( 1- L^{-3} )  \log  b_k  + L^{-3}   \log  (     b_k + bL^{-1} ) \B)  \\
&   -    i \ga_3    b_k^2  \int_0^{\infty   }   \Tr \B[ \sB_{k,y}(\cA) Q_k(\cA)   S_{k,y}(\cA)Q_k^T(-\cA)  \sB_{k,y}(\cA)   \B] dy   \B)
\\
\end{split} 
\ee
where 
\begin{equation}
\begin{split}
\sB_{k,y}(\cA )   =&        \frac{1}{b_k + i \ga_3 y }  (I -  P(\cA) )   +   \frac{1}{ b_k + bL^{-1}+ i \ga_3 y  }  P(\cA)
  \\
 S_{k,y}(\cA  )   =   &   \B( \fD_{\cA}     + \bar  m_k  + \al_{k,y} P_k(\cA)  
   + \beta_{k,y} P_{k+1} (\cA)         \B)^{-1}
      \\   
  &  \al_{k,y}    =       \frac{ b_k i \ga_3 y  }{b_k + i \ga_3 y }  \hs     \beta_{k,y}  =\frac{b_k^2 bL^{-1} }{( b_k + bL^{-1}+i \ga_3 y) (b_k +   i \ga_3 y ) } \\
 \end{split} 
 \ee 
\end{lem}  
\bigskip

\rem   Note that     $ S_{k,y}(\cA  ) $  interpolates between  $ S_{k}(\cA  ) $ at  $y = \infty$ and    $ S^0_{k+1}(\cA  ) $   at  $y=0$    (use (\ref{oprah})).   
\bigskip
  
\pr    By  lemma   \ref{ouch}   
\be 
 \de \sZ_k(\cA)  =  \exp  \B(  \Tr \log   \B( \ (D_k(\cA) + bL^{-1}P(\cA) \B)  \ga_3    \B) 
\ee
where  for any  $R_0$  
\be     \label{singing}  
\begin{split}
 & \log    \B(  (D_k(\cA) + P_k(\cA) \ga_3 \B)  \\ 
   =&   
 \B(D_k(\cA) +\frac{b}{L} P_k(\cA) \B)  \ga_3  \int_{R_0}^{\infty} \frac{dy}{y}  \B(  (D_k(\cA) + bL^{-1} P(\cA))\ga_3 +iy     \B)^{-1}   \\
 &
 -    i \int_{0}^{R_0} dy  \B(  (D_k(\cA) +  bL^{-1}P(\cA))   \ga_3+ i y    \B)^{-1}  + \log  R_0     + \frac{   i\pi  }{2}       \\
  =&   
   \B(D_k(\cA) + bL^{-1}P(\cA) \B)  \int_{R_0}^{\infty} \frac{dy}{y}   \Ga_{k,y}(\cA)  
-  i  \ga_3   \int_{0}^{R_0} dy   \Ga_{k,y}(\cA)  + \log  R_0     + \frac{   i\pi  }{2}    \\
\end{split} 
\ee
Here we defined 
\be
    \Ga_{k,y}(\cA)  =    \B(    D_k(\cA) + bL^{-1}P(\cA)  + i \ga_3 y \B)^{-1} 
\ee
From  appendix  \ref{ID}   we  have the representation   
\be   
\begin{split}
&    \Ga_{k,y}(\cA) 
 =       
\sB_{k,y}(\cA )   +   b_k^2  \sB_{k,y}(\cA ) Q_k(\cA)   S_{k,y}(\cA )Q_k^T(-\cA) \sB_{k,y}(\cA ) \\
\end{split}  
\ee

Now in (\ref{singing})    take the limit  $R_0 \to \infty$.    
We  have        $  \Ga_{k,y}(\cA)   =  \cO ( y^{-1} )$, hence the first term in  (\ref{singing}) goes    to zero.     The   second   term   in (\ref{singing})  
is      
  \be
  \begin{split}
  &    
    \int_0^{R_0}   \frac{  -i\ga_3  dy}{b_k + i \ga_3 y }  (I -  P(\cA) )   +   \int_0^{R_0}  \frac{-i\ga_3  dy }{ b_k + bL^{-1}+ i \ga_3 y  }  P(\cA) \\
    &
      - i \ga_3   b_k^2  \int_0^{R_0} \  \sB_{k,y}(\cA) Q_k(\cA)   S_{k,y}(\cA)Q_k^T(-\cA) \sB_{k,y}(\cA)  \   dy      \\
\end{split}        
\ee
 As     $y \to  \infty $ we have   $B_{k,y}(\cA)  =  \cO ( y^{-1} )$
and   we show  below   that   $S_{k,y} (\cA )  = \one$. 
Hence       last   term  is   $\cO( y^{-2} )$  so we can  take the limit  $R_0 \to \infty$.    
For the first    term  we   compute     
\be  
\begin{split} 
   \int_0^{R_0}    \frac{-i\ga_3 dy  }{b_k + i \ga_3 y } 
    =            \int_0^{R_0}  dy   \frac{-i\ga_3 (b_k -i \ga_3 y)}{b^2_k +y^2 }  dy 
      =& -i\ga_3  \tan^{-1} \B(\frac{R_0}{b_k}\B)  +  \frac12 \B (  \log  b_k^2     -    \log  (b_k^2 + R_0^2)     \B)   \\
\end{split}  
\ee
and similarly for the  second term. 
Now  use   $ \tan^{-1} (R_0/b_k) \to \pi/2$ and  $ -   \frac{1}{2}  \log  (b_k^2 + R_0^2)    + \log R_0 \to 0$.
 and obtain       
 \be 
\begin{split}
  \log    \B( (D_k(\cA) +\frac{b}{L}  P(\cA)) \ga_3 \B)   
= &         \log  b_k   ( I - P(\cA) )     +     \log  (     b_k + bL^{-1} )   P(\cA)    \\
&-     i \ga_3     b_k^2   \int_0^{\infty   }    \sB_{k,y}(\cA)   Q_k(\cA)   S_{k,y}(\cA)Q_k^T(-\cA)  \sB_{k,y} dy  +   i  \frac{\pi }{2}  (1- \ga_3)  
\\
\end{split} 
\ee
For the determinant we  need to take the trace of this and exponentiate.      The trace of the projection  is   
is      
\be 
   \Tr  P(\cA)    =   \Tr ( Q^T(-\cA)Q(\cA) )   =      \Tr (Q(\cA)  Q^T(-\cA) )   =  4 | \bbT^1_{N-k} |  =      4   L^{-3} | \tz  |  
\ee
independent of $\cA$.    
Similarly  $   \Tr  (I  -  P(\cA)  )    = 4 ( 1- L^{-3} )  | \tz  |  $.        Furthermore
\be  
\Tr \frac{i \pi}{2}( 1 - \ga_3)    =   \Tr \frac{i \pi}{2}   =   \frac{i \pi}{2} 4| \tz|    =  2 \pi i |\tz|
\ee
does not contribute when exponentiated.     
 Hence we have the result  (\ref{lankylanky}).  
\bigskip

We will also  need a random walk expansion for   $S_{k,y}(\cA)$.   As  for  $S_k(\cA)$  the main ingredient is control over a  local inverses 
for the modified Dirac   operator.   Instead of lemma  \ref{sweet2} we have:

\begin{lem}  \label{sweet5} 
Under  the hypotheses of  lemma \ref{sweet2}   and for  $\cA$ in the domain  (\ref{listless})  
 there is an operator   $ S_{k,y}( \square, \cA)$  on  functions on    $\tilde  \square^{(5)} $  
 satisfying
\be
\B(   \B( \fD_{\cA}     + \bar  m_k  + \al_{k,y} P_k(\cA)  
   + \beta_{k,y} P_{k+1} (\cA)         \B)   S_{k,y}( \square, \cA) f   \B) (x)   = f(x)   \hs     x \in  \tilde \square
 \ee
 and    
 \be 
| S_{k,y}( \square, \cA) f |, \ \leq 
    C  \|f\|_{\infty} 
\hs
|1_{\De_y} S_{k,y}( \square, \cA)1_{\De_{y'}} f  |, \    \leq      C    e^{  -   \ga  d(y,y') } \|f\|_{\infty}   
\ee
\end{lem}
\bigskip  

The proof is  similar to the analysis of \cite{BOS91}, and  is  postponed to  \cite{Dim17} where multi-scale random walk  expansions are discussed in 
detail.    Assuming   this result we have  instead of lemma \ref{sweet3}:
 
\begin{lem}     \label{sweet6}   Under the hypotheses of   lemma \ref{sweet3}   and for $\cA$ in the domain (\ref{listless}) (or $\tilde \cR_k$) 
there is a    random walk  expansion   
\be   S_{k,y}(\cA  )  =  \sum_{\om}    S_{k,y, \om} (\cA )
\ee
converging to a function  analytic in $\cA$ which satisfies  
\be \label{sweet10}
| S_k( \cA) f  |, \ \leq       C  \|f\|_{\infty} 
  \hs
|1_{\De_y} S_k( \cA)1_{\De_{y'}} f |, \    \leq      C    e^{  -\ga  d(y,y') } \|f\|_{\infty}    
\ee
 \end{lem}
\bigskip

The proof follows the proof  of  lemma  \ref{sweet3}.

 As   in   section  \ref{localgreen} there is an associated polymer 
 expansion  
\be   \label{lounge}  
     S_{k,y}(\cA)   = \sum_X    S_{k,y}(X, \cA) 
 \ee    
with  
\be   \label{lounge2}
  | S_{k,y}(X, \cA)  f |  \leq    C e^{- \ka d_M(X) }  \| f\|_{\infty} 
\ee

 \section{The  main   theorem}

\subsection{the theorem}

 The   starting   density  on $\bbT^0_{N}$ from  (\ref{snow1})       is 
  \be
\rho_0(A_0, \Psi_0)  = \exp\Big( - \frac12 \| dA_0 \|^2  - \blan  \bPsi_0,  (  \fD_{A_0} +  \bar m_0 )  \Psi_0  \bran   - m_0  \blan \bPsi_0, \Psi_0 \bran  - \vep_0     \Big )
\ee
For the full analysis of the model we     define a sequence of densities  $\rho_k(A_k, \Psi_k)     $  for fields  on $\bbT^0_{N-k}$
by successive RG transformations.  Given  $\rho_k$ we first define as in    (\ref{kth})  and   (\ref{funny1})     for fields on  $\bbT^1_{N-k}$
\be  
 \begin{split}    \label{basic1}
&\tilde   \rho_{k+1} (A_{k+1},  \Psi_{k+1} ) 
=      \\   &  \int \ 
  \de\Big( A_{k+1} -  Q A_k \Big)  \   \de( \tau A_{k} )   \de_G\Big( \Psi_{k+1} -  Q(\tilde   \cA_{k +1}) \Psi_k \Big)
\rho_k(A_k, \Psi_k)       D \Psi_k  D  A_k  \\
\end{split}
\ee
We  chose the background field    $\tilde   \cA_{k +1}$ on $\tk$    to be a smeared out version of $A_{k+1}$  defined precisely later on.   
Then    we scale to  fields  on   $\bbT^0_{N-k-1}$ as  in (\ref{scaleddensity}),(\ref{scaleddensity2}) by  
 \begin{equation}   \label{shoe}
 \rho_{k+1} ( A_{k+1},   \Psi_{k+1})  =  \tilde  \rho_{k+1} (A_{k+1,L}, \Psi_{k+1,L}) L^{   \frac12 (b_N - b_{N-k-1})    - \frac{17}{2}(s_N - s_{N-k-1}) }  
 \end{equation}

In this paper  we consider a bounded field approximation in which   (\ref{basic1}) is replaced by  
\be     \label{basic2}
\begin{split}
&\tilde   \rho_{k+1} (A_{k+1},  \Psi_{k+1} ) 
=      \\   &  \int \ 
  \chi_k    \hat \chi_k  \ 
  \de\Big( A_{k+1} -  Q A_k \Big)  \   \de( \tau A_{k} )   \de_G\Big( \Psi_{k+1} -  Q(\tilde  \cA_{k+1}) \Psi_k \Big)
\rho_k(A_k, \Psi_k)      D \Psi_k  D  A_k   \\
\end{split}
\ee
and   scaling is the same.  
New are  the characteristic functions    $ \chi_k    \hat \chi_k $ enforcing bounds on the fields.
The bounds are logarithmic in the coupling constant and depend on the quantities  
\be
p_k =   ( - \log e_k  )^p      \hs    p_{0,k} =   ( - \log e_k  )^{p_0} 
\ee
where   $p, p_0$  are  sufficiently large positive   integers satisfying  $p_0 < p$.     Since  $e_k$ is small these are somewhat large.   
The   bounds are  on  the real minimizer    $\cA_k  =  \cH_k A_k$  on  $\tk$  and    on the  real   fluctuation  field    $( A_k -   H^{\sx}_k A_{k+1} )$
on $\tz$
\be
\begin{split}
 \chi_k  =   &    \chi \B(   |d \cA_k |  \leq   p_k    \B)  \\
\hat \chi_k   =   &  \chi   \B( |   A_k -   H^{\sx}_k A_{k+1} |  \leq   p_{0,k} \B)  \\
\end{split}
\ee
We have   $A_k =  \cQ_k \cA_k$  and $dA_k = \cQ_k^{(2)} d\cA_k$  where $\cQ_k^{(2)}$ is a certain averaging operator on functions on plaquettes.   Hence $\chi_k$ also enforces  that  $  |d   A_k |  \leq   p_k$.
These  restrictions  are natural   
in Balaban's  formulation  of the renormalization group.  Our goal   is to  study the flow of these modified transformations. As   noted   earlier this  is the location of   the  renormalization problem.

We  are going to assert  that after    $k$  steps  for real  $A_k$ with  
 $|d\cA_k| \leq p_k$  we have a density  $\rho_k(A_k, \Psi_k)$    essentially   of the form
\be  \label{snuffit}
\begin{split}
&\rho_k(A_k, \Psi_k)  =  \cN_k  \sZ_k  \sZ_k(0)  \\&
\exp  \Big(   -  \frac12   \| d \cA_k \|^2  -  \fS_k( \cA_k, \psi_k(\cA_k) )
-  m_k \blan \bpsi_k(\cA_k),  \psi_k(\cA_k) \bran - \vep_k \Vol(\bbT_{N-k})  +    E_k(  \cA_k,   \psi^\#_k(\cA_k))  \Big) \\
\end{split}  
\ee
Here    $\psi_k(\cA)  =  \cH_k(\cA) \Psi_k$ and  
\be 
     \psi^\#_k(\cA)  =  \B( \psi_k(\cA), \chi_k(\cA)  \B)  = \B( \psi_k(\cA), \de_{\al, \cA} \psi_k(\cA)\B) 
\ee
The free fermi action    $  \fS_k( \cA_k, \psi_k(\cA_k) )$  is defined in (\ref{freefermi}), the determinant    $\sZ_k(\cA_k)$  is  defined in  (\ref{route66}) and the determinant  $\sZ_k$ is defined in  (\ref{route76}).   The function      $E_k(\cA, \psi^\#(\cA))$  is a sum over polymer  functions   
\begin{equation}
\begin{split}
 E_k(\cA,  \psi^\#(\cA))  =&   \sum_{X \in \cD_k }   E_k(X, \cA,  \psi^\#(\cA))  \\
\end{split}
\end{equation}

These assumptions are    true  for  $k=0$ with   $  \sZ_0=  \sZ_0(\cA) = 1$,  $E_0 = 0$,   and the convention that  $\cA_0 =  A_0$  and  
 and  $\psi_0 (\cA_0 )  =  \Psi_0$ and  $D_0(A_0)     =   \fD_{A_0} +  \bar m_0     $.

\begin{thm}   \label{lanky}
Let   $L$  be   sufficiently large,  let    $M$  be   sufficiently large  (depending on $L$),   and  let $e$  be  sufficiently small   (depending on $L,M$).   Suppose    that   $\rho_k(A_k, \Psi_k)$  has the representation   (\ref{snuffit})     for  $A_k$ such that $| d\cA_k|   \leq  p_k  $.
  Suppose the  polymer function $E_k(X, \cA, \psi^\#_k(\cA_k))$ has kernels  $  E_k(X, \cA)$  defined and analytic
in  $\cA  \in \tilde   \cR_k$   with    all the symmetries  of section \ref{symmetries}.
Suppose  also  that  
  \begin{equation}    \label{gong} 
|m_k|     \leq    e_k^{\frac12}               \hs   \| E_k\|_k   \leq   1   \hs        
 \end{equation}
 
 Then up to a phase shift   $\rho_{k+1}(A_{k+1},\Psi_{k+1}) $  has  a representation of the same form   for  $A_{k+1}$ such
that     $| d\cA_{k+1}|   \leq  p_{k+1}  $,  
  now with   $e_{k+1}=L^{1/2}e_k$.      The  bounds (\ref{gong}) do not necessarily hold for $k+1$,  but we do have  
    \begin{equation}  \label{recursive}
\begin{split}
\vep_{k+1}   =&  L^3 \B( \vep_k    +  \vep(   E_k)  + \vep_k^0)\B) \\
m_{k+1}   =&   L \B(  m_k   + m (    E_k) \B) \\
E_{k+1}   =&  \cL  \B( \cR   E_k  +  E_k^{\det} +    E^\#_k(   m_k,  E_k)  \B) \\
 \end{split}
\end{equation}
where  $\cL: \cK_k \to \cK_{k+1}$ is the linear reblocking and scaling   operator  $\cL E  =  ( \cB   E  )_{L^{-1} }$.      The   polymer function   $ E_k^{\det} = E_k^{\det}(X, \cA)$ depends only on $\cA$, vanishes at  $\cA =0$,  and   satisfies
  $\|  E_k^{\det}  \|_k  \leq  e_k^{\frac14 - \ep}$.     The   polymer function  $E_k^*  = \cL(E_k^{\#}) $ satisfies
\begin{equation}    \label{nuts} 
    \| E^*_k \|_{k+1}     \leq  e_k^{\frac14 - 6 \ep} 
  \end{equation}
Finally       $ \vep_k^0 = \cO(e_k^n)$   for any  $n$. 
\end{thm}
\newpage

\rems
\begin{enumerate} 
\item   The  quantities $\vep( E_k),  m( E_k)$ are the extracted corrections  from $E_k$ to the energy density and  mass as
defined in  (\ref{sundown0}), and $ \cR E_k$ is the remaining irrelevant part.
They are linear in $ E_k$ and satisfy
\be
    | \vep( E_k)| \leq  \one \| E_k\|_k       \hs     m( E_k)\leq  \one e_k^{ \frac12}  \| E_k\|_k  
\ee
For the kernel of  $\cL \cR E_k$ we allow the adjustment of lemma   \ref{adj}  and take  $\cL \cR  E_k  =  ( \cB  (\cR  E_k)^{\nat}  )_{L^{-1}}$. 
Combining the results of lemma \ref{amble}, lemma \ref{adj}, lemma \ref{scalinglem}  we  then   have the key  contactive  estimate  
 \be   \label{lugnuts}
    \|  \cL \cR   E_k  \|_{k+1}  \leq  \one L^{-\frac14 + 2 \ep}   \| E_k  \|_k
\ee

\item
     The reference to the phase shift  means we are actually showing that   $\rho_{k+1}(A_{k+1},e^{i \theta} \Psi_{k+1}) $
     has the stated form for some  real  $\theta  = \theta(A_{k+1})$.      For iterating  the transformation   we redefine  it as   $\rho_{k+1}(A_{k+1}, \Psi_{k+1}) $.  This is allowed since it 
does not change the crucial normalization property   (\ref{lincoln0}).

\item The bound  $|d \cA_k|  \leq  p_k$ and the local estimate  (\ref{localest}) easily imply that  $\cA_k \in  \tilde \cR_k$,
in fact is well inside.
Therefore  $\psi^\#_k(\cA_k) =( \psi_k(\cA_k),  \chi_k(\cA_k))$,  the kernels   $ E_k(X, \cA_k)$,  and hence
$E_k(\cA_k, \psi_k^\#(\cA_k))$ are all well-defined under our assumptions.

\item  The proof  follows strategies developed  in \cite{Bal87}, \cite{Bal88a}, \cite{BIJ88}  \cite{Dim11},   \cite{Dim15}.   
\end{enumerate}

\subsection{proof of the theorem}

\subsubsection{extraction}  First we normalize  by extracting the relevant terms  from  $ E_k$.   We   have   by  (\ref{renorm2})
 \begin{equation}  \label{renorm4}
 \begin{split}
   E_k(  \cA,   \psi_k(\cA),    \chi_k(\cA)  )  = &     -  \vep(   E_k)     \Vol(  \bbT_{ \sN-k} )
    -      m(  E_k)    
  \blan   \bpsi_k(\cA),     \psi_k(\cA) \bran 
+      E'_k( \cA,   \psi_k(\cA),  \chi_k(\cA) )   \\
\end{split}
\end{equation} 
The last term  $E'_k \equiv  \cR E_k$  has a local expansion  $ E'_k (\cA, \psi, \chi) =  \sum_X  E'_k(X,  \cA, \psi , \chi) $   with 
 $  E'_k(X,  \cA)$  which are normalized and   satisfy   
 \be  \label{young}
     \|   E'_k \|_k  \leq  \one  \|  E_k \|_k   \leq   \one 
 \ee 
Then the representation   (\ref{snuffit}) holds  with   $E_k$ replaced  by $E'_k$  and  $\vep_k, m_k$
replaced by  
\be     \vep_k'  = \vep_k   +   \vep(   E_k)    \hs   
m'_k    =  m_k +   m(  E_k)  
\ee
 A renormalized free action is defined as 
 \be    \label{toast}
   \fS^{ +}_{k}( \cA_k, \psi_k(\cA_k) ) 
  =  \fS_k(\cA_k, \psi_k(\cA_k) ) + m'_k \blan \bpsi_k(\cA_k),  \psi_k(\cA_k) \bran + \vep'_k \Vol(\bbT_{N-k})  
  \ee

The block averaging now has the form  
\be   
\begin{split}
& \tilde    \rho_{k+1} (A_{k+1},  \Psi_{k+1} ) \\
=   &  \cN_k   \sZ_k \sZ_{k}(0) 
\int    D  A_k\       D \Psi_k \      \chi_k\      \hat \chi_k \  
\de \Big( A_{k+1} -  Q A_k \Big)  \  \de (  \tau A_k )    \de_G\Big( \Psi_{k+1} -  Q(  \tilde  \cA_{k+1}) \Psi_k \Big)
 \\
&\exp \B(   -  \frac12   \|  d \cA_k \|^2  -   \fS^{ +}_{k}( \cA_k, \psi_k(\cA_k) )   +    E'_k(  \cA_k,    \psi^\#_k(\cA_k))  \B)
\\
\end{split}
\ee

What restrictions should we put  on $A_{k+1}$  on  $\bbT^1_{N-k}$ here?
Later when we scale by  $A_{k+1}=  A'_{k+1,L}$ we will require  $A'_{k+1}$  on  $\bbT^0_{N-k-1}$ to satisfy  
$ |d \cA'_{k+1}|   \leq  p_{k+1}$ where   $\cA'_{k+1} =  \cH_{k+1} A'_{k+1}$.  
So define  an operator  $\cH^0_{k+1}$ on  $\bbT^1_{N-k}$ and a field  $\cA^0_{k+1}$ on $\tk$ by
\be  
 \cA^0_{k+1} = \cH^0_{k+1} A_{k+1} =(\cH_{k+1} A_{k+1,L^{-1}})_L  =\cA'_{k+1, L}
 \ee
Then
\be  \label{singsong}
  | d \cA^0_{k+1}  |   = | d (\cA'_{k+1,L}) |  \leq  L^{-\frac32} \| d \cA'_{k+1} \|_{\infty}  \leq   L^{-\frac32} p_{k+1}
\ee
Conversely this condition will scale to  $ |d \cA'_{k+1}|   \leq  p_{k+1}$.  Thus we impose (\ref{singsong})
as the condition on $A_{k+1}$.   Furthermore the statement  that    $\cA'_{k+1}$   is well inside  $ \tilde \cR_{k+1}$
translates to the statement that   $\cA^0_{k+1} $ is well inside   $\tilde \cR_k$.

\subsubsection{gauge field  translation}

We   translate to the minimum   of     $   \| d \cA_k \|^2$  on the surface   $\cQ A_k =A_{k+1}, \tau A_k =0$    as  before.  
Write    $A_k  =  H^{\sx}_k A_{k+1}  + Z$   and integrate over  $Z$  instead of  $A_k$.   
Then  $\cA_k = \cH_k A_k$ becomes     $\tilde \cA_{k+1}    + \cZ_k  $
where
\be 
    \tilde \cA_{k+1} =    \cH_k  H_k^{\sx} A_{k+1}  \hs    \cZ_k  =  \cH_k Z   
\ee
This is the  $ \tilde \cA_{k+1} $  that  appears in   (\ref{basic1}).   With this translation  
 $   \frac12   \| d \cA_k \|^2 $   becomes $  \frac12   \|  d  \tilde \cA_{k+1} \|^2  +           \frac12 \Big<Z, \De_k  Z  \Big>$
as before and our expression becomes
\be   \label{snore}
\begin{split}
& \tilde    \rho_{k+1} (A_{k+1},  \Psi_{k+1} ) 
= \cN_k  \sZ_{k} \sZ_{k}(0)  \exp \B(  -  \frac12   \|  d \tilde  \cA_{k+1} \|^2 \B)  \int     D Z  \  D \Psi_k  \   
  \chi_k\     \hat \chi_k \      \de  (   QZ     )  \  \de (  \tau Z  )   \\
 &     
\de_G\B( \Psi_{k+1} -  Q( \tilde  \cA_{k+1}  ) \Psi_k \B)
 \exp  \B( -    \frac12 \Big<Z,  \De_k  Z  \Big>  - \fS^{ +}_{k}\B( \cA^0_{k+1}  + \cZ_k ,\psi_k(\cA^0_{k+1}  + \cZ_k  ) \B) \\
  +   &   E'_k\B( \tilde \cA_{k+1}  + \cZ_k      ,   \psi^\#_k(\tilde   \cA_{k+1}  + \cZ_k     )  \Big)   \B) \\
\end{split}
\ee

Next  note that  $\tilde \cA_{k+1}$ differs from   $\cA^0_{k+1}$ by a gauge transformation
To see this     we use  $ \cH^{\sx}_{k}  =     \cH_{k}    +   \pa    \cO_k$  connecting the axial and Landau gauges.
Also define  $\cH^{0,\sx}_{k+1}$ by  scaling $\cH^{\sx}_{k+1}$  and then
   $\cH^{0,\sx}_{k+1}  =  \cH^0_{k+1}  +  \pa \cO^0_{k+1}$.
But by   (\ref{sammy}) we also have   $\cH^{0,\sx}_{k+1}  A_{k+1}= \cH^{\sx}_k   H_k^{\sx} A_{k+1}$
 Using these facts 
\be 
\begin{split}
 \tilde \cA_{k+1}  \equiv   &   \cH_k  H_k^{\sx} A_{k+1}  \\ = &   \cH^{\sx}_k   H_k^{\sx} A_{k+1}  -  \pa    \cO_k   H_k^{\sx} A_{k+1}    \\ 
  = &    \cH^{0, \sx}_{k+1}  A_{k+1}    -  \pa    \cO_k   H_k^{\sx} A_{k+1}    \\ 
  =  &   \cH^{0}_{k+1}  A_{k+1}  -   \pa \B(    \cO_k   H_k^{\sx}A_{k+1}   -   \cO^0_{k+1} A_{k+1}  \B)    \\ 
  \equiv   &    \cA^0_{k+1}  - \pa   \om   \\ 
\end{split}
\ee
where the last line defines   $\om= \om(A_{k+1})$.
 
We use this identity to replace $\tilde \cA_{k+1}$ by  $\cA^0_{k+1}$ in (\ref{snore}).  
If      $\om^{(0)}$ the restriction of  $ \om $  to the unit lattice  $\tz$ then  by (\ref{sync})
\be 
\begin{split}
      \psi_k( \cA -  \pa    \om   )  =  &    \cH_k (  \cA -  \pa    \om   ) \Psi_{k}    
=     e^{ie_k \om  }   \cH_k (  \cA   )   e^{-ie_k \om^{(0)} }  \Psi_{k}    \\
\end{split}
\ee
We  also change variables   by    $\Psi_{k}   \to  e^{ie_k \om^{(0)}}  \Psi_{k}  $.    This  is a rotation so the Jacobian is one.
Then   $\psi_k( \cA -  \pa   \om   ) $  becomes     $e^{ie_k \om}  \psi_k( \cA  )$  and   
\be
\fS^{ +}_{k}\B( \cA  - \pa \om+ \cZ , e^{ie_k \om}  \psi_k(\cA  + \cZ  )
=\fS^{ +}_{k}\B( \cA  + \cZ ,\psi_k(\cA  + \cZ  ) \B)
\ee
    We also  have   by  (\ref{study})  
\be   \de_G\Big( \Psi_{k+1} -  Q( \cA^0_{k+1} -  \pa \om  )  e^{ie_k \om^{(0)} }   \Psi_k \Big)  =   
 \de_G\Big(  \Psi_{k+1} -  e^{ie_k \om^{(1)} }Q( \cA^0_{k+1} )    \Psi_k \Big)   \ee
where          $\om^{(1)}$ is the restriction of $ \om  $  to  $\bbT^{1}_{N-k}$.
We     replace  $\Psi_{k+1} $ by    $ e^{ie_k \om^{(1)} } \Psi_{k+1}$ so the phase factor here
disappears   as well.  The terms   $\| \pa \cA \|^2$
and   $\cZ_k(\cA)$  and  $E'_k(\cA,  \psi^\#(\cA))$    are all gauge invariant, as are the characteristic functions.
 
Finally as      in   section \ref{rgg} we parametrize the integral  replacing      $Z$   by  $C  \tilde   Z$,  and identify a   Gaussian integral by          
\be          
\int f(Z)    \de  (   QZ     )  \  \de (  \tau Z  )   \exp\B(  -    \frac12 \Big<Z,  \De_k  Z  \Big>  \B)  
=      \de \sZ_k \int  f(C\tilde Z)\   d \mu_{C_k} (\tilde Z)       \ee  
So   now  we  understand   $\cZ_k$ as         $\cZ_k  = \cH_k C \tilde Z $.  
The characteristic functions have become
\be
\begin{split}
 \chi_k  =   &    \chi \B(  | d (\cA^0_{k+1} + \cZ_k \B)  |  \leq    p_k   )  \\
\hat \chi_k   =   &  \chi  \B( |C \tilde    Z|  \leq   p_{0,k} \B)  \\
\end{split}
\ee
 But we are assuming $|d \cA^0_{k+1}|      \leq L^{-\frac32}p_{k+1} \leq  L^{-\frac32}p_{k} \leq \frac12 p_k$
 and   by the bounds  (\ref{slavic3})   on  $\cH_k$  the fluctuation field  $\cZ_k= \cH_k C \tilde Z$ satisfies 
  $  | d\cZ_k|       \leq   C \| C\tilde Z \|_{\infty} \leq  C  p_{0,k}. \leq \frac12 p_k$.  Thus the first characteristic functiion is always one
  (and we could have omitted it from the start).

With all these changes:    
\be   \label{snore2}
\begin{split}
& \tilde    \rho_{k+1} (A_{k+1},  e^{ie_k \om^{(1)} }   \Psi_{k+1} ) 
=  \cN_k  \sZ_k \de  \sZ_k    \sZ_k( 0)
 \exp \B(  - \frac12   \|  d \cA \|^2\B)  \int    \    d \mu_{C_k} (\tilde Z)   \  D \Psi_k    \ 
  \\
 &      \hat \chi_k \      
\de_G\B( \Psi_{k+1} -  Q( \cA ) \Psi_k \B)
 \exp  \B(  -    \frac12 \Big<Z,  \De_k  Z  \Big> - \fS^{ +}_{k}\B(\cA  + \cZ_k, \psi_k(\cA  + \cZ_k  ) \B)\\
 &  +    E'_k\B(  \cA  + \cZ_k ,   \psi^\#_k( \cA  + \cZ_k)  \B)  \B|_{\cA =\cA^0_{k+1} }  \\
\end{split}
\ee

Next   separate out   leading terms in an expansion in the fluctuation field    $\cZ_k$.  Split  $\fS^{ +}_{k}$ as in (\ref{toast}) and  define        $E_k^{(1)},E_k^{(2)},E_k^{(3)}  $   by  
\be  
\begin{split}
\label{sanibel}
 E^{(1)}_k(  \cA,   \cZ,     \Psi_k )  = &   \fS_{k}\B(\cA  , \psi_k(\cA)    \B) 
  -  \fS_{k}\B( \cA  + \cZ_k ,\psi_k(\cA + \cZ_k  )  \B)
 \B)  \\
  E^{(2)}_k\B(  \cA,   \cZ,    \psi^\#_k(  \cA  ), \Psi_k  \B)  = &
 E'_k\B(   \cA + \cZ,     \psi^\#_k(  \cA  + \cZ  )  \B)
-   E'_k\B(   \cA,  \psi^\#_k(  \cA   )\B)   \\
  E^{(3)}_k(  \cA,   \cZ,     \psi_k(  \cA  ),\Psi_k  ) 
  =&       m'_k \blan \bpsi_k(\cA    ),  \psi_k(\cA   ) \bran   -  m'_k \blan \bpsi_k(\cA  + \cZ   ),  \psi_k(\cA + \cZ   ) \bran  \\
\end{split}
\ee

These functions    are naturally     functions of  $\Psi_k$  but we  eventually 
change  back  to  functions of  $\psi_k(\cA)$ or  $\psi^\#_k(\cA)$     
 using   the identity   $\Psi_k   =   T_k(\cA)   \psi_k(\cA)
$  where  $T_k(\cA)$ is the left inverse of  $\cH_k(\cA)$:
\be  \label{lunch}
 T_k(\cA)  \psi    \equiv 
  b_k^{-1}   Q_k(\cA)  \B( \fD_{\cA} +  \bar   m_k    + b_k Q^T_k(\cA) Q_k (\cA ) \B)
   \psi 
 \ee   
 In  appendix C in    \cite{Dim15} 
 it is shown that    $     \| Q_k(\cA)  \nabla_{\cA} \cdot  f  \|_{\infty} 
\leq    C  \|f \|_{\infty}  $   and that  
  $ \|  Q_k(\cA)( \eta\De_{\cA} )  f  \|_{\infty}     \leq    C   \eta  \| \pa_{\cA}     f \|_{\infty}   \leq  C   \|      f \|_{\infty} $ and
  hence    $  |  \fD_{\cA} f  |  \leq   C \|f  \|_{\infty}$.  
     Since also  $\| Q_k(\cA)f \|_{\infty}  \leq C \| f \|_{\infty}$ we have  for some constant $C_T$
\be  \label{larch}
   | T_k(\cA) f  |  \leq   C_T \|f  \|_{\infty}  
    \ee

Now    with   
$    E^{(\leq  3)}  =   E^{(1)}  +E^{(2)}+E^{(3)}    $
we have  the representation  
\be     \label{outlier1}
\begin{split}
& \tilde    \rho_{k+1} (A_{k+1}, e^{ie_k \om^{(1)} } \Psi_{k+1} ) 
= \cN_k   \sZ_k\ \de \sZ_k \    \sZ_k( 0 ) \exp \B(  - \frac12   \|  d \cA \|^2 \B)  \int   \    d \mu_{C_k} (\tilde Z)   \  D \Psi_k  
  \\ 
  &       \hat \chi_k \     
 \de_G\Big( \Psi_{k+1} -  Q( \cA  )\Psi_k \B)      \exp\B(     -   \fS^{ +}_{k}\B( \cA   ,\psi_k(\cA   )\B)
  + E'_k\B(  \cA ,   \psi^\#_k( \cA) \B)
   +    E^{(\leq 3)}_k\B(    \cA,   \cZ_k,      \psi^\#_k(\cA)     \B)    \B)  \B|_{\cA =\cA^0_{k+1} }  \\
\end{split}
\ee

\subsubsection{first localization}

We  want to localize  the terms contributing  to   $   E^{ (\leq 3)}_k( \cA,   \cZ,     \psi^\#_k(\cA) )$.     
These  will be treated  in the region 
\be 
  \label{region1}  
\cA     \in  \frac12  \tilde   \cR_k  \hs      |  \cZ|    \leq   e_k^{-3\ep} \ \   \ \  
  |\pa \cZ|  \leq   e_k^{-2\ep}      \ \    \ \  
 \|\de_{\al} \pa \cZ \|    \leq   e_k^{-\ep} 
 \ee
 The conditions on $\cZ$ are stronger than the condition $\cZ \in \frac12 \cR_k$.  Thus we have
 $\cA + \cZ  \in \tilde  \cR_k$ so  the   $E^{(i)}_k( \cA,   \cZ,  \psi,    \Psi_k )$ as given by 
 (\ref{sanibel})      are well-defined.

 We are particularly interested in the case $\cA =  \cA_{k+1}^0$ and $\cZ = \cZ_k$. 
We  have already noted that $\cA_{k+1}^0 \in  \frac12 \tilde \cR_k$, but what about  $\cZ_k= \cH_k  C\tilde Z$?  We
take as a condition on $\tilde  Z$ that $|\tilde Z| \leq e_k^{-2 \ep}$
and consider the domain 
\be 
  \label{region2}  
\cA     \in  \frac12  \tilde   \cR_k  \hs   | \tilde Z| \leq e_k^{-2 \ep}
 \ee
On this domain  $ |\cZ_k|     \leq C e_k^{-2\ep}  \leq e_k^{- \ep} $ and similarly for derivatives 
so we are in the domain (\ref{region1}).   Furthermore 
 the characteristic  function  $\hat \chi_k$ in (\ref{snore2})   enforces that
    $|\tilde  Z |  \leq    p_{0,k} $ which puts us in the domain (\ref{region2}). 
    \bigskip

\begin{lem}   \label{snooze1}  
The function $E^{(1)}_k = E^{(1)}_k(  \cA,   \cZ_k, \Psi_k )$ has a  local expansion   $E^{(1)}_k =  \sum_X     E^{(1)}_k(X)$
where    $ E^{(1)}_k(X,       \cA,  \tilde  Z,    \psi_k(\cA)) $ depends on these fields   only in  $X$,  is 
    analytic in   (\ref{region2})
  and satisfies there
  \be   \label{pinkb}
\begin{split}
\|     E^{(1)}_k\B(X,       \cA, \tilde  Z \B) \|_{h_k} 
\leq   &     e_k^{\frac14- 5\ep} e^{- ( \ka - \ka_0 -2) d_M(X)} \\
\end{split}
\ee
\end{lem}
\bigskip

\pr   First we study   $E^{(1)}_k(     \cA,   \cZ,   \Psi_k   )$  in the region (\ref{region1}). 
 By  (\ref{freefermi}) 
\be     \label{phil} 
\begin{split}
 E^{(1)}_k(     \cA,   \cZ,   \Psi_k   ) =     &     \blan \bPsi_k,\B(   D_k(\cA+ \cZ   )   -   D_k(\cA) \B) \Psi_k\bran  
=    \blan \bPsi_k,\B( M_k(\cA+ \cZ)   -   M_k(\cA) \B)  \Psi_k\bran   \\   
   \end{split}
\ee
where  $M_k$ is the operator on spinors on $\bbT^0_{N-k}$
 \be  
 M_k( \cA)  =-    b_k^2 Q_k(\cA)S_k(  \cA ) Q^T  _k(-\cA)
 \ee
 Next    insert   $S_k(\cA ) = \sum_X S_k(X,\cA )$  from  (\ref{algae})    and get    $M_k(X) = \sum_X M_k(X,\cA )$ 
 where
  \be  
 M_k( X, \cA)  =  -   b_k^2 Q_k(\cA)S_k(X,   \cA ) Q^T  _k(-\cA)
 \ee
 Then        $ E^{(1)}_k = \sum_X  \tilde E^{(1)}_k(X )$
 where
\be
\tilde E^{(1)}_k(  X,   \cA,   \cZ,   \Psi_k   )  
=     \blan \bPsi_k,\B( M_k(X, \cA+ \cZ)   -   M_k(X, \cA) \B)  \Psi_k\bran      
\ee
  This  only depends  on  $  \Psi_k$ in $X$  since   $S_k(X, \cA  )$
 only connects points in $X$  and  $Q_k(\cA)$  is    local on  $M$ scale.

 The matrix  elements   of  $ M_k( X, \cA) $   are
 \be   \label{ttt0} 
 [M_k(X,  \cA)]_{xy}   =       \blan  \de_x,      M_k(X,  \cA,)  \de_y    \bran    
=-   b_k^2  \blan    Q^T_k (\cA)   \de_x,  S_k(X,  \cA)  Q^T_k (-\cA)   \de_y  \bran  
\ee
Then   by (\ref{short})
\be      \label{ttt}
\begin{split}
|  [M_k(X,  \cA)]_{xy}  | 
\leq  &   \|Q^T_k (\cA)   \de_x \|_1 \|  S_k(X,  \cA)  Q^T_k (-\cA)   \de_y  \|_{\infty}  \\
\leq    &   C  \|Q^T_k (\cA)   \de_x \|_1 \|  \| Q^T_k (-\cA)   \de_y   \|_{\infty}         e^{-  \ka   d_M(X) }\\
\leq    &   C   e^{-  \ka   d_M(X) }\\
\end{split}
\ee

To estimate  $ \tilde E^{(1)}_k(X )$
 note that under  our assumptions (\ref{region1})  on $\cZ$   if   we  take complex    $|t|  \leq  \frac12   e_k^{-\frac34+4 \ep}  $
 then    $t \cZ  \in  \frac12 \cR_k$,   
 hence  $\cA  + t \cZ$ is  in  $\tilde   \cR_k$. Hence we are in the analyticity domain of $t \to  M_k(\cA+ t \cZ)$ and can write 
\be   
[ M_k(X, \cA+ \cZ)   -   M_k(X, \cA)]_{xy}  =  
 \frac{1}{2 \pi i}
 \int_{|t|  = \frac12  e_k^{-\frac34+ 4\ep}   }      \frac{dt}{t(t-1)}   [  M_k(\cA+t \cZ)    ]_{xy}
\ee
which yields    
\be   
|[ M_k(X, \cA+ \cZ)   -   M_k(X, \cA)]_{xy}|  \leq          C  e_k^{\frac34- 4 \ep}  e^{-  \ka   d_M(X) }
\ee

  The  kernel  of  $\tilde  E^{(1)}_k(  X,    \cA,   \cZ,   \Psi_k  )$   has the single non-zero entry (suppressing spin indices) 
  \be    
 \tilde   E^{(1)}_{k,2} \B(  X,    \cA,   \cZ, (1,x),(0,y)  \B)  =    [ M_k(X, \cA+ \cZ)   -   M_k(X, \cA)]_{xy} 
   \ee
Since   $ \Vol (X)  =M^3 |X|_M  \leq  \one  M^3  e^{\frac12   \ka d_M(X)}  $  and $h_k^2 = e_k^{-\frac12}$     we
have  
\be    
   \|\tilde  E^{(1)}_{k} (X,       \cA,   \cZ      )   \|_{h_k}   \leq        C   h_k^2  e_k^{\frac34- 4 \ep}  \Vol (X)^2  e^{-  \ka   d_M(X) }   
  \leq    CM^6  e_k^{\frac14- 4 \ep} e^{- ( \ka -1 )   d_M(X) }   
\ee

\bigskip
  
Specialize to the case $\cZ = \cZ_k = \cH_k C \tilde Z$ and take the domain (\ref{region2}). 
The  $  \tilde  E^{(1)}_k(X, \cA, \cZ_{k}, \Psi_k ) $   
is not local in   $ \tilde Z$.   
   To localize we   use the random walk expansion for $\cH_k$    to  introduce weakening parameters $s= \{ s_{\square}  \}$
 and define   $\cZ_k(s) = \cH_k(s, \cA) C \tilde Z$
 and
 \be      \tilde E^{(1)}_k( s, X,   \cA,  \tilde Z,   \Psi_k   )   \equiv  \tilde E^{(1)}_k(  X,   \cA,   \cZ_k(s),   \Psi_k   )
 \ee    
  For $|s_{\square} | \leq M^{\al_0}$ and $\al_0$ sufficiently small  these $s$  dependent quantities satisfy  bounds  of the same form as the original case $s_{\square} =1$
  so  \be    
   \|\tilde  E^{(1)}_{k} (s, X,       \cA,   \tilde Z     )   \|_{h_k}   
  \leq    CM^6  e_k^{\frac14- 4 \ep} e^{- ( \ka -1 )   d_M(X) }   
\ee

In  each variable  $s_{\square}$ for $ \square \subset X$ we  interpolate between  $s_{\square} =1$  and   $s_{\square} =0$  by
\be
   f(s_{\square} =1  )     =  f(s_{\square} =0  )     +   \int_0^1      d  s_{\square}   \frac{\pa  f}{  \pa  s_{\square}}  
    \ee
This yields a new expansion
 $ E^{(1)}_k  =  \sum_Y    \breve   E^{(1)}_k  (Y) $ where 
 \be  \label{summer2}
\begin{split}
 \breve { E }^{(1)}_k(Y,\cA,  \tilde Z ,   \Psi_{k} )   
= & \sum_{ X  \subset  Y  }        \int   ds_{Y-X} 
 \frac { \pa  }{ \pa s_{Y-X}}   \left[  \tilde  E^{(1)}_k(s,X ,  \cA,   \tilde Z,   , \Psi_{k} )  \right]_{s_{Y^c} = 0, s_{X}=1}\\
 \end{split}
 \ee
  and       $s_X =   \{ s_{\square} \}_{\square \subset X}$. 
  Then  $   \breve E^{(1)}_k(Y)   
$     only depends  on  the indicated fields    in $Y$   since there is  no coupling through $Y^c$. 

Now  $ \tilde  E^{(1)}_k(s, X )$ is analytic  in  $|s_{\square} | \leq M^{\al_0}$ and we 
  estimate the derivatives  for $0 \leq s_{\square} \leq 1$ by Cauchy inequalities. Each derivative then contributes a factor $M^{-{\al_0}}$
 and   $M^{-\al_0} \leq   e^{- \ka} $ for  $M$ sufficiently large.     Hence  we gain a factor  $e^{- \ka|Y-X|_M}$  from the derivatives in  $s$  and   have 
    \be      \label{studious} 
\|    \breve{  E } ^{(1)}_k(Y,\cA,  \tilde Z)   \|_{h_k} 
\leq      CM^6    e_k^{\frac14- 4\ep}    \sum_{X \subset Y}  e^{- \ka |Y-X|_M -  ( \ka -1)  d_M(X)  }  
 \ee
But  one can show  that  $   |Y-X|_M     +    d_M(X)   \geq  d_M(Y)  $ (see for example \cite{Dim11}).
Hence one can extract a factor   $ e^{- (\ka  - \ka_0-1)  d_M(X)}$   leaving   a factor   $ e^{ - \ka_0   d_M(X)}$
for the convergence of the sum over $X$.   The sum is bounded by  $\one  |Y|_M \leq   \one e^{ d_M(Y) }$   and so we have  
the announced bound  
\be
     \|    \breve{  E } ^{(1)}_k (Y, \cA, \tilde Z) \|_{h_k}   \leq     CM^6  e_k^{\frac14- 4\ep}   e^{- (\ka  - \ka_0-2 ) d_M(Y)}
\ee

Finally   as in (\ref{lunch})     insert      $\Psi_k(\cA)  =  T_k(\cA) \psi_k(\cA)  $  defining  
\be
     E^{(1)}(X,       \cA,  \tilde Z,   \psi_k(\cA)      )   =    \breve E^{(1)}(X,       \cA,   \tilde Z, T_k(\cA) \psi_k(\cA)     )  
\ee
From (\ref{larch}) we have     $  \|  T_k(\cA)  \|_{1, \infty}  \leq  C_T  $. Then by (\ref{loganberry}) in Appendix \ref{A}, and taking account that the function is quadratic in the fields  we
have 
\be    \label{ladylike}
\begin{split} 
 \|     E^{(1)}_k(X,       \cA,   \tilde Z      )  \|_{h_{k} } \leq &  \|   \breve  E^{(1)}_k(X,       \cA,   \tilde Z      )  \|_{C_Th_{k} }
\leq          C_T^2  \|  \breve    E^{(1)}_k(X,       \cA,   \tilde Z     )   \|_ {h_k}    \\ 
\leq  &  CM^6  e_k^{\frac14- 4\ep}   e^{- (\ka  - \ka_0-2 ) d_M(X)}
     \leq      e_k^{\frac14    - 5\ep  } e^{- (\ka  - \ka_0-2 ) d_M(X)} \\
\end{split}   
\ee
In the last step we used    $CM^6 e_k^{\ep}  \leq  1$.
This completes the proof.

\begin{lem}   \label{snooze2} The function
$E^{(2)}_k=  E^{(2)}_k(  \cA,   \cZ_k,    \psi^\#_k(  \cA  ), \Psi_k  )$ has a  local expansion   $E^{(2)}_k =  \sum_X     E^{(2)}_k(X)$
where    $E^{(2)}_k(X)=  E^{(2)}_k(X,       \cA,  \tilde Z ,    \psi^\#_k(\cA)) $ depends on these fields   only in  $X$, is   analytic in   (\ref{region2})  and satisfies there
\be   \label{pinkb2}
\begin{split}
\|     E^{(2)}_k\B(X,       \cA,   \tilde Z \B) \|_{\frac12  \bh_k} 
\leq   &  \one      e_k^{\frac34- 5 \ep} e^{- ( \ka - \ka_0-1) d_M(X)} \\
\end{split}
\ee
\end{lem}
\bigskip

\pr  
First we study  $E^{(2)}_k(  \cA,   \cZ,    \psi^\#_k(  \cA  ), \Psi_k  )$ in the domain (\ref{region1}).
We have $E^{(2)}_k =  \sum_X   \tilde  E^{(2)}_k(X)$ where
 \be   \label{tingly4}
\begin{split} 
 \tilde  E^{(2)}_k(X, \cA, \cZ,      \psi^\#_k(\cA), \Psi_k    ) =   &   E'_k\B(  X,  \cA + \cZ,    \psi^\#_k( \cA + \cZ  )  \B) 
     -       E'_k\B( X,   \cA,   \psi^\#_k(  \cA   )  \B)     \\
  \equiv    &    E'_k\B(  X,  \cA + \cZ,    \psi^\#_k( \cA  )  +  \cJ^\#_k( \cA,  \cZ) \Psi_k \B)    
    -       E'_k\B( X,   \cA,   \psi^\#_k(  \cA   )  \B)  \\
=    &    \frac{1}{2 \pi i}
 \int_{|t|  = e_k^{-\frac34+5 \ep}   }      \frac{dt}{t(t-1)}    E'_k\B(t,   X, \cA, \cZ,       \psi^{\#}_k(\cA),  \Psi_k \B)  \\
 \end{split}
\ee 
Here we defined
\be   
 \cJ_k ( \cA, \cZ )\Psi_k =       \psi_k(\cA+ \cZ )  -    \psi_k(\cA)  
=(\cH_k(\cA+ \cZ) - \cH_k(\cA) ) \Psi_k   
\ee
and 
\be
    \cJ^\#_k ( \cA, \cZ )\Psi_k =       \psi^\#_k(\cA+ \cZ )  -    \psi^\#_k(\cA)  
= \B(  \cJ_k ( \cA, \cZ )\Psi_k, \de_{\al} \cJ_k ( \cA, \cZ )\Psi_k    \B)
\ee
and
\be  \label{tingly3}
\begin{split}
& E'_k\B(t,   X, \cA,   \cZ,       \psi^{\#}_k(\cA),  \Psi_k \B) 
=      E'_k\B(  X,  \cA + t\cZ,    \psi^\#_k( \cA  )  + t \cJ^\#_k( \cA,  \cZ) \Psi_k \B)      \\
\end{split}
\ee

We need to justify the representation (\ref{tingly4}) and use it for estimates. 
As in the previous lemma   if   we  take complex    $|u|  <   e_k^{-\frac34+4 \ep}  $
 then    $u \cZ  \in  \frac12 \cR_k$ and   $\cA  + u\cZ$ is  in  $\tilde   \cR_k$.   Hence we are in  the analyticity domain of $u \to  \cH(\cA+ u \cZ)$.  
and  can write
\be  \label{leak1}
  \cJ_k ( \cA, \cZ ) f  =   
  \frac{1}{2 \pi i}
 \int_{|u|  =  e_k^{-\frac34+4 \ep} }      \frac{du}{u(u-1)}     \cH_k(\cA+u \cZ)    f
\ee
Using the bounds  (\ref{slavic2}) on  $\cH_k(\cA)$
we get 
\be  
\begin{split}    \label{leak2} 
&  |  \cJ_k ( \cA, \cZ ) f|
 \leq  Ce_k^{\frac34 -4 \ep}  \|f\|_{\infty}   \\
\end{split}    
   \ee
Then  for $ |t|  \leq e_k^{-\frac34+5 \ep}  $ we have $|t|  |  \cJ_k ( \cA, \cZ ) f| \leq Ce_k^{\ep} \| f\|_{\infty} $.  The same
 bound holds with the Holder derivative so  $|t|  |  \cJ^\#_k ( \cA, \cZ ) f| \leq Ce_k^{\ep}\| f\|_{\infty} $. 
Hence by  (\ref{paraguay})   in  appendix  \ref{A} 
  $ E'_k\B(t,   X, \cA, \cZ,       \psi^{\#}_k(\cA),  \Psi_k \B) $ satisfies
  (we take  $C_T h_k$ for later purposes)
\be \label{lion}
\begin{split}
&\|  E'_k\B(t,   X, \cA,\cZ \B) \|_{\frac12 \bh_k, C_T h_k}
\leq  \|  E'_k\B(  X, \cA  + t \cZ \B) \|_{\frac12 \bh_k + C C_Te_k^{\ep}  \bh_k} \\
 & \leq   \|  E'_k\B(  X, \cA  + t \cZ \B) \|_{\bh_k} \leq  \one e^{-\ka d_M(X) }
 \end{split}
\ee
Hence the representation (\ref{tingly4}) holds and  we have the bound 
   \be  
        \label{lingo} 
\|\tilde  E^{(2)}_{k} (X, \cA, \cZ  ) \|_{\frac12 \bh_k, C_T h_k}  \leq     \one   e_k^{\frac34- 5\ep}e^{-\ka d_M(X) } 
\ee  

Now  specialize to the case    $\cZ  =  \cZ_{k}= \cH_k C  \tilde Z$  and study   
the function $  \tilde  E^{(2)}_k(X, \cA, \cZ_k ,    \psi^\#_{k}(\cA) , \Psi_k ) $ on the domain (\ref{region2}). This   
is not local in   $\tilde Z$.      To localize   introduce weakening parameters $s= \{ s_{\square}\}$
based on the random walk expansions and define 
  \be
  \begin{split}
  \cJ_k(s,  \cA, \cZ) \Psi_k  =  &  \B(  \cH_k(s,\cA+ \cZ)  -   \cH_k(s,\cA) \B) \Psi_k\\
   \cJ^\#_k (s, \cA, \cZ )\Psi_k    = & \B(  \cJ_k (s, \cA, \cZ )\Psi_k, \de_{\al} \cJ_k ( s,\cA, \cZ )\Psi_k    \B)\\
   \cZ_{k} (s)   =  &   \cH_{k}(s) C\tilde Z      \\
\end{split}
    \ee 
and \be    \label{tingly45}
\begin{split} 
 \tilde  E^{(2)}_k(s,X, \cA, \tilde Z,      \psi^\#_k(\cA), \Psi_k    ) 
 =    &    \frac{1}{2 \pi i}
 \int_{|t|  = e_k^{-\frac34+5 \ep}   }      \frac{dt}{t(t-1)}   E'_k\B(s, t,   X, \cA, \tilde Z,     \psi^{\#}_k(\cA),  \Psi_k \B)  \\
   E'_k\B(s, t,   X, \cA, \tilde Z,     \psi^{\#}_k(\cA),  \Psi_k \B)
   =  & E'_k\B(  X,  \cA + t   \cZ_k(s),    \psi^\#_k( \cA  )  + t \cJ^\#_k\B(s, \cA,  \cZ_k(s)\B), \Psi_k \B)  
   \\
 \end{split}
\ee 
For complex   $|s_{\square} | \leq M^{\al_0}$  we get bounds of the same form and so
  \be       \label{lingo2} 
\|\tilde  E^{(2)}_{k} (s,  X, \cA, \tilde Z  ) \|_{\frac12 \bh_k, C_T h_k}  \leq     \one   e_k^{\frac34- 5\ep}e^{-\ka d_M(X) } 
\ee  
Again   interpolate between  $s_{\square} =1$  and   $s_{\square} =0$  and  get a new expansion 
 $ E^{(2)}_k 
=  \sum_{Y }  \breve    E^{(2)}_k  (Y) $ where 
 \begin{equation}  \label{summer3}
\begin{split}
 \breve  E^{(2)}_k(Y,\cA, \tilde Z,     \psi^\#_k(\cA) , \Psi_{k} )   
= & \sum_{ X  \subset  Y  }        \int   ds_{Y-X} 
 \frac { \pa  }{ \pa s_{Y-X}}   \left[  \tilde  E^{(2)}_k(s, X,  \cA,   \tilde Z,     \psi^\#_k(\cA)  , \Psi_{k} )  \right]_{s_{Y^c} = 0, s_{X}=1}\\
 \end{split}
 \end{equation}
  and   $   \breve  E^{(2)}_k(Y)   $     only depends  on  the indicated fields    in $Y$. 
The derivatives are again estimated 
 by Cauchy inequalities which gives  a factor  $e^{- \ka|Y-X|_M}$ and then 
    \be      \label{studious5} 
\|   \breve   E^{(2)}_k(Y,\cA, \tilde Z)   \|_{\frac12 \bh_k, C_T h_k} 
\leq    \one        e_k^{\frac34- 5\ep}    \sum_{X \subset Y}  e^{- \ka |Y-X|_M -  \ka  d_M(X)  }
\leq  \one    e_k^{\frac34- 5\ep} e^{  -  (\ka- \ka_0 -1)  d_M(Y)  }
 \ee

Finally  from  (\ref{lunch})   we define 
\be  E^{(2)}_k(Y,\cA, \tilde Z,     \psi^\#_k(\cA) )  = \breve E^{(2)}_k\B(Y,\cA, \tilde Z,     \psi^\#_k(\cA) ,  T_k(\cA)  \psi_k(\cA)  \B)  
\ee
which yields the desired bound 
\be
 \|    E^{(2)}_k (Y, \cA, \tilde Z) \|_{\frac12 \bh_k}   \leq  \|  \breve   E^{(2)}_k (Y, \cA, \tilde Z) \|_{\frac12 \bh_k, C_T h_k}  \leq    \one    e_k^{\frac34- 5\ep}   e^{- (\ka  - \ka_0-1 ) d_M(Y)}
\ee
\bigskip

\begin{lem}   \label{snooze3} The function 
$E^{(3)}_k = E^{(3)}_k(  \cA,   \cZ_k,    \psi_k(  \cA  ), \Psi_k  )$ has a  local expansion   $E^{(3)}_k =  \sum_X     E^{(3)}_k(X)$
where    $ E^{(3)}_k(X,       \cA,  \tilde Z,    \psi_k(\cA)) $ depends on these fields   only in  $X$, is   
   analytic in   (\ref{region2})  
and satisfies there
\be   \label{pinkb3}
\begin{split}
\|     E^{(3)}_k\B(X,       \cA,   \tilde Z \B) \|_{\frac12  h_k} 
\leq   &     e_k^{\frac34- 6 \ep} e^{- ( \ka - \ka_0-1) d_M(X)} \\
\end{split}
\ee
\end{lem}
\bigskip

\pr  
First write   
\be 
 m'_k   \blan  \bpsi_k (\cA ),  \psi_k (\cA ) \bran   =   m'_k \sum_{\square}  \blan  \bpsi_k (\cA ), 1_{\square}  \psi_k (\cA )  \bran 
 \equiv  m_k'    \sum_{\square}    E(\square, \psi_k(\cA) )    
\ee
The   kernel $  E(\square  )$   has the single non-vanishing element 
$    E_{2} (\square, (x,1),(y,0)    )  =m'_k  1_{\square} (y)\de(x-y)  $
which satisfies    $    \|   E_{2}(\square  ) \|  \leq  M^3 m'_k$
and so
 \be   
   \|   E(\square  ) \|_{h_k}  =       h_{k}^2   \|   E_2(\square  ) \| \leq    h_{k}^2   M^3   m'_k 
   \ee
But we are assuming $m_k \leq e_k^{\frac12}$ and by (\ref{sundown}) $m( E_k) \leq  \one e_k^{\frac12}$.  Therefore
$m'_k \leq \one  e_k^{\frac12} $. Since also $h_k^2 \leq e_k^{-\frac12}$ we have
 \be   
   \|   E(\square  ) \|_{h_k}   \leq \one M^3
   \ee 
Now we are in the situation of lemma      \ref{snooze2} except that our starting bound is worse
by the factor   $  M^3 $.   Hence we get the result with a  constant $\  \one  M^3    e_k^{\frac34- 5 \ep} \leq
        e_k^{\frac34- 6 \ep} $.

\subsubsection{fermi    field   translation}
Now  in  (\ref{outlier1})  with  $\cA  =  \cA^0_{k+1}$, we  diagonalize the quadratic form
\be  \label{snuff}
  bL^{-1}\blan  \bPsi_{k+1}-Q(-\cA)\bPsi_k, \Psi_{k+1}-Q(\cA)\Psi_k \bran  +   \fS(\cA, \psi_k(\cA))
\ee
which sits in the exponential. 
   As   in section   \ref{single}  this is   accomplished by   the transformations
   \be
   \begin{split}
    \Psi_k   =  & \Psi^{\crit}_k(\cA) + W \\
  \psi_k(\cA)   = &   \psi^{0}_{k+1}(\cA)  +     \cW_k( \cA)   \\
  \end{split} 
 \ee
  Also  define    $\psi^{0,\#}_{k+1}(\cA) =   ( \psi^{0,\#}_{k+1}(\cA)  ,  \de_{\al} \psi^{0,\#}_{k+1}(\cA)  )$   and  $\cW_k^\#(\cA) =
  ( \cW_k(\cA)  ,  \de_{\al} \cW_k(\cA)  )$ and then  the transformation is 
 \be   
   \psi^{ \#}_{k}(\cA)      =       \psi^{0, \#}_{k+1}(\cA)  +   \cW_k^\# (\cA)
 \ee
  By lemma \ref{samsung} the expression  (\ref{snuff})  becomes   
  \be
 \fS^0_{k+1}\B(\cA, \Psi_{k+1}, \psi^0_{k+1}(\cA)\B)    +     \blan\bar  W,  \B(  D_k(\cA)  +  bL^{-1}   P(\cA)\B) W\bran  
 \ee
Again  we    identify  the Gaussian measure  
 \be        \de \sZ_k (\cA)   \      d \mu_{\Ga _k(\cA)} (W)   =     
  \exp \B( -   \Big<\bar  W,  \B(D_{k} (\cA)+  bL^{-1} P(\cA)  \B)  W  \Big> \B)  D W
\ee
We  also  define $E_k^{(4)},E_k^{(5)}$    with  $\psi   =  \psi^{0}_{k+1}(\cA), \cW  =\cW_k (\cA)$ etc.  by
\be   
\begin{split}   
 E_k^{(4)}( \cA,   \psi^\#,     \cW^\#)  =  &  E'_k(  \cA ,    \psi^\#  +    \cW^\# )  - E'_k(  \cA ,     \psi^\#  )   \\
 E_k^{(5)} ( \cA,  \psi,    \cW)   =  &   
 m'_k \blan \bpsi  , \psi  \bran  -   m'_k \blan \bpsi + \bar \cW , \psi  +  \cW   \bran
\end{split}  
\ee
and   define
\be   
\label{xmas2}   E^{(\leq 5)}_k\B(  \cA,  \tilde Z,   \psi^\#,    \cW ^\#   \B) 
=  \sum_{i=1,2,3} E^{(i)} _k\B(  \cA,  \tilde Z ,    \psi^\#  +  \cW^\#     \B) 
+ E_k^{(4)} ( \cA,  \psi^\#,     \cW^\#)      +    E_k^{(5)}( \cA,  \psi^\#,    \cW^\#)     
\ee
We also define
\be   \fS^{0,+}_{k+1}\B(\cA, \Psi_{k+1}, \psi\B)  =  \fS^0_{k+1}\B(\cA, \Psi_{k+1}, \psi\B)  
+  m'_k \blan \bpsi  , \psi\bran +  \vep' \Vol(\tz)
\ee

With these changes (\ref{outlier1}) becomes 
\be     \label{bong}
\begin{split}
& \tilde    \rho_{k+1} (A_{k+1}, e^{ie_k \om^{(1)} }   \Psi_{k+1} ) = \cN_k  N_k  \sZ_k  \de \sZ_{k+1} \sZ_k( 0)  \de  \sZ_k( \cA)  \\
&\
  \exp \B(  - \frac12   \| d \cA \|^2-   \fS^{0,+}_{k+1}\B(\cA, \Psi_{k+1}, \psi^0_{k+1}(\cA)\B) + E'_k\B( \cA,   \psi^{0, \#}_{k+1}(\cA)   \B)   \B) 
   \Xi_k\B(  \cA,  \psi^{0,\#}_{k+1}( \cA )   \B)    \B|_{\cA = \cA^0_{k+1}}\\
\end{split}
\ee
Here we have isolated a fluctuation integral
\be   \label{fluctuate}
\begin{split}
& \Xi_k\B(  \cA,  \psi^{0, \#}_{k+1}( \cA )   \B) =
  \int    
  \hat \chi_k     \exp \B(   E^{(\leq 5)}_k\B(  \cA,  \tilde Z,   \psi^{0, \#}_{k+1}( \cA ) ,   \cW^\#_k (\cA)  \B)      \B) 
  \  d \mu_{C_k}  (\tilde       Z)   d \mu_{\Ga_k( \cA)} ( W ) \\
  \end{split}
\ee 
  \bigskip

  We re-express the fermion integral in ultralocal form.
  \footnote{Eventually we do this for the gauge field also, but the presence of the characteristic function $\hat \chi$ is
  a temporary obstable here}
  In general  for Gaussian Grassman integrals we have the identity
\be  
\int   F(\bar W, W )  d \mu_{\Ga} (W)   =  \int   F( \bar W',  \Ga  W' )  d \mu_I( W')
\ee
That is we can change to a Gaussian integral with  identity covariance by making the change of variables
$(\bar W, W) =  (\bar W',  \Ga  W')$.    We  make this change in the fluctuation integral   (\ref{fluctuate})  by 
the replacement  
\be  W_{\beta}( \sx)     =  (\tilde     \Ga_k(\cA) W')_{\beta}( \sx) 
\equiv   \begin{cases}  
(\Ga_k(\cA) W')_{\beta}( x)     & \hs   \sx =    (x, \beta, 0)  \\
\bar    W'_{\beta}(x)     &  \hs   \sx =    (x, \beta, 1) \\ 
 \end{cases} 
 \ee
With this change      $  \cW_k(\cA) = \cH_k(\cA)W$  is redefined as  
\be
 \cW_k(\cA) = \cH_k(\cA)\tilde \Ga_k(\cA)W'
 \ee
 and still $\cW_k^\#(\cA) =
  ( \cW_k(\cA)  ,  \de_{\al} \cW_k(\cA)  )$  
The fluctuation integral is   now       
\be 
\begin{split} 
  \Xi_k\B(  \cA,  \psi^{0,\#}_{k+1}( \cA )   \B) = &   \int  \hat  \chi_k
        \exp \B(   E^{(\leq 5)}_k( \cA,  \tilde Z,    \psi^{0,\#}_{k+1}( \cA ) ,     \cW^\#_k(\cA)    \B) 
          d \mu_{C_k} (\tilde  Z)   d \mu_I(W' ) )  \    \\
  \end{split}
\ee

   \subsubsection{second  localization}

To analyze the partition function $\Xi_k$ we need  a polymer expansion for    $ E^{(\leq 5)}_k$ which is local in  $W'$.

 \begin{lem}   \label{sweet7} 
The  function  $ E^{(\leq 5)}_k =  E^{(\leq 5)}_k( \cA,  \tilde Z,    \psi^{0,\#}_{k+1}( \cA ) ,     \cW^\#_k(\cA)    \B) $
   has a polymer expansion $ E^{(\leq 5)}_k  = \sum_X   E^{(\leq 5)}_k(X)$ where  $   E^{(\leq 5)}_k(X) =   E^{(\leq 5)}  _k( X,  \cA,  \tilde Z,   \psi^{0, \#}_{k+1}(\cA),    W'   )  $   depends  fields   only in  $X$, is    analytic in   (\ref{region2}),  
and satisfies   there
\be       \label{archery} 
 \|    E^{( \leq 5)} _k  (X,    \cA,  \tilde Z  )   \|_{\frac14 \bh_k  , 1 }   
 \leq    e_k^{\frac14 -  5\ep } e^{- ( \ka -3 \ka_0  -3)  d_M(X)}
\ee   
 \end{lem}
\bigskip

\pr  There are five terms in $E^{(\leq 5)}_k $ as defined in (\ref{xmas2}).   We first consider the term  $E^{(4)} _k  = \sum_X E_k^{(4)}(X)$
where  $E_k^{(4)}(X)$ can be written 
  \be     E^{(4)} _k \B(X,  \cA,   \psi^{0, \#}_{k+1}( \cA ),    \cW^\#_k( \cA) \B)
 =     \frac{1}{2 \pi i}
 \int_{|t|  =  \frac 14 e_k^{-\frac14   +2\ep}   }     \frac{dt}{t(t-1)}      E^{(4)}_k\Big( t,   X,      \cA,   \psi^{0,\#}_{k+1}( \cA ),     \cW^\#_k( \cA)     \Big)
\ee
where    
\be  \label{spirit}
\begin{split}
  E^{(4)}_k\Big(t,   X,      \cA,  \psi^{\#}( \cA ),   \cW^\#( \cA)      \B)  
    =    &   E^+_k\Big( X,      \cA,  \psi^{\#}( \cA ) ,    t   \cW^\#( \cA)      \B)  \\
     E^+ _k\Big( X,      \cA,  \psi^{\#}( \cA ),   \cW^\#( \cA)     \B) 
       =    &   E'_k\Big( X,      \cA,   \psi^{\#}( \cA )   +  \cW^\#( \cA)    \B) 
  \end{split}  
   \ee
   Define $
  \bh_{0,k} = (e_k^{- \ep},e_k^{- \ep} ) 
  $. 
  If        $|t|   \leq   \frac 14 e_k^{-\frac14   +2\ep}     $  then  $|t | \bh_{0,k}  \leq  \frac14 \bh_k$.   By  (\ref{loganberry}) in Appendix \ref{A}  , then    (\ref{owly}), and then   $\|  E'_k \|_k \leq \one$
     we have 
     \be
 \|    E^{(4)}_k(t,   X,      \cA)  \|_{\frac14 \bh_k,    \bh_{0,k}   }    \leq      \|    E^+_k(X,    \cA)  \|_ { \frac14 \bh_k,  \frac14   \bh_k  } 
 \leq     \|    E'_k(X,    \cA)  \|_ {\frac12  \bh_k}       \leq     \one   e^{ -   \ka  d_M(X)  }
\ee
Hence  (\ref{spirit})  gives us    for  $e_k$ sufficiently small  
\be   \label{spout}
 \|   E^{(4)} _k  (X,  \cA ) \|_{\frac14 \bh_k,    \bh_{0,k}  }    \leq     \one      e_k^{\frac14-  2\ep}         e^{ -   \ka  d_M(X)  }
 \ee

Next define  
 \be     \label{mod} 
\tilde  E^{( 4)}_k  \B( X,  \cA, \tilde   Z,    \psi^{0,\#}_{k+1}(\cA),     W'     \B) 
=  E^{( 4)}_k \B(X,  \cA, \tilde Z,    \psi^{0,\#}_{k+1}(\cA),    \cW^\#_k( \cA)  \B) 
\ee
Now $ \cW_k(\cA) = \cH_k(\cA)\tilde \Ga_k(\cA)W'$  and  and    by      (\ref{slavic}), (\ref{clavicle}) 
  \be   
          | \cH_k (\cA) \tilde    \Ga_k( \cA) f |,   | \de_{\al} \cH_k (\cA)  \tilde  \Ga_k( \cA) f |       \leq       \  C  \| f\|_{\infty}  
  \ee
  Then  by a generalization of    lemma \ref{lunge} in Appendix \ref{A}  to four fields   and (\ref{spout}) we have  
   \be   \label{allowit}
   \begin{split} 
 & \| \tilde   E^{(4)}_k (X,  \cA  )  \|_{\frac14 \bh_k, 1  }     
    \leq       \|     E^{(4)}_k(X,  \cA  )  \|_{\frac14 \bh_k,       (C,C)   }      
    \leq          \|    E^{(4)}_k(X,   \cA )  \|_{\frac14 \bh_k,    \bh_{0,k}}  
 \leq    \one    e_k^{\frac14 -  2\ep  } e^{-  \ka   d_M(X)  }  \\
 \end{split}       
 \ee

To localize in $W'$  we use the random walk expansion described in section \ref{weak}  to introduce weakened operators
 $ \cH_k(s, \cA),  \Ga_k(s, \cA)$ and hence   $ \cW_k(s,\cA) = \cH_k(s,\cA)\tilde \Ga_k(s, \cA)W'$.   Then   define  the   modified version of  (\ref{mod})    
 \be
\tilde  E_k^{(4)} \B(s,  X, \cA, \tilde    Z,   \psi^{0, \# }_{k+1}(\cA),     W'    \B) 
=  E^{(4)}_k \B( X,  \cA, \tilde    Z,    \psi^{0, \#}_{k+1}(\cA),   \cW^\#_k(s,  \cA) \B) 
\ee
which also satsifies for $|s_{\square} | \leq M^{\al_0}$
\be 
\| \tilde   E^{(4)}_k ( sX,  \cA, \tilde Z  )  \|_{\frac14 \bh_k, 1  }     
 \leq     \one    e_k^{\frac14 -  2  \ep  } e^{-  \ka   d_M(X)  }  
\ee
Again   interpolate between  $s_{\square} =1$  and   $s_{\square} =0$  and  get a new expansion 
 $ E^{(4)}_k 
=  \sum_{Y }  \breve    E^{(4)}_k  (Y) $ where 
 \begin{equation}  \label{summer4}
\begin{split}
 \breve  E^{(4)}_k(Y,\cA, \tilde Z,     \psi^\#_k(\cA) , W' )   
= & \sum_{ X  \subset  Y  }        \int   ds_{Y-X} 
 \frac { \pa  }{ \pa s_{Y-X}}   \left[  \tilde  E^{(4)}_k(s, X,  \cA,   \tilde Z,     \psi^\#_k(\cA)  ,  W'  )  \right]_{s_{Y^c} = 0, s_{X}=1}\\
 \end{split}
 \end{equation}
  and   $   \breve  E^{(4)}_k(Y)   $     only depends  on  the indicated fields    in $Y$. 
The derivatives are again estimated 
 by Cauchy inequalities which gives  a factor  $e^{- \ka|Y-X|_M}$ and then 
    \be      \label{studious6} 
\|   \breve   E^{(4)}_k(Y,\cA )   \|_{\frac12 \bh_k, 1} 
\leq    \one        e_k^{\frac14 -  2  \ep  }   \sum_{X \subset Y}  e^{- \ka |Y-X|_M -  \ka  d_M(X)  }
\leq  \one    e_k^{\frac14 -  2  \ep  }e^{  -  (\ka- \ka_0 -1)  d_M(Y)  }
 \ee
This is better than we need for the theorem, and  so $\breve  E^{(4)}_k(X)$ contributes to $E_k^{\leq 5}(X)$.  
\bigskip

The treatment of $E^{(5)}_k$ is similar. 
\bigskip

Now consider the term $ E^{(1),+} _k= \sum_X  E^{(1),+} _k(X)$ where 
\be  E^{(1),+} _k\B(X,   \cA, \tilde Z,   \psi^{0,\#}_{k+1}(\cA ),   \cW^\#_k( \cA)     \B) 
=E^{(1)}_k\B(X,   \cA,  \tilde Z,  \psi^{0,\#}_{k+1}(\cA ) +   \cW^\#_k( \cA)    \B) 
\ee
Since   $\bh_{0,k} \leq \frac 14 \bh_k$  we  have by    (\ref{owly}) in Appendix  \ref{A}  and (\ref{pinkb})
\be  
\begin{split}
 \|    E^{(1),+} _k(X,   \cA,  \tilde Z)  \|_ {\frac14 \bh_k  ,  \bh_{0,k}  }      
\leq  &  \|    E^{(1)} _k(X,   \cA,  \tilde Z)  \|_ { \frac14      \bh_k   +    \bh_{0,k} }    \\
  \leq  &  \|    E^{(1)} _k(X,   \cA,  \tilde Z)  \|_ { \frac12     \bh_k   }  \\
  \leq  &  \one   e_k^{\frac14- 5\ep} e^{- ( \ka - \ka_0 -2) d_M(X)} \\
\end{split}  
\ee
Now we are in the same situation as we were at (\ref{spout}) in the analysis of $E^{(4)}_k$, but with slightly
 weaker bounds.   Localizing in $W'$ as we did then we find a new expansion $E^{(1),+} _k = \sum_Y \breve E^{(1),+} _k(Y)$
 where
  \be      \label{studious7} 
\|   \breve   E^{(1),+}_k(Y,\cA,\tilde Z )   \|_{\frac12 \bh_k, 1} 
\leq  \one    e_k^{\frac14 -  5  \ep  }e^{  -  (\ka- 2\ka_0 -3)  d_M(Y)  }
 \ee
This is again is  better than we need   so $\breve  E^{(4)}_k(X)$ contributes to $E_k^{\leq 5}(X)$.  
\bigskip

The treatment of $E^{(2),+} _k$ and $E^{(3),+} _k$ is similar.

\subsubsection{cluster expansion}

 We study the fluctuation integral which can now be written
  \be
 \Xi_k(\cA,    \psi^{0, \#} _{k+1}( \cA ) )=  
  \int \exp     \Big(    \sum_X     E^{(\leq 5)}_k(X, \cA, \tilde Z,  \psi^{0, \#}_{k+1}( \cA ) ,W')  \Big)  \hat \chi_k(C\tilde Z )      d\mu_{C_k}(\tilde Z)    d\mu_I(W')   
\ee
The cluster  expansion gives this a local structure.   The most straightforward way to proceed would be to mimic the fermion treatment and make a change of variables   $\tilde Z = C_k^{\frac12}Z'$  which changes the  Gaussian measure to  $d\mu_{I}(Z')$.   Then localize    $ E_k^{\leq 5}(X)$ in the new variable $Z'$.   With strictly local $ E^{\leq 5}_k(X)$ and both  ultralocal measures one might now contemplate using a standard cluster expansion.
The trouble is that  the characteristic function $\chi_k$ is messed up by the change of variables.  For the purposes of this paper one could fix this by altering the definition of $\chi_k$.  This is the approach taken in \cite{Dim15} and earlier versions of this paper.  However this does not generalize very  well when we consider the full model in \cite{Dim17}.  Instead we  closely  follow the approach used by Balaban  \cite{Bal88a}.

\bigskip
 
\begin{thm} 
 \label{cluster0}(cluster expansion) 
For   $\cA  \in   \frac12 \tilde  \cR_k$  
\begin{equation}  \label{sunshine}
  \Xi_k(\cA,   \psi^{0, \#}_{k+1}( \cA ) )       = \exp  \Big(  \sum_X      \ E^\#_k(X,\cA,  \psi^{0, \#}_{k+1}( \cA ) )  \Big)
\end{equation}
 where    $ E^ \#_k(X,\cA,  \psi^{0, \#}_{k+1}( \cA ) )$ satisfies
   \begin{equation}  \label{osprey}
\| \hat E^\#_{k}(X,\cA   ) \|_{\frac14    \bh_k}   \leq   \cO(1)  e_k^{\frac14- 5 \ep} 
  e^{ - (  \ka     - 10 \kappa_0 -10    )  d_{M}(X)}  
\end{equation}
\end{thm}
\bigskip

\pr
\textbf{step I}: 
   First make a Mayer expansion  writing
 \be
 \exp \B( \sum_X  E^{(\leq 5)}_k(X) \B) = \prod_X  \B( ( \exp E^{(\leq 5)}_k(X) -1) + 1 \B)
 = \sum_Y K(Y)
 \ee
 Here in the second step we expand the product and classify the terms in the resulting sum by the union of the polymers.  Thus 
 \be \label{lemony}
 K(Y ) 
=   \sum_{ \{ X_i\}:     \cup X_i   =  Y}  \prod_i   ( \exp E^{(\leq 5)}_k(X_i) -1)  
\ee
  where $\{ X_i\}$ is a collection of distinct polymers. 
   Two polymers are connected if they intersect  or   have a face in common.  If $\{ Y_j \}$ are the connected components of $Y$
  then   $K(Y)$ factors as
  \be K(Y) = \prod_j K(Y_j)
  \ee 
  and each $K(Y_j)$ is again given by (\ref{lemony}).
  Instead of  distinct  unordered  $\{ X_i\}$  we  can  write  $K(Y)$  as a sum over  distinct ordered sets  $(X_1,  \dots  X_n)$  by
 \be
 K(Y  ) 
=    \sum_{n=1}^{\infty} \frac{1}{n!}  \sum_{(X_1,\cdots  X_n) : \cup_i   X_i   =    Y}    \prod_i   (e^{E^{(\leq 5)}_k(X_i)} -1)
\ee
The partition function is now
\be. \label{lorna}
\Xi_k = \sum_Y  \int K(Y ) \hat \chi_k(C\tilde Z )      d\mu_{C_k}(\tilde Z)    d\mu_I(W') 
\ee

To estimate $K(Y)$  for  $Y$  connected
we use  the estimate      (\ref{archery})   to obtain 
\be
\| e^{ E^{(\leq 5) }_k(X) }  -1   \|_{\frac14 \bh_k ,1}
  \leq \sum_{n=1}^{\infty}    \|   E^{(\leq 5)}_k(X) \|^n_{\frac14 \bh_k ,1}  
   \leq    2       \|  E^{(\leq 5)}_k(X)   \|_{\frac14 \bh_k ,1}    \leq   \one   e_k^{\frac14 - 5 \ep} e^{ - (\kappa - 3 \ka_0 - 3)   d_M( X)  }
\ee
and so 
\begin{equation}
 \|  K(Y)\|_{\frac14 \bh_k ,1}    
\leq     \sum_{n=1}^{\infty} \frac{1}{n!}  \sum_{(X_1,\cdots  X_n) : \cup_i   X_i   =    Y}   \prod_{i=1}^n   \one   e_k^{\frac14 - 5 \ep} e^{ - (\kappa - 3 \ka_0 - 3)   d_M( X)  }
\end{equation}
Next    extract a factor    $\exp  \Big(- (\ka - \ka_0) ( d_M(Y)  -  (n-1))  \Big)$  and drop    all conditions on
 the  $X_i$  except  $X_i \subset  Y$  to obtain      (see Appendix B in \cite{Dim11} for details)  
\begin{equation}
\|  K(Y)     \|_{\frac14 \bh_k ,1}   \leq    \one   e_k^{\frac14 - 5 \ep} e^{ - (\kappa - 5 \ka_0 - 5)   d_M( X)  }
\end{equation}

\bigskip

  Now  in (\ref{lorna}) do the integral over  $W'$ defining
\be
 K'\B(Y, \cA, \tilde Z,  \psi^{0, \#}_{k+1}( \cA ) \B) 
=  \int K\B(Y, \cA, \tilde Z,  \psi^{0, \#}_{k+1}( \cA ) ,W' \B) d \mu_I (W') 
\ee
Since the integral is strictly local this also factors over its connected components. 
By lemma \ref{skunk} in Appendix \ref{A} we have  for $Y$ connected
\be
\|   K'(Y )  \|_{\frac14 \bh_k} 
\leq   \|  K(Y)  \|_{\frac14 \bh_k ,1} 
\leq  \one   e_k^{\frac14 - 5 \ep} e^{ - (\kappa - 5 \ka_0 - 5)   d_M( X)  }
\ee
The fluctuation integral is now written (relabeling $Y$ as $X$)
\be    \label{sour}
\Xi_k =  \sum_X   \int K'(X )  \ \hat \chi_k(C\tilde Z ) \  d\mu_{C_k}(\tilde Z) 
\ee 
The fermion field $\psi^{0, \#}_{k+1}( \cA )$ is a spectator for the rest of the proof.
\bigskip

\noindent
\textbf{Step II}:  
Next   take  a fixed  $X$  and  remove characteristic functions from $X^c$ as follows.
Define
\be
\hat \chi_k(X) = \prod_{\square \subset X}  \hat \chi_k(\square) \hs \textrm{ where } \hs \hat  \chi_k(\square,C \tilde Z)  =
\chi\B( \sup_{b \cap \square \neq \emptyset} |( C\tilde Z)(b)| \leq p_{0,k} \B)
\ee
Then   define
   $\hat  \zeta_k(\square)   =  1  -  \hat \chi_k(\square) $ and write   
   \be   
       \hat   \chi_k(X^c)  =
      \prod_{\square \subset X^c}     \hat  \chi_k(\square)   
      =       \prod_{\square \subset X^c} \B( 1   -   \hat \zeta_k(\square)  \B)   
  =     \sum_{P   \subset X^c }   (-1)^{|P| }  \hat  \zeta_k(P)         
\ee
where $\hat \zeta(P) = \prod_{\sq \subset P} \hat \zeta (\sq)$. Then 
\be
  \hat    \chi_k  =     \hat    \chi_k( X)     \hat   \chi_k(X^c)  
=  \sum_{P  \subset X^c}    (-1)^{|P| }    \hat   \chi_k( X)    \hat   \zeta_k(P)  
\ee
Insert this back into   (\ref{sour})  and classify the  the terms in the double sum over  $X, P$   by  $Y =  X \cup  P$.  Thus we have
 \be  \label{stacked3} 
 \Xi_{k}  = \sum_{Y }      \int       F(Y, \tilde Z )   d \mu_{ C_k } ( \tilde Z) 
\ee
where   
\be 
 F(Y, \tilde Z )   = 
  \sum_{X, P :   X \cup  P  = Y}      (-1)^{|P| }     \hat  \zeta_k(P)     \hat  \chi_k( X)   
  K'(X)  
\ee
Then $F(Y)$ is  local in its variables and it factors over its connected components. 
The characteristic function $  \hat  \zeta_k(P)  $ forces that there is a point in every cube $\square$ in $P$ where $| C\tilde Z| \geq p_{0,k}$.
Hence for $\square \subset P$  we have $p_{0,k} ^2 \leq \| C \tilde Z \|_{\sq}^2 \leq C_0  \|  \tilde Z \|_{\sq}^2$ and so
\be
\hat \zeta ( \square )  \leq   \exp \B( -p_{0,k}  +  p_{0,k}^{-1} C_0  \|  \tilde Z \|_{\sq}^2   \B)
\ee
If $|P|_M = M^{-3} \Vol(P)$ is the number of $M$ cubes in $P$ then 
\be 
 |   \hat \zeta_k(P) | \leq \exp \B( -p_{0,k}  |P|_M +  p_{0,k}^{-1} C_0  \|  \tilde Z \|_{P}^2   \B)
\ee
Using this we have the estimate  for connected $Y$
\be   \label{fy}
\begin{split}
 \|   \hat F(Y ) \|_{\frac14 \bh_k}    \leq &    
 \one e_k^{\frac14- 5 \ep}   \sum_{X, P :   X \cup  P  = Y}  \exp \B(  p_{0,k}^{-1} C_0  \|  \tilde Z \|_{P}^2    \B)  e^{-(\ka - 5\ka_0 -5)d_M(X)  -p_{0,k}  |P|_M }
 \\
 \leq &    
 \one e_k^{\frac14- 5 \ep}  \exp \B(  p_{0,k}^{-1} C_0  \|  \tilde Z \|_{P}^2    \B) e^{-(\ka - 6\ka_0 -5)d_M(Y)}   
 \sum_{X, P \subset Y }  e^{- \ka_0 d_M(X)}
  e^{ - \frac12 p_{0,k}  |P|_M }\\
  \leq &    
 \one  e_k^{ \frac14- 5 \ep }  \exp \B(  p_{0,k}^{-1} C_0  \|  \tilde Z \|_{Y}^2   \B) e^{ -(\ka - 6\ka_0 -6)d_M(Y) }   
 \\
   \end{split}
 \ee
\bigskip
In the second step we used  first that $\frac12 p_{0,k} \geq (\ka - 6 \ka_0 - 5)$ and then $ d_M(X) +  |P|_M \geq d_M(Y)$.  In the third step the  sum over $Y$ is  bounded by 
$\one |Y|_M \leq \one(d_M(Y) + 1) \leq \one  \exp (\frac12 d_M(Y) )$.    The  sum  over $P$ (which may not be connected) is bounded taking  $e^{-\frac12 p_{0,k} }\leq \la_k^n$
for any integer $n$ and then
\be
\sum_{P \subset Y} \la_k^{n |P|_M} \leq ( 1 + \la_k^{n}  )^{|Y|_M } \leq e^{\la_k^n |Y|_M} \leq \one e^{\frac12 d_M(Y) }
\ee
\bigskip

\noindent
\textbf{Step III}:  
We study the integral in (\ref{stacked3})  by conditioning  on the value of  $\tilde Z$ on  $Y^c$.   This has the form   (see \cite{Bal88a} and  appendix C)
\be  
\begin{split}   
  \int  d \mu_{C_{k}} (\tilde  Z)      F(Y,  \tilde   Z )  
   =& \int  d \mu_{C_{k} } (   Z')   
   \exp \B(   - \frac12 \blan  Z' , \B[ \tilde  \De_kC_{k} (Y)   \tilde \De_k \B]_{Y^c}  Z' \bran    \B)   \\
       &       
\B [ \int  d \mu_{C_{k} (Y) } ( \tilde Z)   \exp \B(   -\blan  Z' , [ \tilde \De_k \tilde  Z]_{Y^c}    \bran  \B)   
  F(Y,  \tilde  Z ) \B] \\
\end{split} 
\ee
Here we have defined $\tilde \De_k = C^T\De_k C$ so that $C_k = \tilde \De_k^{-1}$ and defined
 $C_k(Y) = [\tilde  \De_k ]_{Y}^{-1}$  as in section \ref{loco}.
In the outside integral    make the change of variables   $ Z'  =   (C_{k} )^{\frac12}   Z''$.   Note that this does not 
affect  the characteristic functions  in  $ F(Y,  \tilde  Z )$ which was our goal.
Then we have
\be   
  \Xi_{k}  = \sum_{Y }  \int  d \mu_{ I} (  Z'')  G( Y, Z'' )
\ee
where
\be   \label{souci}
\begin{split}   
G( Y, Z'') =  &
   \exp \B(   - \frac12 \blan  Z'' , C_k ^{\frac12}  [ \tilde  \De_k  C_k (Y)   \tilde \De_k ]_{Y^c} 
    C_k^{\frac12}  Z'' \bran    \B)   \\
       &       
  \int  d \mu_{ C_k(Y) } (\tilde Z)  
   \exp \B(   -\blan  Z'' ,  C_k ^{\frac12} [ \tilde \De_k \tilde  Z]_{Y^c}   \bran    \B)   
  F(Y,  \tilde  Z )  \\
\end{split} 
\ee
\bigskip

\noindent
\textbf{Step IV}:  Next  localize the expression $G( Y; Z'') $ using generalized random walk expansions. 
First consider  $C_k$  which has the expansion (\ref{licorice2}):
 \be \label{signal1}  
  C_k =     \sum_{\bd}    h_{\bd}  C_{k, \bd}h_{\bd} 
 + \sum_{\om: |\om| \geq 1} C_{k, \om}
\ee
where
 $\om = ( X_0,\al_1, X_1,  \dots,\al_n, X_n )$  has   localization domains with $|X_i|_M \leq r_0 = \one$.  Let $r= r_0 +1$.
 We  resum   the expansion so that the  basic units  include  connected components of the enlargement    $\tilde Y^{r}$, denoted $\tilde Y^{r}_{\beta}$.
 This operation is explained in section \ref{newly}.
  Resum the nonleading terms  (resummation of the leading term is optional)
  and get  a similar expansion now with 
 $\om = ( \cX_0, \dots, \cX_n )$ where  each $\cX_i$ is either  a $\tilde Y^{r}_{\beta}$  or an  $(\al_i, X_i)$  such that $X_i$ intersects $(\tilde Y^{r})^c$.
 The latter  satisfy $d(X_i, Y) \geq M$.

We   introduce weakening parameters in $(\tilde Y^{r})^c$ defining
\be \label{signal2}  
  C_k (s)=     \sum_{\bd}    h_{\bd}  C_{k, \bd}h_{\bd} 
 + \sum_{\om: |\om| \geq 1}  s_{\om}C_{k, \om}
\ee
Here as  before  $s= \{ s_{\sq} \}$  are  parameters $0 \leq s_{\sq} \leq 1$  indexed by  a partition into $M$ cubes $\sq$, but now
\be 
s_{\om} = \prod_{\sq \subset X_{\om} \cap (\tilde Y^{r})^c}    \ s_{\sq} \hs   \hs  X_{\om} = \cup_i  \cX_i
\ee
and if   $ X_{\om} \cap (\tilde Y^{r})^c = \emptyset$ then $s_{\om} \equiv 1$. Only walks leaving  $\tilde Y^{r}$ are suppressed.
 If $s=0$ then the only terms in the second sum which contribute are those with   $X_{\om} \subset  \tilde Y^{r}$.  These have  the form 
 $ \om =(\tilde Y^{r}_\beta, \dots, \tilde Y^{r}_\beta)$ for some $\beta$.

Similarly one can use resummed random walks to  define a  weakened version $C^{\frac12}_k(s)$ of  $C^{\frac12}_k$ and a weakened version   $C_k(s,Y)$ of  $C_k(Y)$.
($C_k(Y)$ is an operator on $Y$, but the random walk expansion involves the whole space.)
We also have a weakened version  $ \De_{k}(s ) =\ \cH^T_{k}(s) \de d  \cH_{k}(s)  $ of  $\De_{k} =\ \cH^T_{k} \de d  \cH_{k}$ and  define
$\tilde  \De_{k}(s )  = C^T  \De_{k}(s ) C$.

  Making these substitutions
in    $ G( Y; Z'' ) $  we  get   a new function 
\be       \label{souci2}
\begin{split}   
G( s,Y, Z'') =  &
   \exp \B(   -  \frac12 \blan  Z'' , C_k ^{\frac12}(s)  [ \tilde  \De_k (s) C_k (s,Y)   \tilde \De_k(s) ]_{Y^c} 
    C_k^{\frac12} (s) Z''\bran    \B)   \\
       &       
  \int  d \mu_{ C_k(s,Y) } ( \tilde Z)  
   \exp \B(   -\blan  Z'' ,  C_k ^{\frac12}(s)  [ \tilde \De_k (s)\tilde Z]_{Y^c}   \tilde  Z \bran    \B)   
  F(Y, \tilde Z )  \\
\end{split} 
\ee
The covariance of the Gaussian measure is still positive definite since $C_k(s)$ is  a small $\cO(M^{-1})$ perturbation of $C_k(0) $  which in 
turn is  a small $\cO(M^{-1})$ perturbation of the positive definite $\sum_{\sq}    h_{\sq}  C_{k, \sq}h_{\sq}$.
 
 Expanding around $s_{\sq} =1$ we find 
\be
\sum_{Y} G( Y; Z'' )  = \sum_{Y'} \breve  G( Y'; Z'' )  
\ee
where
\be    \label{guide}
\breve  G( Y'; Z'' )   =      \sum_{Y: \tilde Y^{r} \subset  Y'}    
   \int   ds_{Y'-\tilde Y^{r}} 
 \frac { \pa  }{ \pa s_{Y' -\tilde Y^{r}}}   \left[   G(s,  Y, Z'' ) \right]_{s_{(Y')^c} = 0 }
\ee
Since $s_{(Y')^c} = 0$ no operator in  (\ref{guide}) connects points in different connected components of $Y'$.
Hence  $\breve G( Y'; Z'' )  $ factors over its connected components and only depends on fields in $Y'$.  (To see the measure factors look at the characteristic function). 

Now we can write
\be
 \Xi_k  = \sum_{Y'}  H(Y') 
\ee
where
\be
 H(Y')  \equiv    \int  \breve G( Y, Z'' ) d \mu_{ I} ( \tilde  Z'') 
\ee
also factors over its connected components and depends on fields only in $Y'$.
\bigskip

\noindent
\textbf{Step V}:  We would like to estimate the $s$ derivatives in  (\ref{guide}) by Cauchy bounds.
This requires considering  $ G(s,  Y, Z'' ) $ for complex  $s_{\sq}$.   In particular we need control over the complex covariance $C(s,Y)$.   
We collect some relevant bounds.

First we note that for $\Up,\Up' $ in the basis (\ref{Ubasis}) and  $\Up,\Up'  \in  Y $
(i.e. $ \supp \  \Up, \supp \ \Up' \subset Y $ )
\be \label{dough}
 |C_k (0,Y)(\Up,\Up'  )| \leq C e^{-\ga d_M(\Up,\Up' ) } 
\ee
This follows just as the bound (\ref{succinct}) on $C_k (Y )$. It is also positive definite and then  by (a modification of) Balaban's
lemma on unit lattice operators  \cite{Bal83b}.
 we have a bound of the same form for the inverse:
\be 
 |C_k(0,Y)^{-1} (\Up,\Up'  )| \leq C e^{-\ga d_M(\Up,\Up' ) } 
\ee

Next define $\de C_k (s,Y)$ and $\de C_k^{-1} (s, Y)$ by
\be
\begin{split}
C(s,Y) = &  C(0,Y) + \de C(s,Y) \\
C(s,Y)^{-1} =& C(0,Y) ^{-1}+ \de C^{-1}(s,Y) \\
\end{split}
\ee
We  claim that for $M$ suffciently large, $\al_0$ sufficiently small,  and  $|s_{\sq} | \leq M^{\al_0}$ and  $\Up,\Up'  \in  Y $
\be  \label{many}
\begin{split}
 |  \de C_k (s,Y)(\Up,\Up' ) | \leq   & C M^{-\frac12} e^{-\ga d_M(\Up,\Up' ) }  \\
 |  C_k (s,Y )(\Up,\Up') | \leq   & C e^{-\ga d_M(\Up,\Up' ) }  \\
 |  C^{-1}_k (s,Y)(\Up,\Up' )  | \leq   & C e^{-\ga d_M(\Up,\Up' ) }  \\
 |   \de C_k^{-1}(s,Y)(\Up,\Up')  | \leq   & C M^{-\frac12} e^{-\ga d_M(\Up,\Up' ) }  \\
 \end{split}
 \ee
 Indeed the  operator $ \de C_k (s,Y )$ only involves walks with $|\om| \geq 1$ so if   $|s_{\sq}| \leq 1$ the first bound holds with $M^{-1}$ rather than $M^{-\frac12}$.  
 For $|s_{\sq} | \leq M^{\al_0}$ the bound erodes to $M^{-\frac12}$ just as in lemma \ref{oscar4}  (since $s_{\om}$ only involves the localization domains $X_i$, not the 
 $\tilde Y_{\beta}^{r_0 +1}$).
 The second bound follows from the first and 
(\ref{dough}).  For the third bound make the expansion
\be  \label{kingdom}
 \begin{split}
 C^{-1}_{k }(s,Y)  &  =  ( C_{k }(0,Y)  + \de C_{k }(s,Y) )^{-1} = C(0,Y)^{-1} ( I +  \de C_{k }(s,Y) C(0,Y)^{-1})^{-1} \\
 = &  \sum_{n=0}^{\infty} C(0,Y)^{-1} ( -  \de C_{k }(s,Y) C(0,Y)^{-1})^{n} \\
 \end{split}
 \ee
This converges for $M$ sufficiently large, and the result follows by the bounds on  $\de C_{k }(s,Y)$ and  $ C(0,Y)^{-1} $.  
The expansion also gives the fourth bound.

 Actually the first  and last bounds  in (\ref{many}) can be made even stronger.   We have excluded from $\de C(s, Y)$ walks which stay in $\tilde Y^{r_0 +1}$.  We are starting
 (and finishing)  in $Y$  so some  step must then go to a localization domain $X_i$   which necessarily satisfies  $d(X_i, Y) \geq M$.   The accumulated
 exponent factors in a bound like (\ref{borneo3}) result in any overall factor  $ e^{ -\ga d(X_i, Y)} \leq e^{-\ga M} $.
 Thus  we have the improved bounds
 \be   \label{manyprime}
\begin{split}
 |  \de C_k (s,Y)(\Up,\Up' ) | \leq   & C e^{-\ga M} e^{-\ga d_M(\Up,\Up' ) }  \\
  |  \de C^{-1}_k (s,Y)(\Up,\Up' ) | \leq   & C e^{-\ga M} e^{-\ga d_M(\Up,\Up' ) }  \\
  \end{split}
   \ee
 \bigskip
 
 \noindent
\textbf{Step VI}: 
 We proceed with the estimate on $G( s,Y, Z'') $ for $|s_{\sq}| \leq M^{\al_0}$ and     $s_{(Y')^c } =0$.  Write as
  \be       \label{souci3}
\begin{split}   
G( s,Y, Z'') =  &
   \exp \B(   -  \frac12 \blan  Z'' , \Ga^T_k (s,Y) C_k(s,Y) \Ga_k (s,Y)    Z''\bran    \B)   \\
       &       
  \int  d \mu_{ C_k(s,Y) } ( \tilde Z)  
   \exp \B(   -\blan   \tilde Z ,  \Ga_k (s,Y)    Z''  \bran    \B)   
  F(Y, \tilde Z )  \\
\end{split} 
\ee
where
\be  \Ga_k (s,Y)   =  [\tilde \De_k(s)]_{Y,Y^c} C_k^{\frac12}(s)
\ee
For any $f(\tilde Z) $ 
\be
\begin{split}
&| \int   f(\tilde Z) d \mu_{ C_k(s,Y) } ( \tilde Z) | 
   \leq \left| \frac{  \det \B(  C_k(s,Y)^{-1} \B)}{ \det \B( \re  C_k(s,Y) ^{-1} \B)} \right |^{\frac12} 
\int   |f(\tilde Z)| d \mu_{ (\re C_k(s,Y)^{-1} )^{-1}} ( \tilde Z) 
   \end{split}
  \ee
 Thus we get the bound
  \be       \label{souci4}
\begin{split}   
&|G( s,Y, Z'')|  \leq    
   \exp \B(   -  \frac12 \blan  Z'' ,\re \B( \Ga^T_k (s,Y)   C_k(s,Y) \Ga_k (s,Y)  \B)   Z''\bran    \B)   \\
       &    \leq \left| \frac{  \det \B(  (C_k(s,Y))^{-1} \B)}{\det \B( \re  C_k(s,Y) ^{-1} \B)} \right |^{\frac12}     
  \int  d \mu_{ (\re C_k(s,Y)^{-1} )^{-1}} ( \tilde Z) 
   \exp \B(   -\blan  \tilde  Z ,  \re (\Ga_k (s,Y)    Z'' \bran    \B)   
 | F(Y, \tilde Z ) |   \\
\end{split} 
\ee

To estimate this we start by 
replacing  every object  with  weakening parameters $s$ by the same with $s=0$.  There are several steps.
\bigskip

\noindent(1.)  First consider the ratio of determinants.
 We have
 \be \label{shout}
 \begin{split}
  \det  (C_k(s,Y)^{-1})=  &  \det  C_k(0,Y)^{-1} \det \B( I +  C_k(0,Y)\de C_k^{-1}(s, Y)\B) \\
     \det  (\re C_k(s,Y)^{-1})=  &  \det  C_k(0,Y)^{-1} \det \B( I +  C_k(0,Y) \re\de C_k^{-1}(s, Y)\B) \\
     \end{split}
 \ee
 The terms $ \det  C_k(0,Y)^{-1} $ cancel. 
In general we have  for $\|T\|<1$
\be            e^{ - \one  \|T\|_1} \leq               |\det(I + T)|    \leq      e^{ \one  \|T\|_1}
\ee
where the trace norm is $\|T\|_1 =  \Tr(T^TT)^{\frac12} $.   Therefore
\be
|\det \B( I +  C_k(0,Y)\de C^{-1}_k(s, Y)\B)| \leq \exp \B( \one  \| C_k(0,Y)\de C^{-1}_k(s, Y) \|_1 \B)
\ee
We dominate the trace norm by Hilbert-Schmidt norms and use (\ref{dough}), (\ref{manyprime})
 \be 
 \begin{split}
\one \| C_k(0,Y)\de C^{-1}_k(s, Y) \|_1
\leq &  \one  \| C_k(0,Y)\|_{HS} \|\de C^{-1}_k(s, Y) \|_{HS} \\
=  &\one  \B(\sum_{\Up,\Up' \in Y} |C_k(0, Y)(\Up,\Up' )|^2 \B)^{\frac12} \B(\sum_{\Up,\Up'  \in Y} |\de C^{-1}_k(s, Y)( \Up,\Up' )|^2 \B)^{\frac12} \\
\leq  & Ce^{-\ga M}  \Vol(Y) = Ce^{-\ga M} M^3|Y|_M \leq  \cO(M^{-1}) |Y|_M\\
\end{split}
\ee
 Then we have 
  \be
  |\det \B( I +  C_k(0,Y)\de C^{-1}_k(s, Y)\B)|   \leq e^{ \cO(M^{-1}) |Y|_M}
 \ee
Similarly   $\det ( I +  C_k(0,Y) \re \de C^{-1}_k(s, Y)) ^{-1}$ is bounded  by $e^{ \cO(M^{-1}) |Y|_M}$. Hence  the ratio of determinants  
in (\ref{souci4}) is bounded by $e^{ \cO(M^{-1}) |Y|_M}$
\bigskip 

\noindent(2.) Next consider the term in the first exponential in (\ref{souci4}).
We define a  quadratic form  $R_1$ by 
\be 
   \blan Z'', R_1 Z'' \bran 
=\blan  Z'' , \B[\re \B( \Ga^T_k (s,Y)   C_k(s,Y) \Ga_k (s,Y)  \B) -   \Ga^T_k (0,Y)   C_k(0,Y) \Ga_k (0,Y)    \B] Z''\bran 
\ee
We have the estimate  (\ref{manyprime})   for $ \de C_k(s,Y)$. 
By similar arguments the same holds for   $\Ga_k (s,Y) - \Ga_k (0,Y)$  and so 
 \be \label{sushi}
 \begin{split}
|  \Ga_k (s,Y)( \Up, \Up') - \Ga_k (0,Y)(\Up, \Up'))| \leq &  C e^{-\ga M} e^{- \ga d(\Up, \Up') }\\
\end{split}
\ee
These imply that
 $  |  R_1(\Up, \Up')   | \leq  Ce^{-\ga M}e^{- \ga d(\Up, \Up') }$.
Since  $\Ga_k (s,Y)  $ only connects to $Y'$ this gives
\be 
 | \blan Z'', R_1 Z'' \bran | \leq Ce^{-\ga M} \|Z '' \|^2_{Y'}
 \ee
We also have in the integrand
\be
\| \blan  \tilde  Z ,  \B( \re  \Ga_k (s,Y) - \Ga_k (0,Y)  \B)    Z'' \bran  |
 \leq  Ce^{- \ga M}  \|Z''\|_{Y'} \| \tilde Z \|_Y \leq Ce^{- \ga M} ( \|Z''\| ^2_{Y'}  + \| \tilde Z \|^2_Y   )  
 \ee
We change the Gaussian measure by
\be
\begin{split}
& \int   f(\tilde Z) d \mu_{  (\re C_k(s,Y)^{-1})^{-1} } ( \tilde Z)   
 \\
& =      \left  [ \frac{ \det \B( \re  C_k(s,Y) ^{-1} \B)}{  \det \B(  C_k(0,Y)^{-1} \B)} \right ]^{\frac12}
\int   f(\tilde Z) 
\exp \B(  -\frac12        \blan \tilde Z,  \re   \de C^{-1}_k(s,Y)  \tilde Z \bran \B)      d \mu_{  C_k(0,Y) } ( \tilde Z)   \\
   \end{split}
  \ee
 The ratio of determinants is again bounded by  $e^{ \cO(M^{-1}) |Y|_M}$ and
 \be   
 |      \blan \tilde Z,  \re   \de C ^{-1} _k(s,Y) \tilde Z \bran | \leq Ce^{-\ga M} \| \tilde  Z \|^2_Y
 \ee
 We also bound $F(Y)$ which is a product over its connected components $F(Y)  = \prod_{\al} F(Y_{\al}) $and each  $F(Y_{\al})$ is bounded by 
 (\ref{fy}).   Combining all of this  we have
 \be       \label{souci5}
\begin{split}   
&\|G( s,Y, Z'')\|_{\frac14 \bh_k}    \leq   e^{ \cO(M^{-1}) |Y|_M}  \prod_{\al}   \one  e_k^{ \frac14- 5 \ep }  e^{ -(\ka - 6\ka_0 -6)d_M(Y_{\al}) }  \\
 &  \exp \B(   -  \frac12 \blan  Z'' ,  \Ga^T_k (0,Y)  C_k(0,Y) \Ga_k (0,Y)     Z''\bran   + Ce^{-\ga M}  \|Z''\|^2_{Y'} \B)   \\
       &        
  \int d \mu_{  C_k(0,Y) } ( \tilde Z)   \exp \B(   -\blan \tilde  Z ,  \Ga_k (0,Y)    Z''  \bran   + (C_0 p_{0,k}^{-1} +  Ce^{-\ga M} )  \| \tilde Z \|^2_Y   \B)  
    \\
\end{split} 
\ee
\bigskip

\noindent
(3.) 
Let $\frac12 \beta_k  = C_0 p_{0,k}^{-1} +  Ce^{-\ga M}$ which is tiny.    The  integral is evaluated as  
\be
\begin{split}
&  \int  d \mu_{  C_k(0,Y) } ( \tilde Z) 
   \exp \B(   -\blan \tilde  Z ,  \Ga_k (0,Y)    Z''  \bran + \beta_k \| \tilde Z \|^2_Y   \B) \\
= &\left[ \frac{  \det  \B(  C_{k}(0, Y)^{-1} \B)}{\det  \B(C_{k}(0, Y)^{-1} -\beta_k1_{Y} \B)}  \right]^{\frac12}
\exp \B( \frac 12  \blan   Z'', \Ga^T_k (0,Y)\B(C_{k}(0, Y)^{-1} -\beta_k1_{Y} \B)^{-1}  \Ga_k (0,Y)  Z '' \bran \B) \\
\end{split}
 \ee

The ratio of the determinants is bounded  by 
 \be 
| \B[ \det ( I - \beta_k C_k(0,Y) ) \B] ^{-1} | \leq  \exp\B(  \beta_k \|C_k(0,Y)\|_1 \B)  \leq
  \exp \B(\beta_kCM^3 |Y|_M \B) \leq   e^{ \cO(M^{-1})Y|_M }
\ee
Next consider the quadratic  form
 $\frac 12  \blan   Z'', \Ga^T_k (0,Y)\B(C_{k}(0, Y)^{-1} -\beta_k1_{Y} \B)^{-1}  \Ga_k (0,Y) Z '' \bran$.
Expand it in powers of the small parameter $\al_k$.
The leading term   $\frac 12  \blan  Z'', \Ga^T_k (0,Y)C_{k}(0, Y) \Ga_k (0,Y) Z '' \bran$
is canceled by a corresponding term in (\ref{souci5}).
The remainder has the form
\be 
\blan   Z'', R_2  Z '' \bran =
  \blan  Z'',  \Ga^T_k (0,Y)\B[ \sum_{n=1}^{\infty}  \beta_k^n\  C_{k}(0,Y)^{n+1}  \B]  \Ga_k (0,Y)Z''  \bran   
\ee
Arguing as before this satisfies  $ |R_2(\Up, \Up') | \leq  C \beta_k e^{-\ga d(\Up, \Up') }$ and so 
$ | < Z'', R_2  Z '' >| \leq C \beta_k \|Z'' \|^2 $.
We also take  in (\ref{souci5})
 \be
  \cO(M^{-1}) | Y|_M  =  \sum_{\al} \cO(M^{-1}) | Y_{\al} |_M \leq \sum_{\al} d_M(Y_{\al}) + 1
 \ee
 Let $\beta'_k \equiv  Ce^{-\ga M} +  C\beta_k$ which is tiny.  The bound is now for $Y$ with connected components $\{ Y_{\al} \}$ and   $|s_{\sq} | \leq M^{\al_0}$
 and     $s_{(Y')^c } =0$:
\be       \label{souci6}
\|G( s,Y, Z'')\|_{\frac14 \bh_k}    \leq      e^{\beta'_k  \|Z''\|^2_{Y'} }\prod_{\al}   \one  e_k^{ \frac14- 5 \ep }  e^{ -(\ka - 6\ka_0 -7)d_M(Y) }   
\ee
\bigskip

\noindent (4.)
The function  $\breve G( Y'; Z'')$  is given in terms of  derivatives  of  $G( s,Y, Z'')$ for $|s_{\sq}|  \leq 1$  in (\ref{guide}).   Having established analyticity 
in the larger domain $|s_{\sq} | \leq M^{\al_0}$ we can estimate these derivatives    by  Cauchy bounds.   Each derivative contributes a factor $M^{- \al_0}$.
Using also the bound (\ref{souci6}) yields 
\be \label{larger}
 \|  { \breve G }( Y'; Z'')  \|_{\frac14 \bh_k}     \leq e^{ \beta'_k \| Z''\|^2_{Y'}}  \sum_{Y: \tilde Y^{r} \subset Y'}
 M^{-\al_0 |Y' - \tilde Y^r|_M  } \prod_{\al} \one   e_k^{\frac14- 5 \ep}  e^{-(\ka - 6\ka_0 -7)d_M(Y_{\al})} 
\ee

For $Y'$ connected we want to extract a factor  $ e^{-(\ka - 7\ka_0 -7)d_M(Y')} $ from the exponentials in this expression. 
 A connected component  $\tilde Y_{\beta}^r$  or $\tilde Y^r$ is the union of the  $\tilde Y^r_{\al}$  contained in it
so 
\be 
d_M( \tilde Y_{\beta}^r )  \leq  \sum_{\al: Y_{\al} \subset   \tilde Y_{\beta}^r } (d_M(\tilde Y^r_{\al} ) + 2)
\ee
and therefore
\be
 \sum_{\beta} d_M( \tilde Y_{\beta}^r )  \leq  \sum_{\al} (d_M(\tilde  Y^r_{\al} ) + 2 )
\ee
However for connected $Y$ we have     $  d_M (  \tilde Y) \leq  \one(d_M(Y) +1 )$ (see (364) in \cite{Dim15} ) and 
hence   $  d_M (  \tilde Y^r) \leq  \one(d_M(Y) +1 )$.  Therefore
\be \label{sue1}
 \sum_{\beta} d_M( \tilde Y_{\beta}^r )  \leq  \one   \sum_{\al} (d_M(Y _{\al} ) + 2 )
\ee
The same bound holds with $ \sum_{\beta}( d_M( \tilde Y_{\beta}^r )  +2)$ on the left.
Also suppose we  divide  the connected $Y'$ up  into pieces $Y'_{\beta}$  each  containing exactly one $\tilde Y^r_{\beta}$. 
Then 
\be \label{sue2}
\begin{split}
d_M(Y') \leq & \sum_{\beta} (d_M(Y'_{\beta} )+2 ) \\
\leq &  \sum_{\beta} ( |Y'_{\beta} - \tilde Y^r _{\beta}|_M + d_M( \tilde Y^r_{\beta} ) + 2)\\
= &   |Y' - \tilde Y^r|_M + \sum_{\beta} (d_M( \tilde Y^r_{\beta} ) + 2)\\
\end{split}
\ee
Combining (\ref{sue1}) and (\ref{sue2}) we have that  there is a constant $c_1 = \one, c_1 <1$ such that
\be
  c_1 d_M (Y') \leq |Y' - \tilde Y^r|_M    +  \sum _{\al} (d_M(Y_{\al}) + 2)
\ee
Assuming  $ M^{-\frac12\al_0 } \leq  e^{-(\ka - 7\ka_0 -7)}$ 
the last estimate allows us to pull  a factor  $ e^{-c_1(\ka - 7\ka_0 -7)d_M(Y')}$ out of the sum (\ref{larger}) 
leaving
\be \label{larger2}
 \|  { \breve G }( Y'; Z'')  \|_{\frac14 \bh_k}     \leq           e^{-c_1(\ka - 7\ka_0 -7)d_M(Y')}     e^{ \beta'_k \| Z''\|^2_{Y'}} \sum_{Y: \tilde Y^{r} \subset Y'}
 M^{-\frac12 \al_0 |Y' - \tilde Y^r|_M  } \prod_{\al} \one   e_k^{\frac14- 5 \ep}  e^{- \ka_0 d_M(Y_{\al})} 
\ee  
The sum here can also be written
\be
\sum_{Z \subset Y'} \sum_{Y: \tilde Y^{r} =Z}
M^{-\frac12 \al_0 |Y' - Z|_M  } \prod_{\al} \one   e_k^{\frac14- 5 \ep}  e^{- \ka_0 d_M(Y_{\al})} 
\ee
The sum over $Y = \{ Y_{\al} \}$ is estimated by 
\be
\begin{split}
&   \sum_{  \{ Y_{\al} \}: Y_{\al}  \subset Z}  \prod_{\al} \one   e_k^{\frac14- 5 \ep}  e^{- \ka_0 d_M(Y_{\al})}  
  \leq  \sum_{n=1}^{\infty}   \frac{1}{n!} \sum_{ (Y_1, \dots, Y_n) : Y_{\al} \subset Z}  \prod_{\al} \one   e_k^{\frac14- 5 \ep}  e^{- \ka_0 d_M(Y_{\al})}  \\
&   = \sum_{n=1}^{\infty}   \frac{1}{n!}\B( \sum_{Y \subset Z}\one   e_k^{\frac14- 5 \ep}  e^{- \ka_0 d_M(Y)} \B)^n 
    \leq \sum_{n=1}^{\infty}   \frac{1}{n!}\B( \one   e_k^{\frac14- 5 \ep} |Z|_M \B)^n \\
 &   \leq    \one   e_k^{\frac14- 5 \ep} |Z|_M\exp \B( \one   e_k^{\frac14- 5 \ep} |Z|_M \B) 
      \leq   \one   e_k^{\frac14- 5 \ep}e^{\frac12 c_1d_M(Y')}  \\
\end{split}
\ee
The  sum over $Z$ is estimated by 
\be 
\sum_{Z \subset Y'}  M^{-\frac12 \al_0 |Y' - Z|_M  } \leq  \B( 1 +   M^{-\frac12 \al_0 }\B)^{|Y'|_M}  \leq  \exp \B(  M^{-\frac12 \al_0 }  |Y'|_M \B)
\leq \one e^{\frac12c_1d_M(Y') }
\ee
This yields for $Y'$ connected
\be \label{lunchtime3}
 \|   \breve G ( Y'; Z'')  \|_{\frac14 \bh_k}     \leq  \one   e_k^{\frac14- 5 \ep} e^{ - c_1(\ka  -7 \ka_0 - 8) d_M(Y') }e^{ \beta_k \| Z''\|^2_{Y'}}
\ee  
For   $H(Y') =  \int   \breve G ( Y'; Z'') d \mu_{I} (Z'') $ the integral over $Z''$ is estimated by 
\be
\int  e^{ \beta_k \| Z''\|^2_{Y'} }  d \mu_{I} (Z'')  \leq e^{\beta_k' \Vol |Y'|} \leq  e^{\beta_k'M^3 |Y'|_M} \leq \one e^{c_1d_M(Y')}
\ee
and so for $Y'$ connected we can take
\be
\| H(Y')\|_{\frac14 \bh_k}  \leq  \one   e_k^{\frac14- 5 \ep} e^{ - c_1(\ka  -7 \ka_0 - 9) d_M(Y') }
\ee
\bigskip

\noindent
\textbf{Step VII}: 
Now  $H(Y) $ factors over its connected components, $H(Y) = \prod_i H(Y_i)$, and is small so  we can exponentiate
$ \Xi_{k}  = \sum_{Y } H(Y)    $  by a standard
formula  (see for example \cite{Dim11}). This yields 
\be
    \Xi_k =    \exp   \B(   E_k^\#   \B)  =
 \exp   \B( \sum_X   E_k^\#(X)   \B)  
 \ee 
where the sum is over connected $X$ and 
\be   \label{hstar}
  E_k^\#(X)  =    \sum_{n=1}^{\infty}
 \frac{1}{n!}  \sum_{(Y_1, \dots,  Y_n): \cup_i Y_i  =X}  \rho^T(Y_1, \dots,  Y_n)  \prod_i H( Y_i )
 \ee
 for a certain function $ \rho^T(Y_1, \dots,  Y_n) $  enforcing connectedness. 
This  satisfies
\be
\|   E_k^\#(X)   \|_{\frac14 \bh_k}  \leq  \one  e_k^{\frac14- 5 \ep}  e^{-c_1(\ka - 10\ka_0 -10)d_M(X)}  
\ee

 It is straightforward, but somewhat tedious,      to check  that    $ \ E^\#_k(X,\cA,  \psi )$ still has  all the symmetries. 
This   generally comes down to   a statement about the covariance.   For example for charge conjugation invariance we need
$ \sC^{-1}  \Ga( - \cA)  \sC   =  \Ga(\cA) ^T$.   This follows  from the representation   (\ref{lamp})   and the  same property for
$ S^0_{k+1}(\cA)$  as in     (\ref{superduper}).  This completes the proof of theorem \ref{cluster0}.
\bigskip

We are still working on the proof of theorem \ref{lanky}.  Updating the expression 
  (\ref{bong})  we have 
 \be     \label{bing}
\begin{split}
& \tilde    \rho_{k+1} (A_{k+1}, e^{ie_k \om^{(1)} }   \Psi_{k+1} )  
=  \cN_k N_k   \sZ_k  \de \sZ_{k+1} \sZ_k( 0) \de    \sZ_k( \cA) \\
&  \exp \Big( - \frac12   \|  d \cA \|^2     -   \fS^{0,+}_{k+1}\B(\cA, \Psi_{k+1}, \psi^0_{k+1}(\cA)\B) 
 +   E'_k\B(  \cA,   \psi^{0, \#}_{k+1}( \cA ) \B) +      \ E^\#_k\B( \cA,    \psi^{0, \#}_{k+1}( \cA )  \B)  \B)
 \B|_{\cA =   \cA^0_{k+1}}    \\
\end{split}
\ee

\newpage

\subsubsection{fermion determinant}

We now remove the gauge field from the fermion determinant

 \begin{lem}  For $\cA \in  \tilde \cR_k $
 \be       \label{listless2}  
  \frac{\de  \sZ_k  (\cA )  }{  \de     \sZ_k  (0 ) }   =      \exp  \B (       \sum_X      E_k^{\det}  (X, \cA  )    \B)
\ee
where  $E_k^{\det} (X, \cA) $ vanishes at  $\cA  = 0$    and satisfies  
\be       
|E^{\det}(X,  \cA  )    |  \leq    e_k^{\frac14 - \ep}   e^{-\ka   d_M(X)}
\ee   
\end{lem}
\bigskip

\pr   
First take $\cA$ in the larger domain $e_k^{-\frac14} \cR_k$. This is still in the region of analyticity for propagators $S_k(\cA), S_y(\cA)$.
 Take the   expression for   $\de  \sZ_k  (\cA )  $ from    (\ref{lankylanky})    and insert  the polymer expansion  for  
$ S_{k,y}(\cA)   $ from   (\ref{lounge}).    Then we  have  
\be    
\de  \sZ_k  (\cA )  
=    \exp  \B (    4  | \tz  |    \log  b_k   +   \sum_X    E_k^d (X, \cA )    \B)
\ee
where 
\be   E_k^d (X, \cA   )   = -   i \ga_3     b_k^2  \int_0^{\infty   }   \Tr \B[ \sB_{k,y}(\cA)  Q_k(\cA)   S_{k,y}(X,  \cA)Q_k^T(-\cA)  \sB_{k,y}(\cA) \B] dy
\ee
  The     bound  (\ref{lounge2}) is easily modified to   $ | S_{k,y}(X,\cA) f |
 \leq   e^{-(\ka + 1) d_M(X)}\|f\|_{\infty} $  and then as in (\ref{ttt0}), (\ref{ttt})   
\be  
\begin{split} 
 \B|     \B[  Q_k(\cA) S_{k,y} (X, \cA)Q_k^T(-\cA)   \B]_{xx'} \B|
 \leq    &  C  e^{-(\ka + 1)   d_M(X)}   \\ 
 \end{split}      
\ee
The factors  $\sB_{k,y}(\cA)$  are local  and each supplies a factor  $ \cO(y^{-1} )$ for convergence of the integral.  
 The  trace is only over the region $X$ and  gives a 
factor    $M^3 |X|_M   \leq   M^3  e^  { d_M(X) }$.  Therefore  
\be        \label{lizard}
|E^d(X,  \cA  )    |  \leq   CM^3   e^{-\ka   d_M(X)}
\ee   
For the ratio  $ \de \sZ_k  (\cA )  / \de    \sZ_k  (0 )  $   the volume factors cancel and we  have     the expression (\ref{listless2}) with  
\be        
  E_k^{\det}  (X, \cA  )   = E_k^d  (X, \cA  ) - E_k^d  (X, 0  ) 
\ee
which again satisfies the bound (\ref{lizard}).

Now take the smaller domain $\tilde \cR_k$.  Since  $E_k^{\det}  (X, 0 ) =0$  and $e_k^{-\frac14} \cA$ is in the larger domain     we have
\be
 E_k^{\det}  (X, \cA  ) = \frac{1}{2 \pi i} \int_{|t| = e_k^{-\frac14} }\frac{dt}{t(t-1) } E_k^{\det}  (X, t\cA  ) 
 \ee
 and this gives 
 \be  |E^d(X,  \cA  )    |  \leq   CM^3 e_k^{\frac14}  e^{-\ka   d_M(X)} \leq e_k^{\frac14 - \ep}  e^{-\ka   d_M(X)}
 \ee
 to complete the proof. 
 \bigskip 
 
 Now we have      
 \be     \label{bing2}
\begin{split}
& \tilde    \rho_{k+1} (A_{k+1}, e^{ie_k \om^{(1)} }   \Psi_{k+1} )  
=  \cN_k N_k   \sZ_k  \de \sZ_{k+1} \sZ_k( 0) \de    \sZ_k( 0) \\
&  \exp \Big( - \frac12   \|  d \cA \|^2     -   \fS^{0}_{k+1}\B(\cA, \Psi_{k+1}, \psi^0_{k+1}(\cA)\B) 
  - ( \vep'_k  +  \vep_k^0  )\Vol(  \bbT^0_{ N -k}) 
  \Big) \\
&   \exp  \Big(    - m'_k \blan \bpsi^{0}_{k+1}(\cA), \psi^{0}_{k+1}(\cA)  \bran  
 +   E'_k\B(  \cA,   \psi^{0, \#}_{k+1}( \cA ) \B) +  E^{\det}_k(\cA)   +         \ E^\#_k\B( \cA,    \psi^{0, \#}_{k+1}( \cA )  \B)  \B)
 \B|_{\cA =   \cA^0_{k+1}}    \\
\end{split}
\ee

\subsubsection{scaling}    \label{loomis} 
 The last step is scaling.   
We   reblock  $E'_k   =  \cR E_k$  to    $\cB \cR E_k$,    $E_k^{det}$  to   $\cB E_k^{det}$,   and     $E^\#_k  $  to   $\cB E_k^\#$,  all    defined on  $LM$-polymers.  
Then  we scale  (\ref{bing})    according to  (\ref{shoe}).     Taking account that  $\cA^0_{k+1}$ scales to  $\cA_{k+1}$, that
  $\psi^0_{k+1}(\cA_{k+1}^0)$ scales to  $\psi_{k+1}(\cA_{k+1})$, and (\ref{sundry})   we obtain  the desired form   
 \be     \label{bing3}
\begin{split}
&    \rho_{k+1} (A_{k+1}, e^{i\theta }   \Psi_{k+1} )  
= \cN_{k+1}  \sZ_{k+1}  \sZ_{k+1} ( 0)  \\
&  \exp \Big( - \frac12   \|  d \cA_{k+1} \|^2     - \fS_{k+1}\B(\cA, \Psi_{k+1}, \psi_{k+1}(\cA)\B)   
 -    \vep_{k+1}  \Vol(  \bbT^0_{ N -k-1 })   \Big) \\
&   \exp  \Big(  -  m_{k+1} \B \langle   \bpsi_{k+1}( \cA_{k+1}),    \psi_{k+1}( \cA_{k+1}) \B \rangle
 +  E_{k+1}   \B(  \cA_{k+1},   \psi^\#_{k+1}(\cA_{k+1} ) \B)  \B)
\end{split}
\ee
Here  we have made the identification from  (\ref{z}), (\ref{sammy})
\be 
\begin{split}   
& \B( \cN_k N_k  \sZ_k( 0) \de    \sZ_k( 0) L^{ -8 (s_N- s_{N-k-1}) }  \B) 
 \B(   \sZ_k    \de \sZ_{k+1}    L^{\frac12 (b_N- b_{N-k-1} )  -  \frac12 (s_N- s_{N-k-1})  } \B)  \\   & \hs  =  \cN_{k+1}  \sZ_{k+1} ( 0)  \sZ_{k+1}  \\
\end{split}
\ee
As  announced in   the theorem we have  defined $  \vep_{k+1}  =     L^3  (  \vep'_k   +  \vep_k^0   )
  =  L^3  (  \vep_k    +   \vep_k ( E_k)  +  \vep_k^0   )$,   and    
         $m_{k+1}    = Lm'_k  = L( m_k + m( E_k))$,    and  with   $  \cL E    =  (\cB  E)_{L^{-1}}   $ 
  \be
  E_{k+1}  =       \cL \B( \cR E_k  +   E_k^{\det} +     E_k^\#    \B)
  \ee

 We already have a bound on  $\vep_k^0$  so all that remains is a bound on the kernel of   $E_k^* = \cL E_k^\# $.  
We  would like to use  the bound   (\ref{L4})   and   $  \|  E^\#_k \|_{k}    \leq    \one  e_k^{\frac14  - 5 \ep}  $ to   obtain  
\be   \label{hut3}
    \| \cL  E_k^\#\|_{k+1}     \leq    
 \cO(1)  L^3       \|  E^\#_k \|_{k}    \leq    \one L^3  e_k^{\frac14  - 5 \ep}    \leq   e_k^{\frac14  - 6 \ep} 
  \ee 
 But this   is not quite correct since   our bound  (\ref{osprey})  does not give   the  bound on   $  \|  \hat E^\#_k \|_{k} $  but on a somewhat different  quantity.  We have to revisit the proof of  (\ref{L4}) using the actual bound   (\ref{osprey}).  
    If   $\cA \in  \tilde \cR_{k+1}$ then     $\cA_L  \in  \frac12 \tilde  \cR_k$ so  after scaling we are in the domain needed for  (\ref{osprey}).   
 The  fermion field parameter  in  (\ref{osprey})   is $\frac14 \bh_k$  not   $\bh_k$.    But this does not affect the  derivation of  (\ref{L4})
 since we can take   $\bh_{k+1}  \leq  \frac14  \bh_k$ rather  than    $\bh_{k+1}  \leq    \bh_k$.   Finally  the decay factor 
 in   (\ref{osprey}) is   $e^{-c_1(\ka - 10\ka_0 -10)d_M(X)}  $
 rather than    $ e^{ -   \ka    d_{M}(X)}$.    This  means  that in  the bound  (\ref{xmas})   we  get    $\one L^3   e^{-Lc_1(\ka - 10\ka_0 -10)d_M(X)}  $
   instead of   $\one L^3   e^{ -L (  \ka     - \kappa_0 -1   ) d_{M}(Y)} $.   For $L$ large this is still dominated by     $\one L^3   e^{ - \ka   d_M(Y)} $.   Thus   the conclusion   of  (\ref{hut3}) is
 still valid.
  
 This completes the proof of theorem \ref{lanky}.

\section{The flow}

We    seek  well-behaved    solutions of   the    RG   equations  (\ref{recursive}).  
Thus we  study
 \begin{equation}  \label{recursive3}
\begin{split}
\vep_{k+1}   =&  L^3 \B( \vep_k    +  \vep(  E_k)  + \vep_k^0)\B) \\
m_{k+1}   =&   L \B(  m_k   + m (   E_k) \B) \\
E_{k+1}   =&  \cL\B(\cR  E_k  +     E_k^{\det}   +  E^\#_k(   m_k,  E_k) \B)  \\
 \end{split}
\end{equation}
This can be iterated at most up  to $k= K \equiv N-m$ since at this point  we are on the torus  $\bbT^0_{N-K} =  \bbT^0_m$ which consists of
a single $M= L^m$ cube. 
Our  goal is to show  that for   any    $N$   we can choose  the  initial   point  so  that the solution exists   for  $k=0,1,  \dots,  K$
and finishes  at  preassigned values  $(\vep_K,  m_K)   =    (\vep^N_K,  m^N_K)       $  independent of  $N$.   This procedure  is  nonperturbative  renormalization -  the initial  values for  $(\vep_0,  m_0)  =   (\vep^N_0,  m^N_0)  $  will depend  $N$.  
Our proof follows the analysis  in  \cite{Dim11}, \cite{Dim15}.

  Arbitrarily  fixing the final values at  zero,    and  starting with  $E_0 =0$ as dictated by the model,   
we look for solutions  $\vep_k,  m_k,  E_k$    for    $k = 0,1,2,  \dots  , K$
satisfying  
\begin{equation} \label{bc}
\vep_K  = 0       \hs      m_K  = 0     \hs     E_0  =0    
  \end{equation}
  This makes the effective mass  $ \bar m_K + m_K= \bar m_K  = L^{-m} \bar m$.  
At  this point we  temporarily drop the equation for the   energy density  $\vep_k$ and just  study  
 \begin{equation}  \label{recursive4}
\begin{split}
m_{k+1}   =&   L \B( m _k   + m(  E_k)  \B) \\
E_{k+1}   =&  \cL\B( \cR E_k  +     E_k^{\det}   +  E^\#_k(   m_k,  E_k) \B)  \\
 \end{split}
\end{equation}

Let    $\xi_k  =  (m_k, E_k)$  be  an  element of   the complete metric space   
  $\bbR \times   \cK_k  $  where     $   \cK_k$  is the Banach space  defined in section \ref{polymersection}.   
            Consider sequences      
\begin{equation}
\underline{ \xi }  =  ( \xi_0,  \dots  ,   \xi_K)
 \end{equation}
 Let    $\sB$   be the    space of all   such   sequences
 with norm   
  \begin{equation}
\|  \underline{\xi } \|  =  \sup_{0 \leq  k   \leq  K}       \{  e_k^{- \frac34 + 8 \ep }  | m_k|, \
 e_k^{-\frac14 + 7 \ep}  \| E_k \|_{k}   \}  
\end{equation}
  Let  $\sB_0$   be the   subset  of all sequences satisfying the  boundary 
 conditions.
 Thus   
 \begin{equation}
 \sB_0  =  \{\underline{ \xi}   \in  \sB:    m_K  =  0,        E_0  =0\} 
 \end{equation}
  This is a complete metric space  with distance  
  $\| \underline {  \xi } - \underline {\xi' }  \|$.
 Finally   let  
 \begin{equation}
 \sB_1  = \sB_0 \cap    \{\underline{ \xi}   \in  \sB:    \| \underline{\xi} \|   < 1  \}   
 \end{equation}
Note that  the condition   $\|\xi  \|  \leq   1$  implies       
\be  
|m_k|   <  e_k^{\frac34 - 8 \ep  }    \hs    \|E_k  \|_k   <    e_k^{\frac14 - 7 \ep}   
\ee
   so   we  are well   within  the domain of validity for the main theorem.

 Next  define an    operator  $\un{\xi' } =T \un{\xi } $    by      
  \begin{equation}  \label{recursive5}
\begin{split}
m'_k   =&   L^{-1}m_{k+1}  -  m(E_k)    \\
E'_k   =&      \cL\B(  \cR  E_{k-1}  +     E_{k-1}^{\det}   +  E^\#_{k-1}(   m_{k-1},  E_{k-1}) \B)  \\
 \end{split}
\end{equation}
Then   $\underline{  \xi }$ is a solution  of  (\ref{recursive4}) and the boundary conditions iff  it is a fixed point for  $T$ on 
$\sB_0$.     We    look for  such   fixed points  in $\sB_1$.

\begin{lem}   Let  $L$ be sufficiently large and $e$    sufficiently small.  Then   for all  $N$  
\begin{enumerate}
\item   The  transformation   $T$     maps   the set
  $\sB_1$  to itself.  
\item   There is a unique fixed point   $T\underline{ \xi}  = \underline{ \xi }$  in this  set. 
\end{enumerate}
\end{lem}
\bigskip

\pr 
We  use the bound from  (\ref{sundown})  
\be  \label{tingle1}
  |m(E_k)|  \leq  \one  e_k^{\frac12 }\| E_k \|_k   \hs
\ee
We also use    
\be  \label{tingle2} 
\begin{split}
\|  \cL \cR E_{k-1}  \|_{k}   \leq  & \one  L^{-\frac14 +2 \ep}   \| E_{k-1}  \|_{k-1}  
  \hs     \| \cL E_{k-1}^{\det}  \|_{k} <   e_k^{\frac14- 2\ep} \\
    \| E^*_{k-1} \|_k  =&    \|\cL  E^\#_{k-1}  \|_{k}  \leq   e_{k-1}^{ \frac14 - 6 \ep}\\
 \end{split}
 \ee
 The first is the key contractive estimate   (\ref{lugnuts}).   The second follows since  $ \| E_{k-1}^{det} \|_{k-1}  \leq  e_k^{\frac14- \ep}$  so
 $  \| \cL E^{\det}_{k-1} \|_{k}  \leq  \one L^3e_k^{\frac14 - \ep} \leq e_k^{\frac14 -2 \ep} $. 
  The third is (\ref{nuts}).   
 
\bigskip 

\noindent  (1.) To   show the the map sends  $\sB_1$ to itself
we  estimate 
using  $e_{k+1} = L^{\frac12} e_k$
\begin{equation}  \label{jelly}
\begin{split}
e_k^{- \frac 34 + 8 \ep   } |  m'_k| 
 \leq   &
  e_{k}^{- \frac 34 + 8 \ep    }  \B(   L^{-1} | m_{k+1}|    +  \one   e_k^{\frac12}\| E_k \|_k   \B)\\
  \leq   &
   L^{- \frac 58 }\B[e_{k+1}^{- \frac34 + 8 \ep   }  | m_{k+1}|\B] + \one   e_{k}^{\ep }  \B[  e_k^{-\frac14 + 7 \ep} \| E_k \|_k \B] \\
    \leq  & 
     \|  \underline  \xi  \|   <  1 \\
\end{split}
\end{equation}
We   also   
have 
\begin{equation}
\begin{split}
 e_{k}^{-\frac14 + 7 \ep }\|E'_k\|_{k}  
   \leq   & e_{k}^{-\frac14 + 7 \ep } \B(\one  L^{-\frac14 +2\ep}    \|  E_{k-1} \|_{k-1}    + e_k^{\frac14 - 2\ep} +    e_{k-1}^{ \frac14 - 6 \ep}\B)  \\
   \leq    &  \one   L^{  -\frac38 + 6\ep }  \Big[ e_{k-1}^{-\frac14 + 7 \ep } \| E_{k-1} \|_{k-1} \Big]  +  e_k^{\ep} \\
       \leq  & \frac 12  \|  \underline  \xi  \|  +   \frac12    <  1  \\
     \end{split}
\end{equation}
 Hence  $ \|  T (\underline{ \xi  } ) \|   <  1$  as  required.
\bigskip

\noindent  (2.)
  It suffices  
 to show   that   the mapping is a contraction.
 We show  that   under our assumptions
\begin{equation}
\|  \underline{ \xi'_1}- \underline{ \xi'_2} \| =
\| T( \underline{ \xi_1}) -   T( \underline{ \xi_2}) \|  \leq  \frac12
\|  \underline{ \xi_1}- \underline{ \xi_2} \| 
\end{equation}

First consider the  $m$  terms.  Since  $m(E)$ is linear  
 $|m(  E_{1,k}) -  m(  E_{2,k})| \leq  \one  e_k^{\frac12 } \| E_{1,k}-  E_{2,k}\|_k$
  so 
   \begin{equation}  \label{filet1}
\begin{split}
  e_k^{- \frac 34 + 8 \ep   }|m'_{1,k}-m'_{2,k}| 
  \leq &  
  e_k^{- \frac 34 + 8 \ep   }\B(L^{-1}    | m_{1,k+1}-  m_{2, k+1} | + \B| m(  E_{1,k}) -  m(  E_{2,k}) \B|  \B)   \\
 & \leq  
   L^{- \frac58 } \B[e_{k+1}^{- \frac 34 + 8 \ep   }       | m_{1,k+1}-  m_{2, k+1} | \B]
  + 
   \one e_k^{\ep} \B[  e_k^{-\frac14 + 7 \ep}  \|  E_{1,k} -E_{2,k}\|_k  \B]   \\
  & \leq  \frac12  \|  \underline{ \xi_1}- \underline{ \xi_2} \| \\  
  \end{split}
\end{equation}

Now consider the $E$ terms. The  term  $\cL E_{k-1}^{\det}$ cancels and we have  with  $E_{k-1}^*  = \cL E_{k-1}^\#$
\begin{equation}
\begin{split}
E'_{1,k}-E'_{2,k}   = &  \cL \cR (E_{1,k-1}  -  E_{2,k-1})  
  +  ( E^*_{k-1}(m_{1, k-1}, E_{1, k-1})  -  E^*_{k-1}(m_{2, k-1}, E_{2, k-1})  )  \\     
\end{split}
\end{equation}
Then
\begin{equation}  \label{slippery}  
\begin{split}
e_k^{-\frac14 + 7 \ep} \|E'_{1,k}-E'_{2,k}\|_{k} 
\leq   
 &
   \one e_k^{-\frac14 + 7 \ep} 
L^{-\frac14 + 2 \ep }   \|  E_{1,k-1}  -      E_{2,k-1}\|_{k-1}   \\ 
  +  &  
 e_k^{-\frac14 + 7 \ep}  \|  E^*_{k-1}(m_{1, k-1}, E_{1, k-1})  - E^*_{k-1}(m_{2, k-1}, E_{2, k-1})   \|_k    \B)   \\
\end{split}  
\end{equation}
For the    first  term  in (\ref{slippery})  
we have   
\be \label{greenie}
      \one e_k^{-\frac14 + 7 \ep} 
L^{-\frac14 + 2 \ep }   \|  E_{1,k-1}  -      E_{2,k-1}\|_{k-1}  \leq  \one  L^{-\frac38 + 6 \ep  } \B[ e_{k-1}^{-\frac14 + 7 \ep} \|  E_{1,k-1}  -      E_{2,k-1}\|_{k-1} \B]
\leq \frac16  \|  \underline{ \xi_1}- \underline{ \xi_2}\| 
\ee
 For the   second    term  in (\ref{slippery}) we  write  
\be   \label{slippery2}
\begin{split}
&  e_k^{-\frac14 + 7 \ep} \|  E^*_{k-1}(m_{1,k-1}, E_{1,k-1}) - E^*_{k-1}(m_{2,k-1}, E_{2,k-1}) \|  \\
&\leq   e_k^{-\frac14 + 7 \ep} \|\B( E_{k-1}^*(m_{1,k-1}, E_{1,k-1}) - E_{k-1} ^*(m_{2,k-1}, E_{1,k-1})\B)\|\\
&+
 e_k^{-\frac14 + 7 \ep}  \| \B( E_{k-1}^*(m_{2,k-1}, E_{1,k-1}) - E_{k-1} ^*(m_{2,k-1}, E_{2,k-1}) \B)\|  \\
\end{split}
\ee
Now   $E^*_k(m,E) $ is actually an analytic function of its arguments so  
for the first  term in  (\ref{slippery2}) we can  write for         $r>1$   
  \begin{equation}
\begin{split}
 &   E_{k-1}^*(m_{1,k-1}, E_{1,k-1}) - E_{k-1} ^*(m_{2,k-1}, E_{1,k-1})   \\
  & =      \frac{1}{2 \pi i}
 \int_{|t|  =r }     \frac{dt}{t(t-1)}      E^*_{k-1}   \Big(m_{2,k-1} +  t( m_{1,k-1}- m_{2,k-1}), E_{1,k-1})  \Big)   \\
 \end{split}
\end{equation}
We  can assume  $ m_{1,k-1}\neq   m_{2,k-1}$   and   take    $r  =3 e_{k-1}^{ \frac34 - 8 \ep    } | m_{1,k-1}- m_{2,k-1}|^{-1}$.    This is greater than one since   $|m_{1,k-1}- m_{2,k-1}| \leq  e_{k-1}^{ \frac34 - 8 \ep    }  \|  \underline{ \xi_1}- \underline{ \xi_2} \|  \leq   2  e_{k-1}^{ \frac34 - 8 \ep    } $.  Also it keeps  us well inside the domain for    
$ E^*_{k-1}  $ as given by the main theorem.   Hence we can use the estimate $ \|E^*_{k-1}   \|_{k}  \leq   e_{k-1}^{ \frac14 - 6 \ep}$ from  (\ref{tingle2}).   
 Hence the first term in (\ref{slippery2}) is 
bounded by   
\be   \label{greenie1} 
 \one e_k^{-\frac14 + 7 \ep}    \B[  e_{k-1}^{-\frac34 + 8 \ep   } |m_{1,k-1}-m_{2,k-1}|  \B ]   e_{k-1} ^{ \frac14- 6 \ep } 
      \leq    \one e_k^{\ep}    \|  \underline{ \xi_1}- \underline{ \xi_2} \|  \leq  \frac16 \|  \underline{ \xi_1}- \underline{ \xi_2} \| 
\ee
For the second term  in   (\ref{slippery2})   we write  for  $r>1$
\begin{equation}
\begin{split}
&   E^*_{k-1}(m_{2,k-1}, E_{1,k-1}) -E^*_{k-1}(m_{2,k-1}, E_{2,k-1})  \\
  &=    \frac{1}{2 \pi i}
 \int_{|t|  =r }     \frac{dt}{t(t-1)}      E^*_{k-1}   \Big(m_{2,k-1},  E_{2,k-1} +  t( E_{1,k-1}- E_{2,k-1})  \Big)   \\
 \end{split}
\end{equation}
Now we       take   $r= 3 e_{k-1}^{\frac14-7 \ep}  \|E_{1,k-1} - E_{2,k-1} \|^{-1}_{k-1}  $
which is bigger than one since   $\|E_{1,k-1} - E_{2,k-1}  \|_{k-1}   \leq    2  e_{k-1}^{\frac14-7 \ep}  $,  and it keeps us 
well within the domain of    $E^*_{k-1}$.  
Again using  $ \|E^*_{k-1}   \|_{k}  \leq   e_{k-1}^{ \frac14 - 6 \ep}$   this term is bounded by       
\be   \label{greenie2} 
 \one e_k^{-\frac14 + 7 \ep}    \B[  e_{k-1}^{-\frac14 + 7 \ep   } \|E_{1,k-1} - E_{2,k-1} \|_{k-1}   \B ]   e_{k-1} ^{ \frac14- 6 \ep } 
      \leq    \one  e_k^{\ep}    \|  \underline{ \xi_1}- \underline{ \xi_2} \|  \leq  \frac16 \|  \underline{ \xi_1}- \underline{ \xi_2} \| 
\ee
Combining (\ref{greenie}), (\ref{greenie1}), (\ref{greenie2})  yields 
$
e_k^{-\frac14 + 7 \ep}   \| E'_{1,k}-E'_{2,k}  \|_{k} 
\leq   \frac12    \|  \underline{ \xi_1}- \underline{ \xi_2} \|  
$ which together with        (\ref{filet1}) gives the desired result   $    \|  \underline{ \xi'_1}- \underline{ \xi'_2} \|  \leq
 \frac12    \|  \underline{ \xi_1}- \underline{ \xi_2} \|  $.
\bigskip

 Now  we  can state:

 \begin{thm}  \label{gsf}  Let  $L$ be  sufficiently large and  $e$ be  sufficiently  small.     Then  for  each    $N $  there is a unique sequence
  $\vep_k,  m_k,  E_k$    for    $k = 0,1,2,  \dots, K=N-m$   
satisfying of  the dynamical equation   (\ref{recursive3}) and   the boundary conditions  (\ref{bc}),     
and    with    $e_k =  L^{-\frac12(N-k)}e$
\begin{equation}  \label{somewhat}
  | m_k|  \leq     e_k^{\frac34 - 8 \ep  }  \hs
      \| E_k \|_{k}   \leq  e_k^{\frac14- 7 \ep}  
\end{equation}
Furthermore  
 \begin{equation}  \label{eg}
|\vep_{k}|   \leq    2 e_k^{\frac14- 7 \ep}  
 \end{equation}
\end{thm}
\bigskip

\pr   This  solution  $(m_k, E_k)$  is the fixed point from the previous  lemma and  the bounds  (\ref{somewhat})  are
a consequence.  

The energy density   $\vep_k$ is determined   by   
\be      \vep_k =  L^{-3}  \vep_{k+1}   -  \vep(  E_k)  - \vep_k^0
\ee
with the final condition  $\vep_K=0$,  which however we  now treat as a initial condition.   From  
(\ref{sundown})  we have   $ |\vep(  E_k)|  \leq  c \|E_k \|_k  \leq  c e_k^{\frac14- 7 \ep}  $ for some constant  $c  =  \one$,
and we   also have     $|\vep_k^0|    \leq     e_k^{n_0} $.   We claim that
\be      |\vep_k| \leq      2c e_k^{\frac14- 7 \ep}  
\ee
It is true for  $k=K$. We suppose it is true for  $k+1$ and prove it for $k$.   This follows for $L$ large by    
\be    
  | \vep_k|   \leq  L^{-3} 2c e_{k+1}^{\frac14- 7 \ep}     +  c  e_{k}^{\frac14- 7 \ep}    + e_k^{n_0}      \leq    2c  e_{k}^{\frac14- 7 \ep}
\ee
This completes the proof. 
\bigskip

\rem  The    treatment  of   scalar QED$_3$  in  \cite{Dim15}    featured a more awkward
treatment of the  matter   determinant  (as did  an earlier version of this paper).  This could be improved by adopting the present strategy 
for determinants.      Also the iteration in  
that paper should have been stopped at $K=N-m$  rather than $N$. 
\bigskip

\begin{appendix}

 \section{Grassman integrals}    \label{A}  We  develop facts about Grassman algebras and associated integrals.  General 
   references are   \cite{Sal99},  \cite{FKT02}.

 \subsection{basic  estimates}   \label{A1}

Consider  the Grassman algebra generated by  $\{\Psi(s) \}_{s \in S}$    where $(S, \mu)$ is    a finite  measure space.  The general element has the form   
\be  
E=  E(\Psi)  =      \sum_n   \frac{1}{n!}   \sum_{s_1, \dots, s_n }   E_n(s_1,  \dots,  s_n)  \Psi(s_1)  \cdots  \Psi (s_n)   \mu(s_1) \cdots  \mu(s_n)
\ee 
 where   $ E_n(s_1,  \dots,  s_n) $ is totally anti-symmetric in its  arguments and  $\Psi(s) \Psi(s') = -  \Psi(s') \Psi(s)$.  Pick a fixed 
 ordering for  $S$.   Then we   can also write this as a sum over ordered   subsets  $I  =  (s_1, \dots,  s_n)$ of variable length      as 
 \be   \label{standard}
    E (\Psi)=     \sum_I     E(I)  \Psi(I ) \mu(I)
 \ee
 where  
 \be   E(I) =       E(s_1,  \dots,  s_n) \hs   \Psi(I)   =   \Psi(s_1)  \cdots  \Psi (s_n)  \hs   \mu(I)  = \mu(s_1) \cdots  \mu(s_n)
\ee

We define a norm   
\be
\|  E \|_h   =     \sum_n   \frac{h^n }{n!}   \sum_{s_1, \dots, s_n }  | E_n(s_1,  \dots,  s_n) |  \mu(s_1) \cdots  \mu(s_n)
\ee
which can also be written 
\be
\|  E \|_h   =       \sum_I  h^{|I|}   |E(I)| \mu(I)
\ee
 
 \begin{lem}  \label{shalala}
 \be
 \|  E F\|_h    \leq  \|  E \|_h  \|  F \|_h  
 \ee
 \end{lem}  
 \bigskip
 
 \pr   We have   
 \be    E F=     \sum_{I,J}      E(I)F(J) \Psi(I) \Psi(J)  \mu(I)  \mu(J)    \ee
Only terms  with  $I \cap J = \emptyset$ contribute   and we  classify them by   $I \cup J$ so   
\be   
  E F=  \sum_K   \left[  \sum_{  I \cup J = K,  I \cap J = \emptyset }      E(I)F(J)   \sgn((I, J) \to K)  \right] \Psi(K)   \mu(K)  
    \ee
  where      $ \sgn((I, J) \to K)  $ is the sign of the permutation taking  $(I,J)$ to  $K$. 
 Then   dropping the condition  $ I \cap J = \emptyset$
 \be 
 \begin{split}
   \| EF \|_h   =   &  \sum_K  h^{|K|}  \left[ \sum_{  I \cup J = K,  I \cap J = \emptyset }      E(I)F(J)   \sgn(I, J \to K)  \right]  \mu(K)  \\
   \leq     &  \sum_K   \sum_{  I \cup J = K,  I \cap J = \emptyset }     h^{|I| + |J| }     E(I)F(J)     \mu(I)   \mu(J)  \\
     \leq   &   \sum_{  I ,J }  h^{|I| + |J| }      E(I)F(J)    \mu(I)   \mu(J)  =  \|E \|_h  \|F \|_h \\
 \end{split}
 \ee
\bigskip

Next  consider  the Grassman algebra generated by   $\Psi$  indexed by  $(S_1, \mu_1)$  and  $\chi$   indexed by  $(S_2, \mu_2)$. 
 The general element can be  written
  \be  \label{sudsy} 
E  =      E(\Psi, \chi ) =   \sum_{I \subset  S_1,  J \subset S_2}   E(I,J)  \Psi(I) \chi(J)  \mu_1(I)   \mu_2(J)
  \ee
  An associated  norm  depends on two parameters   $h,k$   and  is  
  \be  
  \|    E    \|_{h,k}=   \sum_{I,J} h^{|I|}  k^{|J|}   |E(I,J)|  \mu_1(I)   \mu_2(J)
\ee

 \begin{lem}    \label{G}  Let   $\Psi,  \Psi'$  be  indexed by the same  $(S, \mu)$  and  define  
 \be     
    E^+(  \Psi, \Psi' )   =  E( \Psi + \Psi' )
 \ee
 Then   
 \be     \| E^+  \|_{h, h' }     \leq    \| E \|_{h+h' }  
 \ee
 \end{lem}
 \bigskip
 
 \pr  With  $E(\Psi)$ of the form   (\ref{standard}) 
 \be   
 \begin{split}
 E^+(\Psi, \Psi' )   = &   \sum_K   E(K)( (\Psi + \Psi')(K) )\mu(K)     \\ 
  =   &\sum_K     E(K)\B[ \sum_{I\cup J =  K,  I \cap J = \emptyset}     \sgn((I, J)  \to   K )   \Psi(I)   \Psi'(J)  \B] \mu(K)    \\
    =   &\sum_{  I \cap J = \emptyset}  \B[ E(I \cup J)    \sgn(I, J ) \B]  \Psi(I)  \Psi'(J)\mu(I) \mu(J)   \\
 \equiv     &   \sum_{  I,J}   E^+(I, J )   \Psi(I)  \Psi'(J) \mu(I) \mu(J)  \\
\end{split}  
 \ee
 where    $ \sgn(I, J )$  is the sign of the permutation that  puts  $(I,J)$ is  standard order  and   
 $ E^+(I, J ) =    E(I \cup J)    \sgn(I, J )$ if  $I \cap J = \emptyset$ and is zero otherwise.
 Therefore
 \be   
 \begin{split}
\|  E^+ \|_{h,h'} 
=  &    \sum_{I,J } h^{|I|}  (h')^{|J|}   |    E^+(I,J)| \mu(I) \mu(J)\\ 
  \leq     &     \sum_{ I \cap J = \emptyset}  h^{|I|} ( h')^{|J|} | E(I \cup J) |\mu(I) \mu(J)    \\
  =   &   \sum_K        \sum_{I \cup J = K,   I \cap J = \emptyset}  h^{|I|}  (h')^{|J|} | E(K) |  \mu(K) \\
   =   &   \sum_K     (h+h')^{|K|  }    | E(K) |    \mu(K)  \\
   = &    \| E\|_{h  + h' }    \\
  \end{split}  
 \ee

\subsection{dressed Grassman variables}  \label{A2}
Now  suppose  we  introduce  dressed fields $\psi(t)$   defined on a new measure space  $(T,  \nu)$  and defined by 
\be     \label{snookie0}
 \psi (t)  =   (\cH  \Psi)(t)   \equiv  \sum_s \cH (t,s) \Psi (s)  \mu(s)  
\ee
We  consider    elements of the algebra of the form  
\be      \label{snookie1}
  E= E(\psi)   =       \sum_n   \frac{1}{n!}   \sum_{t_1, \dots, t_n }   E_n(t_1,  \dots,  t_n)  \psi(t_1)  \cdots  \psi (t_n)   \nu(t_1) \cdots  \nu(t_n)
\ee
or for ordered  subsets  $I \subset  T$    in some fixed ordering  
 \be    \label{snookie2}
    E (\psi) =     \sum_I     E(I)  \psi(I ) \nu(I)
 \ee
 We  define a norm on the kernel    by     
\be
\|  E \|_h   =     \sum_n   \frac{h^n }{n!}   \sum_{t_1, \dots, t_n }  | E_n(t_1,  \dots,  t_n) |  \nu(t_1) \cdots  \nu(t_n)
\ee
which    can also be written  
\be   
\|  E \|_h   =       \sum_I  h^{|I|}   |E(I)| \nu(I)
\ee
Since   $\psi(t) \psi(t')  =  -   \psi(t') \psi(t)$  we have as in  lemma  \ref{shalala}     that if  $G = EF$ then  the 
kernels satisfy   
\be      \label{enron}
 \| G \|_h    \leq  \|  E \|_h  \|  F \|_h  
 \ee

Next  we estimate  the basic norm in terms of the kernel norm.  We  use the norm
\be 
  \| \cH  \|_{1, \infty}   \equiv    \sup_{t \in T }  \sum_{s \in S} | \cH (t, s ) | \mu(s)  
\ee

\begin{lem}    \label{H}  If   $E' ( \Psi) =  E( \cH \Psi)$
as in   (\ref{snookie0}), (\ref{snookie1})    then   
\be
  \| E' \|_h   \leq   \| E  \|_{  \|\cH  \|_{1, \infty}  h}
\ee
\end{lem} 
\bigskip

\pr  
Start with
  \be      \label{snookie4}
  E'(\Psi)=     \sum_n   \frac{1}{n!}   \sum_{s_1, \dots, s_n } E'_n(s_1,  \dots,  s_n)   \Psi (s_1)   \cdots    \Psi (s_n)   \mu(s_1) \cdots  \mu(s_n)    
\ee
where
\be  
E'_n(s_1,  \dots,  s_n)   =     \sum_{t_1, \dots, t_n }  E_n(t_1,  \dots,  t_n) \prod_{i=1}^n  \nu(t_i)\cH (t_i,s_i)
\ee
 is totally     anti-symmetric under permutations.   Then
  \be
 \begin{split}
    \| E'\|_h  \leq    &   \sum_{n=0}^{\infty}    \frac{h^n}{n!}    \sum_{s_1, \dots, s_n}  \B[   \sum_{t_1, \dots, t_n } | E_n(t_1,  \dots,  t_n) |\prod_{i=1}^n | \nu(t_i)\cH (t_i,s_i)|  \B]  \mu(s_1) \cdots  \mu(s_n)  \\
  \leq    &       \sum_{n=0}^{\infty}    \frac{( \|\cH  \|_{1, \infty} h)  ^n}{n!}      \sum_{t_1, \dots, t_n }  |E_n(t_1,  \dots,  t_n)|  \nu(t_1)\cdots  \nu(t_n)     \\
  =   &    \| E \|_{ \cH  \|_{1, \infty} h}  \\
 \end{split}
 \ee

\subsection{several dressed Grassman variables  }   \label{A3}

 The  previous results   generalize directly to the case of  two or more    dressed fields.   Again consider the Grassman algebra generated
 by  $\Psi$  on  measure spaces    $(S, \mu)$  
 Define  new fields   $\psi_1, \psi_2 $  on  measure spaces   $(T_1, \nu_1),   (T_2, \nu_2)     $ by   
 
 \be
     \psi_i(t_i)   =   ( \cH_{i}  \Psi )(t_i)   =   \sum_{s \in S}  \cH_{i}(t_i, s)  \Psi(s )  \mu(s)  \hs     i=1,2 
 \ee  
We  consider    elements of the algebra of the form  
\be   \label{lanky1} 
E=   E( \psi_1,  \psi_2)   
=    \sum_{I,J}  E(I,J)  \psi_1(I)  \psi_2(J)   \nu_1( I)   \nu_2(J)
\ee
where   $I$ is an ordered subset from   $T_1$  and    $J$ is an ordered subset from   $T_2$.   
This can also be written  without ordering   as
\be      \label{slumber 1}
\begin{split}
  E(\psi_1, \psi_2)  =   &  \sum_{n, m}   \frac{1}{n! m!}   \sum_{t_1, \dots, t_{n},  t'_1,  \dots,  t'_{m} }   E_{nm}(t_1,  \dots,  t_{n}, t'_1, \dots  t'_{m}) \\
&    \psi_1(t_1)  \cdots  \psi_1 ( t_{n})       \psi_2(t'_1)  \cdots  \psi_2 ( t'_{m})     \prod_{i=1}^{n}   \nu_1(t_i)  \prod_{i=1}^{m}   \nu_2(t'_i)
      \\
\end{split}  
\ee
The associated norm on the  kernels  depends on parameters   $\bh  =(h_1, h_2)$
and  is  
\be
\| E   \|_{\bh}    
=    \sum_{I,J }h_1^{|I|}   h_2^{|J|}   | E(I,J ) |   \nu_1( I)   \nu_2(J )
\ee
which   is also written  
\be      \label{snookie8}
\begin{split}
 \| E \|_{\bh}   =   &  \sum_{n,m}\frac{h_1^{n} h_2^{m}}{n!m!}  
   \sum_{t_1, \dots, t_{n},  t'_1,  \dots,  t'_{m} }  | E_{nm}(t_1,  \dots,  t_{n}, t'_1, \dots  t'_{m})|     
     \prod_{i=1}^{n}   \nu_1(t_i)  \prod_{i=1}^{m}   \nu_2(t'_i)
      \\
\end{split}  
\ee
As in lemma   \ref{shalala} we  have again that 
 if  $G = EF$ then 
 \be   \label{slough}
 \| G \|_{ \bh}     \leq  \|  E \|_{ \bh}  \|  F \|_{ \bh} 
 \ee
 
 \begin{lem}  \label{lunge}  If   $ E' (\Psi) =  E( \cH_1 \Psi, \cH_2 \Psi)$  as above    then
\be
\| E' \|_h   \leq     \|   E  \|_ { \| \cH_1 \|_{1, \infty}h,  \| \cH_2  \|_{1, \infty} h }
\ee
\end{lem}       
\bigskip

\pr
We have
\be  \label{soapy}
\begin{split}
E'( \Psi)   = &  \sum_{n,m} \frac{1}{n!m!} \sum_{ s_1, \dots, s_n, s'_1, \dots  s'_n}E'_{nm}(s_1, \dots, s_n, s'_1, \dots  s'_n)\\
&\Psi(s_1) \cdots  \Psi(s_n)\Psi(s'_1) \cdots  \Psi(s'_n)  \prod_i \mu(s_i)  \prod_j \mu(s'_j)\\
\end{split}
\ee
where
\be
\begin{split}
& E'_{nm}(s_1, \dots, s_n, s'_1, \dots  s'_n)  \\
 & =   
    \sum_{t_1, \dots, t_n, t_1' \dots,  t_n' }  E_{nm}(t_1,  \dots,  t_n, t_1' \dots,  t_n' )
     \prod_{i=1}^n  \nu(t_i)\cH_1 (t_i,s_i)  \prod_{j=1}^m  \nu(t'_i)\cH_2 (t'_j,s'_j)\\
\end{split}     
\ee
 is      anti-symmetric under permutations within each group.
We can rewrite  (\ref{soapy})  as 
\be
\begin{split}
E'(\Psi ) = &  \sum_{I \cap J  = \emptyset}   E'(I,J) \Psi(I)  \Psi(J) \mu(I) \mu(J)  \\
=  &    \sum_K  \left[  \sum_{I \cup J  = K,  I \cap J = \emptyset}   E'(I,J) \sgn(I,J \to K)   \right]   \Psi(K) \mu(K)\\
\equiv &  \sum_k   E'(K)   \psi(K)  \mu(K)\\
\end{split} 
\ee
Then  
\be   
\begin{split}
\|   E' \|_h    = &  \sum_K  h^{|K|} |E'(K) |  \mu(K)  \\
\leq  &   \sum_K    \sum_{I \cup J  = K,  I \cap J = \emptyset} h^{|I|}   h^{|J|} E'(I,J) |   \mu(I) \mu(J)\\
\leq  &    \sum_{I, J } h^{|I|}   h^{|J|}| E'(I,J) |   \mu(I) \mu(J)\\
\end{split}
\ee
which  can be rewritten as
\be
\begin{split}
\| E' \|_h 
\leq
& \sum_{n,m} \frac{h^{n+m}}{n!m!}\B[ \sum_{s_1, \dots, s_n, s'_1, \dots  s'_n} |E'(s_1, \dots, s_n, s'_1, \dots  s'_n)|
  \prod_i \mu(s_i)  \prod_j \mu(s'_j) \B] \\
\leq     &\sum_{n,m} \frac{h^{n+m}}{n!m!} \| E_{nm}  \|  \|  \cH_1  \|_{1, \infty}^n    \| \cH_2 \|_{1, \infty}^m  \\
=  & \|   E  \|_ { \| \cH_1 \|_{1, \infty}h,  \| \cH_2  \|_{1, \infty} h }\\
\end{split}
\ee

 \subsection{further results} 
  We  list some further  results.     Theses are  mostly variations of earlier results,  but now involve fields $\psi_1, \psi_2$ which
can both be dressed fields.    They  are straightforward to check.

  \begin{itemize}

 \item 
If  $\psi, \psi' $  are defined on the same space $(T, \nu )$  and    
$  E^+(\psi, \psi')  =     E(\psi+ \psi')$  then the kernels satisfy
   \be 
    \| E^+  \|_{h, h'}    \leq     \| E  \|_{h+h'}
\ee
More generally if  $(\psi_1, \psi_2)$ and $(\psi'_1, \psi_2')$  are fields such that 
$\psi_1, \psi'_1$ are defined on the same space and   $\psi_2, \psi_2'$ are defined on the 
same space and   $E^+(   (\psi_1, \psi_2),  (\psi'_1, \psi_2')  )=
E(   \psi_1 +, \psi'_1,  \psi_2+ \psi_2'  )$
then with   $\bh =  (h_1,h_2)$ and      $\bh' =  (h'_1,h'_2)$ the kernels satisfy  
\be     \label{owly}
\| E^{+} \|_{\bh ,\bh'}   \leq   \|  E   \|_{\bh + \bh'}
\ee

 \item  Let  $A$ be   an operator  from      functions on  $(T_2, \nu_2)$ to  $(T_1, \nu_1)$. 
Let   $E$  be  defined on  fields indexed by  $(T_1, \nu_1)$  and   define       $ E'$   on fields indexed
by  $(T_2, \nu_2)$  by       
   $  E'(  \psi )   = E ( A  \psi  ) $.
 Then   the kernels satisfy  
 \be    \label{loganberry}
  \|   E'  \|_h    \leq  \| E \|_{   \| A \|_{1, \infty}  h}
 \ee
 More  generally   let    
  $\psi_1,  \psi_2$ be indexed by    $(T_1, \nu_1)$   and   $(T_2, \nu_2)$
 and suppose   
 \be    E'( \psi_1, \psi_2)   =  E(A_{11} \psi_1  + A_{12} \psi_2,  A_{21}   \psi_1 + A_{22}  \psi_2 )  
  \ee
  where  $A_{ij}$ is an operator mapping functions on   $(T_j, \nu_j)$   to  functions on  $(T_i, \nu_i)$.
If 
\be    \|  A_{ij}   \|_{1, \infty}   \leq  C_{ij}  
\ee
then  the kernels satisfy
  \be        \label{paraguay}
  \| E'  \|_{h_1,h_2 }     \leq     \| E  \|_{ C_{11}h_1 + C_{12}   h_2,  C_{21}h_1 + C_{22} h_2}      
 \ee
 \end{itemize}

 \subsection{Gaussian integrals}
 
 Unitl now we have implicitly treated  $\Psi, \bPsi$ as different components of the same field.   Now we distinguish them 
 and consider our Grassman algebra as generated by $\Psi, \bPsi$ each   indexed by $(S, \mu)$.
 The general element now   has the form 
  \be    
 \begin{split}
& E(\Psi, \bPsi)  \\
=  &    \sum_{nm}   \frac{1}{n!m!}   \sum_{s_1, \dots, s_n,t_1, \dots, t_n  }   E(  s_1, \dots, s_n,t_1, \dots, t_m         ) 
 \Psi(s_1)  \cdots  \Psi (s_n)   \bPsi(t_1)  \cdots  \bPsi (t_m)   \prod_i  \mu (s_i)   \prod_j  \mu(t_j)  \\
 \end{split}
\ee 
which  can also be written
\be  
 E(\Psi, \bPsi)=  \sum_{I,\bar I}    E(I, \bar I)  \bPsi(I ) \Psi(\bar I)  \mu(I) \mu(\bar I)
\ee      
where   $I, \bar I$  are ordered sequences of points.  
There is an associated norm   
\be      
\begin{split}
\|E  \|_{h}  =   &     \sum_{nm}   \frac{h^{n+m} }{n!m!}   \sum_{s_1, \dots, s_n,t_1, \dots, t_m  }  | E(  s_1, \dots, s_n,t_1, \dots, t_m         ) |
  \prod_i  \mu (s_i)   \prod_j  \mu(t_j)  \\
\end{split}
\ee  or
\be      
\begin{split}
\|E  \|_{h} =   &   \sum_{I,\bar I}   h^{|I| + |\bar I|}  | E(I, \bar I)|   \mu(I) \mu(\bar I)
   \\
\end{split}
\ee      
This norm agrees with the norm  used when   treating   $\Psi, \bPsi$ on the same footing.

The    Gaussian integral    with covariance  $\Ga$   satisfies  
\be
\int   \Psi(s_1)  \cdots  \Psi (s_n)   \bPsi(t_1)  \cdots  \bPsi (t_m)   
 d \mu_{\Ga}(\Psi) 
=    \begin{cases}        \det   \{  \Ga(s_i, t_j ) \}    &       n=m  \\
0   &      n \neq m   \\
\end{cases}
\ee
If   we have the identity covariance
\be
\int     \Psi(s_1)  \cdots  \Psi (s_n)   \bPsi(t_1)  \cdots  \bPsi (t_m)   
 d \mu_{I}(\Psi) 
=    \begin{cases} \det\{  \de_{s_i, t_j}  \}  &       n=m  \\
0   &      n \neq m   \\
\end{cases} 
\ee
which is  also  written  
\be
\int      \Psi (I)   \bPsi(\bar I) 
 d \mu_{I}(\Psi) 
=    \begin{cases} 1  &       I=\bar I  \\
0   &     I \neq \bar I   \\
\end{cases} 
\ee
It follows that 
\be
\int   E(\Psi, \bPsi)  d \mu_I(\Psi) =     \sum_{I}    E(I, I)    \mu(I)^2 
\ee
and so   
\be
|\int   E(\Psi, \bPsi)  d \mu_I(\Psi)|   \leq    \| E \|_1   
  \ee

Here is a variation.    Suppose  that in addition to $\Psi, \bPsi$ there  are independent fields  $\psi, \bpsi$  indexed by 
$(T, \nu)$.   (Or there could be more extra fields).  Consider elements of the form    
\be   \label{artichoke}
    E(\psi,  \bpsi,   \Psi, \bPsi ) =   \sum_{I,\bar I, J, \bar J  }   E(I,\bar I,  J, \bar J)  \psi(I) \bpsi(\bar I) \Psi(J)  \bPsi(\bar J) 
      \nu(I) \nu(\bar I)  \mu(J) \mu(\bar J)
\ee
and   define
\be     E^\#(\psi,  \bpsi    )    =    \int       E(\psi,  \bpsi,   \Psi, \bPsi )    d \mu_{I }(\Psi)
\ee
Note that if contributions to    $E$ have equal numbers of  $(\psi, \Psi)$ and  $(\bpsi, \bPsi)$ variables, the integral 
selects terms with equal numbers of   $\Psi, \bPsi$ variables and hence  $E^\#$ must hade equal numbers of  
$\psi, \bpsi$ variables.

\bigskip

\begin{lem}   \label{skunk}   
\be
    \|    E^\#      \|_{h}  \leq   \|  E  \|_{h, 1} 
\ee
\end{lem}
\bigskip  

\pr   We have 
\be
\begin{split}
  E^\#   (   \psi, \bpsi    )   =    &       \sum_{I,\bar I, J, \bar J  }    E(I,\bar I,  J, \bar J)  \psi(I) \bpsi(\bar I)
   \nu(I) \nu(\bar I)  \mu(J) \mu(\bar J)
 \int  \Psi(J)  \bPsi(\bar J)  d \mu_I(\Psi) \\
    =  &    \sum_{I,\bar I }   \B[  \sum_J E(I,\bar I,  J,  J)  \mu(J)^2 
 \B] \psi(I) \bpsi(\bar I)  \nu(I) \nu(\bar I) \\
  \equiv   &    \sum_{I,\bar I }  \B[  E^\#(I,\bar I)  \B] \psi(I) \bpsi(\bar I)  \nu(I) \nu(\bar I) \\
   \end{split}
 \ee 
Then  
\be
\begin{split}
   \|  E^\#  \|_h   \equiv    &    \sum_{I, \bar I}      h^{|I| +| \bar I| }     | E^\#(I,\bar I)   | \nu(I) \nu(\bar I)    \\
   \leq   &    \sum_{I, \bar I,J }      h^{|I| +| \bar I| }     | E(I,\bar I, J,J )   |    \nu(I) \nu(\bar I)  \mu(J)^2 
  \\
   \leq   &    \sum_{I, \bar I,J, \bar J }      h^{|I| +| \bar I| }     | E(I,\bar I, J,\bar  J )   |    \nu(I) \nu(\bar I)  \mu(J) \mu(\bar J)
  \\
   \equiv    &   \|E \|_{h, 1} \\
   \end{split} 
\ee

\section{an identity} \label{ID}

\begin{lem}   $ \Ga_{k,y}(\cA)   =(  D_k(\cA)  +  bL^{-1} P(\cA)  +  i\ga_3 y  )^{-1}$
has the representation   
\begin{equation}
  \Ga_{k,y}(\cA)  =   
\sB_{k,y}(\cA)   +   b_k^2  \sB_{k,y}(\cA) Q_k(\cA)   S_{k,y}(\cA)Q_k^T(-\cA) \sB_{k,y}(\cA)
\end{equation}
where 
\begin{equation}    \label{stinger}  
\begin{split}
\sB_{k,y}(\cA)   =&     \Big(   b_k  +  bL^{-1} P(\cA) + i\ga_3y  \Big) ^{-1}=      \frac{1}{b_k+ i\ga_3y}  (I -  P(\cA) )   +   \frac{1}{ b_k + bL^{-1}+ i\ga_3y }  P(\cA)  \\
 S_{k,y} (\cA)=   &   \Big( \fD_{\cA}     + \bar  m_k  +  \frac{b_k i\ga_3y}{b_k +  i\ga_3y}P_k(\cA)   +     \frac{b_k^2 bL^{-1} }{ (b_k +  i\ga_3y) (b_k + b L^{-1}+ i\ga_3y) }  P_{k+1}(\cA)  \Big)^{-1}  \\
  \\
 \end{split}
 \end{equation}
\end{lem}
\bigskip

\pr  
 Start with 
\be
\begin{split}
& \exp  \B(   < \bar J,  \Ga_k(\cA)  J  >  \B)  \\
=  &  \const \int   d\Psi   \exp \left(  \blan \bPsi,  J \bran +  \blan \bar J, \Psi \bran     -  \blan  \bPsi, \B( bL^{-1}P(\cA)+  i\ga_3y  \B) \Psi \bran 
  -   \blan \bPsi,  D_k(\cA) \Psi \bran  \right)   \\
\end{split}
\ee
and   from section \ref{minimizers1}
\begin{equation}
\exp \left(  -  \blan\bPsi,  D_k(\cA) \Psi\bran  \right)  
=   \const   \int   \exp \left(   - b_k \blan   \bPsi    -   Q_k(-\cA)  \bpsi,     \Psi    -   Q_k(\cA)  \psi \bran      -   \blan \bpsi,  ( \fD_{\cA}  + \bar m_k  )  \psi  \bran   \right)    d \psi
\end{equation}
Insert the second into the first   and  do the integral  over  $\Psi$  which is 
\begin{equation}
\begin{split}
&  \int   d\Psi   \exp \left(  \blan\bPsi,  J\bran +  \blan\bar J, \Psi\bran     -  \blan  \bPsi, \B( bL^{-1}P(\cA)+  i\ga_3y  \B) \Psi \bran    - b_k \blan   \bPsi    -   Q_k(-\cA)  \bpsi,     \Psi    -   Q_k(\cA)  \psi \bran    \right) \\
=&     \int   d\Psi   \exp \B(  \blan\bPsi, J +b_k Q_k(\cA) \psi\bran   +    \blan\bar  J  +b_k Q_k(-\cA) \bpsi  , \Psi   \bran  \\
&
\hs   \hs   \hs   \hs  - \blan \bPsi,  \Big(   b_k  +   bL^{-1}P(\cA) +  i\ga_3y \Big)  \Psi  \bran    -   b_k \blan    \bpsi,    P_k(\cA)  \psi \bran   \B) \\
=&  \const    \exp   \Big(   \blan (\bar  J  +b_k Q_k(-\cA) \bpsi) ,  \sB_{k,y} (\cA)  (J +b_k Q_k(\cA) \psi) \bran 
 -   b_k \blan    \bpsi,    P_k(\cA)  \psi \bran   \Big) \\
\end{split} 
\end{equation}
This gives    
\begin{equation}
\begin{split}
&\exp  \B(  < \bar J,  \Ga_{k}(\cA) J  >  \B)\\
=&   \const
\int   \exp   \Big(       \blan (\bar  J  +b_k Q_k(-\cA) \bpsi) ,  \sB_{k,y} (\cA)  (J +b_k Q_k(\cA) \psi)    \bran  
-   \blan \bpsi,  ( \fD_{\cA}     + \bar m _k  + b_k P_k(\cA))  \psi  \bran  \Big)   \ d \psi
 \\
= &     
 \const \exp   \Big(   <  \bar  J ,  \sB_{k,y}(\cA)   J  >    \Big)
 \\
& \int    \exp   \Big(   \blan b_k Q_k^T(\cA)  \sB^T_{k,y}(\cA)\bar  J  ,   \psi \bran  
+   \blan \bpsi ,b_k  Q_k^T(-\cA) \sB_{k,y}(\cA)   J \bran   -  \blan \bpsi,  S_{k,y}(\cA)^{-1}  \psi  \bran   \Big)
  \ d \psi
 \\
=& \const \exp \Big(    <  \bar  J ,  \sB_{k,y} (\cA)  J  > 
+   b_k^2  \blan   \bar J  , \sB_{k,y}(\cA)Q_k(\cA)  S_{k,y}(\cA) Q_k^T(-\cA) \sB_{k,y}(\cA)  J    \bran   \Big)  \\
\end{split}
  \end{equation}
  where  we defined    
\be  
   S_{k,y}(\cA)  =  \B(     \fD_{\cA}     + \bar m _k  + b_k P_k(\cA)  -    b_k^2 Q_k^T(-\cA) \sB_{k,y}(\cA)  Q_k(\cA)  \B) ^{-1}  
  \ee
  This is the same as  the definition   in  (\ref{stinger}) 
  since     we can write       
  \be 
 \sB_{k,y}(\cA)   =           \frac{1}{b_k + i\ga_3y }     -  \B( \frac{bL^{-1} }{( b_k + bL^{-1}+ i\ga_3y) (b_k + i\ga_3y ) }\B) P(\cA) \\
\ee
and   this yields       
\be
\begin{split}  
&  b_kP_k(\cA)   -   b_k^2  Q^T_k(-  \cA)  \sB_{k,y}(\cA )      Q_k(  \cA)   \\
  =&   \B( b_k   -  \frac{b_k^2}{b_k + i\ga_3y } \B)   P_k(\cA)  
   +  \B( \frac{b_k^2 bL^{-1} }{( b_k + bL^{-1} +  i\ga_3y ) (b_k + i\ga_3y ) }\B) P_{k+1} (\cA) \\
     =&   \B(   \frac{b_k  i\ga_3y  }{b_k + i\ga_3y} \B)   P_k(\cA)  
   +  \B( \frac{b_k^2 bL^{-1} }{( b_k + bL^{-1}+ i\ga_3y) (b_k + i\ga_3y ) }\B) P_{k+1} (\cA) \\
\end{split}
\ee

  \section{another identity} \label{C}

\begin{thm} \cite{Bal84b}
For $F$ on bonds on $\bbT^1_{N-k}$   we have $\cQ^T F$ on $\tz$ and  $\cQ_{k+1}^T F$ on $\tk$ and the identify
\be
\blan \cQ^T_{k+1}F,\  \cG^0_{k+1} \cQ^T_{k+1}F \bran =  \blan [\cQ^TF]^{\sx},\tilde   C_k [\cQ^TF]^{\sx} \bran
\ee
Here  $\tilde C_k$ is the operator on the subspace $\{ Z \textrm{ on }\bbT^0_{N-k}:   \tau Z =0 \}$ defined by
\be    e^{ \frac12 < J,  \tilde C_k J >} =  \int  DZ\  \de ( \tau  Z )  
\exp \B(- \frac 12 \blan Z,  \B[ \De_k + a\cQ^T \cQ \B] Z \bran  \B)
e^{<  Z,  J >}       / \{ J = 0\}
\ee
and  $ [\cQ^TF]^{\sx}$ is  the projection of  $\cQ^TF$  onto $\ker \tau $.
\end{thm} 
\bigskip

\pr We sketch the proof and refer to \cite{Bal84a}, \cite{Bal84b} for more details. Let  $J =\cQ^T_{k+1}F$. 
Start with 
\be
\begin{split}
&  e^{\frac12  < J,\  \cG^0_{k+1} J >}= \int D\cA \B( -\frac12 \blan \cA, \ \B( \de d + d R_{k+1} \de + a \cQ_{k+1}^T \cQ_{k+1} \B) \cA \bran  \B)  e^{ <\cA,J>} / \{ J = 0\}
  \\
\end{split}
\ee
Let $\De = \de d + d \de$ and write
   $\de d + d R_{k+1} \de = \De - d P_{k+1} \de $  where $P_{k+1} = I - R_{k+1}$ is a projection.  We change from $P_{k+1}$ to $P_k$
using the identities for $\la, \la'$ on $\tk$ 
\be 
\begin{split}
\exp \B( - \frac 12  \blan \de\cA,  P_{k}\de \cA \bran \B)
= &  \int  D\la'\   \de (Q_{k} \la') \exp \B(  - \frac12 \| \de \cA - \De \la' \|^2 \B)/ \{ \de \cA = 0\} \\
\exp \B( - \frac 12  \blan \de\cA,  P_{k+1}\de \cA \bran \B)
= &  \int  D\la\   \de (Q_{k+1} \la) \exp \B(  - \frac12 \| \de \cA - \De \la \|^2 \B)/ \{ \de \cA = 0\} \\
\end{split}
\ee
Also insert  for $\om$ on $\tz$  insert the basic axial gauge  identity
\be
 \int D \om\  \de ( Q \om) \de \B( \tau ( \cQ_k \cA + d \om) \B)= \const
\ee
 This yields 
\be
\begin{split}
    e^{\frac12  < J,\  \cG^0_{k+1} J >} =& \const   \int D \om \ \de ( Q \om) 
\int D\cA  \exp \B( -\frac12 \blan \cA, \ \B( \De - d P_{k} \de + a \cQ_{k+1}^T \cQ_{k+1} \B) \cA \bran  \B)
 \\
& \de \B( \tau ( \cQ_k \cA + d \om) \B)
\frac{   \int  D\la'\   \de (Q_{k} \la') \exp \B(  - \frac12 \| \de \cA - \De \la' \|^2 \B)}
{  \int  D\la\   \de (Q_{k+1} \la) \exp \B(  - \frac12 \| \de \cA - \De \la \|^2 \B)} e^{ <\cA,J>}
\\
\end{split}
\ee

Next define $\la  = \cH_k'  \mu$ to be the minimizer of     $ \| \De \la \|^2 $  subject to  $  Q_k \la = \mu$
and  make the change of variables $
\cA \to \cA - d \cH'_k \om $.
This is constructed to leave the quadratic form in $\cA$  invariant.  It can be written 
$
 < \cA, \ ( \de d + d R_{k} \de + a \cQ_{k+1}^T \cQ_{k+1} ) \cA > 
$.
To see the invariance note that $\cQ_k d \cH_k' \om = d Q_k \cH_k' \om = d \om$ so $\cQ_k\cA \to \cQ_k \cA- d \om$.
Then  $\cQ_{k+1} \cA \to \cQ_{k+1}\cA -\cQ d \om$, but $ \cQ d \om =  d Q \om =0$ so $\cQ_{k+1} \cA$ is invariant.
Furthermore $\de \cA \to \de \cA - \De \cH_k' \om$ since $\de d = \De$ on scalars.  But one has the explicit representations
 $\cH_k' = \De^{-2}Q^T_k (Q_k \De^{-2} Q_k^T)^{-1}$ 
as well as $P_k = \De^{-1} Q_k^T (Q_k \De^{-2} Q_k^T)^{-1}Q_k \De^{-1}$
and these combine to give $P_k \De \cH_k' \om = \De \cH_k' \om$ and hence  $ R_k \De \cH_k' \om = 0 $
and so $R_k \de \cA$ is invariant.  And of course $ d\cA$ is invariant so the form is invariant.

Other changes are that $\de( \tau ( \cQ_k \cA + d \om) )$ becomes $\de ( \tau ( \cQ_k \cA ) )$
and  that  $<  \cA, J > =  < \cQ_{k+1} \cA, F >  $  is invariant. 
  The  numerator in the big fraction becomes
\be
  \int  D\la' \   \de (Q_{k} \la') \exp \B(  - \frac12 \| \de \cA -  \De ( \cH'_k \om  +\la')  \|^2 \B)
\ee
Make  make the further change of variables $    \la' \to \la' -  \cH'_k \om   $
which gives
\be
  \int  D\la' \   \de (Q_{k} \la'- \om) \exp \B(  - \frac12 \| \de \cA -  \De \la'  \|^2 \B)
\ee
 Similarly the denominator in the big fraction is invariant since the change in $Q_{k+1} \la$ is  $Q_{k+1} \cH_k' \om =Q \om  =0$.
 
Now  we have 
\be
\begin{split}
&   e^{\frac12  < J,\  \cG^0_{k+1} J >}=  \const \int D \om \ \de ( Q \om) \int D\cA 
 \exp \B( -\frac12 \blan \cA, \ \B( \de d + d R_{k} \de + a \cQ_{k+1}^T \cQ_{k+1} \B) \cA \bran  \B)
\\
& \de \B( \tau  \cQ_k \cA  \B)
\frac{     \int  D\la'\   \de (Q_{k} \la'- \om ) \exp \B(  - \frac12 \| \de \cA - \De \la' \|^2 \B)}
{  \int  D\la\   \de (Q_{k+1} \la) \exp \B(  - \frac12 \| \de \cA - \De \la \|^2 \B)}   e^{ <\cA,J>} \\
=  & \const \int D\cA  \exp \B( -\frac12 \blan \cA, \ \B(\de d + d R_{k} \de  + a\cQ_{k+1}^T \cQ_{k+1} \B) \cA \bran  \B)
 \de \B( \tau  \cQ_k \cA  \B) e^{ <\cA,J>} 
\\
\end{split}
\ee
In the second step we did the integral over $\om$ and canceled out the big fraction.

Next  insert 
$
1 =  \int \de ( \cQ_k \cA - Z) DZ
$
and obtain 
\be
\begin{split}
&   e^{\frac12  < J,\  \cG^0_{k+1} J >} \\
=  &   \const \int D\cA\ DZ\  \de ( \cQ_k \cA - Z) \de ( \tau  Z )
\exp \B( -\frac12 \blan \cA, \ \B( \de d + d R_{k} \de   \B) \cA \bran   -\frac12 a  \|\cQ Z \|^2 \B)
e^{ <\cA,J>} 
 \\
\end{split}
\ee
Let  $\hat \cH_k$ be the minimizer in $\cA$  of $<\cA, ( \de d + d R_{k} \de ) \cA >$ subject to to $\cQ_k \cA = Z$.
In  the integral over $\cA$ make   the translation $\cA \to \cA +\hat  \cH_k Z$.
Then the integral factors into 
\be
\begin{split}
&   e^{\frac12  < J,\  \cG^0_{k+1} J >} 
=  \const  \int D\cA \   \de ( \cQ_k \cA )   \exp \B( -\frac12 \blan \cA, \ \B( \de d + d R_{k} \de  \B) \cA \bran  e^{ <\cA,J>} \\
 & \int  DZ\  \de ( \tau  Z )  
\exp \B( -\frac12 \blan \hat \cH_k Z, \ \B(\de d + d R_{k} \de  \B) \hat \cH_k Z \bran   -\frac12 a  \|\cQ Z \|^2 \B)
e^{  < \hat \cH_k Z, J > }
 \\
\end{split}
\ee
But in the first integral $< \cA, J > = < \cQ_{k+1} \cA, F >  =0$ so this integral  is  a constant. 
In the second integral we have  $ < \hat  \cH_k Z, J > = < \cQ_{k+1} \hat \cH_kZ,F> =  < \cQ Z, F >$.
Also in  the second integral we  change from exponential gauge fixing to delta function gauge fixing by a Fadeev-Popov 
procedure and  identify the Landau gauge minimizer $\cH_k$ to obtain 
\be 
\begin{split}
&\exp \B( -\frac12 \blan \hat \cH_k Z, \ \B(\de d + d R_{k} \de   \B) \hat  \cH_k Z \bran    \B)\\
=&  \const   \int  D\cA\ \de ( Q_k\cA - Z )   \exp \B( -\frac12 \blan \cA, \ \B( \de d + d R_{k} \de   \B) \cA \bran    \B)\\
=&  \const    \int  D\cA\ \de ( Q_k\cA - Z ) \de (R_k \de \cA )  \exp \B( -\frac12 \| d\cA \|^2  \B)\\
= & \exp \B( -\frac12 \| d\cH_k Z \|^2  \B) = \exp \B( - \frac 12 \blan Z, \De_k Z \bran \B)\\
\end{split}
\ee
This yields
\be
\begin{split}
   e^{\frac12  < J,\  \cG^0_{k+1} J >}= & \const
  \int  DZ\  \de ( \tau  Z )  
\exp \B(- \frac 12 \blan Z,  \B[ \De_k + a\cQ^T \cQ \B] Z \bran  \B)
\exp \B(\blan  Z, \cQ^TF \bran  \B)
 \\
 = &  \const  \exp\B ( \frac12 \blan [\cQ^TF]^{\sx},  \tilde C_k [\cQ^T F]^{\sx}  \bran \B) \\
\end{split}
\ee
Setting $F=J=0$ we see that the constant must be one.  This gives the result.

  \section{an estimate on $\cQ^T$}   \label{D}

\begin{lem}
 For $A$ on any $L$-lattice $\bbT^1_{N-k}$
\be
\| \cQ^T A \|^2 \geq  L^{-1} \|A \|^2
\ee
Also if  $[\cQ^T A]^{\sx} $ is the projection of $\cQ^T A$ onto $\ker  \tau$ 
\be
\| [\cQ^T A]^{\sx} \|^2 \geq  L^{-1} \|A \|^2
\ee
\end{lem}
\bigskip

\pr  Let $\cQ_s$ be the operator the averages over surface bonds of the cubes $B(y)$, see \cite{BIJ85} for the precise definition. 
This satisfies  $\cQ_s \cQ_s^T  =L$   so  $L^{-1} \cQ_s^T \cQ_s$ is an orthogonal projection.  Also   
since $\cQ \cQ^T_s =I$ implies $\cQ_s \cQ^T =I$  we have 
\be 
\| \cQ^T A \|^2 \geq L^{-2} \| \cQ_s^T \cQ_s\cQ^T A \|^2  =  L^{-2} \| \cQ_s^T A \|^2 = L^{-1} \|A\|^2
\ee
The projection $L^{-1} \cQ_s^T \cQ_s$ annihilates functions on the interiors of the cubes $B(y)$.  The projection
 $\cQ^T A  \to [\cQ^T A]^{\sx} $ only affects the functions on the interiors.   Thus $ L^{-1} \cQ_s^T \cQ_s [\cQ^T A]^{\sx}
 =    L^{-2}  \cQ_s^T \cQ_s\cQ^T A $ and the same proof can be carried out.

 \section{conditioning}
 
 Consider integrals of the form   
 \be  
 I   =    \int        d \mu_{T^{-1}}(A)
F(A_{\La})   \ee
  where the integral is over functions   $A$  on bonds in  a unit lattice,  $\mu_{T^{-1}}$ is a Gaussian measure with covariance $T^{-1}$,  $\La$ is a subset of
  the lattice,   and $A_{\La}= 1_{\La} A$ is the restriction   to  $\La$, i.e. to bonds intersecting $\La$.   We  want to express the integral in terms of a conditional expectation  of  $ F(A_{\La}) $
  given the values  $A_{\La^c}$.    This comes in two  different forms.   In general we define $T_{\La} = 1_{\La}T 1_{\La}$ and
  $T_{\La, \La^c} = 1_{\La}T 1_{\La^c}$.
  \bigskip

  \begin{lem} 
  The integral  $I$ can be expressed as  
\be  \label{linus1}
  I   =     \int      d \mu_{T^{-1}}(A')  \B[  \int   d \mu_{T^{-1}_{\La}}(A_{\La})     F\B(A_{\La} -  T_{\La}^{-1} T_{\La \La^c}  A'_{\La^c}  \B)\B]
\ee  
or as    
\be  \label{linus2}
  I   =     \int      d \mu_{T^{-1}}(A') 
  \exp  \B(  -\frac12 \blan   A'_{\La^c} ,  T_{\La^c \La}     T_{\La}^{-1} T_{\La \La^c}   A'_{\La^c} \bran  \B) 
    \B[  \int   d \mu_{T^{-1}_{\La}}(A_{\La})  \exp  \B(  - \blan     A'_{\La^c},   T_{\La^c \La}  A_{\La} \bran  \B)    F(A_{\La} )   \B]
\ee  
 \end{lem}  
 \bigskip
  
  \rem      These formulas were first used in \cite{Bal82b} and  \cite{Bal88a} respectively.  
  \bigskip
  
 \pr    We have
 \be 
  I  =    Z^{-1}  \int   DA    \exp  \B(  - \frac12  \blan   A,TA \bran    \B)    F(A_{\La})     \hs      Z  =   \int   DA    \exp  \B(  - \frac12  \blan   A,TA \bran    \B) 
 \ee
 We write 
 \be
 \frac12   \blan   A,TA \bran   =    \frac12   \blan   A_{\La^c},TA_{\La^c} \bran  +      \blan   A_{\La^c},TA_{\La} \bran +   \frac12   \blan   A_{\La},TA_{\La} \bran 
 \ee
 Diagonalize the quadratic form by the change of variables       $A_{\La}   \to   A _{\La}   -   T_{\La}^{-1} T_{\La \La^c}  A_{\La^c}$.   This yields
 \be
 \begin{split}
   I   = &  Z^{-1}     \int D A _{\La^c}  \exp  \B( - \frac12   \blan     A_{\La^c} , \B(T_{\La} -  T_{\La^c \La}     T_{\La}^{-1} T_{\La \La^c} \B)  A_{\La^c} \bran  \B)  \\
   &   \B[       \int  DA_{\La}  \exp  \B(  - \frac12   \blan   A_{\La},TA_{\La} \bran   \B)      F\B(A_{\La} -  T_{\La}^{-1} T_{\La \La^c}  A'_{\La^c} \B) \B]   \\
 \end{split}  
 \ee
 Now relabel     $A_{\La^c}$ as    $A'_{\La^c}$  and insert  
 \be   1  =  Z_{\La}^{-1}    \int   DA'_{\La}    \exp  \B(  - \frac12  \blan   A'_{\La},TA'_{\La} \bran    \B) 
 \ee
 Then  the  change of variables      $A'_{\La}   \to   A' _{\La^c}   +  T_{\La}^{-1} T_{\La \La^c}  A'_{\La^c}$ restores the quadratic form 
  $ \frac12   \blan   A',TA' \bran  $   and does not affect the interior integral.   Thus 
  \be
   I   = Z^{-1}     \int D A'   \exp  \B(-\frac12   \blan   A',TA' \bran  \B) 
 \B[   Z_{\La}^{-1}         \int    DA_{\La} \exp  \B(  - \frac12   \blan   A_{\La},TA_{\La} \bran   \B)       F\B(A_{\La} -  T_{\La}^{-1} T_{\La \La^c}  A'_{\La^c} \B) \B]
 \ee
  This is   (\ref{linus1}).    In  the interior integral make the inverse  change of variables   
   $A_{\La}   \to   A _{\La}   +   T_{\La}^{-1} T_{\La \La^c}  A'_{\La^c}$  to regain  $F(A_{\La})$.   The  quadratic form     $\frac12   \blan   A_{\La},TA_{\La} \bran  $  becomes
  \be   
\frac12     \blan   A_{\La},TA_{\La} \bran       
   +    \blan   A'_{\La^c},T_{\La^c \La}  A_{\La} \bran  + \frac12  \blan   A'_{\La^c} ,  T_{\La^c \La}     T_{\La}^{-1} T_{\La \La^c}   A'_{\La^c} \bran
 \ee 
  which gives  (\ref{linus2}).

  \end{appendix}

\end{document}